\documentclass[a4paper, 12pt]{scrartcl}
\DeclareOldFontCommand{\bf}{\normalfont\bfseries}{\mathbf}
\usepackage{geometry}
\usepackage{titling}
\usepackage[ansinew]{inputenc} 
\usepackage{graphicx} 
\usepackage[font = footnotesize]{caption}
\usepackage{natbib} 
\usepackage{amsmath, amsthm, amssymb} 
\usepackage{bm} 
\usepackage{dsfont} 
\usepackage{tabularx} 
\usepackage[plain]{algorithm} 
\usepackage[noend]{algorithmic} 
\usepackage{multirow} 
\usepackage{xcolor}
\usepackage{makeshape} 
\usepackage{flowchart}
\usetikzlibrary{arrows} 
\usepackage{booktabs, colortbl} 
\usepackage{siunitx} 
\usepackage{pgfplotstable} 
\usepackage{array}
\pgfplotsset{compat=1.14}
\usepackage{caption}
\usepackage{etoolbox}
\usepackage{hyperref} 
\makeatletter 
\renewcommand*{\@fnsymbol}[1]{\ensuremath{\ifcase#1\or *\or \dagger\or \mathsection\or
    \ddagger\or \mathparagraph\or \|\or **\or \dagger\dagger
    \or \ddagger\ddagger \else\@ctrerr\fi}}
\makeatother
\DeclareMathOperator*{\argmin}{arg\,min}
\DeclareMathOperator*{\argmax}{arg\,max}
\makeindex
\begin{document}
\renewcommand{\baselinestretch}{1.2} 
\normalsize
\theoremstyle{plain}
\newtheorem{T}{Theorem}
\newtheorem*{GMt}{Gauss-Markov theorem}
\newtheorem*{GMAt}{Gauss-Markov-Aitken theorem}
\newtheorem*{IRLS}{IRLS algorithm}
\newtheorem*{PIRLS}{PIRLS algorithm}
\newtheorem*{ML}{ML algorithm}
\newtheorem{proposition}{Proposition} 
\newtheorem{corollary}{Corollary}
\newtheorem*{Acknowledgements}{Acknowledgements}
\pagenumbering{Alph}
\begin{titlingpage}
\centering
\title{\Large Machine Learning in Least-Squares Monte Carlo Proxy Modeling of Life Insurance Companies} 
\author{Anne-Sophie Krah\thanks{Department of Mathematics, TU Kaiserslautern, Erwin-Schr\"{o}dinger-Stra{\ss}e, Geb\"{a}ude 48, {67653 Kaiserslautern}, Germany; anne-sophiekrah@web.de.}\and Zoran Nikoli\'{c}\thanks{Mathematical Institute, University Cologne, Weyertal 86-90, 50931 Cologne, Germany; znikolic@uni-koeln.de.} \and Ralf Korn\footnotemark[1]${\,\,}^{,}$\thanks{Department Financial Mathematics, Fraunhofer ITWM, Fraunhofer-Platz 1, 67663 Kaisers-lautern, Germany; korn@mathematik.uni-kl.de.}}
\date{\today}
\maketitle
\vspace{1cm}
{\normalsize\bfseries Abstract\par}
\begin{abstract}
Under the Solvency II regime, life insurance companies are asked to derive their solvency capital requirements from the full loss distributions over the coming year. Since the industry is currently far from being endowed with sufficient computational capacities to fully simulate these distributions, the insurers have to rely on suitable approximation techniques such as the least-squares Monte Carlo (LSMC) method. The key idea of LSMC is to run only a few wisely selected simulations and to process their output further to obtain a risk-dependent proxy function of the loss. In this paper, we present and analyze various adaptive machine learning approaches that can take over the proxy modeling task. The studied approaches range from ordinary and generalized least-squares regression variants over GLM and GAM methods to MARS and kernel regression routines. We justify the combinability of their regression ingredients in a theoretical discourse. Further, we illustrate the approaches in slightly disguised real-world experiments and perform comprehensive out-of-sample tests.
\end{abstract}
\end{titlingpage}
\pagenumbering{roman}
\tableofcontents
\newpage
\pagenumbering{arabic}
\section{Introduction}\label{sec:introduction}
\subsection*{LSMC Framework under Solvency II}
By the Solvency II directive of the \cite{european2009}, life insurance companies are asked to derive their solvency capital requirements (SCRs) from their full loss probability distributions over the coming year if they do not want to rely on the much simpler standard formula. In order to obtain reasonably accurate full loss distributions via a nested simulations approach as described in \cite{bauer2012}, their cash-flow-projection (CFP) models would need to be simulated several hundred thousand times. But the insurers are currently far from being endowed with sufficient computational capacities to perform such expensive simulation tasks. By applying suitable approximation techniques like the least-squares Monte Carlo (LSMC) approach of \cite{bauer2015}, the insurers are able to overcome these computational hurdles though. For example, they can implement the LSMC framework formalized by \cite{krah2018} and applied by e.g. \cite{bettels2014} to derive their full loss distributions. The central idea of this framework is to carry out a comparably small number of wisely chosen Monte Carlo simulations and to feed the simulation results into a supervised machine learning algorithm that translates the results into a proxy function of the insurer's loss (output) with respect to the underlying risk factors (input). To guarantee a certain approximation quality, the proxy function has to pass an additional validation procedure before it can finally be used for the full loss distribution forecast.

\subsection*{Machine Learning Calibration Algorithm}
Apart from the calibration and validation steps, we adopt the LSMC framework from \cite{krah2018} without any changes. Therefore, we neither repeat the simulation setting nor the procedure for the full loss distribution forecast and SCR calculation here in detail. The purpose of this exposition is to introduce different machine learning methods that can be applied in the calibration step of the LSMC framework and other high-dimensional variable selection applications, to point out their similarities and differences and to compare their out-of-sample performances in a slightly disguised real-world LSMC example. We describe the data basis used for calibration and validation in Section \ref{sec:fitting_and_validation_points}, the structure of the calibration algorithm in Section \ref{sec:calibration_algorithm} and our validation approach in Section \ref{sec:validation_figures}. Our focus lies on out-of-sample performance rather than computational efficiency as the latter becomes only relevant if the former gives reason for it. We analyze a very realistic data basis with $15$ risk factors and validate the proxy functions based on a very comprehensive and compuationally expensive nested simulations test set comprising the SCR estimate.

The idea is to combine different regression methods with an adaptive algorithm, in which the proxy functions are built up of basis functions in a stepwise fashion. In a four risk factor LSMC example, \cite{teuguia2014} applied a full model approach, forward selection, backward elimination and a bidirectional approach as e.g. discussed in \cite{hocking1976} with orthogonal polynomial basis functions. They stated that only forward selection and the bidirectional approach were feasible when the number of risk factors or polynomial degree exceeded $7$ as the other models exploded then. Life insurance companies covering a wide range of contracts in their portfolio are typically exposed to even more risk factors like e.g. $15$. In complex business regulation frameworks such as in Germany, they furthermore require polynomial degrees of at least $4$. In these cases, even the standard forward selection and bidirectional approaches become infeasible as the sets of candidate terms from which the basis functions are chosen will explode then as well. We therefore follow the suggestion of \cite{krah2018} to implement the so-called principle of marginality, an iteration-wise update technique of the set of candidate terms that lets the algorithm get along with comparably few carefully selected candidate terms.

\subsection*{Regression Methods \& Model Selection Criteria}
Our main contribution is to identify, explain and illustrate a collection of regression methods and model selection criteria from the jungle of regression design options that provide suitable proxy functions in the LSMC framework when applied in combination with the principle of marginality. After some general remarks in Section \ref{sec:general_remarks_theory}, we describe ordinary least-squares (OLS) regression in Section \ref{sec:OLStheory}, generalized linear models (GLMs) by \cite{nelder1972} in Section \ref{sec:GLMtheory}, generalized additive models (GAMs) by \cite{hastie1986} and \cite{hastie1990} in Section \ref{sec:GAMtheory}, feasible generalized least-squares (FGLS) regression in Section \ref{sec:FGLStheory}, multivariate adaptive regression splines (MARS) by \cite{friedman1991} in Section \ref{sec:MARStheory}, and kernel regression by \cite{watson1964} and \cite{nadaraya1964} in Section \ref{sec:KRtheory}. While some regression methods such as OLS and FGLS regression or GLMs can immediately be applied in conjunction with numerous model selection criteria such as Akaike information criterion (AIC), Bayesian information crierion (BIC), Mallow's $C_P$ or generalized cross-validation (GCV), other regression methods such as GAMs, MARS, kernel, ridge or robust regression require thought-through modifications thereof or work only with non-parametric alternatives such as $k$-fold or leave-one-out cross-validation. For adaptive approaches of FGLS, ridge and robust regression in life insurance proxy modeling, see also \cite{hartmann2015}, \cite{krah2015} and \cite{nikolic2017}, respectively. 

In the theory sections, we present the models with their assumptions, important properties and popular estimation algorithms and demonstrate how they can be embedded in the adaptive algorithm by proposing feasible implementation designs and combinable model selection criteria. While we shed light on the theoretical basic concepts of the models to lay the groundwork for the application and interpretation of the later following numerical experiments, we forego to describe in detail technical enhancements or peculiarities of the involved algorithms and instead refer the interested reader here and there to some further sources. Additionally we provide the practicioners with R packages containing useful implementations of the presented regression routines. We complement the theory sections by practice sections \ref{sec:general_remarks_experiment}, \ref{sec:OLSexperiment}, \ref{sec:GLMexperiment}, \ref{sec:GAMexperiment}, \ref{sec:FGLSexperiment}, \ref{sec:MARSexperiment} and \ref{sec:KRexperiment}, respectively, throughout which we perform the same Monte Carlo approximation task to make the performance of the various methods comparable. We measure the approximation quality of the resulting proxy functions by means of aggregated validation figures on three out-of-sample test sets.

\subsection*{Further Machine Learning Alternatives}
Conceivable alternatives to the entire adaptive algorithm are other typical machine learning techniques such as artificial neural networks (ANNs), decision tree learning or support vector machines. In particular, the classical feed forward networks proposed by \cite{hejazi2017} and applied in various ways by \cite{kopczyk2018}, \cite{castellani2018}, \cite{born2018} and \cite{schelthoff2019} were shown to capture the complex nature of CFP models well. A major challenge here is not only to find reliable hyperparameters such as the numbers of hidden layers and nodes in the network, batch size, weight initializer probability distribution, learning rate or activation function but also the high dependence on the random seeds. Future research should therefore be dedicated to hyperparameter search algorithms and stabilization methods such as ensemble methods. As an alternative to feed forward networks, \cite{kazimov2018} suggested to use radial basis function networks albeit so far none of the tested approaches worked out well.

In decision tree learning, random forests and tree-based gradient boosting machines were considered by \cite{kopczyk2018} and \cite{schoenenwald2019}. While random forests were outperformed by feed forward networks but did better than the least absolute shrinkage and selection operator (LASSO) by \cite{tibshirani1996} in the example of the former author, they generally performed worse than the adaptive approaches by \cite{krah2018} with OLS regression in numerous examples of the latter author. The gradient boosting machines, requiring more parameter tuning and thus being more versatile and demanding, came overall very close to the adaptive approaches. The tree-based methods belong by definition to the aforementioned ensemble methods, a modeling concept transferrable to arbitrary regression techniques, mitigating random model artefacts through averaging.

\cite{castellani2018} compared support vector regression (SVR) by \cite{drucker1997} to ANNs and the adaptive approaches by \cite{teuguia2014} in a seven risk factor example and found the performance of SVR placed somewhere inbetween the other two approaches with the ANNs getting closest to the nested simulations benchmark. As some further non-parametric approaches, \cite{sell2019} tested least-squares support-vector machines (LS-SVM) by \cite{suykens1999} and shrunk additive least-squares approximations (SALSA) by \cite{kandasamy2016} in comparison to ANNs and the adaptive approaches by \cite{krah2018} with OLS regression. In his examples, SALSA was able to beat the other two approaches whereas LS-SVM was left far behind. The analyzed machine learning alternatives have in common that they require at least to some degree a fine-tuning of some model hyperparameters. Since this is often a non-trivial but crucial task for generating suitable proxy functions, finding efficient search algorithms should become a subject of future research.

\section{Calibration \& Validation in the LSMC Framework}
\subsection{Fitting \& Validation Points}\label{sec:fitting_and_validation_points}
\subsubsection*{Outer Scenarios \& Inner Simulations}
In the LSMC approach, the proxy function of the economic variable (e.g. the loss, available capital or best estimate of liabilities) is calibrated conditional on the \textit{fitting points} which have been generated besides the \textit{validation points} by the Monte Carlo simulations of the CFP model in the step before. The fitting and validation points describe relationships between the economic variable and the different financial and actuarial risk factors the insurance company is exposed to such as the interest rate, equity, property, credit, mortality, morbidity, lapse or expense stresses. By an \textit{outer scenario} we refer to a specific stress level combination of these risk factors, and by an \textit{inner simulation} to a stochastic path of an outer scenario in the CFP model under the given risk-neutral probability measure. The fitting values of the economic variable are defined as the mean values over only few inner simulations of the same outer fitting scenario whereas the validation values of the economic variable are defined as the mean values over many inner simulations of the same outer validation scenario.

\subsubsection*{Different Trade-off Requirements}
According to the law of large numbers, this construction makes the fitting values very volatile and the validation values comparably stable. Typically, the very limited fitting and validation simulation budgets are of similar sizes. Hence the few inner simulations in the case of the fitting points allow a great diversification among the outer scenarios whereas the many inner simulations in the case of the validation points let the validation values be quite close to their expectations but at the cost of only little diversification among the outer scenarios. These opposite ways to deal with the trade-off between the numbers of outer scenarios and inner simulations reflect the different requirements for the fitting and validation points in the LSMC approach. While the fitting scenarios should cover the domain of the real-world scenarios well to serve as a good regression basis, the validation values should approximate the expectations of the economic variable at the validation scenarios well to provide appropriate target values for the proxy functions.

\subsection{Calibration Algorithm}\label{sec:calibration_algorithm} 
\subsubsection*{Five Major Components}
\begin{figure}[htbp]
  \centering
  \begin{tikzpicture}[>=latex',font={\sffamily \footnotesize}]
    \def\smbwd{2cm}
    \node (terminal1) at (0,0.65) [draw, terminal, align=center, minimum width=\smbwd, minimum height=0.75cm] {START\\ PROXY};
    \node (process0) at (0,-0.7) [draw, process, align=center, minimum width=\smbwd, minimum height=0.75cm] {$k = 0$};
    \node (process1) at (0,-2.15) [draw, process, align=center, minimum width=\smbwd, minimum height=0.75cm] {Regression\\ Model Selection Criterion};
    \node (process2) at (0,-3.5) [draw, process, align=center, minimum width=\smbwd, minimum height=0.75cm] {$\text{MSC}_{\text{min,old}} := \text{MSC}_{\text{min}} := \text{MSC}_{0}$};
    \node (process3) at (0,-4.65) [draw, process, align=center, minimum width=\smbwd, minimum height=0.75cm] {$k = 1$};
    \node (process4) at (0,-6) [draw, process, align=center, minimum width=\smbwd, minimum height=0.75cm] {Get Candidate Terms\\$c=1,\ldots,C$};
    \node (process5) at (0,-8) [draw, process, align=center, minimum width=\smbwd, minimum height=0.75cm] {$c = 1$};
    \node (process6) at (0,-13.5) [draw, process, align=center, minimum width=\smbwd, minimum height=0.75cm] {Regression with $c$\\ Model Selection Criterion};
    \node (decide1) at (0,-16) [draw, decision, align=center, minimum width=\smbwd, minimum height=0.75cm] {$\text{MSC}_{c} <$\\ $\text{MSC}_{\text{min}}$?};
    \node (process7) at (0,-18.5) [draw, process, align=center, minimum width=\smbwd, minimum height=0.75cm] {$\text{MSC}_{\text{min}} := \text{MSC}_{c}$\\ $c_{\text{min}} := c$};
    \node (process8) at (3.8,-16) [draw, process, align=center, minimum width=\smbwd, minimum height=0.75cm] {$c = c + 1$};
    \node (decide2) at (3.8,-13.5) [draw, decision, align=center, minimum width=\smbwd, minimum height=0.75cm] {$c \leq C$?};
    \node (decide3) at (7.6,-13.5) [draw, decision, align=center, minimum width=\smbwd, minimum height=0.75cm] {$\text{MSC}_{\text{min}} <$\\ $\text{MSC}_{\text{min,old}}$?};
    \node (terminal2) at (11.6,-13.5) [draw, terminal, align=center, minimum width=\smbwd, minimum height=0.75cm] {FINAL\\ PROXY};
    \node (process9) at (7.6,-10.75) [draw, process, align=center, minimum width=\smbwd, minimum height=0.75cm] {UPDATED PROXY\\ by adding term $c_{\text{min}}$};
    \node (process10) at (7.6,-9.25) [draw, process, align=center, minimum width=\smbwd, minimum height=0.75cm] {$\text{MSC}_{\text{min,old}} := \text{MSC}_{\text{min}}$};
    \node (process11) at (7.6,-8) [draw, process, align=center, minimum width=\smbwd, minimum height=0.75cm] {$k = k + 1$};
    \node (decide4) at (7.6,-6) [draw, decision, align=center, minimum width=\smbwd, minimum height=0.75cm] {$k \leq K_{\text{max}}$?};
    \draw[->] (terminal1) -- (process0);
    \draw[->] (process0) -- (process1);
    \draw[->] (process1) -- (process2);
    \draw[->] (process2) -- (process3);  
    \draw[->] (process3) -- (process4);
    \draw[->] (process4) -- (process5);
    \draw[->] (process5) -- (process6);
    \draw[->] (process6) -- (decide1);
    \draw[->] (decide1) --node[left]{YES} (process7);
    \draw[->] (decide1) --node[above]{NO} (process8);
    \draw[->] (process7) -| (process8);
    \draw[->] (process8) -- (decide2);
    \draw[->] (decide2) --node[above]{YES} (process6);
    \draw[->] (decide2) --node[above]{NO} (decide3);
    \draw[->] (decide3) --node[left]{YES} (process9);
    \draw[->] (decide3) --node[above]{NO} (terminal2);
    \draw[->] (process9) -- (process10);
    \draw[->] (process10) -- (process11);
    \draw[->] (process11) -- (decide4);
    \draw[->] (decide4) --node[above]{YES} (process4);
    \draw[->] (decide4) -|node[right]{NO} (terminal2);
  \end{tikzpicture}
  \begin{center}
    $\underbrace{\ \ \ \ \ \ \ \ \ \ \ \ \ \ \ \ \ \ \ \ \ \ \ \ \ \ \ \ \ \ \ \ \ \ \ \ \ \ \ \ \ \ \ \ \ \ \ \ \ \ \ }_{\text{Finds the best candidate $c$ in iteration $k$ if there is one}} \underbrace{\ \ \ \ \ \ \ \ \ \ \ \ \ \ \ \ \ \ \ \ \ \ \ \ \ \ \ \ \ \ \ \ \ \ \ \ \ \ \ \ \ \ \ \ \ \ \ \ \ \ \ \ \ \ \ \ \ \ \ }_{\text{Finds the best proxy function in the adaptive algorithm}}$
  \end{center}
	\caption{Flowchart of the calibration algorithm}
	\label{fig:algorithm}
\end{figure}
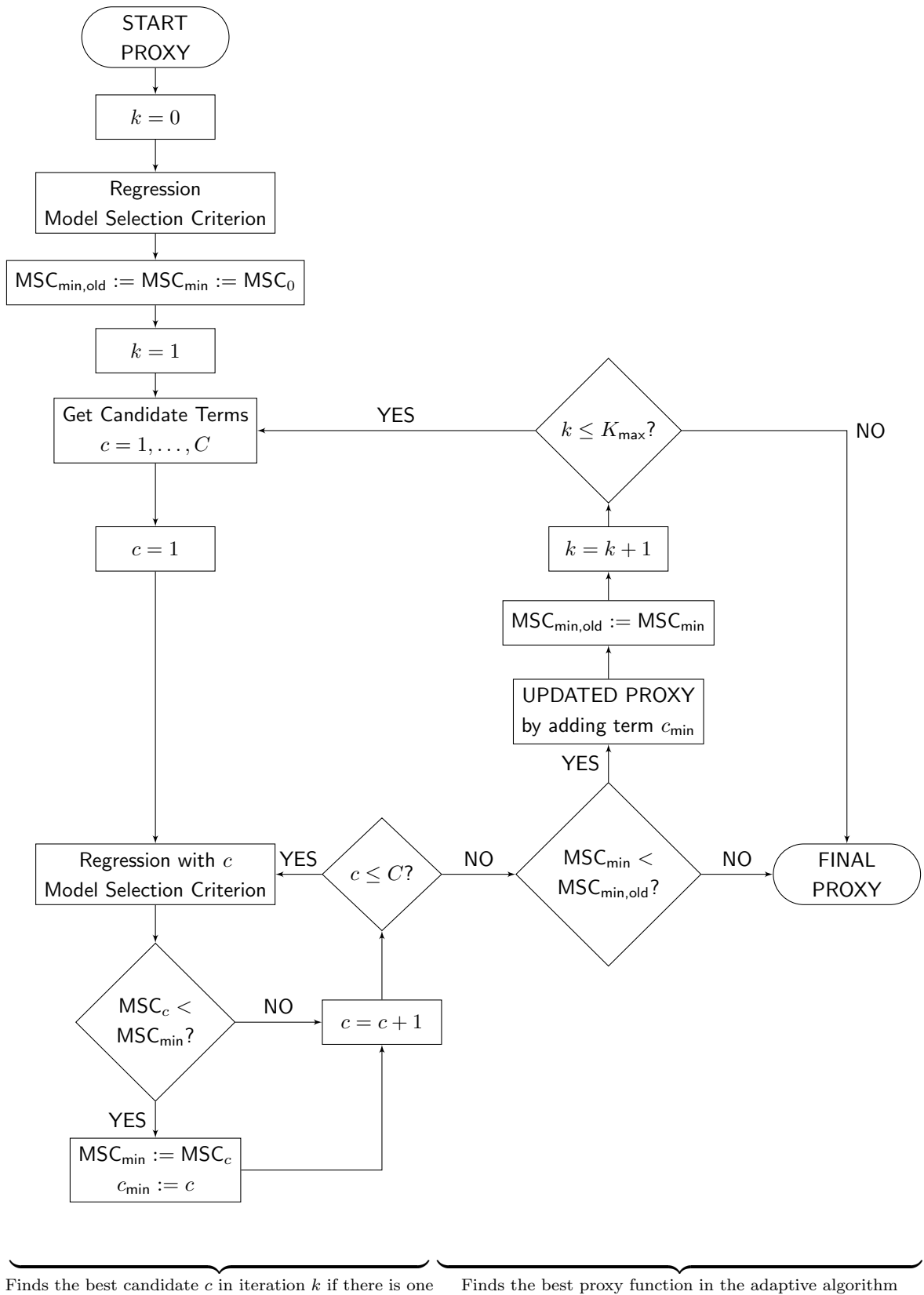

The calibration of the proxy function is performed by an adaptive algorithm that can be decomposed into the following five major components: (1) a set of allowed basis function types for the proxy function, (2) a regression method, (3) a model selection criterion, (4) a candidate term update principle, and (5) the number of steps per iteration and the directions of the algorithm. For illustration, we adopt the flowchart of the adaptive algorithm from \cite{krah2018} and depict it in Figure \ref{fig:algorithm}. While components (1) and (5) enter the flowchart implicitly through the start proxy, candidate terms and the order of the processes and decisions in the chart, components (2), (3) and (4) are explicitly indicated through the labels ``Regression'', ``Model Selection Criterion'' and ``Get Candidate Terms''.

Let us briefly recapitulate the adaptive algorithm under some standard choices of components (1), (2), (3), (4) and (5) which have already been successfully applied in the insurance industry. As the function types for the basis functions (1), let only monomials be allowed. Let the regression method (2) be ordinary least-squares (OLS) regression and the model selection criterion (3) Akaike information criterion (AIC) from \cite{akaike1973}. Let the set of candidate terms (4) be updated by the principle of marginality to which we will return in greater detail below. Lastly, when building up the proxy function iteratively, let the algorithm make only one step per iteration in the forward direction (5) meaning that in each iteration exactly one basis function is selected which cannot be removed anymore (adaptive forward stepwise selection).

\subsubsection*{Iterative Procedure}
The algorithm starts in the upper left side of Figure \ref{fig:algorithm} with the specification of the start proxy basis functions. We specify only the intercept so that the first regression ($k = 0$) reduces to averaging over all fitting values. In order to harmonize the choices of OLS regression and AIC, we assume that the errors are normally distributed and homoscedastic because then the OLS estimator coincides with the maximum likelihood estimator. AIC is a relative measure for the goodness-of-fit of the proxy function and is defined as twice the negative of the maximum log-likelihood plus twice the number of degrees of freedom. The smaller the AIC score, the better is the fit and thus the trade-off between a too complex (overfitting) and too simple model (underfitting).

At the beginning of each iteration ($k = 1,\ldots,K-1$), the set of candidate terms is updated by the principle of marginality which is compatible with the choice of a monomial basis. Using such a principle saves computational costs by selecting the basis functions conditionally on the current proxy function structure. According to the principle of marginality, a monomial basis function becomes a candidate if and only if all its derivatives are already included in the proxy function. In the first iteration ($k = 1$), all linear monomials of the risk factors become candidates as their derivatives are constant values which are represented by the intercept.

The algorithm proceeds on the lower left side of the flowchart with a loop in which all candidate terms are separately added to the proxy function structure and tested with regard to their additional explanatory power. With each candidate, the fitting values are regressed against the fitting scenarios and the AIC score is calculated. If no candidate reduces the currently smallest AIC score, the algorithm is terminated, and otherwise, the proxy function is updated by the one which reduces AIC most. Then the next iteration ($k + 1$) begins with the update of the set of candidate terms, and so on. As long as no termination occurs, this procedure is repeated until the prespecified maximum number of terms $K_{\text{max}}$ is reached.

\subsection{Validation Figures}\label{sec:validation_figures}
\subsubsection*{Validation Sets}
Since it is the objective of this paper to propose suitable regression methods for the proxy function calibration in the LSMC framework, we introduce several validation figures serving as indicators for the approximation quality of the proxy functions. We measure the out-of-sample performance of each proxy function on three different validation sets by calculating five validation figures per set.

The three validation sets are a Sobol set, a nested simulations set and a capital region set. Unlike the Sobol set, the nested simulations and capital region sets do not serve as feasible validation sets in the LSMC routine as they require massive computational capacities but can be regarded as the natural benchmark for the LSMC-based method and are thus very valuable for this analysis. The Sobol set consists of e.g. between $L = 15$ and $L = 200$ Sobol validation points, of which the scenarios follow a Sobol sequence covering the fitting space uniformly. Thereby is the fitting space the cube on which the outer fitting scenarios are defined. It has to cover the space of real-world scenarios used for the full loss distribution forecast sufficiently well. For interpretive reasons, sometimes the Sobol set is extended by points with e.g. one-dimensional risk scenarios or scenarios producing a risk capital close to the SCR ($= 99.5\%$ value-at-risk) in previous risk capital calculations.

The nested simulations set comprises the e.g. $L = 820$ to $L = 6554$ validation points of which the scenarios correspond to the e.g. highest $2.5\%$ to $5\%$ losses from the full loss distribution forecast made by the proxy function that had been derived under the standard calibration algorithm choices described in Section \ref{sec:calibration_algorithm}. Like in the example of Ch. 5.2 in \cite{krah2018}, the order of these losses - which scenarios lead to which quantiles - following from the forth and last step of the LSMC approach is very similar to the order following from the nested simulations approach. Therefore the scenarios of the nested simulations set are simply given by the order of the losses resulting from the LSMC approach. Several of these scenarios consist of stresses falling out of the fitting space. Few points with severe outliers due to extreme stresses far beyond the fitting space should be excluded from the set. The capital region set is a subset of the nested simulations set containing the nested simulations SCR estimate and the e.g. $64$ losses above and below, which makes in total e.g. $L = 129$ validation points.

\subsubsection*{Validation Figures}\label{sec:valfig}
The five validation figures reported in our numerical experiments comprise two normalized mean absolute errors (MAEs), one with respect to the magnitude of the economic variable itself and one with respect to the magnitude of the corresponding market value of assets. Further, they comprise the mean of the residuals, the normalized MAE of the deviation of the economic variable from the base value (see the definition of the base value below) with respect to the magnitude of that deviation and last but not least the mean of the residuals of these deviations. The smaller the normalized MAEs are, the better the proxy function approximates the economic variable. However, the validation values are afflicted with Monte Carlo errors so that the normalized MAEs serve only as meaningful indicators as long as the proxy functions do not become too precise. The means of the residuals should be possibly close to zero since they indicate systematic deviations of the proxy functions from the validation values. While the first three validation figues measure how well the proxy function reflects the economic variable in the CFP model, the latter two address the approximation effects on the SCR, compare Ch. 3.4.1 of \cite{krah2018}.

Let us write the absolute value as $\left|\cdot\right|$ and let $L$ denote the number of validation points. Then we can express the MAE of the proxy function $\widehat{f}\left(x^i\right)$ evaluated at the validation scenarios $x^i$ versus the validation values $y^i$ as $\frac{1}{L} \sum_{i=1}^{L} {\left|y^i - \widehat{f}\left(x^i\right)\right|}$. After normalizing the MAE with respect to the mean of the absolute values of the economic variable or the market value of assets, i.e. $\frac{1}{L} \sum_{i=1}^{L} {\left|d^i\right|}$ with $d^i \in \left\{y^i, a^i\right\}$, we obtain the first two validation figures, i.e.
\begin{equation}
  \text{mae} = \frac{\sum_{i=1}^{L} {\left|y^i - \widehat{f}\left(x^i\right)\right|}}{\sum_{i=1}^{L} {\left|d^i\right|}}.
  \label{eq:mae}
\end{equation}
In the following, we will refer to (\ref{eq:mae}) with $d^i = y^i$ as the MAE with respect to the \textit{relative metric}, and to (\ref{eq:mae}) with $d^i = a^i$ as the MAE with respect to the \textit{asset metric}. The mean of the residuals is given by
\begin{equation}
  \text{res} = \frac{1}{L}\sum_{i=1}^{L} {\left(y^i - \widehat{f}\left(x^i\right)\right)}.
  \label{eq:res}
\end{equation}

Let us refer by the base value $y^0$ to the validation value corresponding to the base scenario $x^0$ in which no risk factor has an effect on the economic variable. In analogy to (\ref{eq:mae}) but only with respect to the relative metric, we introduce another normalized MAE by
\begin{equation}
  \text{mae}^0 = \frac{\sum_{i=1}^{L} {\left|\left(y^i-y^0\right) - \left(\widehat{f}\left(x^i\right)-\widehat{f}\left(x^0\right)\right)\right|}}{\sum_{i=1}^{L} {\left|y^i-y^0\right|}}.
  \label{eq:mae0}
\end{equation}
The mean of the corresponding residuals is given by
\begin{equation}
  \text{res}^0 = \frac{1}{L}\sum_{i=1}^{L} {\left(\left(y^i-y^0\right) - \left(\widehat{f}\left(x^i\right)-\widehat{f}\left(x^0\right)\right)\right)}.
  \label{eq:res0}
\end{equation}

In addition to these five validation figures, let us define the base residual which can be used as a substitute for (\ref{eq:res0}) depending on personal taste. The base residual can easily be extracted from (\ref{eq:res}) and (\ref{eq:res0}) by
\begin{equation}
  \text{res}^{\text{base}} = y^0 - \widehat{f}\left(x^0\right) = \text{res} - \text{res}^0.
  \label{eq:resbase}
\end{equation}

\section{Machine Learning Regression Methods}
\subsection{General Remarks}\label{sec:general_remarks_theory}
As the main part of our work, we will compare various types of machine learning regression approaches for determining suitable proxy functions in the LSMC framework. The methods we present in this section range from ordinary and generalized least-squares regression variants over GLM and GAM approaches to multivariate adaptive regression splines and kernel regression approaches. The performance of the newly derived proxy functions when applied to the validation sets is one way of how to judge the different methods. Their compatibility with the principle of marginality and a suitable model selection criterion such as AIC to compare iteration-wise the candidate models inside the approaches is another way.

We make two approximations to express the expected value of the economic variable $Y(X)$ under the risk-neutral probability measure $\mathbb{Q}$ by a proxy function with respect to the risk factors $X$. The approximations are necessary since the sets of basis functions ($K$ basis functions) and fitting points (sample size $N$) are finite in practice. Unlike \cite{bauer2015} and \cite{krah2018}, who denote the conditional expectation of the economic variable relative to the outer scenario $X$ by $\text{AC}(X)$, we use the notation $Y(X)$ to account for the fact that the economic variable does not have to be the available capital but can instead be e.g. the best estimate of liabilites or the market value of assets. Note that the expectation operator is included in the notation of $Y(X)$. In accordance with this reduced notation, we will refer to the conditional expectation of the economic variable only as the economic variable. Only when we will use the term economic variable in the context of data realizations such as of the fitting or validation values, we will not mean its expectation.

Let the $D$-dimensional fitting scenarios be distributed under the physical probability measure $\mathbb{P}'$ on the fitting space which itself is a subspace of $\mathbb{R}^D$.

\subsection{Ordinary Least-Squares (OLS) Regression}\label{sec:OLStheory}
\subsubsection*{Classical Linear Regression Model}
In iteration $K-1$ of the adaptive forward stepwise algorithm, we can write the linear predictor $f(X)$ for $Y(X)$, containing the first approximation, as a linear combination of suitable linear independent basis functions $e_k \left( X \right) \in L^2\left( \mathbb{R}^D, \mathcal{B}, \mathbb{P}' \right),\ k=0,1,\ldots,K-1,$ i.e.
\begin{equation}
   Y(X) \stackrel{K<\infty}{\approx} f(X) = \sum_{k=0}^{K-1} {\beta_k e_k \left( X \right)}.
  \label{eq:f}
\end{equation}

With the fitting points $\left( x^i, y^i \right),\ i=1,\ldots,N,$ and uncorrelated errors $\epsilon^i$ having the same variance $\sigma^2 > 0$ (= homoscedastic errors), we obtain the classical linear regression model
\begin{equation}
   y^i = \sum_{k=0}^{K-1} {\beta_k e_k \left( x^i \right)} + \epsilon^i,
  \label{eq:y}
\end{equation}
where $e_0 \left( x^i \right) = 1$ and $\beta_0$ is the intercept. Then the ordinary least-squares (OLS) estimator $\bm{\widehat{\beta}}_{\text{OLS}}$ of the coefficients is given by
\begin{equation}
   \bm{\widehat{\beta}}_{\text{OLS}} = \argmin_{\bm{\beta} \in \mathbb{R}^K} \left\{ \sum_{i=1}^{N} {\left( y^i - \sum_{k=0}^{K-1} {\beta_k e_k \left( x^i \right)} \right)^2 } \right\}.
  \label{eq:betaOLSreg}
\end{equation}
Since the residuals corresponding to the OLS solution are $\widehat{\epsilon}^i = y^i - \sum_{k=0}^{K-1} {\widehat{\beta}_{\text{OLS},k} e_k \left( x^i \right) }$, the OLS estimator minimizes by definition the residual sum of squares. By substituting $\bm{\widehat{\beta}}_{\text{OLS}}$ for $\bm{\beta}$ in (\ref{eq:f}), we account for the second approximation and arrive at the proxy function $\widehat{f}\left(X\right)$ for the economic variable $Y(X)$ conditional on any outer scenario $X$, i.e.
\begin{equation}
    Y(X) \stackrel{K,N<\infty}{\approx} \widehat{f}\left(X\right) = \sum_{k=0}^{K-1} {\widehat{\beta}_{\text{OLS},k} e_k \left( X \right)}.
  \label{eq:fhat}
\end{equation}

If we use the notation $z_{ik} = e_k \left( x^i \right)$, we can replace the minimization problem (\ref{eq:betaOLSreg}) by the closed-form expression of the OLS estimator in which $Z = \left( z_{ik} \right)_{\substack{i=1,\ldots,N\ \ \\k=0,\ldots,K-1}}$ denotes the design matrix and $\textbf{\text{y}} = \left(y^1,\ldots,y^N \right)^T$ the response vector, i.e.
\begin{equation}
   \bm{\widehat{\beta}}_{\text{OLS}} = \left( Z^T Z \right)^{-1} Z^T \textbf{\text{y}}.
  \label{eq:betaOLS}
\end{equation}
The system $\left(Z^T Z\right) \bm{\widehat{\beta}}_{\text{OLS}} = Z^T \textbf{\text{y}}$ equivalent to (\ref{eq:betaOLS}) is in practice often solved via a QR or singular value decomposition of $Z$ to increase numerical stability. For a practical implementation see e.g. function \textit{lm($\cdot$)} in R package \textit{stats} of \cite{stats2018}. The sample variance is obtained by $s_{\text{OLS}} = \frac{1}{N-K} \bm{\widehat{\epsilon}}^T \bm{\widehat{\epsilon}}$ where $\bm{\widehat{\epsilon}} = \textbf{\text{y}}-Z\bm{\widehat{\beta}}_{\text{OLS}}$ is the residual vector. With $\textbf{z} = \left(e_0 \left( X \right),\ldots,e_{K-1} \left( X \right)\right)^T$, (\ref{eq:fhat}) becomes in matrix notation $Y(X) \stackrel{K,N<\infty}{\approx} \widehat{f}\left(X\right) = \textbf{z}^T\bm{\widehat{\beta}}_{\text{OLS}}$.

\subsubsection*{Gauss-Markov Theorem, ML Estimation \& AIC}
We formulate the Gauss-Markov theorem in our setting conditional on the fitting scenarios and in line with \cite{hayashi2000} under the assumptions of strict exogeneity $E\left[ \bm{\epsilon} \mid Z \right] = \bm{0}$ (A1), a spherical error variance $V \left[ \bm{\epsilon} \mid Z \right] = \sigma^2 I_N$, where $I_N$ is the $N$-dimensional identity matrix (A2), and no multicollinearity, that is, linear independent basis functions (A3).
\begin{GMt} The OLS estimator is the best linear unbiased estimator (BLUE) of the coefficients in the classical linear regression model (\ref{eq:y}) under Assumptions (A1)-(A3).\end{GMt}

Akaike information criterion (AIC) needs to be evaluated at the maximum likelihood (ML) estimators of the coefficients and variance of the errors. For this purpose, we have to make an assumption about the distribution of the economic variable, or equivalently the errors. In order to make AIC and OLS regression easily combinable we assume in addition to (A1), (A2) and (A3) that the errors are normally distributed conditional on the fitting scenarios (A4) because then Proposition 1.5 of \cite{hayashi2000} states the following.
\begin{T} The ML coefficient estimator coincides with the OLS coefficient estimator and the ML estimator of the error variance $\widehat{\sigma}^2$ can be expressed as $\frac{N-K}{N}$ times the OLS sample variance $s_{\emph{\text{OLS}}}$, i.e. $\widehat{\sigma}^2 = \frac{1}{N} \bm{\widehat{\epsilon}}^T \bm{\widehat{\epsilon}}$, under Assumptions (A1)-(A4).\label{th:1}\end{T}
Furthermore, the OLS estimator is the efficient estimator under these assumptions according to \cite{greene2002}.

According to \cite{krah2018}, AIC has the form of a suitably weighted sum of the calibration error and the number of basis functions under Assumption (A4), i.e.
\begin{align}
  \text{AIC} &= -2 l\left(\bm{\widehat{\beta}}_{\text{OLS}},\widehat{\sigma}^2\right) + 2 \left(K+1\right) \label{eq:AICOLS} \\
  &=N \left( \log \left( 2\pi\widehat{\sigma}^2 \right) + 1 \right) + 2 \left(K+1\right). \nonumber
\end{align}
More generally, the calibration error corresponds to twice the negative of the log-likelihood $l\left(\cdot\right)$ of the model and the number of basis functions corresponds to the degrees of freedom of the model. The smaller the AIC score is, the better is the fitted model supposed to approximate the underlying data. AIC penalizes both a small log-likelihood and a high model complexity and helps thus select a possibly simple model with a possibly high goodness-of-fit. However, since AIC is only a relative measure of the goodness-of-fit, the final proxy function has to pass an additional out-of-sample validation procedure in the LSMC algorithm.

\subsection{Generalized Linear Models (GLMs)}\label{sec:GLMtheory}
\subsubsection*{Random Component, Systematic Component \& Link Function}
\cite{nelder1972} developed the class of generalized linear models (GLMs) as a generalization of the classical linear model in (\ref{eq:y}). A GLM consists of a random component, a systematic component and a link function. From the perspective of a GLM and maximum likelihood (ML) estimation, one has to assume that the economic variable under the risk-neutral probability measure follows a certain distribution. We have already seen with Theorem \ref{th:1} that the OLS estimator of the coefficients equals the ML estimator if the economic variable, or equivalently the errors, are normally distributed conditional on the fitting scenarios (A4).

By following Ch. 2.2 in \cite{mccullagh1989}, we generalize the linear model in the adaptive algorithm. In a GLM, the economic variable is allowed to come from any distribution of the exponential family conditional on the outer scenario, for instance from the normal, gamma, or inverse gaussian distribution. The distribution of the economic variable in a GLM is reflected by the random component. Its canonical form with canonical parameter $\theta$ is given by the density function
\begin{equation}
  \pi(y \mid \theta, \phi) = \exp {\left( \frac{y \theta - b(\theta)}{a(\phi)} + c(y,\phi) \right)},
  \label{eq:pi}
\end{equation}
where $a(\phi)$, $b(\theta)$ and $c(y,\phi)$ are specific functions. While the canonical parameter $\theta$ is related to the expected value of the distribution $E\left[y\right] = \mu = b'\left(\theta\right)$, the dispersion parameter $\phi$ only affects the variance $V\left[y\right] = b''\left(\theta\right)a(\phi) = V\left[\mu\right] a(\phi)$, whereby we refer to $V\left[\mu\right]$ as the variance function. Since we consider only equal prior weights, we can set $a(\phi) = \phi$ constant over all observations. For example, a normally distributed economic variable with mean $\mu$ and variance $\sigma^2$ is given by $a(\phi) = \phi$, $b(\theta) = \frac{\theta^2}{2}$ and $c(y,\phi) = -\frac{1}{2} \left( \frac{y^2}{\sigma^2} + \log {\left(2\pi \sigma^2\right)} \right)$ with $\theta = \mu$ and $\phi = \sigma^2$ because then (\ref{eq:pi}) becomes $\pi(y \mid \theta, \phi) = \frac{1}{\sqrt {2\pi \sigma^2}}\exp {\left(-\frac{\left(y-\mu\right)^2}{2\sigma^2}\right)}$. Here, the distribution of the errors is obtained by shifting the mean to $\mu = 0$. The equivalence between the distribution assumption of the economic variable and raw errors $\epsilon = y - \mu$ persists in GLMs.

The systematic component of a GLM is exactly the linear predictor $\eta = f(X)$ as defined in the linear model in (\ref{eq:f}). However, the first equality in (\ref{eq:f}) does not generally hold anymore. Instead a monotonic link function $g(\cdot)$ relates now the economic variable to the linear predictor, in literature usually formalized by $g(\mu) = \eta$, here by
\begin{equation}
  g(\underbrace{Y(X)}_{=\ \mu}) \stackrel{K<\infty}{\approx} \underbrace{f(X)}_{=\ \eta} = \sum_{k=0}^{K-1} {\beta_k z_k} = \textbf{z}^T \bm{\beta}
  \label{eq:g}
\end{equation}
with $\textbf{z} = \left(e_0 \left( X \right),\ldots,e_{K-1} \left( X \right)\right)^T$. When the link function is the identity as in the normal model this extension disappears, i.e. $\mu=\eta$.

Applying a link function to the economic variable is especially appealing when the range of the linear predictor may deviate substantially from that of the economic variable. For example, an economic variable capturing service times that follow a gamma distribution can only be positive but the linear predictor may also take on negative values. With e.g. $g\left(\cdot\right) = \log{\left(\cdot\right)}$ such a potential inconsistency can be eliminated. Another popular choice are the canonical link functions $\widetilde{g}\left(\cdot\right)$ which express the canonical parameter $\theta = \theta(X)$ with regard to the expected value $\mu = Y(X)$ if the variance is known, i.e. $\widetilde{g}\left(\mu\right) = \theta$, hence due to (\ref{eq:g}) also $\theta \stackrel{K<\infty}{\approx} f(X)$ with $\widetilde{g}\left(\cdot\right)$. For instance, the canonical link functions are $g\left(\mu\right) = \text{id}\left(\mu\right)$ for the normal, $g\left(\mu\right) = \frac{1}{\mu}$ for the gamma, and $g\left(\mu\right) = \frac{1}{\mu^2}$ for the inverse gaussian distribution.

The log-likelihood of a single observation is the logarithm of (\ref{eq:pi}), i.e. $l^i\left(\bm{\beta}, \phi^i\right) = \log \pi(y^i \mid \theta^i, \phi^i)$ with the dependence $\theta^i = \theta^i\left(\mu^i\left(\eta^i\left(\bm{\beta}, x^i\right)\right)\right)$ due to the equality $\mu = b'\left(\theta\right)$ and (\ref{eq:g}). Thus, with constant dispersion $a(\phi^i) = \phi^i = \phi,\ i=1,\ldots,N,$ the GLM estimator $\bm{\widehat{\beta}}_{\text{GLM}}$ of the coefficients is given as the maximizer of $l\left(\bm{\beta}, \phi\right) = \sum_{i=1}^{N} {\log \pi\left(y^i \mid \theta^i, \phi\right)}$, that is as the ML estimator, i.e.
\begin{equation}
   \bm{\widehat{\beta}}_{\text{GLM}} = \argmax_{\bm{\beta} \in \mathbb{R}^K} \left\{ \sum_{i=1}^{N} {\left(\frac{y^i \theta^i - b(\theta^i)}{\phi} + c(y^i,\phi)\right)} \right\}.
  \label{eq:betaGLMreg}
\end{equation}
While for the Poisson or binomial distribution the dispersion is taken as $1$, for the other distributions from the exponential family the dispersion $\phi$ is unknown. Assuming a random component of the form (\ref{eq:pi}) with a constant dispersion (A5), or in other words, with equal prior weights, makes the factors $a(\phi^i)$ disappear in the first-order ML condition. Using unequal prior weights might be beneficial, however it is not clear how they should be selected in the adaptive algorithm, and they would make the estimation procedure more complicated. Since the dispersion is not required for the derivation of $\bm{\widehat{\beta}}_{\text{GLM}}$ in our setting, it is omitted in the IRLS algorithm described below. Once $\bm{\widehat{\beta}}_{\text{GLM}}$ is known, the dispersion is estimated with the aid of the Pearson residual chi-squared statistic.

\subsubsection*{GLM Estimation via IRLS Algorithm}
Under (A5), there generally does not exist a closed-form solution for the GLM coefficient estimator (\ref{eq:betaGLMreg}). In Ch. 2.5, \cite{mccullagh1989} apply Fisher's scoring method, a standard approach in log-likelihood maximization, to obtain an approximation to the GLM estimator, i.e.
\begin{equation}
  \bm{\widehat{\beta}^{(t+1)}} = \bm{\widehat{\beta}^{(t)}} + I^{-1} \frac{\partial l}{\partial \bm{\beta}}.
  \label{eq:FisherScoring}
\end{equation}
Here, $\bm{\widehat{\beta}^{(t)}}$ is the coefficient estimator in iteration $t$, $\frac{\partial l}{\partial \bm{\beta}}$ the score function, and $I = E\left[ -\frac{\partial^2 l}{\partial\bm{\beta}\partial\bm{\beta}^T}\right]$ the Fisher information matrix (equal to the negative of the expected value of the Hessian matrix) with the expectation being taken with respect to the random component. While $\frac{\partial l}{\partial \bm{\beta}}$ depends on the regressors and response values, $I$ depends only on the regressors due to the expectation operator. Both have to be evaluated at $\bm{\widehat{\beta}^{(t)}}$. Further, \cite{mccullagh1989} justify how Fisher's scoring method can be cast in the form of the iteratively reweighted least squares (IRLS) algorithm. As an alternative, they suggest the Newton-Raphson method, which coincides with Fisher's scoring method if canonical link functions are used since the actual value of the Hessian matrix equals its expected value then.

The IRLS algorithm works in our context as follows. Let the dependent variable in the iterative procedure be
\begin{equation}
  \widehat{s}^i\left(\bm{\widehat{\beta}^{(t)}}\right) = \widehat{\eta}_{(t)}^i + \left(y^i - \widehat{\mu}_{(t)}^i\right)\left(\frac{d\eta}{d\mu}\left(\widehat{\mu}_{(t)}^i\right)\right),
  \label{eq:s}
\end{equation}
where $\widehat{\eta}_{(t)}^i = \widehat{f}\left(x^i\right)$ is the estimate for the linear predictor or the proxy function evaluated at fitting scenario $x^i$, compare (\ref{eq:g}), where $\widehat{\mu}_{(t)}^i = g^{-1}\left(\widehat{\eta}_{(t)}^i\right)$ derived from $\widehat{\eta}_{(t)}^i$ is the estimate for the economic variable, and $\frac{d\eta}{d\mu}\left(\widehat{\mu}_{(t)}^i\right) = g'\left(\widehat{\mu}_{(t)}^i\right)$ is the first derivative of the link function with respect to the economic variable evaluated at $\widehat{\mu}_{(t)}^i$. Let $\widehat{\textbf{\text{s}}}^{\bm{(t)}} = \left(\widehat{s}^1\left(\bm{\widehat{\beta}^{(t)}}\right),\ldots,\widehat{s}^N\left(\bm{\widehat{\beta}^{(t)}}\right)\right)^T$ denote the vector of the dependent variable over all fitting points.

Furthermore, the (quadratic) weight in the iterative procedure is given by
\begin{equation}
  \widehat{w}^i\left(\bm{\widehat{\beta}^{(t)}}\right) = \left(\frac{d\eta}{d\mu}\left(\widehat{\mu}_{(t)}^i\right)\right)^{-2} V\left[\widehat{\mu}_{(t)}^i\right]^{-1},
  \label{eq:w}
\end{equation}
where $V\left[\widehat{\mu}_{(t)}^i\right]$ is the variance function from above evaluated at $\widehat{\mu}_{(t)}^i$. Then the (quadratic) weight matrix is defined by $W^{(t)}=\text{diag}\left(w^1\left(\bm{\widehat{\beta}^{(t)}}\right),\ldots,w^N\left(\bm{\widehat{\beta}^{(t)}}\right)\right)$.

\begin{IRLS} Perform the following iterative approximation procedure with e.g. an initialization of $\widehat{\mu}_{(0)}^i = y^i + 0.1$ and $\widehat{\eta}_{(0)}^i = g\left(\widehat{\mu}_{(0)}^i\right)$ as proposed by \cite{dutang2017} until convergence:
\begin{align}
  \bm{\widehat{\beta}^{(t+1)}} &= \argmin_{\bm{\beta} \in \mathbb{R}^K} \left\{ \sum_{i=1}^{N} {w^i \left(\bm{\widehat{\beta}^{(t)}}\right)^{-1} \left(\widehat{s}^i\left( \bm{\widehat{\beta}^{(t)}}\right) - \sum_{k=0}^{K-1} {\beta_k z_{ik}}\right)^2}\right\} \nonumber \\
  &= \left(Z^T W^{(t)} Z\right)^{-1} Z^T W^{(t)} \widehat{\textbf{\emph{\text{s}}}}^{\bm{(t)}}.
  \label{eq:IRLS}
\end{align}
After convergence, we have $\bm{\widehat{\beta}}_{\emph{\text{GLM}}} = \bm{\widehat{\beta}^{(t+1)}}$.\end{IRLS}
For example, \cite{green1984} proposes to solve system $\left(Z^T W^{(t)} Z\right) \bm{\widehat{\beta}^{(t+1)}} = Z^T W^{(t)} \widehat{\textbf{\emph{\text{s}}}}^{\bm{(t)}}$ equivalent to (\ref{eq:IRLS}) via a QR decomposition to increase numerical stability. For a practical implementation of GLMs using the IRLS algorithm, see e.g. function \textit{glm($\cdot$)} in R package \textit{stats} of \cite{stats2018}.

By inserting (\ref{eq:s}), (\ref{eq:w}) and the GLM estimator into (\ref{eq:IRLS}) and by using (\ref{eq:g}), we arrive at the property
\begin{align}
  \bm{\widehat{\beta}}_{\text{GLM}} &= \argmin_{\bm{\beta} \in \mathbb{R}^K} \left\{ \sum_{i=1}^{N} {V\left[\widehat{\mu}_{\text{GLM}}^i\right]\left(y^i-\widehat{\mu}_{\text{GLM}}^i\right)^2}\right\},
  \label{eq:IRLSprop}
\end{align}
that is, the GLM estimator minimizes the squared sum of raw residuals scaled by the estimated individual variances of the economic variable. The Pearson residuals are defined as the raw residuals divided by the estimated individual standard deviations, i.e.
\begin{equation}
  \widehat{\epsilon}^i = \frac{y^i-\widehat{\mu}_{\text{GLM}}^i}{\sqrt{V\left[\widehat{\mu}_{\text{GLM}}^i\right]}}.
  \label{eq:pearsonres}
\end{equation}
For example, in the normal model from above with mean $\mu$ and variance $\sigma^2$, we have $b(\theta)=\frac{\theta^2}{2}$ and thus constant estimated individual variances across all observations $V\left[\mu\right] = b''(\theta)=1$ so that no actual weighting takes place.

\subsubsection*{AIC \& Dispersion Estimation}
Since AIC depends on the ML estimators, it is combinable with GLMs in the adaptive algorithm. Here, it has the form
\begin{equation}
   \text{AIC} = -2 l\left(\bm{\widehat{\beta}}_{\text{GLM}}, \widehat{\phi}\right) + 2 \left(K+p\right),
  \label{eq:AICGLM}
\end{equation}
where $K$ is the number of coefficients and $p$ indicates the number of the additional model parameters associated with the distribution of the random component. For instance, in the normal model, we have $p=1$ due to the error variance/dispersion. A typical estimate of the dispersion in GLMs is the Pearson residual chi-squared statistic divided by $N-K$ as described by \cite{zuur2009} and implemented e.g. in function \textit{glm($\cdot$)} belonging to R package \textit{stats}, i.e.
\begin{equation}
   \widehat{\phi} = \frac{1}{N-K}\sum_{i=1}^{N} {\left(\widehat{\epsilon}^i\right)^2},
  \label{eq:phihat}
\end{equation}
with $\widehat{\epsilon}^i$ given by (\ref{eq:pearsonres}). Even though this is not the ML estimator, it is a good estimate because, if the model is specified correctly, the Pearson residual chi-squared statistic divided by the dispersion is asymptotically $\chi_{N-K}^2$ distributed and the expected value of a chi-squared distribution with $N-K$ degrees of freedom is $N-K$.

\subsection{Generalized Additive Models (GAMs)}\label{sec:GAMtheory}
\subsubsection*{Richly Parameterized GLM with Smooth Functions}
\begin{figure}[thbp]
	\centering
		\includegraphics[width=\textwidth]{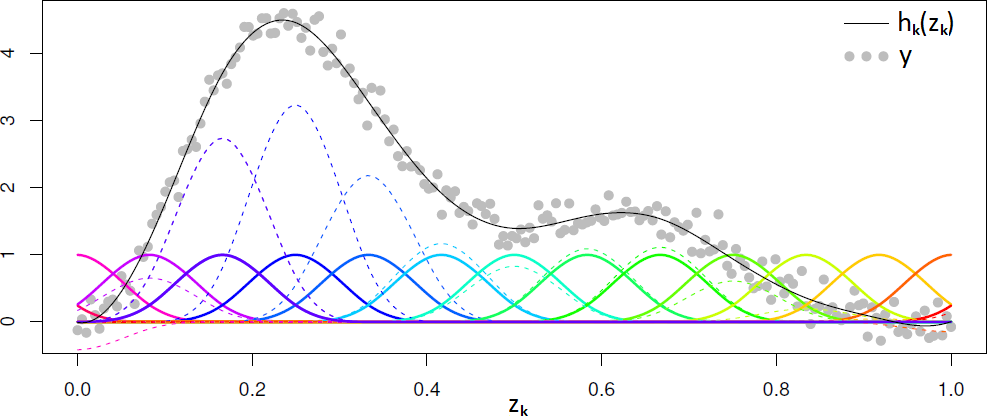}
	\caption{GAM with a basis expansion in one dimension}
	\label{fig:basis_expansion_Wood_2006}
\end{figure}

The class of generalized additive models (GAMs) was invented by \cite{hastie1986} and \cite{hastie1990} to unite the properties of GLMs and additive models. Based on \cite{wood2006}, we introduce one of the most obvious applications of GAMs in the adaptive algorithm of the LSMC framework. It is conceivable that other varieties of GAMs allowing e.g. only linear basis functions of single risk factors are the more natural approach from a theoretical point of view. While GAMs inherit from GLMs the random component (\ref{eq:pi}) and the link function (\ref{eq:g}), they inherit from the additive models of \cite{friedman1981} the linear predictor with the smooth functions. In the adaptive algorithm, we apply GAMs of the form
\begin{equation}
  g(\underbrace{Y(X)}_{=\ \mu}) \stackrel{K<\infty}{\approx} \underbrace{f(X)}_{=\ \eta} = \beta_0 + \sum_{k=1}^{K-1} {h_k\left(z_k\right)},
  \label{eq:gam}
\end{equation}
where $z_k = e_k \left( X \right)$, $\beta_0$ is the intercept and $h_k\left(\cdot\right),\ k=1,\ldots,K-1,$ are the smooth functions to be estimated. In addition to the smooth functions, GAMs could also include simple linear terms of the basis functions as they appear in the linear predictor of GLMs. A smooth function $h_k\left(\cdot\right)$ can be written as a basis expansion
\begin{equation}
  h_k\left(z_k\right) = \sum_{j=1}^{J} {\beta_{kj} b_{kj}\left(z_k\right)},
  \label{eq:basis}
\end{equation}
with coefficients $\beta_{kj}$ and known basis functions $b_{kj}\left(z_k\right),\ j=1,\ldots,J,$ which should not be confused with their arguments, namely the first-order basis functions $z_k=e_k \left( X \right),\ k=0,\ldots,K-1$. Slightly adjusted Figure \ref{fig:basis_expansion_Wood_2006} from \cite{wood2006} depicts an exemplary approximation of $y$ by a GAM with a basis expansion in one dimension $z_k$ without an intercept. The solid colorful curves represent the pure basis functions $b_{kj}\left(z_k\right),\ j=1,\ldots,J,$ the dashed colorful curves show them after scaling with the coefficients $\beta_{kj}b_{kj}\left(z_k\right),\ j=1,\ldots,J,$ and the black curve is their sum (\ref{eq:basis}).
Typical examples for basis functions are thin plate regression splines, duchon splines, penalized cubic regression splines or Eilers and Marx style P-splines. See e.g. function \textit{gam($\cdot$)} in R package \textit{mgcv} of \cite{mgcv2018} for a practical implementation of GAMs admitting these types of basis functions and using the PIRLS algorithm, which we present below.

In vector notation, we can write $\bm{\beta} = \left(\beta_0,\bm{\beta}_1^T,\ldots,\bm{\beta}_{K-1}^T\right)^T$ with $\bm{\beta}_k = \left(\beta_{k1},\ldots,\beta_{kJ}\right)^T$ and $\textbf{a} = \left(1,\textbf{b}_1\left(z_1\right)^T,\ldots,\textbf{b}_{K-1}\left(z_{K-1}\right)^T\right)^T$ with $\textbf{b}_k\left(z_k\right) = \left(b_{k1}\left(z_k\right),\ldots,b_{kJ}\left(z_k\right)\right)^T$, hence (\ref{eq:gam}) becomes
\begin{equation}
  g(\underbrace{Y(X)}_{=\ \mu}) \stackrel{K<\infty}{\approx} \underbrace{f(X)}_{=\ \eta} = \textbf{a}^T \bm{\beta}.
  \label{eq:gamvec}
\end{equation}
This parameterization is a richer version of (\ref{eq:g}) so that a GAM having a random component from the exponential family (\ref{eq:pi}) can be viewed as a richly parameterized GLM. In order to make the smooth functions $h_k\left(\cdot\right),\ k=1,\ldots,K-1,$ identifiable, identifiability constraints $\sum_{i=1}^{N} {h_k\left(z_{ik}\right)}=0$ with $z_{ik} = e_k\left(x^i\right)$ can be imposed. According to \cite{wood2006} this can be achieved by modification of the basis functions $b_{kj}\left(\cdot\right)$ with one of them being lost.

Let the deviance corresponding to observation $y^i$ be $D^i\left(\bm{\beta}\right) = 2\left(l_{\text{sat}}^i-l^i\left(\bm{\beta}, \phi\right)\right)\phi$ where $D^i\left(\bm{\beta}\right)$ is independent of dispersion $\phi$, where $l_{\text{sat}}^i = \max_{\bm{\beta}^i} l^i\left(\bm{\beta}^i, \phi\right)$ is the saturated log-likelihood and $l^i\left(\bm{\beta}, \phi\right)$ the log-likelihood. Then the model deviance can be written as $D\left(\bm{\beta}\right) = \sum_{i=1}^{N}{D^i\left(\bm{\beta}\right)}$. It is a generalization of the residual sum of squares for ML estimation. For instance, in the normal model the unit deviance is $\left(y^i - \mu^i\right)^2$. For given smoothing parameters $\lambda_k > 0,\ k=1,\ldots,K-1$, the GAM estimator $\bm{\widehat{\beta}}_{\text{GAM}}$ of the coefficients is defined as the minimizer of the penalized deviance
\begin{align}
  \bm{\widehat{\beta}}_{\text{GAM}} = \argmin_{\bm{\beta} \in \mathbb{R}^{(K-1)J+1}} \left\{ D\left(\bm{\beta}\right) + \sum_{k=1}^{K-1} {\lambda_k \int {h_k''\left(z_{k}\right)^2 \textrm{d} z_{k}}}\right\}\text{, where} \label{eq:betaGAMreg} \\
  \int {h_k''\left(z_{k}\right)^2 \textrm{d} z_{k}} = \bm{\beta}_k^T \left( \int {\textbf{b}_k''\left(z_{k}\right) \textbf{b}_k''\left(z_{k}\right)^T \textrm{d} z_{k}} \right) \bm{\beta}_k = \bm{\beta}_k^T \mathcal{S}_k \bm{\beta}_k \nonumber
\end{align}
are the smoothing penalties and the smoothing parameters $\lambda_k$ control the trade-off between a too wiggly model (overfitting) and a too smooth model (underfitting). The larger the $\lambda_k$ values are, the more pronounced is the wiggliness of the basis functions reflected by their second derivatives in the minimization problem (\ref{eq:betaGAMreg}), and the higher is thus the penalty associated with the coefficients and the smoother is the estimated model. Similar to how we have defined the GAM estimator as the minimizer of the penalized deviance, we could have defined the GLM estimator (\ref{eq:betaGLMreg}) as the minimizer of the unpenalized deviance.

\subsubsection*{GAM Estimation via PIRLS Algorithm}
\cite{buja1989} proposed to estimate GAMs by a backfitting procedure which can be shown to be the Gauss-Seidel iterative method for solving a set of normal equations associated with the additive model. Their backfitting procedure works for any scatterplot smoother so that the random component does no longer have to come from the exponential family, in fact, non-parametric models such as running-mean, running-line or kernel smoothers are possible as well. However, their suggestions to select the degree of smoothness through e.g. graphical analyses or cross-validation are for practitioners still difficult to implement. Therefore, GAMs have recently been increasingly defined in the form of (\ref{eq:gam}) with basis expansions (\ref{eq:basis}) of which the degree of smoothness is controlled by the smoothing penalties (\ref{eq:betaGAMreg}). A major advantage of this definition is its compatibility with information criteria and other model selection criteria such as generalized cross-validation. Besides, the resulting penalty matrix favors numerical stability in the PIRLS algorithm.

Since the saturated log-likelihood is a constant for a fixed distribution and set of fitting points, we can turn the minimization problem (\ref{eq:betaGAMreg}) into the maximization task of the penalized log-likelihood, i.e.
\begin{align}
  \bm{\widehat{\beta}}_{\text{GAM}} = \argmax_{\bm{\beta} \in \mathbb{R}^{(K-1)J+1}} \left\{ l\left(\bm{\beta}, \phi\right) - \frac{1}{2}\sum_{k=1}^{K-1} {\lambda_k \bm{\beta}_k^T \mathcal{S}_k \bm{\beta}_k}\right\}. \label{eq:betaGAMreg2}
\end{align}
\cite{wood2000} points out that Fisher's scoring method can be cast in a penalized version of the iteratively reweighted least squares (PIRLS) algorithm when being used to approximate the GAM coefficient estimator (\ref{eq:betaGAMreg2}). This derivation is very similar to the one of the IRLS algorithm in the GLM context with the constant dispersion $\phi$ disappearing in the first-order condition. We formulate the PIRLS algorithm based on \cite{marx1998} who indicate the iterative solution explicitly.

Let $\bm{\widehat{\beta}^{(t)}}$ now be the GAM coefficient approximation in iteration $t$. Then the vector of the dependent variable $\widehat{\textbf{\text{s}}}^{\bm{(t)}} = \left(\widehat{s}^1\left(\bm{\widehat{\beta}^{(t)}}\right),\ldots,\widehat{s}^N\left(\bm{\widehat{\beta}^{(t)}}\right)\right)^T$ and the weight matrix given by $W^{(t)}=\text{diag}\left(w^1\left(\bm{\widehat{\beta}^{(t)}}\right),\ldots,w^N\left(\bm{\widehat{\beta}^{(t)}}\right)\right)$ have the same form as in the IRLS algorithm, see (\ref{eq:s}) and (\ref{eq:w}). Additionally, let $S = \text{blockdiag} \left(0, \lambda_1\mathcal{S}_1, \ldots, \lambda_{K-1}\mathcal{S}_{K-1}\right)$ with $S_{11} = 0$ belonging to the intercept be the penalty matrix.

\begin{PIRLS} Perform the following iterative approximation procedure with e.g. an initialization of $\widehat{\mu}_{(0)}^i = y^i + 0.1$ and $\widehat{\eta}_{(0)}^i = g\left(\widehat{\mu}_{(0)}^i\right)$ in analogy to the IRLS algorithm until convergence:
{\footnotesize\begin{align}
  \bm{\widehat{\beta}^{(t+1)}} &= \argmin_{\bm{\beta} \in \mathbb{R}^{(K-1)J+1}} \left\{ \sum_{i=1}^{N} {w^i \left(\bm{\widehat{\beta}^{(t)}}\right)^{-1} \left(\widehat{s}^i\left( \bm{\widehat{\beta}^{(t)}}\right) - \beta_0 - \sum_{k=1}^{K-1} {\sum_{j=1}^{J} {\beta_{kj} b_{kj}\left(z_{ik}\right)}}\right)^2} + \sum_{k=1}^{K-1} {\lambda_k \bm{\beta}_k^T \mathcal{S}_k \bm{\beta}_k}\right\} \nonumber \\
  &= \left(Z^T W^{(t)} Z + S\right)^{-1} Z^T W^{(t)} \widehat{\textbf{\emph{\text{s}}}}^{\bm{(t)}}.
  \label{eq:PIRLS}
\end{align}}
After convergence, we have $\bm{\widehat{\beta}}_{\emph{\text{GAM}}} = \bm{\widehat{\beta}^{(t+1)}}$.\end{PIRLS}

\subsubsection*{Smoothing Parameter Selection, AIC \& GCV}
The smoothing parameters $\lambda_k$ can be selected such that they minimize a suitable model selection criterion, for the sake of consistency preferrably the one used in the adaptive algorithm for basis function selection. The GAM estimator (\ref{eq:betaGAMreg2}) does not exactly maximize the log-likelihood, therefore AIC has another form for GAMs than for GLMs. The degrees of freedom need to be adjusted with respect to the smoothing effects of the penalties on the coefficients. The reasoning behind this adjustment is that high smoothing parameters restrict the coefficients more than low smoothing parameters and need therefore be associated with less effective degrees of freedom.

\cite{hastie1990} propose a widely used version of AIC for GAMs, which uses effective degrees of freedom $\text{df}$ in place of the number of coefficients $(K-1)J+1$. This is
\begin{equation}
   \text{AIC} = -2 l\left(\bm{\widehat{\beta}}_{\text{GAM}}, \widehat{\phi}\right) + 2 \left(\text{df}+p\right),
  \label{eq:AICGAM}
\end{equation}
where 
\begin{equation}
   \text{df} = \text{tr}\left(\left(I+S\right)^{-1}I\right).
  \label{eq:dfGAM}
\end{equation}
The expression $I = Z^T W Z$ for the Fisher information matrix with the weight matrix $W$ evaluated at the GAM estimator is obtained as a by-product when casting Fisher's scoring method in the form of the PIRLS algorithm. Without the penalty matrix $S$, we have $\text{df} = \text{tr}\left(I^{-1}I\right) = (K-1)J+1$. If we follow \cite{wood2006} by denoting the unpenalized GAM estimator by $\bm{\widehat{\beta}}_{\text{GAM}}^0$ and the so-called shrinkage matrix by $F = \left( Z^T W Z + S \right)^{-1} Z^T W Z$ with $\bm{\widehat{\beta}}_{\text{GAM}} = F \bm{\widehat{\beta}}_{\text{GAM}}^0$, we arrive at the equality $\text{df} = \text{tr}\left(F\right)$ revealing the shrinkage effects on the effective degrees of freedom. After convergence of the PIRLS algorithm, the dependent variable is constant, i.e. $\widehat{\textbf{\emph{\text{s}}}} = \widehat{\textbf{\emph{\text{s}}}}^{\bm{(t)}}$, and the hat matrix $H$ satisfies $\left(\widehat{\eta}^1,\ldots,\widehat{\eta}^N\right)^T = Z \bm{\widehat{\beta}}_{\text{GAM}} = H \widehat{\textbf{\emph{\text{s}}}}$ so that $H = Z \left( Z^T W Z + S \right)^{-1} Z^T W$. Due to the cyclic property of the trace, the effective degrees of freedom can thus also be written as $\text{df} = \text{tr}\left(H\right)$. For GAMs, an estimate of the dispersion $\widehat{\phi}$ is obtained similarly to GLMs by (\ref{eq:phihat}). The parameter $p$ is defined as in (\ref{eq:AICGLM}). For a refinement of (\ref{eq:AICGAM}) accounting for the uncertainty of the smoothing parameters and tending to select models less prone to overfitting, see \cite{wood2016}.

Another popular and effective smoothing parameter selection criterion invented by \cite{craven1979} is generalized cross-validation (GCV), i.e.
\begin{equation}
   \text{GCV} = \frac{N D\left(\bm{\widehat{\beta}}_{\text{GAM}}\right)}{\left(N - \text{df}\right)^2},
  \label{eq:GCVGAM}
\end{equation}
with the model deviance $D\left(\bm{\widehat{\beta}}_{\text{GAM}}\right)$ evaluated at the GAM estimator and the effective degrees of freedom defined just like for AIC.

\subsubsection*{Adaptive Forward Stagewise Selection \& Performance}
In situations where the economic variable depends on many risk factors and where large sample sizes are required to derive reliable proxy functions, the adaptive forward stepwise algorithm depicted in Figure \ref{fig:algorithm} can become computationally infeasible with GAMs as opposed to e.g. GLMs. In iteration $k$, a GAM has $(K-1)J+1$ coefficients which need to be estimated while a GLM has only $K$ coefficients. This difference in the estimation effort is increased further due to the iterative nature of the IRLS and PIRLS algorithms. Moreover, GAMs involve the task of optimal smoothing parameter selection. Thereby entails each smoothing parameter constellation of which the goodness-of-fit is assessed in terms of AIC or GCV not only a full coefficient estimation stream but also a quite costly evaluation of the degrees of freedom so that the estimation effort for GAMs is scaled once more tremendously.

\cite{wood2000} has found a way to make smoothing parameter selection more efficient. Furthermore, \cite{wood2015} and \cite{wood2017} have developed practical GAM fitting methods for large data sets. These methods also involve e.g. iterative update schemes which require only subblocks of the design matrix to be recomputed and parallelization. The suitable application of these methods in the adaptive algorithm is beyond the scope of this analysis though since our focus does not lie on computational performance. Besides parallelizing the candidate loop on the lower left side of Figure \ref{fig:algorithm}, we achieve the necessary performance gains in GAMs by replacing the stepwise algorithm by a stagewise algorithm. This means that in each iteration, a predefined number $L$ or proportion of candidate terms is selected simultaneously until a termination criterion is fulfilled. Thereby we select in one stage those basis functions which reduce the model selection criterion of our choice most when added separately to the current proxy function structure. When there are not at least as many basis functions as targeted, the algorithm shall be terminated after the ones which lead to a reduction in the model selection criterion have been selected.

\subsection{Feasible Generalized Least-Squares (FGLS) Regression}\label{sec:FGLStheory}
\subsubsection*{Generalized Regression Model}
While in the classical linear regression model the errors are assumed to be uncorrelated and have the same unknown variance $\sigma^2 > 0$, in the generalized regression model, they are assummed to have the covariance matrix $\Sigma = \sigma^2 \Omega$ where $\Omega$ is positive definite and known and $\sigma^2 > 0$ is unknown. We transform the generalized regression model according to \cite{hayashi2000} to obtain a model (*) which satisfies Assumptions (A1), (A2) and (A3) of the classical linear regression model. As $\Omega$ is by construction symmetric and positive definite, there exists an invertible matrix $H$ such that $\Omega^{-1} = H^T H$. The matrix $H$ is not unique but this is not important since any choice of $H$ works. The generalized response vector $\textbf{\text{y}}^*$, design matrix $Z^*$ and error vector $\bm{\epsilon}^*$ are then given by
\begin{equation}
  \textbf{\text{y}}^* = H\textbf{\text{y}},\ \ \ Z^* = H Z,\ \ \ \bm{\epsilon}^* = \textbf{\text{y}}^* - Z^*\bm{\beta} = H \left(\textbf{\text{y}}-Z\bm{\beta}\right) = H \bm{\epsilon}.
  \label{eq:generalized}
\end{equation}
Strict exogeneity (A1) is satisfied by the transformed regression model (*) as $E\left[ \bm{\epsilon}^* \mid Z^* \right] = H E\left[ \bm{\epsilon} \mid Z \right] = \bm{0}$, the error variance is spherical (A2) because of $\Sigma^* = V \left[ \bm{\epsilon}^* \mid Z^* \right] = H V \left[ \bm{\epsilon} \mid Z \right] H^T = H \left[ \sigma^2 \Omega \right] H^T = H \left[ \sigma^2 \left(H^T H\right)^{-1} \right] H^T = \sigma^2 I_N $ with the $N$-dimensional identity matrix $I_N$ and the no-multicollinearity assumption (A3) holds as $\Omega$ is positive definite.

In analogy to the OLS estimator, the generalized least-squares (GLS) estimator $\bm{\widehat{\beta}}_{\text{GLS}}$ of the coefficients is given as the minimizer of the generalized residual sum of squares, i.e.
\begin{equation}
   \bm{\widehat{\beta}}_{\text{GLS}} = \argmin_{\bm{\beta} \in \mathbb{R}^K} \left\{ \sum_{i=1}^{N} {\left(\epsilon^{*,i}\right)^2} \right\}.
  \label{eq:betaGLSreg}
\end{equation}
The closed-form expression of the GLS estimator is
\begin{equation}
   \bm{\widehat{\beta}}_{\text{GLS}} = \left(Z^{*T} Z^*\right)^{-1} Z^{*T} \textbf{\text{y}}^* = \left(Z^T \Omega^{-1} Z\right)^{-1} Z^T \Omega^{-1}\textbf{\text{y}},
  \label{eq:betaGLSreg2}
\end{equation}
and the proxy function becomes
\begin{equation}
   \widehat{f}\left(X\right) = \textbf{z}^T\bm{\widehat{\beta}}_{\text{GLS}},
  \label{eq:fGLShat}
\end{equation}
where $\textbf{z} = \left(e_0 \left( X \right),\ldots,e_{K-1} \left( X \right)\right)^T$. The scalar $\sigma^2$ can be estimated in analogy to OLS regression by $s_{\text{GLS}} = \frac{1}{N-K} \bm{\widehat{\epsilon}}^{*T} \bm{\widehat{\epsilon}}^*$ where $\bm{\widehat{\epsilon}}^* = \textbf{\text{y}}^* - Z^* \bm{\widehat{\beta}}_{\text{GLS}}$ is the residual vector.

\subsubsection*{Gauss-Markov-Aitken Theorem \& ML Estimation}
We formulate the Gauss-Markov-Aitken theorem conditional on the fitting scenarios in line with \cite{huang1970} and \cite{hayashi2000} under the assumptions of strict exogeneity (A1), no multicollinearity (A3) and a covariance matrix $\Sigma = \sigma^2 \Omega$ of which $\Omega$ is positive definite and known (A6).
\begin{GMAt} The GLS estimator is the BLUE of the coefficients in the generalized regression model (\ref{eq:y}) under Assumptions (A1), (A3) and (A6).\end{GMAt}

In order to make AIC and GLS regression combinable, we assume additionally to (A1), (A3) and (A6) that the economic variable, or equivalently the errors, are jointly normally distributed conditional on the fitting scenarios (A7). The transformation (*) transfers to the ML function of the generalized regression model so that we can state the following theorem in analogy to Theorem \ref{th:1}, see e.g. \cite{hartmann2015}.
\begin{T} The ML coefficient estimator coincides with the GLS coefficient estimator and the ML estimator of the scalar $\widehat{\sigma}^2$ can be expressed as $\frac{N}{N-K}$ times $s_{\emph{\text{GLS}}}$, i.e. $\widehat{\sigma}^2 = \frac{1}{N} \bm{\widehat{\epsilon}}^{*T} \bm{\widehat{\epsilon}}^*$, under Assumptions (A1), (A3), (A6) and (A7).\label{th:2}\end{T}
Moreover, the GLS estimator is the efficient estimator under these assumptions according to \cite{greene2002}.

\subsubsection*{FGLS Estimation via ML Algorithm}
In the LSMC framework, $\Omega$ is unknown. If a consistent estimator $\widehat{\Omega}$ exists, we can apply feasible generalized least-squares (FGLS) regression, of which the estimator
\begin{equation}
   \bm{\widehat{\beta}}_{\text{FGLS}} = \left(Z^T \widehat{\Omega}^{-1} Z\right)^{-1} Z^T \widehat{\Omega}^{-1}\textbf{\text{y}}
  \label{eq:betaFGLSreg}
\end{equation}
has asymptotically the same properties as the GLS estimator (\ref{eq:betaGLSreg2}). \cite{greene2002} remarks furthermore that the asymptotic efficiency of the FGLS estimator does not carry over to finite samples. In small sample studies with no severe heteroscedasticity, the OLS estimator has been shown to be sometimes more efficient than the FGLS estimator. However, if heteroscedasticity is more severe, the FGLS estimator has been shown to outperform the OLS estimator. With $\textbf{z} = \left(e_0 \left( X \right),\ldots,e_{K-1} \left( X \right)\right)^T$ the FGLS proxy function is then given as
\begin{equation}
   \widehat{f}\left(X\right) = \textbf{z}^T\bm{\widehat{\beta}}_{\text{FGLS}}.
  \label{eq:fFGLShat}
\end{equation}

Without loss of generality, we set $\sigma^2 = 1$ so that $\Sigma = \Omega$ and refer to $\Sigma$ in the following. Hereby, any specification of $\sigma^2 > 0$ is possible as long as $\Omega$ is rescaled accordingly so that $\Sigma = \sigma^2 \Omega$ is satisfied since the GLS and FGLS coefficient estimators are invariant to a scaling of $\Omega$ and $\widehat{\Omega}$, respectively. Furthermore, we assume in addition to (A1), (A3) and (A7) that the elements of the covariance matrix $\Sigma$ are twice differentiable functions of parameters $\bm{\alpha} = \left(\alpha_0,\ldots,\alpha_{M-1}\right)^T$ with $K + M \leq N$ so that we can write $\Sigma = \Sigma \left(\bm{\alpha}\right)$ (A8). Theorem 1 of \cite{magnus1978} characterizes the ML estimators $\bm{\widehat{\beta}}_{\text{ML}}$ and $\bm{\widehat{\alpha}}_{\text{ML}}$ under these assumptions. We will relate the FGLS coefficient estimator to the ML coefficient estimator later in this section.
\begin{T} The generalized regression model (\ref{eq:y}) under Assumptions (A1), (A3), (A7) and (A8) has the following first-order ML conditions:
\begin{align}
  \bm{\widehat{\beta}}_{\emph{\text{ML}}} &= \left(Z^T \widehat{\Sigma}^{-1} Z\right)^{-1} Z^T \widehat{\Sigma}^{-1}\textbf{\emph{\text{y}}}, \label{eq:betafglsFOC} \\
  \frac{\partial l}{\partial \alpha_m} &= \frac{1}{2}\emph{\text{tr}}\left(\frac{\partial \Sigma^{-1}}{\partial \alpha_m}\Sigma\right)_{\bm{\alpha}=\bm{\widehat{\alpha}}_{\emph{\text{ML}}}} - \frac{1}{2}\bm{\widehat{\epsilon}}^T \left(\frac{\partial \Sigma^{-1}}{\partial \alpha_m}\right)_{\bm{\alpha}=\bm{\widehat{\alpha}}_{\emph{\text{ML}}}} \bm{\widehat{\epsilon}} = 0, \label{eq:OmegafglsFOC}
\end{align}
where $m=0,\ldots,M-1$, $\widehat{\Sigma} = \Sigma\left(\bm{\widehat{\alpha}}_{\emph{\text{ML}}}\right)$ and $\bm{\widehat{\epsilon}} = \textbf{\emph{\text{y}}}-Z\bm{\widehat{\beta}}_{\emph{\text{ML}}}$.
\label{th:3}\end{T}
Since the system in (\ref{eq:betafglsFOC}) and (\ref{eq:OmegafglsFOC}) typically does not have a closed-form solution, we suggest to solve it iteratively, e.g. according to \cite{magnus1978}. We start the procedure with $\bm{\beta^{(0)}}$ instead of with an arbitrary admissable vector $\bm{\alpha^{(0)}}$ though. The vector $\bm{\widehat{\alpha}^{(t+1)}}$ can be estimated iteratively conditional on $\bm{\widehat{\epsilon}^{(t+1)}}$, that is, $\bm{\beta^{(t)}}$ e.g. by using PORT optimization routines as described in \cite{gay1990} and implemented in function \textit{nlminb($\cdot$)} belonging to R package \textit{stats} of \cite{stats2018}. In this iterative routine, $\bm{\widehat{\alpha}^{(t+1)}}$ can be initialized e.g. by random numbers from the standard normal distribution.
\begin{ML} Perform the following iterative approximation procedure with e.g. an initialization of $\bm{\widehat{\beta}^{(0)}} = \bm{\widehat{\beta}}_{\emph{\text{OLS}}}$ until convergence:
\begin{enumerate}
  \item Calculate the residual vector $\bm{\widehat{\epsilon}^{(t+1)}} = \textbf{\emph{\text{y}}}-Z\bm{\widehat{\beta}^{(t)}}$.
  \item Substitute $\bm{\widehat{\epsilon}^{(t+1)}}$ into the $M$ equations in $M$ unknowns $\alpha_m$ given by (\ref{eq:OmegafglsFOC}) and solve them. If an explicit solution exists, set $\bm{\widehat{\alpha}^{(t+1)}} = \bm{\alpha}\left(\bm{\widehat{\epsilon}^{(t+1)}}\right)$. Otherwise, select the maximum likelihood solution $\bm{\widehat{\alpha}^{(t+1)}}$ iteratively, e.g. by using PORT optimization routines.
  \item Calculate \begin{align}\widehat{\Sigma}^{(t+1)} &= \Sigma \left(\bm{\widehat{\alpha}^{(t+1)}}\right), \nonumber \\
  \bm{\widehat{\beta}^{(t+1)}} &= \left(Z^T \left(\widehat{\Sigma}^{(t+1)}\right)^{-1} Z\right)^{-1} Z^T \left(\widehat{\Sigma}^{(t+1)}\right)^{-1}\textbf{\emph{\text{y}}}.
  \label{eq:betaFGLS}
  \end{align}
  Continue with the next iteration.
\end{enumerate}
After convergence, we have $\bm{\widehat{\beta}}_{\emph{\text{ML}}} = \bm{\widehat{\beta}^{(t+1)}}$ and $\bm{\widehat{\alpha}}_{\emph{\text{ML}}} = \bm{\widehat{\alpha}^{(t+1)}}$.
\label{th:ML}\end{ML}

Theorem 5 of \cite{magnus1978} states some regularity conditions guaranteeing the consistency of the ML estimators. Then $\widehat{\Sigma} = \Sigma\left(\bm{\widehat{\alpha}}_{\text{ML}}\right)$ is a consistent estimator so that the ML coefficient estimator (\ref{eq:betafglsFOC}) provides the FGLS coefficient estimator (\ref{eq:betaFGLSreg}).
\begin{T} The FGLS coefficient estimator can be derived as the ML coefficient estimator by the ML algorithm under Assumptions (A1), (A3), (A7) and (A8) and some further regularity conditions stated in Theorem 5 of \cite{magnus1978}.
\label{th:4}\end{T}

\subsubsection*{Heteroscedasticity \& Breusch-Pagan Test}
Besides Assumption (A8) about the structure of the covariance matrix, we assume that the errors are uncorrelated with further on different variances (= heteroscedastic errors), i.e. $\Sigma = \text{diag}\left(\sigma_1^2,\ldots,\sigma_N^2\right)$. We model each variance $\sigma_i^2,\ i=1,\ldots,N$, by a twice differentiable function in dependence of parameters $\bm{\alpha} = \left(\alpha_0,\ldots,\alpha_{M-1}\right)^T$ and a suitable set of linear independent basis functions $e_m \left( X \right) \in L^2\left( \mathbb{R}^D, \mathcal{B}, \mathbb{P}' \right),\ m=0,1,\ldots,M-1$, with $\textbf{\text{v}}^{i} = \left(e_0 \left(x^i\right),\ldots,e_{M-1} \left(x^i\right)\right)^T$, i.e.
\begin{equation}
  \sigma_i^2 = \sigma^2 V\left[\bm{\alpha}, \textbf{\text{v}}^{i}\right],
  \label{eq:varmod}
\end{equation}
where $V\left[\bm{\alpha},\textbf{\text{v}}^{i}\right]$ is referred to as the variance function in analogy to $V\left[\mu\right]$ for GLMs and GAMs. Without loss of generality, we set again $\sigma^2 = 1$.

\cite{hartmann2015} has already applied FGLS regression with different variance models in the LSMC framework. In her numerical examples, variance models with multiplicative heteroscedasticity led to the best performance of the proxy function in the validation. Therefore, we restrict our analyis on these kinds of structures, compare e.g. \cite{harvey1976}, i.e.
\begin{equation}
  V\left[\bm{\alpha}, \textbf{\text{v}}^{i}\right] = \exp\left(\textbf{\text{v}}^{iT}\bm{\alpha}\right).
  \label{eq:multvarmod}
\end{equation}

We should only apply FGLS regression as a substitute of OLS regression if heteroscedasticity prevails. If the variance function has the structure
\begin{equation}
  V\left[\bm{\alpha}, \textbf{\text{v}}^{i}\right] = h\left(\textbf{\text{v}}^{i,T}\bm{\alpha}\right),
  \label{eq:bpstruct}
\end{equation}
where the function $h(\cdot)$ is twice differentiable and the first element of $\textbf{\text{v}}^i$ is $\text{v}_0^i = 1$, the Breusch-Pagan test of \cite{breusch1979} can be used to diagnose heteroscedasticity under the assumption of normally distributed errors. We use it in the numerical computations to check if heteroscedasticity still prevails during the iterative procedure.

\subsubsection*{Variance Model Selection \& AIC}
Like the proxy function, the variance function (\ref{eq:multvarmod}) has to be calibrated to apply FGLS regression, which means that the variance function has to be composed of suitable basis functions. Again, such a composition can be found with the aid of a model selection criterion. We stick to AIC but have to take care of the fact that the covariance matrix has now $M$ unknown parameters instead of only one as in the OLS case (the same variance for all observations). Under Assumption (A7), AIC is given as
\begin{align}
  \text{AIC} &= -2 l\left(\bm{\widehat{\beta}}_{\text{FGLS}},\widehat{\Sigma}\right) + 2 \left(K+M\right) \label{eq:AICFGLS} \\
  &=N \log\left(2\pi\right) + \log\left(\det\widehat{\Sigma}\right) + \left(\textbf{\text{y}} - Z\bm{\widehat{\beta}}_{\text{FGLS}}\right)^T\widehat{\Sigma}^{-1}\left(\textbf{\text{y}} - Z\bm{\widehat{\beta}}_{\text{FGLS}}\right) + 2 \left(K+M\right). \nonumber
\end{align}
When using a variance model with multiplicative heteroscedasticity, AIC becomes
\begin{equation}
  \text{AIC} = N \log\left(2\pi\right) + \left(\sum_{i=1}^{N} {\textbf{\text{v}}^{iT}}\right)\bm{\widehat{\alpha}} + \sum_{i=1}^{N} {\exp\left(-\textbf{\text{v}}^{iT}\bm{\widehat{\alpha}}\right) \left(\widehat{\epsilon}^i\right)^2} + 2 \left(K+M\right).
  \label{eq:AICFGLShetero}
\end{equation}
As an alternative or complement, the basis functions of the variance model can be selected with respect to their correlations with the final OLS residuals or based on graphical residual analysis.

A difficulty of variance model selection poses its potential interdependency with proxy function selection because the basis functions minimizing the model selection criterion when being added to the proxy function might depend on the selected basis functions of the variance model and vice versa. There are multiple ways to tackle the interdependency difficulty, compare \cite{hartmann2015}, of which we implement two variants with rather short run times and promising out-of-sample validation performances. Our type I variant starts with the derivation of the proxy function by the standard adaptive OLS regression approach and then selects the variance model adaptively from the set of proxy basis functions of which the exponents sum up to at most two. The type II variant builds on the type I algorithm by taking the resulting variance model as given in its adaptive proxy basis function selection procedure with FGLS regression in each iteration.

\subsection{Multivariate Adaptive Regression Splines (MARS)}\label{sec:MARStheory}
\subsubsection*{OLS Regression/GLM with Hinge Functions}
\begin{figure}[thbp]
	\centering
		\includegraphics[width=\textwidth]{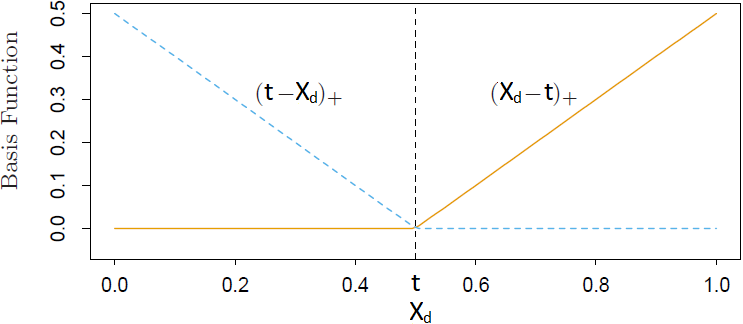}
	\caption{Reflected pair of piecewise linear functions with a knot at $t$}
	\label{fig:basis_function_pair_Hastie_et_al_2017}
\end{figure}

The multivariate adaptive regression splines (MARS) were introduced by \cite{friedman1991}. We describe the standard MARS algorithm in the LSMC routine by Ch. 9.4 of \cite{hastie2017}. The building blocks of MARS proxy functions are reflected pairs of piecewise linear functions with knots $t$ as depicted in Figure \ref{fig:basis_function_pair_Hastie_et_al_2017}, i.e.
\begin{align}
  \left(X_d - t\right)_{+} = \max\left(X_d - t, 0\right), \nonumber \\
  \left(t - X_d \right)_{+} = \max\left(t - X_d, 0\right),
  \label{eq:hinge}
\end{align}
where the $X_d,\ d=1,\ldots,D$, represent the risk factors which form together the outer scenario $X = \left(X_1,\ldots,X_D\right)^T$.

For each risk factor, reflected pairs with knots at each fitting scenario stress $x_d^i,\ i=1,\ldots,N$, are defined. All pairs are united in the following collection serving as the initial candidate term set of the MARS algorithm, i.e.
\begin{equation}
  C_1 = \left\{ \left(X_d - t\right)_{+}, \left(t - X_d \right)_{+} \right\}_{t \in \left\{x_d^1,x_d^2,\ldots,x_d^N\right\}\ \mid\  d=1,\ldots,D}.
  \label{eq:candset}
\end{equation}
We call the elements of such a collection hinge functions and write them as functions $h\left(X\right)$ over the entire input space $\mathbb{R}^D$. The initial set $C_1$ contains in total $2DN$ basis functions.

The classical MARS model is a form of the classical linear regression model (\ref{eq:y}), where the basis functions $e_k\left(x^i\right)$ are hinge functions. Therefore, the theory of OLS regression applies in this context. However, the theory about AIC cannot be transferred without any adjustments since the notion of the degrees of freedom has to be reconsidered due to the knots in the hinge functions acting as additional degrees of freedom. Since GLMs (\ref{eq:g}) are generalizations of the classical linear regression model, they can also be applied in conjunction with MARS models which we refer to in the following as generalized MARS models. In these cases, the theory of GLMs applies but again with the exception of the AIC part.

An especially fast MARS algorithm was developed by \cite{friedman1993} and is implemented e.g. in function \textit{earth($\cdot$)} of R package \textit{earth} provided by \cite{earth2018}.

\subsubsection*{Adaptive Forward Stepwise Selection \& Forward Pass}
The forward pass of the MARS algorithm can be viewed as a variation of the adaptive forward stepwise algorithm depicted in Figure \ref{fig:algorithm}. The start proxy function consists only of the intercept, i.e. $h_0\left(X\right) = 1$. In the classical MARS model, the regression method of choice is the standard OLS regression approach with the estimator (\ref{eq:betaOLSreg}), where in each iteration a reflected pair of hinge functions is selected instead of $e_k\left(x^i\right)$. Similarly, the regression method of choice in the generalized MARS model is the IRLS algorithm (\ref{eq:IRLS}). Let us denote the MARS coefficient estimator by $\bm{\widehat{\beta}}_{\text{MARS}}$. As the model selection criterion serves the residual sum of squares, or equivalently, the negative of R squared.

After each iteration, the set of candidate terms is extended by the products of the last two selected hinge functions with all hinge functions in $C_1$ depending on risk factors the two selected hinge functions do not depend on. Let the reflected pair selected in the first iteration ($k = 1$) be
\begin{align}
  h_1\left(X\right) = \left(X_{d_1} - t_1\right)_{+}, \nonumber \\
  h_2\left(X\right) = \left(t_1 - X_{d_1} \right)_{+}.
  \label{eq:hinge1}
\end{align}
Further, let $C_{1,-} = C_1 \setminus \left\{h_1\left(X\right),h_2\left(X\right)\right\}$. Then, the set of candidate terms is updated at the beginning of the second iteration ($k = 2$) such that
\begin{align}
  C_2 = C_{1,-} &\cup \left\{ \left(X_d - t\right)_{+}h_1\left(X\right), \left(t - X_d \right)_{+}h_1\left(X\right) \right\}_{t \in \left\{x_d^1,x_d^2,\ldots,x_d^N\right\}\ \mid \ d=1,\ldots,D,\ d\neq d_1} \nonumber \\
   &\cup \left\{ \left(X_d - t\right)_{+}h_2\left(X\right), \left(t - X_d \right)_{+}h_2\left(X\right) \right\}_{t \in \left\{x_d^1,x_d^2,\ldots,x_d^N\right\}\ \mid \ d=1,\ldots,D,\ d\neq d_1}.
  \label{eq:candset2}
\end{align}
The second set $C_2$ contains thus $2\left(DN-1\right) + 4\left(D-1\right)N$ basis functions. Often, the order of interaction is limited to improve the interpretability of the proxy functions. Besides the maximum allowed number of terms, a minimum threshold for the decrease in the residual sum of squares can be employed as a termination criterion in the forward pass. Typically, the proxy functions generated in the forward pass overfit the data since model complexity is only penalized conservatively by stipulating a maximum number of basis functions and a minimum threshold.

\subsubsection*{Backward Pass \& GCV}
Due to the overfitting tendency of the proxy function generated in the forward pass, a backward pass is executed afterwards. Apart from the direction and slight differences, the backward pass works like the forward pass. In each iteration, the hinge function of which the removal causes the smallest increase in the residual sum of squares is removed and the backward model selection criterion for the resulting proxy function evaluated. By this backward procedure, we generate the ``best'' proxy functions of each size in terms of the residual sum of squares. Out of all these best proxy functions, we finally select the one which minimizes the backward model selection criterion. As a result, the final proxy function will not only contain reflected pairs of hinge functions but also single hinge functions of which the complements have been removed. Optionally, the backward pass can be omitted or alternatives to the pure backward adaptive algorithm such as combinations with forward steps can be implemented.

Let the number of basis functions in the MARS model be $K$, the number of knots $T$ and the smoothing parameter $c$. The standard choice for the backward model selection criterion is GCV, compare its definition (\ref{eq:GCVGAM}) for GAMs, i.e.
\begin{equation}
   \text{GCV} = \frac{N D\left(\bm{\widehat{\beta}}_{\text{MARS}}\right)}{\left(N - \text{df}\right)^2},
  \label{eq:GCVMARS}
\end{equation}
with the effective degrees of freedom $\text{df} = K + cT$. For cases in which no interaction terms are allowed, \cite{friedman1989} give a mathematical argument for using $c = 2$. For the other cases, \cite{friedman1991} concludes from a wide variety of simulation studies that a parameter of $c = 3$ is fairly effective. Across all these studies, $2 \leq c \leq 4$ was found to give the best value of $c$. Alternatively, but with significantly higher computational costs, $c$ could be estimated by resampling techniques such as bootstrapping by \cite{efron1983} or cross-validation by \cite{stone1974}. Since comparably few basis functions are selected in the forward passes of our numerical MARS experiments, we set $c = 0$.

\subsection{Kernel Regression}\label{sec:KRtheory}
\subsubsection*{One-Dimensional LC \& LL Regression}
Independently from each other, \cite{nadaraya1964} and \cite{watson1964} both proposed to estimate a conditional expectation of a variable relative to another variable by a non-parametric regression approach using a kernel as a weighting function. In the following, we describe at first local constant (LC) regression and local linear regression (LL) in one dimension by Ch. 6 of \cite{hastie2017}. In the next sections, we refer to Ch. 2 of \cite{li2007} for LC and LL regression in more dimensions and suitable model selection criteria.

\begin{figure}[thbp]
	\centering
		\includegraphics[width=\textwidth]{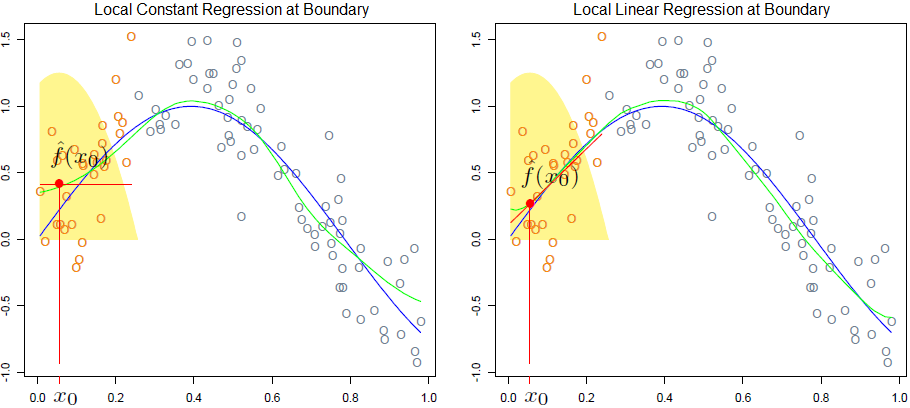}
	\caption{LC and LL kernel regression using the Epanechnikov kernel with $\lambda=0.2$ in one dimension}
	\label{fig:lc_ll_regression_Hastie_et_al_2017}
\end{figure}

We start with LC and LL regression in one dimension to carve out the idea of kernel regression, which generalizes very naturally to more dimensions. For now, let the target scenario be $x_0 \in \mathbb{R}$ and let the univariate kernel with given bandwidth $\lambda > 0$ be 
\begin{equation}
   K_{\lambda}\left(x_0,x^i\right) = D \left( \frac{\left|x^i - x_0\right|}{\lambda} \right),
  \label{eq:kernel}
\end{equation}
where $D \left(\cdot\right)$ denotes the specified kernel function. While e.g. the Epanechnikov (see the yellow shaded areas of Figure \ref{fig:lc_ll_regression_Hastie_et_al_2017} from \cite{hastie2017}), tri-cube and uniform kernels are commonly used kernel functions with bounded support, the gaussian kernel is one with infinite support. Moreover, the kernels can be defined with different orders, often the second order kernels are used, see e.g. \cite{li2007}. The LC kernel estimator or Nadaraya-Watson kernel smoother is given as the kernel-weighted average at each $x_0$, i.e.
\begin{equation}
   \widehat{f}_{\text{LC}}\left(x_0\right) = \widehat{\beta}_{\text{LC}}\left(x_0\right) = \frac{\sum_{i=1}^{N} {K_{\lambda}\left(x_0,x^i\right) y^i}}{\sum_{i=1}^{N} {K_{\lambda}\left(x_0,x^i\right)}}.
  \label{eq:betaHatLC}
\end{equation}
It is a continuous function since the weights die off smoothly with increasing distance from $x_0$. As this locally constant function varies over the domain of the target scenarios $x_0$, it needs to be estimated separately at all of them.
Due to the asymmetry of the kernels at the boundaries of the domain, the LC kernel estimator (\ref{eq:betaHatLC}) can be severly biased in that region, see the left panel of Figure \ref{fig:lc_ll_regression_Hastie_et_al_2017}.

We can overcome this problem by fitting locally linear functions instead of locally constant functions, see the right panel of Figure \ref{fig:lc_ll_regression_Hastie_et_al_2017}. At each target $x_0$, the LL kernel estimator is defined as the minimizer of the kernel-weighted residual sum of squares, i.e.
\begin{equation}
  \bm{\widehat{\beta}}_{\text{LL}}\left(x_0\right) = \argmin_{\bm{\beta}\left(x_0\right) \in \mathbb{R}^2} \left\{ \sum_{i=1}^{N} {K_{\lambda}\left(x_0,x^i\right)\left( y^i - \beta_0\left(x_0\right) - \beta_1\left(x_0\right) x^i \right)^2} \right\},
  \label{eq:betaHatLL}
\end{equation}
with $\bm{\beta}\left(x_0\right) = \left(\beta_0\left(x_0\right),\beta_1\left(x_0\right)\right)^T$. If we omit the linear term in (\ref{eq:betaHatLL}) by setting $\beta_1 = 0$, the intercept $\widehat{\beta}_{\text{LL},0}\left(x_0\right)$ of the LL kernel estimator becomes the LC kernel estimator (\ref{eq:betaHatLC}). The proxy function at $x_0$ is given by
\begin{equation}
  \widehat{f}_{\text{LL}}\left(x_0\right) = \widehat{\beta}_{\text{LL},0}\left(x_0\right) + \widehat{\beta}_{\text{LL},1}\left(x_0\right) x_0.
  \label{eq:fHatLL2}
\end{equation}

In analogy to LC regression, the minimization problem (\ref{eq:betaHatLL}) must be solved separately for all target scenarios so that the coefficients of the proxy function vary across their domain. Each proxy function is only evaluated at the target scenario it has been derived for. Since (\ref{eq:betaHatLL}) is a weighted least-squares (WLS) problem with weights $K_{\lambda}\left(x_0,x^i\right)$, its solution is the WLS estimator
\begin{equation}
  \bm{\widehat{\beta}}_{\text{LL}}\left(x_0\right) = \left(Z^T W\left(x_0\right)Z\right)^{-1} Z^T W\left(x_0\right)\textbf{\text{y}},
  \label{eq:betaHatLL2}
\end{equation}
where $\textbf{\text{y}}$ is the response vector, $W\left(x_0\right) = \text{diag}\left(K_{\lambda}\left(x_0,x^1\right),\ldots,K_{\lambda}\left(x_0,x^N\right)\right)$ the weight matrix and $Z$ the design matrix which contains row-wise the vectors $\left(1,x^i\right)^T$. We call $H$ the hat matrix if $\bm{\widehat{\textbf{\text{y}}}} = H \textbf{\text{y}}$ such that $\bm{\widehat{\textbf{\text{y}}}} = \left(\widehat{f}_{\text{LL}}\left(x^1\right),\ldots,\widehat{f}_{\text{LL}}\left(x^N\right)\right)^T$ contains the proxy function values at their target scenarios.

When we use proxy functions in LL regression that are composed of polynomial basis functions with exponents greater than one, we could also speak of local polynomial regression. 

\subsubsection*{Multidimensional LC \& LL Regression}
We generalize LC regression to $\mathbb{R}^K$ by expressing the kernel with respect to the basis function vector $\textbf{\text{z}} = \left(e_0\left(X\right),\ldots,e_{K-1}\left(X\right)\right)^T$ following from the adaptive forward stepwise selection with OLS regression and small $K_{\text{max}}$. At each target scenario vector $\textbf{\text{z}}_0 \in \mathbb{R}^K$ with elements $z_{0k}$, basis function vector $\textbf{\text{z}}^i \in \mathbb{R}^K$ with elements $z_{ik}$ evaluated at fitting scenario $x^i$ and given bandwidth vector $\bm{\lambda} = \left(\lambda_0,\ldots,\lambda_{K-1}\right)^T$, the multivariate kernel is defined as the product of univariate kernels, i.e.
\begin{equation}
  K_{\bm{\lambda}}\left(\textbf{\text{z}}_0,\textbf{\text{z}}^i\right) = \prod_{k=0}^{K-1} {D \left( \frac{\left|z_{ik} - z_{0k}\right|}{\lambda_k} \right)}.
\label{eq:kernelmulti}
\end{equation}
The LC kernel estimator or Nadaraya-Watson kernel smoother in $\mathbb{R}^K$ is defined at each $\textbf{\text{z}}_0$ as
\begin{equation}
   \widehat{f}_{\text{LC}}\left(\textbf{\text{z}}_0\right) = \widehat{\beta}_{\text{LC}}\left(\textbf{\text{z}}_0\right) = \frac{\sum_{i=1}^{N} {K_{\bm{\lambda}}\left(\textbf{\text{z}}_0,\textbf{\text{z}}^i\right) y^i}}{\sum_{i=1}^{N} {K_{\bm{\lambda}}\left(\textbf{\text{z}}_0,\textbf{\text{z}}^i\right)}}.
  \label{eq:betaHatLCmulti}
\end{equation}
Since we let $e_0\left(X\right)$ represent the intercept so that $z_{i0} = z_{00} = 1$, the corresponding univariate kernel $D \left( \frac{\left|z_{i0} - z_{00}\right|}{\lambda_0}\right) = D \left(0\right)$ is constant over all fitting points, cancels thus out in (\ref{eq:betaHatLCmulti}) and can be omitted in (\ref{eq:kernelmulti}).

The LL kernel estimator in $\mathbb{R}^K$ is given as the multidimensional analogue of (\ref{eq:betaHatLL}) at each $\textbf{\text{z}}_0$, i.e.
\begin{equation}
  \bm{\widehat{\beta}}_{\text{LL}}\left(\textbf{\text{z}}_0\right) = \argmin_{\bm{\beta}\left(\textbf{\text{z}}_0\right) \in \mathbb{R}^K} \left\{ \sum_{i=1}^{N} {K_{\bm{\lambda}}\left(\textbf{\text{z}}_0,\textbf{\text{z}}^i\right)\left( y^i - \textbf{\text{z}}^{i,T} \bm{\beta}\left(\textbf{\text{z}}_0\right) \right)^2} \right\},
  \label{eq:betaHatLLmulti}
\end{equation}
with $\bm{\beta}\left(\textbf{\text{z}}_0\right) = \left(\beta_0\left(\textbf{\text{z}}_0\right),\ldots,\beta_{K-1}\left(\textbf{\text{z}}_0\right)\right)^T$ and the proxy function at $\textbf{\text{z}}_0$ is given by
\begin{equation}
  \widehat{f}_{\text{LL}}\left(\textbf{\text{z}}_0\right) = \textbf{\text{z}}_0^T \bm{\widehat{\beta}}_{\text{LL}}\left(\textbf{\text{z}}_0\right).
  \label{eq:fHatLLmulti}
\end{equation}
The LL kernel estimator can again be computed by WLS regression, i.e.
\begin{equation}
  \bm{\widehat{\beta}}_{\text{LL}}\left(\textbf{\text{z}}_0\right) = \left(Z^T W\left(\textbf{\text{z}}_0\right)Z\right)^{-1} Z^T W\left(\textbf{\text{z}}_0\right)\textbf{\text{y}},
  \label{eq:betaHatLL2multi}
\end{equation}
where $W\left(\textbf{\text{z}}_0\right) = \text{diag}\left(K_{\bm{\lambda}}\left(\textbf{\text{z}}_0,\textbf{\text{z}}^1\right),\ldots,K_{\bm{\lambda}}\left(\textbf{\text{z}}_0,\textbf{\text{z}}^N\right)\right)$ is the weight matrix and $Z$ the design matrix containing row-wise the vectors $\textbf{\text{z}}^{i,T}$. The hat matrix $H$ satisfies $\bm{\widehat{\textbf{\text{y}}}} = H \textbf{\text{y}}$ with $\bm{\widehat{\textbf{\text{y}}}} = \left(\widehat{f}_{\text{LL}}\left(\textbf{\text{z}}^1\right),\ldots,\widehat{f}_{\text{LL}}\left(\textbf{\text{z}}^N\right)\right)^T$ containing the proxy function values at their target scenario vectors.

\subsubsection*{Bandwidth Selection, AIC \& LOO-CV}
The bandwidths $\lambda_k$ in kernel regression can be selected similarly to the smoothing parameters in GAMs by minimization of a suitable model selection criterion. In fact, kernel smoothers can be interpreted as local non-parametric GLMs with identity link functions. More precisely, at each target scenario the kernel smoother can be viewed as a GLM (\ref{eq:g}) where the parametric weights $V\left[\widehat{\mu}_{\text{GLM}}^i\right]$ in (\ref{eq:IRLSprop}) are the non-parametric kernel weights $K_{\bm{\lambda}}\left(\textbf{\text{z}}_0,\textbf{\text{z}}^i\right)$ in (\ref{eq:betaHatLLmulti}). Since GLMs are special cases of GAMs and the bandwidths in kernel regression can be understood as smoothing parameters, kernel smoothers and GAMs are sometimes lumped together in one category. If the numbers $N$ of the fitting points and $K$ of the basis functions are large, from a computational perspective it might be beneficial to perform bandwidth selection based on a reduced set of fitting points.

\cite{hurvich1998} propose to select the bandwidths $\lambda_1,\ldots,\lambda_{K-1}$ based on an improved version of AIC which works in the context of non-parametric proxy functions that can be written as linear combinations of the observations. It has the form
\begin{equation}
  \text{AIC} = \log \left(\widehat{\sigma}^2\right) + \frac{1+\text{tr}\left(H\right)/N}{1-\left(\text{tr}\left(H\right) + 2\right)/N},
\label{eq:AICkr}
\end{equation}
where $\widehat{\sigma}^2 = \frac{1}{N}\left(\textbf{\text{y}} - \bm{\widehat{\textbf{\text{y}}}}\right)^T \left(\textbf{\text{y}} - \bm{\widehat{\textbf{\text{y}}}}\right)$ and $H$ is the hat matrix.

As an alternative, non-parametric leave-one-out cross-validation (LOO-CV) is suggested by \cite{li2004} for bandwidth selection. Let us refer to
\begin{equation}
  \bm{\widehat{\beta}}_{\text{LL},-j}\left(\textbf{\text{z}}_0\right) = \argmin_{\bm{\beta}\left(\textbf{\text{z}}_0\right) \in \mathbb{R}^K} \left\{ \sum_{i\neq j, i=1}^{N} {K_{\bm{\lambda}}\left(\textbf{\text{z}}_0,\textbf{\text{z}}^i\right)\left( y^i - \textbf{\text{z}}^{i,T} \bm{\beta}\left(\textbf{\text{z}}_0\right) \right)^2} \right\}
  \label{eq:LOOCVkr}
\end{equation}
as the leave-one-out LL kernel estimator and to $\widehat{f}_{\text{LL},-j}\left(\textbf{\text{z}}_0\right) = \textbf{\text{z}}_0^T \bm{\widehat{\beta}}_{\text{LL},-j}\left(\textbf{\text{z}}_0\right)$ as the leave-one-out proxy function at $\textbf{\text{z}}_0$. The objective of LOO-CV is to choose the bandwidths $\lambda_1,\ldots,\lambda_{K-1}$ which minimize
\begin{equation}
  \text{CV} = \frac{1}{N} \sum_{i=1}^{N} {\left(y^i - \widehat{f}_{\text{LL},-i}\left(\textbf{\text{z}}_0\right)\right)^2}.
  \label{eq:CVkr}
\end{equation}

\subsubsection*{Adaptive Forward Stepwise OLS Selection}
A practical implementation of kernel regression can be found e.g. in the combination of functions \textit{npreg($\cdot$)} and \textit{npregbw($\cdot$)} from R package \textit{np} of \cite{np2018}.

In the other sections, basis function selection depends on the respective regression methods. Since the crucial process of bandwidth selection in kernel regression takes a very long time in the implementation of our choice, it would be infeasible to proceed here in the same way. Therefore, we derive the basis functions for LC and LL regression by adaptive forward stepwise selection based on OLS regression, by risk factor wise linear selection or a combination thereof. Thereby, we keep the maximum allowed number of terms $K_{\text{max}}$ rather small as we aim to model the subtleties by kernel regression.

\section{Numerical Experiments}
\subsection{General Remarks}\label{sec:general_remarks_experiment}
\subsubsection*{Data Basis}
In our slightly disguised real-world example, the life insurance company has a portfolio with a large proportion of traditional annuity business. In order to challenge the regression techniques, the traditional annuity business features by construction very high interest rate guarantees so that the insurer suffers huge losses in low interest rate environments. We let the insurance company be exposed to $D = 15$ relevant financial and actuarial risk factors. For the derivation of the fitting points, we run its CFP model conditional on $N = 25,000$ fitting scenarios with each of these outer scenarios entailing two antithetic inner simulations. The Sobol validation set is generated based on $L = 51$ validation scenarios with $1,000$ inner simulations, where the 51 scenarios comprise 26 Sobol scenarios, 16 one-dimensional risk scenarios and 9 scenarios that turned out to be capital region scenarios in the previous year risk capital calculations. The nested simulations set which is due to its high computational costs not available in the regular LSMC approach reflects the highest $5\%$ real-world losses and is based on $L = 1,638$ outer scenarios with respectively $4,000$ inner simulations. From the $1,638$ real-world scenarios, $14$ exhibit extreme stresses far beyond the bounds of the fitting space and are therefore excluded from the analysis. The capital region set consists of the $L = 129$ nested simulations points which correspond to the nested simulations SCR estimate ($= 99.5 \%$ highest loss) and the $64$ losses above and below ($= 99.3 \%$ to $99.7 \%$ highest losses).

\subsubsection*{Validation Figures}
We will output validation figure (\ref{eq:mae}) with respect to the relative and asset metric, and figures (\ref{eq:res}), (\ref{eq:mae0}) and (\ref{eq:res0}). While figures (\ref{eq:mae0}) and (\ref{eq:res0}) are evaluated with respect to a base value resulting from $1,000$ inner simulations on the Sobol set, i.e. $\text{v.mae}^0$, $\text{v.res}^0$, they are computed with respect to a base value resulting from $16,000$ inner simulations on the nested simulations set, i.e. $\text{ns.mae}^0$, $\text{ns.res}^0$, and capital region set, i.e. $\text{cr.mae}^0$, $\text{cr.res}^0$. The latter base value is supposed to be the more reliable validation value since it is the one associated with a lower standard error. Therefore it is worth noting here that figure $\text{v.res}^0$ can easily be transformed such that it is also evaluated with respect to the latter base value by subtracting from it the difference of $14$ which the two different base values incur. We will not explicitly state the base residual (\ref{eq:resbase}) as it is just (\ref{eq:res}) minus (\ref{eq:res0}).

\subsubsection*{Economic Variables}
We derive the OLS proxy functions for two economic variables, namely for the best estimate of liabilities (BEL) and the available capital (AC) over a one-year risk horizon, i.e. $Y(X) \in \left\{\text{BEL}(X),\ \text{AC}(X)\right\}$. Their approximation quality is assessed by validation figures (\ref{eq:mae}) with respect to the relative and asset metric and (\ref{eq:res}). Essentially, AC is obtained as the market value of assets minus BEL, which means that AC reflects the negative behavior of BEL. Therefore, we will only derive BEL proxy functions with the other regression methods. The profit resulting from a certain risk constellation captured by an outer scenario $X$ can be computed as $\text{AC}(X)$ minus the base AC. Validation figures (\ref{eq:mae0}) and (\ref{eq:res0}) address the approximation quality of this difference. Taking the negative of the profit yields the loss and evaluating the loss at all real-world scenarios the real-world loss distribution from which the SCR is derived as the $99.5 \%$ value-at-risk. The out-of-sample performances of two different OLS proxy functions of BEL on the Sobol, nested simulations and capital region sets serve as the benchmark for the other regression methods.

\subsubsection*{Numerical Stability}
Let us discuss the subject of numerical stability of QR decompositions in the OLS regression design under a monomial basis. If the weighting in the weighted least-squares problems associated with GLMs, heteroscedastic FGLS regression and kernel regression is good-natured, similar arguments apply as they can also be solved via QR decompositions according to \cite{green1984} where the weighting is just a scaling. However, the weighting itself raises additional numerical questions that need to be taken into consideration when making the regression design choices. In GLMs, these choices are the random component (\ref{eq:pi}) and link function (\ref{eq:g}), in FGLS regression it is the functional form of the heteroscedatic variance model (\ref{eq:varmod}) and in kernel regression it is the kernel function (\ref{eq:kernelmulti}). The following arguments do not apply to GAMs and MARS models as these are constructed out of spline functions, see (\ref{eq:basis}) and (\ref{eq:hinge}), respectively. In GAMs, the penalty matrix increases numerical stability.

\cite{mclean2014} justifies that from the perspective of numerical stability performing a QR decomposition on a monomial design matrix $Z$ is asymptotically equivalent to using a Legendre design matrix $Z'$ and transforming the resulting coefficient estimator into the monomial one. 
Under the assumption of an orthonormal basis, \cite{weiss2018} have derived an explicit upper bound for the condition number of non-diagonal matrix $\frac{1}{N}(Z')^T (Z')$ for $N < \infty$, where the factor $\frac{1}{N}$ is used for technical reasons. This upper bound increases in (1) the number of basis functions, (2) the Hardy-Krause variation of the basis, (3) the convergence constant of the low-discrepancy sequence, and (4) the outer scenario dimension. Our previously defined type of restriction setting controls aspect (1) through the specification of $K_{\text{max}}$ and aspect (2) through the limitation of exponents $d_1 d_2 d_3$. Aspects (3) and (4) are beyond the scope of the calibration and validation steps of the LSMC framework and therefore left aside here.

\subsubsection*{Interpolation \& Extrapolation}
In the LSMC framework, let us refer by interpolation to prediction inside the fitting space and by extrapolation to prediction outside the fitting space. \cite{runge1901} found that high-degree polynomial interpolation at equidistant points can oscillate toward the ends of the interval with the approximation error getting worse the higher the degree is. In a least-squares problem, Runge's phenomenon was shown by \cite{dahlquist1974} not to apply to polynomials of degree $d$ fitted based on $N$ equidistant points if the inequality $d < 2 \sqrt{N}$ holds. With $N = 25,000$ fitting points the inequality becomes $d < 316$ so that we clearly do not have to impose any further restrictions in OLS, FGLS and kernel regression as well as in GLMs to keep this phenomenon under control. Splines as they occur in GAMs and MARS models do not suffer from this oscillation issue by construction.

Since Runge's phenomenon concerns the ends of the interval and the real-world scenarios for the insurer's full loss distribution forecast in the forth step of the LSMC framework partly go beyond the fitting space, its scope comprises the extrapolation area as well. High-degree polynomial extrapolation can worsen the approximation error and play a crucial role if many real-world scenarios go far beyond the fitting space.

\subsubsection*{Principle of Parsimony}
Another problem that can occur in an adaptive algorithm is overfitting. \cite{burnham2002} state that overfitted models often have needlessly large sampling variances which means that their precision of the predictions is poorer than that of more parsimonious models which are also free of bias. In cases where AIC leads to overfitting, implementing restriction settings of the form $K_{\text{max}}$ - $d_1 d_2 d_3$ becomes relevant for adhering to the principle of parsimony.

\subsection{Ordinary Least-Squares (OLS) Regression}\label{sec:OLSexperiment}
\subsubsection*{Settings}
We build the OLS proxy functions (\ref{eq:fhat}) of $Y(X) \in \left\{\text{BEL}(X),\ \text{AC}(X)\right\}$ with respect to an outer scenario $X$ out of monomial basis functions that can be written as $e_k\left(X\right) = \prod_{l=1}^{15} {X_l^{r_k^l}}$ with $r_k^l \in \mathbb{N}_0$ so that each basis function can be represented by a $15$-tuple $\left(r_k^1,\ldots,r_k^{15}\right)$. The final proxy function depends on the restrictions applied in the adaptive algorithm. The purpose of setting restrictions is to guarantee numerical stability, to keep the extrapolation behavior under control and the proxy functions parsimonious. In order to illustrate the impact of restrictions, we run the adaptive algorithm for BEL under two different restriction settings with the second one being so relaxed that it will not take effect in our example. Additionally, we run the adaptive algorithm under the first restriction setting for AC to give an example of how the behavior of BEL can transfer to AC. As the first ingredient of our restriction setting acts the maximum allowed number of terms $K_{\text{max}}$. Furthermore, we limit the exponents in the monomial basis. Firstly we apply a uniform threshold to all exponents, i.e. $r_k^l \leq d_1$. Secondly we restrict the degree, i.e. $\sum_{l=1}^{15} {r_k^l} \leq d_2$. Thirdly we restrict the exponents in interaction basis functions, i.e. if there are some $l_1 \neq l_2$ with $r_k^{l_1},\ r_k^{l_2} > 0$, we require $r_k^{l_1},\ r_k^{l_2} \leq d_3$. Let us denote this type of restriction setting by $K_{\text{max}}$ - $d_1 d_2 d_3$.

As the first and second restriction settings, we choose $150$-$443$ and $300$-$886$, respectively, motivated by \cite{teuguia2014} who found in their LSMC example in Ch. 4 with four risk factors and $50,000$ fitting scenarios entailing two inner simulations that the validation error computed based on $14$ validation scenarios started to stabilize at degree $4$ when using monomial or Legendre basis functions in different adaptive basis function selection procedures. Furthermore, they pointed out that the LSMC approach becomes infeasible for degrees higher than $12$.

We apply R function \textit{lm($\cdot$)} implemented in R package \textit{stats} of \cite{stats2018}.

\subsubsection*{Results}
Table \ref{tab:BGROOLS1} contains the final BEL proxy function derived under the first restriction setting $150$-$443$ with the basis function representations and coefficients. Thereby reflect the rows the iterations of the adaptive algorithm and depict thus the sequence in which the basis functions are selected. Moreover, the iteration-wise AIC scores and out-of-sample MAEs (\ref{eq:mae}) with respect to the relative metric in $\%$ on the Sobol, nested simulations and capital region sets are reported, i.e. $\text{v.mae}$, $\text{ns.mae}$ and $\text{cr.mae}$. Table \ref{tab:PGROOLS1} contains the AC counterpart of the BEL proxy function derived under $150$-$443$ and Table \ref{tab:BGROOLS3008861} the final BEL proxy function derived under the more relaxed restriction setting $300$-$886$. Tables \ref{tab:BGROOLSvalfig} and \ref{tab:PGROOLSvalfig} indicate respectively for the BEL and AC proxy functions derived under $150$-$443$ the AIC scores and all five previously defined validation figures evaluated on the Sobol, nested simulations and capital region sets after each tenth iteration. Similarly, Table \ref{tab:BGROOLS300886valfig} reports these figures for the BEL proxy function derived under $300$-$886$. Here the last row corresponds to the final iteration.

Lastly, we manipulate the validation values on all three validation sets twice insofar as we subtract respectively add pointwise $1.96$ times the standard errors from respectively to them (inspired by $95\%$ confidence interval of gaussian distribution). We then evaluate the validation figures for the final BEL proxy functions under both restriction settings on these manipulated sets of validation value estimates and depict them in Table \ref{tab:BGROOLSstderrvalfig} in order to assess the impact of the Monte Carlo error associated with the validation values.

\subsubsection*{Improvement by Relaxation}
Tables \ref{tab:BGROOLS1} and \ref{tab:PGROOLS1} state that the adaptive algorithm terminates under $150$-$443$ for both BEL and AC when the maximum allowed number of terms is reached. This gives reason to relax the restriction setting to e.g. $300$-$886$ which eventually lets the algorithm terminate due to no further reduction in the AIC score without hitting restrictions $886$, compare Table \ref{tab:BGROOLS3008861} for BEL. In fact, only restrictions $224$-$464$ are hit. Except for the already very small figures $\text{cr.mae}$, $\text{cr.mae}^a$ and $\text{cr.res}$ all validation figures are further improved by the additional basis functions, see Tables \ref{tab:BGROOLSvalfig} and \ref{tab:BGROOLS300886valfig}. The largest improvement takes place between iterations $180$ and $190$. The result that at maximum degrees $464$ are selected is consistent with the result of \cite{teuguia2014} who conclude in their numerical examples of Ch. 4 that under a monomial, Legendre or Laguerre basis the optimum degree is probably $4$ or $5$. Furthermore, \cite{bauer2015} derive a similar result in their one risk factor LSMC example of Ch. 6 when using $50,000$ fitting scenarios and Legendre, Hermite, Chebychev basis functions or eigenfunctions.

According to our Monte Carlo error impact assessment in Table \ref{tab:BGROOLSstderrvalfig}, the slight deterioration at the end of the algorithm is not sufficient to indicate a slight overfitting tendency of AIC. Under the standard choices of the five major components, compare Section \ref{sec:calibration_algorithm}, the adaptive algorithm manages thus to provide a numerically stable and parsimonious proxy function even without a restriction setting. Here, allowing a priori unlimited degrees of freedom is thus beneficial to capturing the complex interactions in the CFP model.

\subsubsection*{Reduction of Bias}
\begin{figure}[t]
	\centering
		\includegraphics[width=\textwidth]{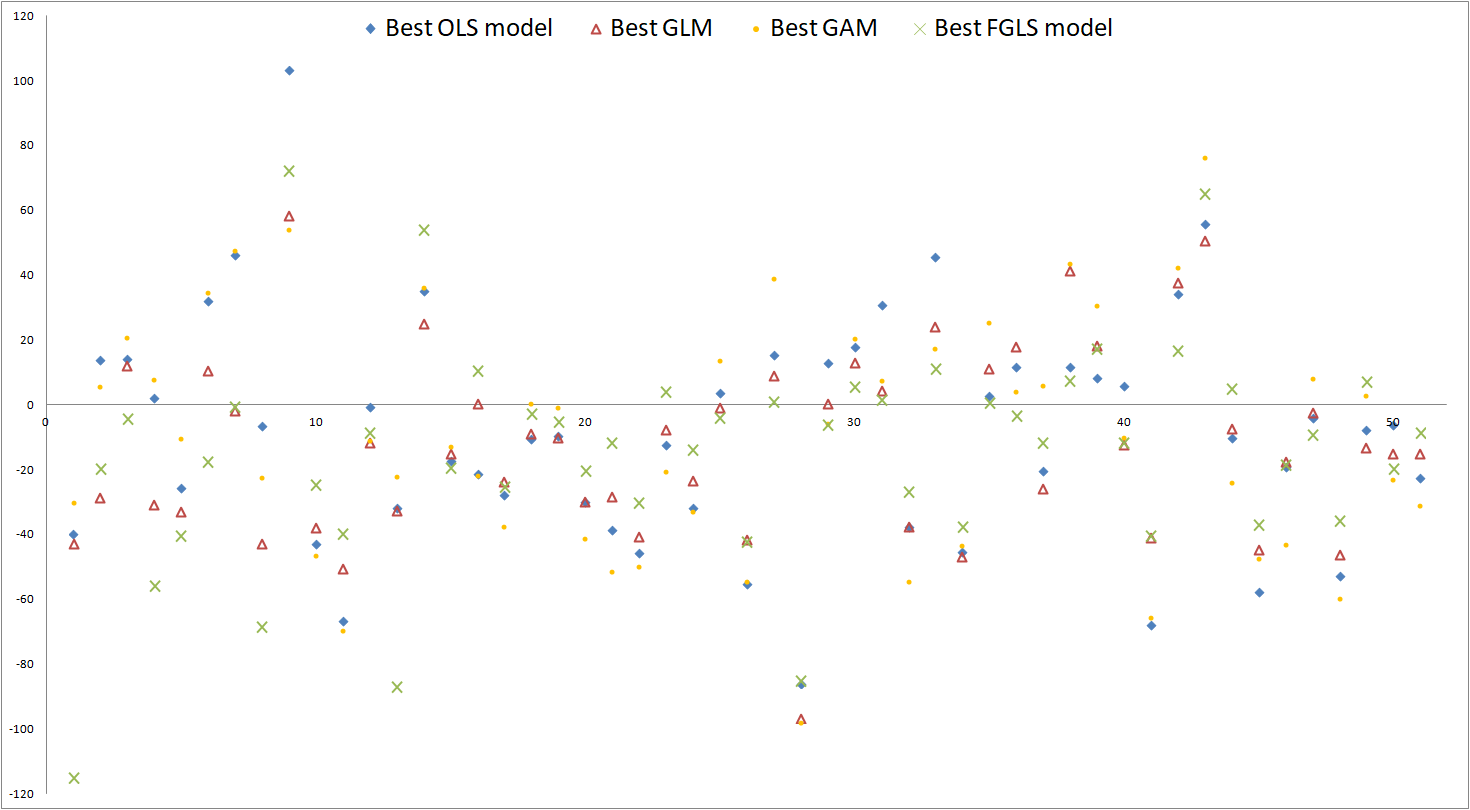}
	\caption{Residual Plots on Sobol Set}
	\label{fig:ValidationResiduals}
\end{figure}

\begin{figure}[t]
	\centering
		\includegraphics[width=\textwidth]{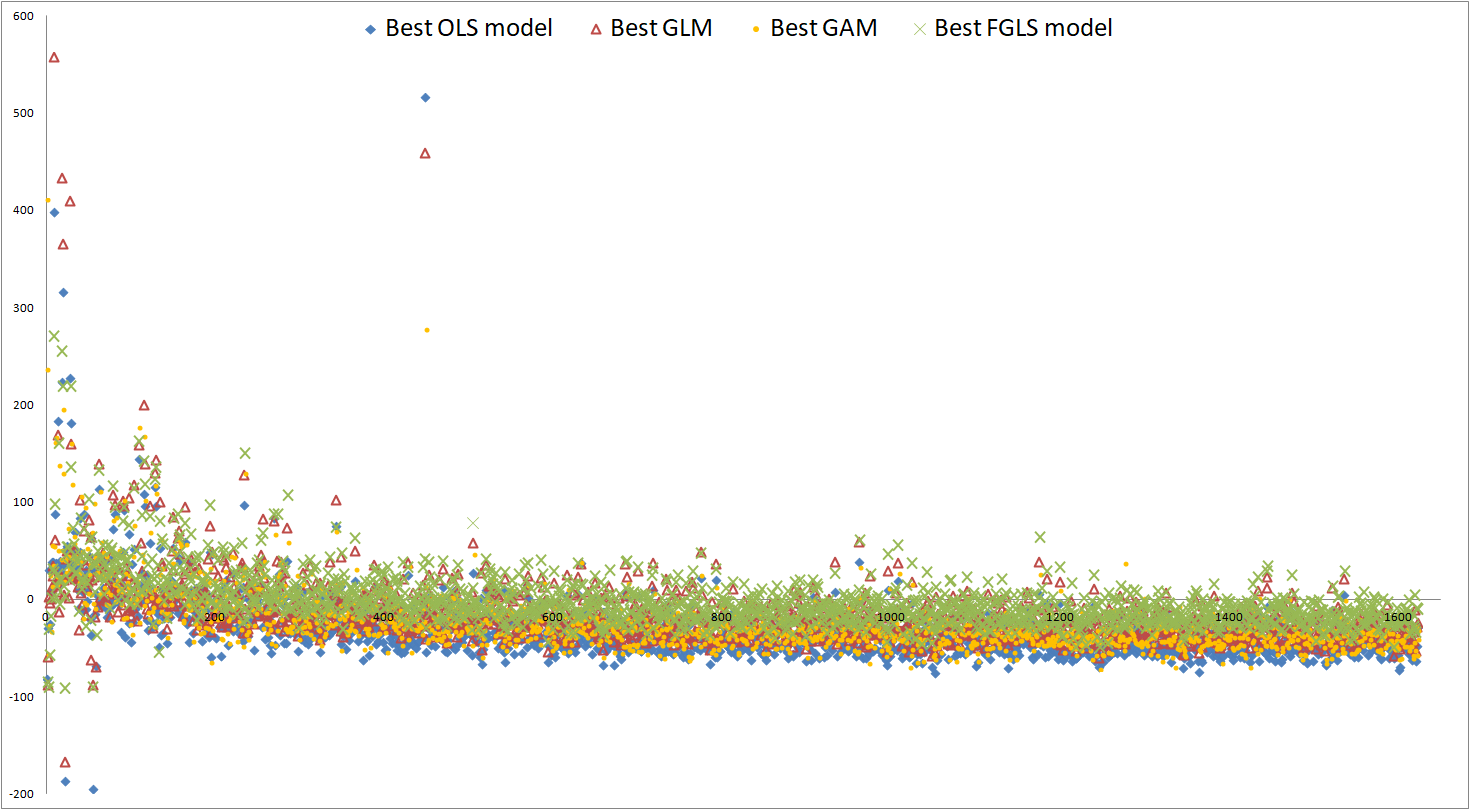}
	\caption{Residual Plots on Nested Simulations Set}
	\label{fig:NestedStochasticsResiduals}
\end{figure}

\begin{figure}[t]
	\centering
		\includegraphics[width=\textwidth]{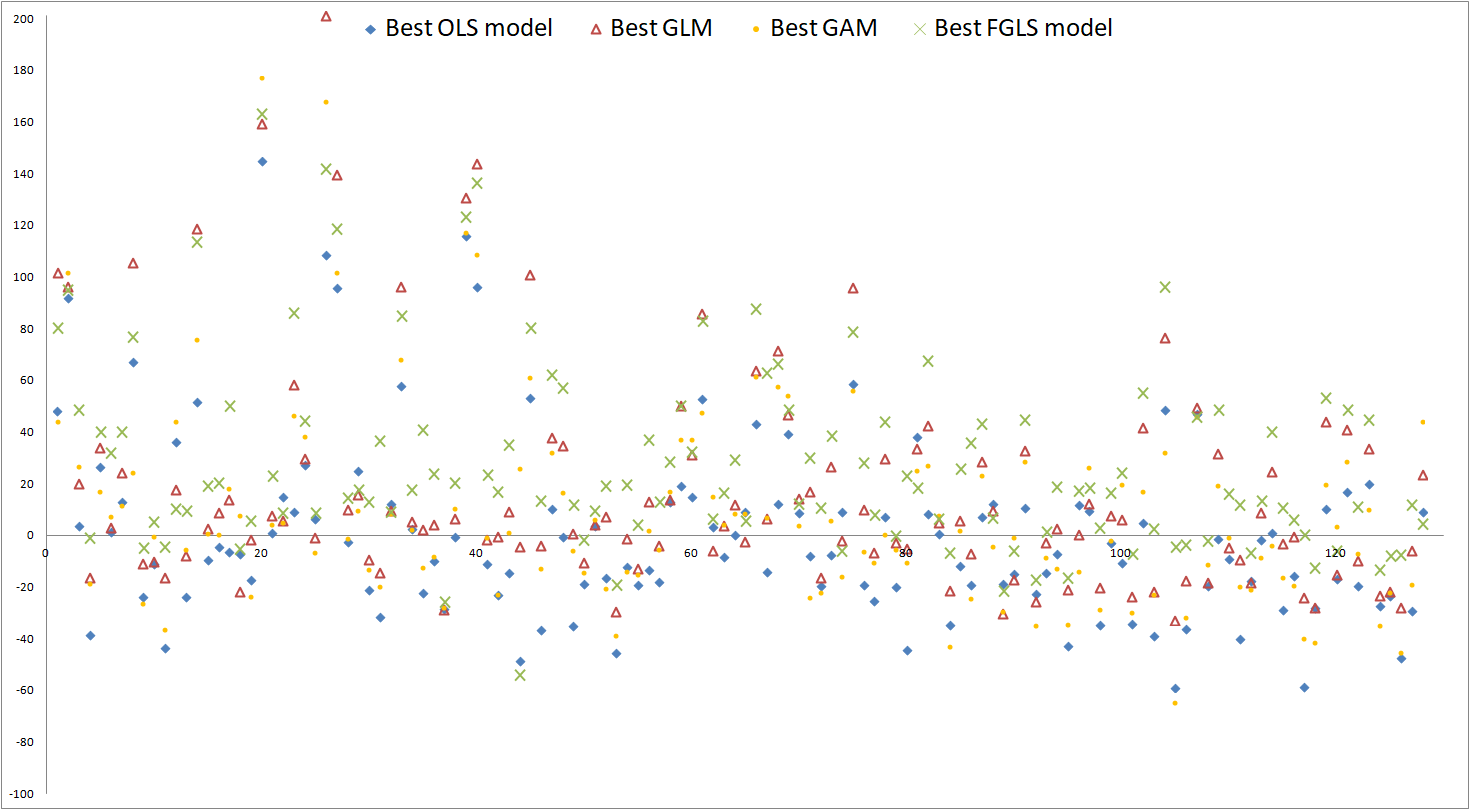}
	\caption{Residual Plots on Capital Region Set}
	\label{fig:CapitalRegionResiduals}
\end{figure}

Overall, the systematic deviations indicated by the means of residuals (\ref{eq:res}) and (\ref{eq:res0}) are reduced significantly on the three validation sets by the relaxation but not completely eliminated. For the $300$-$886$ OLS residuals on the three sets, see the blue residuals in Figures (\ref{fig:ValidationResiduals}), (\ref{fig:NestedStochasticsResiduals}) and (\ref{fig:CapitalRegionResiduals}), respectively. While the reduction of the bias comes along with the general improvement stated above, the remainder of the bias indicates that sample size is not sufficiently large or that the functional form still has some flaws. Note that if the functional form is correctly specified, Proposition 3.2 of \cite{bauer2015} states that if sample size is not sufficiently large, the AC proxy function will on average be positively biased in the tail reflecting the high losses and the BEL proxy function will thus be negatively biased there. Since Propositions 1 and 2 of \cite{gordy2010} state that this result holds for the nested simulations estimators as well, the validation values of the nested simulations and capital region sets need to be more accurate in order to serve for bias detection in this case. For an illustration of such as bias, see Figures 5 and 6 of \cite{bauer2015}. The bias in our one sample example is in the opposite systematic direction.

Unlike figures (\ref{eq:mae}) and (\ref{eq:res}), figures (\ref{eq:mae0}) and (\ref{eq:res0}) do not forgive a bad fit of the base value if the validation values are well approximated by a proxy function. Contrariwise, if a proxy function shows the same systematic deviation from the validation values and the base value, (\ref{eq:mae0}) and (\ref{eq:res0}) will be close to zero whereas (\ref{eq:mae}) and (\ref{eq:res}) will be not. The comparisons $\left|\text{v.res}\right| <  \left|\text{v.res}^0\right|$, $\left|\text{cr.res}\right| < \left|\text{cr.res}^0\right|$ but $\left|\text{ns.res}\right| > \left|\text{ns.res}^0\right|$, holding under both restrictions settings, indicate that on the Sobol and capital region sets primarily the base value is not approximated well whereas on the nested simulations set not only the base value but also the validation values are missed. The MAEs capture this result, too, i.e. $\text{v.mae}, \text{cr.mae} < \text{ns.mae}$ but $\text{ns.mae}^0 < \text{v.mae}^0, \text{cr.mae}^0$.

\subsubsection*{Relationship between BEL \& AC}
The MAEs with respect to the relative metric for BEL are much smaller than for AC since the two economic variables are subject to similar absolute fluctuations with e.g. in the base case BEL being approximately $20$ times the size of AC. The similar absolute fluctuations are reflected by the iteration-wise very similar MAEs with respect to the asset metric of BEL and AC, compare $\text{v.mae}^a$, $\text{ns.mae}^a$ and $\text{cr.mae}^a$ given in $\%$ in Tables \ref{tab:BGROOLSvalfig} and \ref{tab:PGROOLSvalfig}. Furthermore, they manifest themselves in the iteration-wise opposing means of residuals $\text{v.res}$, $\text{v.res}^0$, $\text{ns.res}$ and $\text{cr.res}$ as well as in the similar-sized MAEs $\text{v.mae}^0$, $\text{ns.mae}^0$ and $\text{cr.mae}^0$.

\subsection{Generalized Linear Models (GLMs)}\label{sec:GLMexperiment}
\subsubsection*{Settings}
We derive the GLMs (\ref{eq:g}) of BEL under restriction settings $150$-$443$ and $300$-$886$ which we also employed for the derivation of the OLS proxy functions. Thereby, we run each restriction setting with the canonical choices of random components for continuous (non-negative) response variables, that is, the gaussian, gamma and inverse gaussian distributions, compare \cite{mccullagh1989}. In cases where the economic variable can also attain negative values (e.g. AC), a suitable shift of the response values in a preceding step would be required. We combine each of the three random component choices with the commonly used identity, inverse and log link functions, i.e. $g\left(\mu\right) \in \left\{ \text{id}\left(\mu\right), \frac{1}{\mu}, \log\left(\mu\right)\right\}$, compare \cite{hastie1992glm}. In combination with the inverse gaussian random component, we consider additionally link function $\frac{1}{\mu^2}$. Further choices are conceivable but go beyond this first shot.

We take R function \textit{glm($\cdot$)} implemented in R package \textit{stats} of \cite{stats2018}.

\subsubsection*{Results}
While Tables \ref{tab:BGROGLMgaussian}, \ref{tab:BGROGLMgamma} and \ref{tab:BGROGLMinvgauss} display the AIC scores and five previously defined validation figures after each tenth iteration for the just mentioned combinations under $150$-$443$, Tables \ref{tab:BGROGLMgaussian300886}, \ref{tab:BGROGLMgamma300886} and \ref{tab:BGROGLMinvgauss3008861} do so under $300$-$886$ and include furthermore the final iterations. Table \ref{tab:BGROGLMgaussiangammainvgauss} gives an overview of the AIC scores and validation figures corresponding to all considered final GLMs and highlights in green and red respectively the best and worst values observed per figure.

\subsubsection*{Improvement by Relaxation}
The OLS regression is the special case of a GLM with gaussian random component and identity link function which is why the first sections of Tables \ref{tab:BGROGLMgaussian} and \ref{tab:BGROGLMgaussian300886} coincide respectively with Tables \ref{tab:BGROOLSvalfig} and \ref{tab:BGROOLS300886valfig}. The adaptive algorithm terminates under $150$-$443$ not only for this combination but also for all other ones when the maximum allowed number of terms is reached. Under $300$-$886$ termination occurs due to no further reduction in the AIC score without hitting the restrictions - the different GLMs stop between $208$-$454$ and $250$-$574$.

For all GLMs except for the one with gamma random component and identity link, the AIC scores and eight most significant validation figures for measuring the approximation quality, namely leftmost figure $\text{v.mae}$ to rightmost figure $\text{ns.res}$ in the tables, are improved through the relaxation as can be seen in Table \ref{tab:BGROGLMgaussiangammainvgauss}. For gamma random component with identity link, the deteriorations are negligible. Overall, figures $\text{ns.mae}^0$ and $\text{cr.mae}^0$ are deteriorated by at maximum $0.5\%$ points and figures $\text{ns.res}^0$ and $\text{cr.res}^0$ by at maximum $4$ units. Figures $\text{cr.mae}$ and $\text{cr.mae}^a$ are especially small under $150$-$443$ so that slight deteriorations by at maximum $0.05\%$ points under $300$-$886$ towards the levels of $\text{v.mae}$ and $\text{v.mae}^a$ or $\text{ns.mae}$ and $\text{ns.mae}^a$ are not surprising. Similar arguments apply to the acceptability of the maximum deterioration of $\text{cr.res}$ by $13$ to $17$ units for inverse gaussian with $\frac{1}{\mu^2}$ link. We conclude that the more relaxed restriction setting $300$-$886$ performs better than $150$-$443$ for all GLMs in our numerical example. This result appears plausible in comparison with the OLS result from the previous section and hence also compared to the OLS results of \cite{teuguia2014} and \cite{bauer2015}.

AIC cannot be said to show an overfitting tendency according to Tables \ref{tab:BGROGLMgaussian300886}, \ref{tab:BGROGLMgamma300886} and \ref{tab:BGROGLMinvgauss3008861} and also Table \ref{tab:BGROOLSstderrvalfig} since the validation figures do not deteriorate in the late iterations more than they underly Monte Carlo fluctuations, compare the OLS interpretation. Using GLMs instead of OLS regression in the standard adaptive algorithm, compare Section \ref{sec:calibration_algorithm}, lets the algorithm thus maintain its property to yield numerically stable and parsimonious proxy functions even without restriction settings.

\subsubsection*{Reduction of Bias}
According to Table \ref{tab:BGROGLMgaussiangammainvgauss}, inverse gaussian with $\frac{1}{\mu^2}$ link shows the most significant decrease in $\text{v.mae}$ by $-0.088\%$ points when moving from $150$-$443$ to $300$-$886$. Under $300$-$886$ this combination even outperforms all other ones (highlighted in green) whereas under $150$-$443$ it is vice versa (highlighted in red). Hence, the performance of a random component link combination under $150$-$443$ does not generalize to $300$-$886$. On the Sobol and nested simulations sets, the MAEs (\ref{eq:mae}) are not only considerably lower for inverse gaussian with $\frac{1}{\mu^2}$ link than for all others but also the closest together even when the capital region set is included. This speaks for a great deal of consistency.

In fact, the systematic overestimation of $81\%$ of the points on the nested simulations set by inverse gaussian with $\frac{1}{\mu^2}$ link is certainly smaller than e.g. that of $89\%$ by gaussian with identity link but still very pronounced. On the capital region set, the overestimation rates for these two combinations are $41\%$ and $56\%$, respectively, meaning that here the bias is negligibe. Surprisingly, for most GLMs the bias is here smaller than for inverse gaussian with $\frac{1}{\mu^2}$ link but since this result does not generalize to the nested simulations set, we regard it as a chance event and do not question the rather mediocre performance of inverse gaussian with $\frac{1}{\mu^2}$ link here further. Interpreting the mean of residuals (\ref{eq:res}) provides similar insights.

In particular, for inverse gaussian $\frac{1}{\mu^2}$ link GLM the reduction of the bias comes along with the general improvement by the relaxation. The small remainder of the bias indicates not only that this GLM is a promising choice here but also that identifying suitable regression methods and functional forms is crucial to further improving the accuracy of the proxy function. For the residuals on the three sets, see the red residuals in Figures (\ref{fig:ValidationResiduals}), (\ref{fig:NestedStochasticsResiduals}) and (\ref{fig:CapitalRegionResiduals}), respectively. 

\subsubsection*{Major \& Minor Role of Link Function \& Random Component}
Apart from the just considered case, for all three random components, the relaxation to $300$-$886$ yields the largest out-of-sample performance gains in terms of $\text{v.mae}$ with identity link (between $-0.047\%$ and $-0.058\%$ points), closely followed by log link (between $-0.033\%$ and $-0.047\%$ points), and the least gains with inverse link (between $-0.017\%$ and $-0.020\%$ points). While with identity link the largest improvements before finalization take place for gaussian, gamma and inverse gaussian random components between iterations $180$ to $190$, $170$ to $180$, and $150$ to $160$, respectively, with log link they occur much sooner between iterations $120$ to $130$, $110$ to $120$, and $110$ to $120$, respectively, see Tables \ref{tab:BGROGLMgaussian300886}, \ref{tab:BGROGLMgamma300886} and \ref{tab:BGROGLMinvgauss3008861}. As a result of this behavior, under $150$-$443$ log link performs better than identity link for gaussian and inverse gaussian whereas under $300$-$886$ it is vice versa. Inverse link always performs worse than identity and log links, in particular under $300$-$886$.

Applying the same link with different random components does not bring much variation under $300$-$886$ with gamma and inverse gaussian being slightly better than gaussian for all considered links though. A possible explanation is that the distribution of BEL is slightly skewed conditional on the outer scenarios. Thereby results the skewness in the inner simulations from an asymmetric profit sharing mechanism in the CFP model: While the policyholders are entitled to participate at the profits of an insurance company, see e.g. \cite{mourik2003}, the company has to bear its losses fully by itself. Since gaussian performs only slightly worse than the skewed distributions, it should still be considered for practical reasons because it has a closed-form solution and a great deal of statistical theory has been developed for it, compare e.g. \cite{dobson2002}. By conclusion, the choice of the link is more important than that of the random component so that trying alternative link functions might be beneficial.

\subsection{Generalized Additive Models (GAMs)}\label{sec:GAMexperiment}
\subsubsection*{Settings}
For the derivation of the GAMs (\ref{eq:gamvec}) of BEL, we apply only restriction settings $K_{\text{max}}$-$443$ with $K_{\text{max}} \leq 150$ in the adaptive algorithm since we use smooth functions (\ref{eq:basis}) constructed out of splines that may already have exponents greater than $1$ to which the monomial first-order basis functions are raised. As the model selection criterion we take GCV (\ref{eq:GCVGAM}) used by our chosen implementation by default. We vary different ingredients of GAMs while holding others fixed to carve out possible effects of these ingredients on the approximation quality of GAMs in adaptive algorithms and our application.

We rely on R function \textit{gam($\cdot$)} implemented in R package \textit{mgcv} of \cite{mgcv2018}.

\subsubsection*{Results}
Table \ref{tab:BGROGAMsplinedim} contains the validation figures for GAMs with varying number of spline functions per smooth function, i.e. $J \in \left\{4, 5, 8, 10\right\}$, after each tenth and the finally selected smooth function. In the case of adaptive forward stepwise selection the iteration numbers coincide with the numbers of selected smooth functions. In contrast, table sections with adaptive forward stagewise selection results do not display the iteration numbers in the smooth function column $k$. In Table \ref{tab:BGROGAMsplinedimdfk4k10part1}, we display the effective degrees of freedom, p-values and significance codes of each smooth function of the $J=4$ and $J=10$ GAMs from the previous table at stages $k \in \left\{50, 100, 150\right\}$. The p-values and significance codes are based on a test statistic of \cite{marra2012} having its foundations in the frequentist properties of Bayesian confidence intervals analyzed in \cite{nychka1988}. Tables \ref{tab:BGROGAMbfdim5} and \ref{tab:BGROGAMbfdim10} report the validation figures respectively for GAMs with numbers $J = 5$ and $J = 10$, where the types of the spline functions are varied. Thin plate regression splines, penalized cubic regression splines, duchon splines and Eilers and Marx style P-splines are considered. Thereafter, Tables \ref{tab:BGROGAMrclinkdim4} and \ref{tab:BGROGAMrclinkdim8} display the validation figures respectively for GAMs with numbers $J = 4$ and $J = 8$ and different random component link function combinations. As in GLMs, we apply the gaussian, gamma and inverse gaussian distributions with identity, log, inverse and $\frac{1}{\mu^2}$ (only inverse gaussian) link functions.

Table \ref{tab:BGROGAMadalg} compares by means of two exemplary GAMs the effects of adaptive forward stagewise selection of length $L = 5$ and adaptive forward stepwise selection. Last but not least, Table \ref{tab:BGROGAMchallengePS} contains a mixture of GAMs challenging the results which we will have deduced from the other GAM tables. Table \ref{tab:BGROGAMall} gives an overview of the validation figures corresponding to all derived final GAMs and highlights in green and red respectively the best and worst values observed per figure.

\subsubsection*{Efficiency \& Performance Gains by Tailoring the Spline Function Number}
Table \ref{tab:BGROGAMsplinedim} indicates that the MAEs (\ref{eq:mae}) and (\ref{eq:mae0}) of the exemplary GAMs built up of thin plate regression splines with gaussian random component and identity link tend to increase with the number $J$ of spline functions per dimension until $k=100$. Running more iterations reverses this behavior until $k=150$. Hence, as long as comparably few smooth functions have been selected in the adaptive algorithm fewer spline functions tend to yield better out-of-sample performances of the GAMs whereas many smooth functions tend to perform better with more spline functions. A possible explanation of this observation is that an omitted-variable bias due to too few smooth functions is aggravated here by an overfitting due to too many spline functions. For more details on an omitted-variable bias, see e.g. \cite{pindyck1998}, and for the needlessly large sampling variances and thus low estimation precision of overfitted models, see e.g. \cite{burnham2002}. Differently, the absolute values of the means of residuals (\ref{eq:res}) and (\ref{eq:res0}) tend to become smaller with increasing $J$ regardless of $k$.

According to Table \ref{tab:BGROGAMsplinedimdfk4k10part1}, the components of the effective degrees of freedom (\ref{eq:dfGAM}) associated with each smooth function tend to decrease for $J=4$ and $J=10$ slightly in $k$. This is plausible as the explanatory power of each additionally selected smooth term is expected to decline by trend in the adaptive algorithm. Conditional on $\text{df} > 1$, that is for proportions of at least $40\%$ of all smooth terms, the averages of the effective degrees of freedom belonging to $k \in \left\{50, 100, 150\right\}$ amount for $J=4$ and $J=10$ to $\left\{2.494, 2.399, 2.254\right\}$ and $\left\{5.366, 4.530, 4.424\right\}$, respectively. The values are by construction smaller than $J-1$ since one degree of freedom per smooth function is lost to the identifiability constraints. Hence, for at least $40\%$ of the smooth functions, on average $J = 6$ is a reasonable choice to capture the CFP model properly while maintaining computational efficiency, compare \cite{wood2017b}. The other side of the coin here is that up to $60\%$ of the smooth functions are supposed to be replacable by simple linear terms without losing accuracy so that here tremendous efficiency gains can be realized by making the GAMs more parsimonious. Furthermore, setting $J$ individually for each smooth function can help improve computational efficiency (if $J$ should be set below average) and out-of-sample performance (if $J$ should be set above average). However, such a tailored approach entails the challenge that the optimal $J$ per smooth function is not stable across all $k$, compare row-wise the degrees of freedom in the table for $J=4$ and $J=10$.

\subsubsection*{Dependence of Best Spline Function Type}
According to Tables \ref{tab:BGROGAMbfdim5} and \ref{tab:BGROGAMbfdim10}, the adaptive algorithm terminates only due to no further decrease in GCV when the GAMs are composed of duchon splines discussed in \cite{duchon1977}. Whether GCV has an overfitting tendency here cannot be deduced from this example since only restriction settings with $K_{\text{max}} \leq 150$ are tested. The thin plate regression splines of \cite{wood2003} and penalized cubic regression splines of \cite{wood2017b} perform similarly and significantly better than the duchon splines for both $J=5$ and $J=10$. For $J=5$ the Eilers and Marx style P-splines proposed by \cite{eilers1996} perform by far best when $K_{\text{max}} = 100$ smooth functions are allowed. However, for $J=10$ they are outperformed by both the thin plate regression splines and penalized cubic regression splines when between $K_{\text{max}} = 125$ and $150$ smooth functions are allowed. This result illustrates well that the best choice of the spline function type varies with $J$ and $K_{\text{max}}$, meaning that it should be selected together with these parameters.

\subsubsection*{Minor Role of Link Function \& Random Component}
For GLMs, we have seen that varying the random component barely alters the validation results whereas varying the link function can make a noticeable impact. While this result mostly applies to the earlier compositions of GAMs as well, it certainly does not to the later ones. See for instance early composition $k=40$ in Table \ref{tab:BGROGAMrclinkdim4}. Here identity link GAMs with gamma and inverse gaussian random components perform more similar to each other than identity and log link GAMs with gamma random component or identity and log link GAMs with inverse gaussian random component do. Log link GAMs with gamma and inverse gaussian random components show such a behavior as well. However identity link GAM with the less flexible gaussian random component (no skewness) does not show at all a behavior similar to that of identity link GAMs with gamma or inverse gaussian random components. Now see later compositions $k \in \left\{70, 80\right\}$ to verify that all available GAMs in the table produce very similar validation results.

For another example see Table \ref{tab:BGROGAMrclinkdim8}. For early composition $k=50$, identity link GAMs with gaussian and gamma random components behave very similar to each other just like log link GAMs with gaussian and gamma random components do. For later compositions $k \in \left\{100, 110\right\}$, again all available GAMs produce very similar validation results. A possible explanation of this result is that the impact of the link function and random component decreases with the number of smooth functions as the latter take the modeling over. By conclusion, the choices of the random component and link function do not play a major role when the GAM is built up of many smooth functions.

\subsubsection*{Consistency of Results}
Table \ref{tab:BGROGAMadalg} shows based on two exemplary GAMs constructed out of $J=8$ thin plate regression splines per dimension varying in the random component and link function that the adaptive forward stagewise selection of length $L = 5$ and adaptive forward stepwise selection lead to very similar GAMs and validation results. As a result, stagewise selection should be preferred due to its considerable run time advantage. As we will see in the following, the run time can be further reduced without any drawbacks by dynamically selecting even more than $5$ smooth functions per iteration.

The purpose of Table \ref{tab:BGROGAMchallengePS} is to challenge the hypotheses deduced above. Like Table \ref{tab:BGROGAMsplinedim}, this table contains the results of GAMs with varying spline function number $J \in \left\{5, 8, 10\right\}$ and fixed spline function type. Instead of thin plate regression splines, now Eilers and Marx style P-splines are considered. Since adaptive forward stepwise and stagewise selection do not yield significant differences in the examples of Table \ref{tab:BGROGAMadalg}, we do not expect that permutations thereof affect the results much here as well. This allows us to randomly assign three different adaptive forward selection approaches to the three exemplary proxy function derivation procedures. As one of these approaches, we choose a dynamic stagewise selection approach in which $L$ is determined in each iteration as the proportion $0.25$ of the size of the candidate term set. Again we see that as long as only $k \in \left\{90, 100\right\}$ smooth functions have been selected, $J = 5$ performs better than $J = 8$ and $J = 8$ better than $J = 10$. However, $k=150$ smooth functions are not sufficient this time for $J = 10$ to catch up with the performance of $J = 5$. The observed performance order is consistent with the hypotheses of a high stability of the GAMs with respect to the adaptive selection procedure and random component link function combination.

\subsubsection*{Potential of Improved Interaction Modeling}
Table \ref{tab:BGROGAMall} presents as the most suitable GAM the one with highest allowed maximum number of smooth functions $K_{\text{max}} = 150$ and highest number of spline functions $J = 10$ per dimension. The slight deterioration after $k=130$ reported by Table \ref{tab:BGROGAMsplinedim} indicates that at least one of the parameters is already comparably high. According to Table \ref{tab:BGROGAMsplinedimdfk4k10part1}, there are a few smooth terms which might benefit from being composed of more than ten spline functions and increasing $K_{\text{max}}$ might be helpful to capturing the interactions in the CFP model more appropriately, particularly in the light of the fact that the best GLM, having $250$ basis functions, outperforms the best GAM on both the Sobol and nested simulations set, compare Table \ref{tab:BGROGLMgaussiangammainvgauss}, with the best GAM showing a comparably low bias across the three validation sets though, see the orange residuals in Figures (\ref{fig:ValidationResiduals}), (\ref{fig:NestedStochasticsResiduals}) and (\ref{fig:CapitalRegionResiduals}), respectively. Variations in the random component link function combination and adaptive selection procedure are not expected to change the performance much. By conclusion, we recommend the fast gaussian identity link GAMs (several expressions in the PIRLS algorithm simplify) with tailored spline function numbers per smooth function and simple linear terms under stagewise selection approaches of suitable lengths $L \geq 5$ and more relaxed restriction settings where $K_{\text{max}} > 150$.

\subsection{Feasible Generalized Least-Squares (FGLS) Regression}\label{sec:FGLSexperiment}
\subsubsection*{Settings}
Like the OLS proxy functions and GLMs, we derive the FGLS proxy functions (\ref{eq:fFGLShat}) under restriction settings $150$-$443$ and $300$-$886$. For the performance assessment of FGLS regression, we apply type I and II algorithms with variance models of different complexity, where type I results are obtained as a by-product of type II algorithm since the latter algorithm builds upon the former one. We control the complexity through the maximum allowed numbers of variance model terms $M_{\text{max}} \in \left\{2; 6; 10; 14; 18; 22\right\}$.

We combine R functions \textit{nlminb($\cdot$)} and \textit{lm($\cdot$)} implemented in R package \textit{stats} of \cite{stats2018}.

\subsubsection*{Results}
Tables \ref{tab:BGROGLSvarmod} and \ref{tab:BGROGLS300886varmod} display respectively the adaptively selected FGLS variance models of BEL corresponding to maximum allowed numbers of terms $M_{\text{max}}$ based on final $150$-$443$ and $300$-$886$ OLS proxy functions given in Tables \ref{tab:BGROOLS1} and \ref{tab:BGROOLS3008861}. For reasons of numerical stability and simplicity, only basis functions with exponents summing up to at max two are considered as candidates. Additionally, the AIC scores and MAEs with respect to the relative metric are reported in the tables. By construction, these results are also the type I algorithm outcomes. Tables \ref{tab:BGROGLSvarmodvalfig} and \ref{tab:BGROGLSvarmod300886valfig} summarize respectively under $150$-$443$ and $300$-$886$ all iteration-wise out-of-sample test results. The results of type II algorithm after each tenth and the final iteration of adaptive FGLS proxy function selection are respectively displayed by Tables \ref{tab:BGROGLSvalfig1} and \ref{tab:BGROGLS300886valfig1}. Table \ref{tab:BGROGLSallvalfig} gives an overview of the AIC scores and validation figures corresponding to all final FGLS proxy functions and highlights as in the previous overview tables in green and red respectively the best and worst values observed per figure.

\subsubsection*{Consistency Gains by Variance Modeling}
By looking at Tables \ref{tab:BGROGLSvarmod} and \ref{tab:BGROGLS300886varmod} we see similar out-of-sample performance patterns during adaptive variance model selection based on the basis function sets of $150$-$443$ and $300$-$886$ OLS proxy functions. In both cases, the p-values of Breusch-Pagan test indicate that heteroscedasticity is not eliminated but reduced when the variance models are extended, i.e. when $M_{\text{max}}$ is increased. In fact, in a more good-natured LSMC example \cite{hartmann2015} shows that a type I alike algorithm manages to fully eliminate heteroscedasticity. While the MAEs (\ref{eq:mae}) barely change on the Sobol set, they decrease significantly on the nested simulations set and increase noticeably on the capital region set. Under $300$-$886$ the effects are considerably smaller than under $150$-$443$ since the capital region performance of $300$-$886$ OLS proxy function is less extraordinarily good than that of $150$-$443$ OLS proxy function. The three MAEs approach each other under both restriction settings. Hence the reductions in heteroscedasticity lead to consistency gains across the three validation sets.

Tables \ref{tab:BGROGLSvarmodvalfig} and \ref{tab:BGROGLSvarmod300886valfig} complete the just discussed picture. The remaining validation figures on the Sobol set improve through type I FGLS regression slightly compared to OLS regression. Like $\text{ns.mae}$, figure $\text{ns.res}$ and the base residual improve a lot with increasing $M_{\text{max}}$ under $150$-$443$ and a little less under $300$-$886$ but $\text{ns.mae}^0$ and $\text{ns.res}^0$ do not alter much as the aforementioned two figures cancel each other out here. On the capital region set, the figures deteriorate or remain comparably high in absolute values. The type I FGLS figures converge fast so that increasing $M_{\text{max}}$ successively from $10$ to $22$ barely affects the out-of-sample performance anymore. As a result of heteroscedasticity modeling, the proxy functions are shifted such that overall approximation quality increases. Unfortunately, this does not guarantee an improvement in the relevant region for SCR estimation as our example illustrates well.

\subsubsection*{Monotonicity in Complexity}
Let us address the type II FGLS results under $150$-$443$ in Table \ref{tab:BGROGLSvalfig1} now. For $M_{\text{max}} = 2$, figures (\ref{eq:mae0}) and (\ref{eq:res0}) are improved on all three validation sets significantly compared to OLS regression with the type I figures lying inbetween. The other validation figures are similar for OLS, type I and II FGLS regression, which traces the performance gains in (\ref{eq:mae0}) and (\ref{eq:res0}) back to a better fit of the base value. For $M_{\text{max}} = 6$ to $22$, the type II figures show the same effects as the type I ones but more pronouncedly, see the previous two paragraphs. These effects are by trend the more distinct the more complex the variance model becomes. The type II figures stabilize less than the type I ones because of the additional variability coming along with adaptive FGLS proxy function selection. \cite{hartmann2015} shows in terms of Sobol figures in her LSMC example that increasing the complexity while omitting only one regressor from the simpler variance model can deteriorate the out-of-sample performance dramatically. Intuitively, it is plausible that the FGLS validation figures are the farther from the OLS figures away the more elaborately heteroscedasticity is modeled.

Now let us relate the type II FGLS results under $300$-$886$ in Table \ref{tab:BGROGLS300886valfig1} to the other FGLS results. Under $300$-$886$ for $M_{\text{max}} = 2$, figures (\ref{eq:mae0}) and (\ref{eq:res0}) are already at a comparably good level with both OLS and type I FGLS regression so that they do not alter much or even deteriorate with type II FGLS regression. Like under $150$-$443$ for $M_{\text{max}} = 6$ to $22$, the type II figures show the effects of the type I ones more pronouncedly. Under both restriction settings, $\text{ns.mae}$ and $\text{ns.res}$ decrease thereby significantly. While this barely causes $\text{ns.res}^0$ to change under $150$-$443$, it lets $\text{ns.res}^0$ increase in absolute values under $300$-$886$. The slight improvements on the Sobol set and the deteriorations on the capital region set carry over to $300$-$886$. When $M_{\text{max}}$ is increased up to $22$, the type II FGLS validation figures under $300$-$886$ do not stop fluctuating. The variability entailed by adaptive FGLS proxy function selection intensifies thus through the relaxation of the restriction setting in this numerical example. According to Breusch-Pagan test, heteroscedasticity is neither eliminated by the type II algorithm here nor by a type II alike approach of \cite{hartmann2015} in her more good-natured example.

\subsubsection*{Improvement by Relaxation}
Among all FGLS proxy functions listed in Table \ref{tab:BGROGLSallvalfig}, we consider type II with $M_{\text{max}} = 14$ in variance model selection under $300$-$886$ as the best performing one. Apart from nested simulations validation under type I algorithm, $300$-$886$ performs better than $150$-$443$. Since on the other hand type II algorithm performs better than type I algorithm under the respective restriction settings, $300$-$886$ and type II algorithm are the most promising choices here. Differently $M_{\text{max}} = 14$ does not constitute a stable choice due to the high variability coming along with $300$-$886$ and type II algorithm.

While all type I FGLS proxy functions are by definition composed of the same basis functions as the OLS proxy function, the compositions of type II FGLS proxy functions vary with $M_{\text{max}}$ because of their renewed adaptive selection. Consequently, under $300$-$886$ all type I FGLS proxy functions hit the same restrictions $224$-$464$ as the OLS proxy function does, whereas the restrictions hit by type II FGLS proxy functions vary between $224$-$454$ and $258$-$564$. This variation is consistent with the OLS and GLM results from the previous sections and hence the OLS results of \cite{teuguia2014} and \cite{bauer2015}.

AIC does not have an overfitting tendency according to Tables \ref{tab:BGROGLSvarmodvalfig}, \ref{tab:BGROGLSvarmod300886valfig}, \ref{tab:BGROGLSvalfig1} and \ref{tab:BGROGLS300886valfig1} as the validation figures do not deteriorate in the late iterations more than they underly Monte Carlo fluctuations, compare the OLS and GLM interpretations. Using FGLS instead of OLS regression in the standard adaptive algorithm, compare Section \ref{sec:calibration_algorithm}, lets the algorithm thus yield numerically stable and parsimonious proxy functions without restriction settings as well.

\subsubsection*{Reduction of Bias}
The type II $M_{\text{max}} = 14$ FGLS proxy function under $300$-$886$ reaches with $258$ terms the highest observed number across all numerical experiments and not only outperforms all derived GLMs and GAMs in terms of combined Sobol and nested simulations validation, it also shows by far the smallest bias on these two validation sets and approximates the base value comparably well. This observation speaks for a high interaction complexity of the CFP model. The reduction of the bias comes again along with the general improvement by the relaxation. Given the fact that the capital region set presents the most extreme and challenging validation set in our analysis, the still mediocre performance here can be regarded as acceptable for now. Nevertheless, especially the bias on this set motivates the search for even more suitable regression methods and functional forms. For the residuals of the $300$-$886$ FGLS proxy function on the three sets, see the green residuals in Figures (\ref{fig:ValidationResiduals}), (\ref{fig:NestedStochasticsResiduals}) and (\ref{fig:CapitalRegionResiduals}), respectively.

\subsection{Multivariate Adaptive Regression Splines (MARS)}\label{sec:MARSexperiment}
\subsubsection*{Settings}
We undertake a two-step approach to identify suitable generalized MARS models out of numerous possibilities. In the first step, we vary several MARS ingredients over a wide range and obtain in this way a large number of different MARS models. To be more specific, we vary the maximum allowed number of terms $K_{\text{max}} \in \left\{50, 113, 175, 237, 300\right\}$ and the minimum threshold for the decrease in the residual sum of squares $t_{\text{min}} \in \left\{0, 1.25, 2.5, 3.75, 5\right\} \cdot 10^{-5}$ in the forward pass, the order of interaction $o \in \left\{3, 4, 5, 6\right\}$, the pruning method $p \in  \left\{\text{'n'}, \text{'b'}, \text{'f'}, \text{'s'}\right\}$ with $\text{'n'}=\text{'none'}$, $\text{'b'}=\text{'backward'}$, $\text{'f'}=\text{'forward'}$ and $\text{'s'}=\text{'seqrep'}$ in the backward pass, as well as the random component link function combination of the GLM extension. In addition to the $10$ random component link function combinations applied in the numerical experiments of the GLMs, compare e.g. Table \ref{tab:BGROGLMgaussiangammainvgauss}, we use poisson random component with identity, log and squareroot link functions. We work with the default fast MARS parameter $\text{fast.k} = 20$ of our chosen implementation.

We use R function \textit{earth($\cdot$)} implemented in R package \textit{earth} of \cite{earth2018}.

\subsubsection*{Results}
In total, these settings yield $4 \cdot 5 \cdot 5 \cdot 4 \cdot 13 = 5,200$ MARS models with a lot of duplicates in our first step. We validate the $5,200$ MARS models on the Sobol, nested simulations and capital region sets through evaluation of the five validation figures. 
Then we collect the five best performing MARS models in terms of each validation figure per set which gives us in total $5 \cdot 5 = 25$ best performing models per first step validation set. Since the MAEs (\ref{eq:mae}) with respect to the relative and asset metric entail the same best performing models, only $5 \cdot 4 = 20$ of the collected models per first step set are potentially different. Based on the ingredients of each of these $20$ MARS models per first step set, we define $5 \cdot 5 = 25$ new sets of ingredients varying only with respect to $K_{\text{max}}$ and $t_{\text{min}}$ and derive the corresponding new but similar MARS models in the second step. As a result, we obtain in total $20 \cdot 25 = 500$ new MARS models per first step set. Again, we assess their out-of-sample performances through evaluation of the five validation figures on the three validation sets. Out of the $500$ new MARS models per first step set, we collect then the best performing ones in terms of each validation figure per second step set. Now this gives us in total $5 \cdot 3 = 15$ best MARS models per first step set, or taking into account that the MAEs (\ref{eq:mae}) with respect to the relative and asset metric entail once more the same best performing models, $4 \cdot 3 = 12$ potentially different best models per first step set. In total, this makes $12 \cdot 3 = 4 \cdot 9 = 36$ best MARS models, which can be found in Table \ref{tab:BGROMARSRun2} sorted by first and second step validation sets.

\subsubsection*{Poor Interaction Modeling \& Extrapolation}
In Table \ref{tab:BGROMARSRun2}, the out-of-sample performances of all MARS models derived in our two-step approach are sorted using the first step validation set as the primary and the second step validation set as the secondary sort key. Let us address the first step second step validation set combinations by the headlines in Table \ref{tab:BGROMARSRun2}. By construction, the combinations $\text{Sobol set}^2$, $\text{Nested simulations set}^2$ and $\text{Capital region set}^2$ yield respectively the MARS models with the best validation figures (\ref{eq:mae}), (\ref{eq:res}), (\ref{eq:mae0}) and (\ref{eq:res0}) on the Sobol, nested simulations and capital region sets. See that in the table all corresponding diagonal elements are highlighted in green. But the best MAEs (\ref{eq:mae}) and (\ref{eq:mae0}) are not even close to what OLS regression, GLMs, GAMs and FGLS regression achieve. Finding small residuals (\ref{eq:res}) and (\ref{eq:res0}) regardless of the other validation figures is not sufficient. The performances on the nested simulations and capital region sets, comprising several scenarios beyond the fitting space, are especially poor. All these results indicate that MARS models do not seem very suitable for our application. Despite the possibility to select up to $300$ basis functions, the MARS algorithm selects only at maximum $148$ basis functions, which suggests that without any alterations, the algorithm is not able to capture the behavior of the CFP model properly, in particular extrapolation behavior is comparably poor.

The MARS model with the set of ingredients $K_{\text{max}} = 50$, $t_{\text{min}} = 0$, $o = 4$, $p = \text{'b'}$, inverse gaussian random component and identity link function is selected as the best one six times out of $36$, or once for each Sobol and nested simulations first step validation set combination. Furthermore, this model performs best in terms of $\text{v.res}^0$, $\text{ns.mae}^0$ and $\text{ns.mae}^a$. Since there is no other MARS model with a similar high occurrence and performance, we consider it the best performing and most stable one found in our two-step approach. For illustration of a MARS model, see this one in Table \ref{tab:BGROMARSbest}. The fact that this best MARS model performs worse than other ones in terms of several validation figures stresses the infeasibility of MARS models in this application.

\subsubsection*{Limitations}
Table \ref{tab:BGROMARSRun2} suggests that, up to a certain upper limit, the higher the maximum allowed number of terms $K_{\text{max}}$ the higher tends the performance on the Sobol set to be. However, this result does not generalize to the nested simulations and capital region sets. Since at maximum $148$ basis functions are selected here even if up to $300$ basis functions are allowed, extending the range of $K_{\text{max}}$ in the first step of this numerical experiment would not affect the output in this regard. The threshold $t_{\text{min}}$ is an instrument controlling the number of basis functions selected in the forward pass up to $K_{\text{max}}$ which cannot be extended below zero, meaning that its variability has already been exhausted here as well. For the interaction order $o$ similar considerations as for $K_{\text{max}}$ apply. The pruning method $p$ used in the backward pass does not play a large role compared to the other ingredients as it only helps reduce the set of selected basis functions. In terms of Sobol validation, inverse gaussian random component with identity link performs best, whereas in terms of nested simulations and capital region validation, inverse gaussian random component with any link or log link with gaussian or poisson random component perform best. We conclude that if there was a suitable MARS model for our application, our two-step approach would have found it.

\subsection{Kernel Regression}\label{sec:KRexperiment}
\subsubsection*{Settings}
We make a series of adjustments affecting either the structure or the derivation process of the multidimensional LC and LL proxy functions (\ref{eq:betaHatLCmulti}) and (\ref{eq:fHatLLmulti}) to get as broad a picture of the potential of kernel regression in our application as possible. Our adjustments concern the kernel function and its order, the bandwidth selection criterion, the proportion of fitting points used for bandwidth selection, and the sets of basis functions of which the local proxy functions are composed of. Thereby we combine in various ways the gaussian, Epanechnikov and uniform kernels, orders $o \in \left\{2, 4, 6, 8\right\}$, bandwidth selection criteria LOO-CV and AIC, and between $2,500$ (proportion $\text{bw} = 0.1$) and $25,000$ (proportion $\text{bw} = 1$) fitting points for bandwidth selection.

We work with R functions \textit{npregbw($\cdot$)} and \textit{npreg($\cdot$)} implemented in R package \textit{np} of \cite{np2018}.

\subsubsection*{Results}
Furthermore, we alternate the four basis function sets contained in Tables \ref{tab:KRregressors1} and \ref{tab:KRregressors2}. The first two basis function sets with $K_{\text{max}} \in \left\{16, 27\right\}$ are derived by adaptive forward stepwise selection based on OLS regression, the third one with $K_{\text{max}} = 15$ by risk factor wise linear selection and the last one with $K_{\text{max}} = 22$ by a combination thereof. All combinations including their out-of-sample performances can be found in Table \ref{tab:KR}. Again, the best and worst values observed per validation figure are highlighted in green and red, respectively.

\subsubsection*{Poor Interaction Modeling \& Extrapolation}
We draw the following conclusions based on the validation results in Table \ref{tab:KR}. The comparisons of LC and LL regression applied with gaussian kernel and $16$ basis functions or Epanechnikov kernel and $15$ basis functions suggest that LL regression performs better than LC regression. However, even the best Sobol, nested simulations and capital region results of LL regression are still outperformed by OLS regression, GLMs, GAMs and FGLS regression. Possible explanations for this observation are that kernel regression is not able to model the interactions of the risk factors equally well with its few basis functions and that local regression approaches perform rather poorly close to and especially beyond the boundary of the fitting space because of the thinned out to missing data basis in this region. While the first explanation applies to all three validation sets, the latter one applies only to the nested simulations and capital region sets on which the validation figures are indeed worse than on the Sobol set. While LC regression produces interpretable results with the sets of $22$ and $27$ basis functions, the more complex LL regression does not in most cases.

\subsubsection*{Limitations}
On the Sobol and capital region sets, both LC and LL regression show similar behaviors when relying on gaussian kernel and $16$ basis functions compared to Epanechnikov kernel and $15$ basis functions. But on the nested simulations set, gaussian kernel and $16$ basis functions are the superior choices. Using a uniform kernel with LC regression deteriorates the out-of-sample performance. The results of LC regression indicate furthermore that an extension of the basis function sets from $15$ to $27$ only slightly affects the validation performance. With gaussian kernel switching from $16$ to $27$ basis functions barely has an impact and with Epanechnikov kernel only the nested simulations and capital region validation performance improve when using $27$ as opposed to $15$, $16$ or $22$ basis functions. While increasing the order of the gaussian or Epanechnikov kernel deteriorates the validation figures dramatically, for the uniform kernel the effects can go in both directions. AIC performs worse than LOO-CV when used for bandwidth selection of the gaussian kernel in LC regression. For LC regression, increasing the proportion of fitting points entering bandwidth selection improves all validation figures until a specific threshold is reached. But thereafter the nested simulations and capital region figures are deteriorated. For LL regression no such deterioration is observed.

Overall we do not see much potential in kernel regression for our practical example compared to most of the previously analyzed regression methods. Nonetheless in order to achieve comparably good kernel regression results, we consider LL regression more promising than LC regression due to the superior but still poor modeling close to and beyond the boundary of the fitting space. We would apply it with gaussian, Epanechnikov or other similar kernel functions. A high proportion of fitting points for bandwidth selection is recommended and it might be worth trying alternative comparably small basis function sets reflecting e.g. the risk factor interactions better than in our examples.

\section{Conclusion}
\subsection*{General Remarks}
For high-dimensional variable selection applications such as the calibration step in the LSMC framework, we have presented various machine learning regression approaches ranging from ordinary and generalized least-squares regression variants over GLM and GAM approaches to multivariate adaptive regression splines and kernel regression approaches. At first we have justified the combinability of the ingredients of the regression routines such as the estimators and proposed model selection criteria in a theoretical discourse. Afterwards we have applied numerous configurations of these machine learning routines to the same slightly disguised real-world example in the LSMC framework. With the aid of different validation figures, we have analyzed the results, compared the out-of-sample performances and adviced to use certain routine designs.

In this conclusion, we recap the assumptions, properties and estimation algorithms of the analyzed routines in conjunction with their results in the numerical experiments. Furthermore, we give an outlook for possible future research streams.

\subsection*{OLS Regression}
The OLS regression algorithm in Section \ref{sec:OLStheory} requires the assumptions of strict exogeneity, homoscedastic errors and linear independent basis functions for the coefficient estimator to be the best linear unbiased estimator by Gauss-Markov theorem. The OLS estimator minimizes the residual sum of squares by definition and has a closed-form expression. For AIC to be evaluable at the OLS estimator, the errors also have to be normally distributed according to Theorem \ref{th:1}.

We applied the OLS regression algorithm in Section \ref{sec:OLSexperiment} under suitable restriction settings and found that relaxing the setting from $150$-$443$ to $300$-$886$ (= no actual restriction) improved out-of-sample performance considerably. Thereby the bias indicated by the means of residuals on the three validation sets was reduced, see Tables \ref{tab:BGROOLSvalfig} and \ref{tab:BGROOLS300886valfig}, but not eliminated so that we stated that the functional form of the proxy function still had some flaws, see Figure \ref{fig:NestedStochasticsResiduals}. We concluded that overall the adaptive algorithm managed to provide a numerically stable and parsimonious proxy function even without imposing a restriction setting and that the a priori unlimited degrees of freedom served capturing the complex CFP model better. Furthermore we pointed out that BEL and AC were subject to similar absolute fluctuations.

\subsection*{GLMs}
The GLM algorithm in Section \ref{sec:GLMtheory} is a generalization of the OLS regression algorithm insofar as the errors are now allowed to come from an arbitrary distribution of the exponential family and the economic variable is related to the linear predictor by a monotonic link function. The GLM estimator maximizes the log-likelihood and can be derived by an IRLS algorithm. Without more ado, the GLM estimator can be fed into AIC.

Like in the OLS regression algorithm, we observed in all applied GLM algorithms in Section \ref{sec:GLMexperiment} that relaxing the setting from $150$-$443$ to $300$-$886$ (= no actual restriction) helped improve out-of-sample performance and reduce the bias. From the small remainder of the bias we deduced that identifying suitable regression methods and functional forms is crucial to further improving the accuracy of the proxy function. We concluded that the adaptive algorithm maintained its property to yield numerically stable and parsimonious proxy functions without requiring restriction settings in the GLM context. The performance of a random component link combination under $150$-$443$ did not generalize to $300$-$886$. Moreover, we saw in the variation of the results that the choice of the link was more important than that of the random component so that regarding additional link functions might be beneficial. While continuous skewed random components led to slightly advantageous out-of-sample performances, the use of the gaussian random component had practical advantages. Compared to the OLS regression routine, there were GLM routine designs with better out-of-sample performances. While performing best on both the Sobol and nested simulations set, $300$-$886$ inverse gaussian $\frac{1}{\mu^2}$ link GLM showed only a mediocre performance on the capital region set. For an overview of these results, see Table \ref{tab:BGROGLMgaussiangammainvgauss}.

\subsection*{GAMs}
The GAM algorithm in Section \ref{sec:GAMtheory} acts as a generalization of the GLM algorithm and brings in the additive models with the smooth functions as the new component. The GAM estimator maximizes the penalized log-likelihood and can be derived by a PIRLS algorithm. The penalization takes place with respect to smoothing parameters controlling the trade-off between a too wiggly and too smooth model. For AIC to be evaluable at the GAM estimator, the degrees of freedom are generalized such that they account for the smoothing. As an alternative to AIC, generalized cross-validation GCV is introduced. The smoothing parameters are selected such that they minimize the chosen model selection criterion. For reasons of computational efficiency, adaptive forward stagewise selection is suggested.

We ran the different GAM algorithms in Section \ref{sec:GAMexperiment} only under restriction setting $150$-$443$. Whether GCV had an overfitting tendency in the adaptive algorithm could therefore not be assessed. We saw that as long as comparably few smooth functions had been selected fewer spline functions performed better whereas many smooth functions did better with more spline functions, compare Table \ref{tab:BGROGAMsplinedim}. We gave a possible explanation of these effects by arguing that an omitted-variable bias due to too few smooth functions might have been aggravated here by an overfitting due to too many spline functions. In order to realize the efficiency and performance gains incentivized by Table \ref{tab:BGROGAMsplinedimdfk4k10part1} by making the GAMs more parsimonious, we proposed to set the spline function numbers individually for each smooth function and to use linear terms where sufficient. Another result was that the spline function type should be selected conditional on the spline function number(s) and number of smooth functions, see Tables \ref{tab:BGROGAMbfdim5} and \ref{tab:BGROGAMbfdim10}. As soon as the GAM had been composed of many smooth functions, the choices of both the link and random component turned out to be less crucial which made us recommended the fast gaussian identity GAMs in the exemplary application, compare Tables \ref{tab:BGROGAMrclinkdim4} and \ref{tab:BGROGAMrclinkdim8}. Since adaptive forward stagewise selection of length $L = 5$ and adaptive forward stepwise selection led to very similar GAMs according to Table \ref{tab:BGROGAMadalg}, we suggested to use the former selection approach due to its run time advantage. From the fact that the best found GLM had $250$ terms and outperformed the best found GAM reported in Table \ref{tab:BGROGAMall}, we deduced that using more than $150$ smooth functions might improve the results.

\subsection*{FGLS Regression}
The FGLS regression algorithm in Section \ref{sec:FGLStheory} is another generalization of the OLS regression algorithm insofar as the errors are here allowed to have any positive definite covariance matrix. For the GLS estimator to be the best linear unbiased estimator by Gauss-Markov-Aitken theorem, the assumptions of strict exogeneity, linear independent basis functions and a known covariance matrix are required. The GLS estimator minimizes the generalized residual sum of squares. When the covariance matrix is unknown but can be estimated consistently, the FGLS estimator serves as a substitute for the GLS estimator that has asymptotically the same properties. If furthermore the errors are jointly normally distributed, the FGLS estimator can be derived by a maximum likelihood algorithm and fed into AIC according to Theorem \ref{th:3}. Suitable implementations are multiplicative heteroscedasticity, adaptive variance model selection procedures and Breusch-Pagan test for heterogeneity diagnosis.

Among the applied FGLS algorithms in Section \ref{sec:FGLSexperiment}, the type I algorithms led to consistency gains across the three validation sets. According to Breusch-Pagan test, they induced at least a reduction in heteroscedasticity in the generalized least-squares problem, which tended to be the more pronounced the more complex the variance models became but converged fast, compare Tables \ref{tab:BGROGLSvarmod} and \ref{tab:BGROGLS300886varmod}. Despite the overall improvement in out-of-sample performance and the base approximation, they led to a deterioration in the relevant region for SCR estimation. The type II algorithms showed the effects of the type I algorithms in an amplified and more volatile way. While the type II routines under $300$-$886$ (= no actual restriction) constituted systematically the best choices except for on the extreme and challenging capital region set where their performance was still acceptable, there was no systematically best choice of variance model complexity due to the high variability accompanied by the type II routines under $300$-$886$. The best found FGLS routine reached with $258$ terms the highest observed number across all numerical experiments and outperformed all considered GLM and GAM routines in terms of combined Sobol and nested simulations validation. Furthermore, it reduced the bias on these two validation sets by far the most. This result spoke once more for a high interaction complexity of the CFP model. We concluded that the adaptive algorithm maintained its property to yield numerically stable and parsimonious proxy functions without requiring restriction settings in the FGLS context. Nonetheless, the bias of the best FGLS routine on the capital region set motivated the search for even more suitable regression methods and functional forms, see Figure \ref{fig:CapitalRegionResiduals}. For an overview of these results, see Table \ref{tab:BGROGLSallvalfig}.

\subsection*{MARS}
The classical and generalized MARS algorithms in Section \ref{sec:MARStheory} are special cases of respectively the OLS regression algorithm and GLM algorithm, in which the basis functions are hinge functions and variable selection is carried out subsequently in a forward and backward pass. While in the forward pass the proxy functions are built up with respect to the residual sum of squares as model selection criterion, in the backward pass they are cut back with respect to GCV where the degrees of freedom are modified to account for the knots in the hinge functions.

By applying a great variety of MARS algorithms in Section \ref{sec:MARSexperiment} in a two-step approach, we ensured that no comparably well suited MARS model would have been missed in our analysis. All tested MARS algorithms selected at maximum $148$ basis functions and showed rather poor out-of-sample performances as well as a weak extrapolation behavior compared to the previously discussed routines, see Table \ref{tab:BGROMARSRun2}. The conclusion was that MARS routines were not able to model the complex interactions in the CFP model appropriately.

\subsection*{Kernel Regression}
The kernel regression algorithm in Section \ref{sec:KRtheory} is a non-parametric local regression approach using a kernel as a weighting function. While at each target point the LC kernel estimator is given as the kernel-weighted average, the LL kernel estimator minimizes there the kernel-weighted residual sum of squares. For AIC to be evaluable at a kernel estimator, a non-parametric version accounting for the bandwidths is presented. As an alternative to AIC, non-parametric leave-one-out cross-validation LOO-CV is introduced. The bandwidths are selected such that they minimize the chosen model selection criterion. For reasons of computational efficiency, the adaptive basis function selection procedures need to be performed prior to the kernel regression approach.

Like we did with MARS, we applied numerous variants of kernel regression algorithms in Section \ref{sec:KRexperiment}. We found that the LL regression algorithms performed better than the LC ones but still worse than the previously discussed routines, see Table \ref{tab:KR}. We traced the rather poor out-of-sample performances back to an insufficient interaction modeling by too few basis functions and a poor behavior of local regression approaches close to and beyond the boundary of the fitting space.

\subsection*{Outlook}
In our slightly disguised real-world example and given LSMC setting, the adaptive OLS regression, GLM, GAM and FGLS regression algorithms turned out to be suitable machine learning methods for proxy modeling of life insurance companies with potential for both performance and computational efficiency gains by fine-tuning model hyperparameters and implementation designs. For recommendations of specific hyperparameter settings and designs, see the aforementioned suggestions. Differently, the MARS and kernel regression algorithms were not found to be convincing in our application. In order to study the robustness of our results, the approaches can be repeated in multiple other LSMC examples.

After all, none of our tested approaches was able to completely eliminate the bias observed in the validation figures and to yield consistent results across the three validation sets though. Investigations on whether these observations are systematic for the approaches, a result of the Monte Carlo error or a combination thereof help further narrow down the circle of recommended regression techniques. In order to assess the variance and bias of the proxy estimates conditional on an outer scenario, seed stability analyses in which the sets of fitting points are varied and convergence analyses in which sample size is increased need to be carried out. While such analyses would be computationally very costly, they would provide valuable insights into how to further improve approximation quality, that is, whether additional fitting points are necessary to reflect the underlying CFP model more accurately, whether more suitable functional forms and estimation assumptions are required for a more appropriate proxy modeling, or whether both aspects are relevant. Furthermore, one could deduce from such an analysis the sample sizes needed by the different regression algorithms to meet certain validation criteria. Since the generation of large sample sizes is currently computationally expensive for the industry, algorithms getting along with comparably few fitting points should be striven for.

Picking a suitable calibration algorithm is most important from the viewpoint of capturing the CFP model and hence the SCR appropriately. Therefore, if the bias observed in the validation figures indicates indeed issues with the functional forms of our approaches, doing further research on techniques not entailing such a bias or at least a smaller one is vital. On the one hand, one can fine-tune the approaches of this exposition and try different configurations thereof, and on the other hand, one can analyze further machine learning alternatives such as the ones mentioned in the introduction and already used in other LSMC applications. Ideally, various approaches like adaptive OLS regression, GLM, GAM and FGLS regression algorithms, artificial neural networks, tree-based methods and support vector machines would be fine-tuned and compared based on the same realistic and comprehensive data basis. Since the major challenges of machine learning calibration algorithms are hyperparameter selection and in some cases their dependence on randomness, future research should be dedicated to efficient hyperparameter search algorithms and stabilization methods such as ensemble methods.
\vspace{1cm}

\begin{Acknowledgements}\normalfont The first author would like to thank Christian Wei{\ss} for his valuable comments which greatly helped improve the paper. Furthermore, she is grateful to Magdalena Roth, Tamino Meyh\"{o}fer and her colleagues who have been supportive by providing her with academic time and computational resources.\end{Acknowledgements}
\newpage

\apptocmd{\thebibliography}{\csname References\endcsname\addcontentsline{toc}{section}{References}}{}{}
\bibliographystyle{agsm}
\bibliography{Stardust}
\newpage

\appendix
\section*{Appendix}
\addcontentsline{toc}{section}{Appendix}
\renewcommand{\baselinestretch}{1.2}

\newcolumntype{C}{>{\centering\arraybackslash}p{4mm}}
\begin{table}[htb]
  \tiny
  \begin{center}
  \tabcolsep=0.106cm
  \renewcommand{\arraystretch}{0.90}
      \begin{filecontents*}{BGROOLSadalgwre1.csv}
k;r1;r2;r3;r4;r5;r6;r7;r8;r9;r10;r11;r12;r13;r14;r15;beta;AIC;v.wre.y;ns.wre.y;cr.wre.y
0;0;0;0;0;0;0;0;0;0;0;0;0;0;0;0;14718.23756;437251.0284;4.556602;3.23053;4.02699629
1;0;0;0;0;0;0;0;1;0;0;0;0;0;0;0;7850.170275;386721.8775;2.4738689;0.845084;0.91309099
2;1;0;0;0;0;0;0;0;0;0;0;0;0;0;0;-269.3262983;375143.8396;2.0650148;2.139015;1.83147434
3;0;0;0;0;0;1;0;0;0;0;0;0;0;0;0;145.2104436;366567.289;1.6562468;0.443792;0.49630791
4;0;0;0;0;0;0;0;0;0;0;0;0;0;0;1;-5.364114744;358893.518;1.6467626;1.006377;0.55626101
5;0;0;0;0;0;0;1;0;0;0;0;0;0;0;0;434.035817;355731.9286;1.6347756;0.853357;0.46869164
6;1;0;0;0;0;0;0;1;0;0;0;0;0;0;0;1753.395453;354317.7979;1.6790248;0.956304;0.37400957
7;0;0;0;0;0;0;0;2;0;0;0;0;0;0;0;19145.78013;349758.5786;1.2340493;0.491057;0.62820691
8;2;0;0;0;0;0;0;0;0;0;0;0;0;0;0;33.32989246;347795.9634;0.9985358;0.339591;0.59410883
9;0;0;0;0;0;1;0;1;0;0;0;0;0;0;0;868.2484473;346443.9203;0.9119805;0.357314;0.60191958
10;0;0;0;0;0;0;0;1;0;0;0;0;0;0;1;30.58863404;345045.2474;0.8386126;0.388718;0.64994998
11;1;0;0;0;0;0;0;0;0;0;0;0;0;0;1;1.651426285;341082.8124;0.7587438;0.397868;0.46541559
12;0;1;0;0;0;0;0;0;0;0;0;0;0;0;0;86.78831694;339360.374;0.7178702;0.394231;0.38998224
13;1;0;0;0;0;1;0;0;0;0;0;0;0;0;0;33.35112791;337730.9855;0.5735332;0.653032;0.51187199
14;0;0;0;0;0;0;0;0;1;0;0;0;0;0;0;49.58931116;336842.8235;0.5889452;0.658094;0.51788433
15;0;0;0;0;0;0;0;0;0;0;0;1;0;0;0;71.25322367;335980.0391;0.6276231;0.678399;0.51220045
16;0;0;0;0;0;0;1;1;0;0;0;0;0;0;0;2667.92004;335351.1441;0.6085193;0.671499;0.50343614
17;1;0;0;0;0;0;1;0;0;0;0;0;0;0;0;96.43176931;334875.5029;0.5791571;0.700678;0.54495249
18;1;0;0;0;0;0;0;1;0;0;0;0;0;0;1;-6.308129423;334412.7937;0.5933914;0.719882;0.53111425
19;0;0;0;0;0;0;0;2;0;0;0;0;0;0;1;-47.09165961;333904.2185;0.5619575;0.620595;0.47362037
20;0;0;0;0;0;0;0;0;0;0;0;0;0;1;0;48.93239255;333447.4362;0.5648041;0.59708;0.45439558
21;1;0;0;0;0;0;0;2;0;0;0;0;0;0;0;-3412.683377;333116.2226;0.5534635;0.543166;0.40661664
22;0;0;0;0;0;0;0;0;0;0;0;0;0;0;2;0.018134234;332805.7679;0.5621878;0.477606;0.35839493
23;2;0;0;0;0;0;0;0;0;0;0;0;0;0;1;-0.117639258;332547.2224;0.5497891;0.449694;0.38124004
24;0;0;0;0;0;0;0;0;0;0;0;0;1;0;0;43.76801991;332293.8077;0.5445717;0.46758;0.37847101
25;0;0;1;0;0;0;0;0;0;0;0;0;0;0;0;118.939158;332042.1939;0.5300925;0.46364;0.36198325
26;0;0;1;0;0;0;0;1;0;0;0;0;0;0;0;-1288.446437;331686.6184;0.5217526;0.453302;0.35476415
27;1;0;1;0;0;0;0;0;0;0;0;0;0;0;0;-44.72140407;331404.843;0.5247251;0.443834;0.34290691
28;0;0;0;0;0;0;0;3;0;0;0;0;0;0;0;-24908.9923;331136.4147;0.4985978;0.404711;0.32693856
29;2;0;0;0;0;0;0;1;0;0;0;0;0;0;0;-86.88148988;330562.055;0.5037983;0.348409;0.26837872
30;0;0;0;0;0;1;0;0;0;0;0;0;0;0;1;0.550792762;330361.1506;0.5184664;0.41797;0.26442846
31;0;0;0;0;0;1;1;0;0;0;0;0;0;0;0;77.26068633;330163.0398;0.5122091;0.442838;0.27185191
32;1;0;0;0;0;0;0;0;1;0;0;0;0;0;0;24.77718347;329988.2028;0.5082752;0.442784;0.26411044
33;0;0;0;0;0;2;0;0;0;0;0;0;0;0;0;14.32669183;329834.2063;0.4766377;0.491447;0.2856491
34;0;1;0;0;0;0;0;0;0;0;0;0;0;0;1;-0.391536685;329687.797;0.4769757;0.499703;0.29008603
35;0;0;0;0;0;0;0;0;0;0;1;0;0;0;0;28.35812098;329550.3981;0.4759476;0.502168;0.29069359
36;0;1;0;0;0;0;0;1;0;0;0;0;0;0;0;-370.9160146;329441.9896;0.4721628;0.498611;0.28832454
37;1;1;0;0;0;0;0;0;0;0;0;0;0;0;0;-17.89944165;329146.5443;0.4624657;0.505377;0.30108613
38;0;0;0;1;0;0;0;0;0;0;0;0;0;0;0;8574.531304;329042.672;0.4722863;0.518349;0.29975566
39;0;0;0;0;0;1;0;1;0;0;0;0;0;0;1;-2.172057731;328934.797;0.4740133;0.510304;0.29532844
40;0;0;0;0;0;0;0;1;1;0;0;0;0;0;0;223.9053198;328832.2073;0.4749207;0.509415;0.29103128
41;0;0;0;0;0;1;0;2;0;0;0;0;0;0;0;-1801.730708;328732.9494;0.4551975;0.444528;0.24823832
42;1;0;0;0;0;1;0;1;0;0;0;0;0;0;0;-102.1033956;327927.0187;0.3719387;0.34547;0.23732419
43;0;0;0;0;0;0;1;0;0;0;0;0;0;0;1;0.696377645;327857.5891;0.3681123;0.353053;0.23519227
44;0;0;0;0;0;0;0;0;1;0;0;0;0;0;1;0.55601722;327792.2809;0.3664425;0.352132;0.23280702
45;1;0;0;1;0;0;0;0;0;0;0;0;0;0;0;-3034.323719;327728.9214;0.365056;0.356124;0.2283351
46;0;0;0;1;0;0;0;1;0;0;0;0;0;0;0;-13127.80808;327659.061;0.3675825;0.36443;0.22679792
47;1;0;0;0;0;0;0;0;0;0;0;1;0;0;0;-17.53908064;327603.2303;0.3682433;0.366201;0.2259735
48;0;0;0;0;0;0;0;1;0;0;0;1;0;0;0;-187.072963;327537.0309;0.374316;0.36744;0.22628398
49;0;0;0;0;0;1;1;1;0;0;0;0;0;0;0;-300.537927;327482.8636;0.3688262;0.366659;0.23044851
50;1;0;0;0;0;1;0;0;0;0;0;0;0;0;1;-0.089229959;327432.0902;0.3684989;0.390834;0.22138584
51;0;0;0;0;0;2;0;1;0;0;0;0;0;0;0;-60.83894181;327381.5018;0.3589157;0.389841;0.22771688
52;0;0;1;0;0;1;0;0;0;0;0;0;0;0;0;-20.90549361;327331.0188;0.3518843;0.389732;0.22502614
53;1;0;0;0;0;0;0;0;0;0;0;0;0;0;2;-0.003742602;327286.7993;0.3462766;0.376815;0.20560833
54;0;0;0;0;0;0;0;1;0;0;0;0;0;0;2;-0.085356514;327148.6298;0.338558;0.357072;0.18545478
55;2;0;0;0;0;1;0;0;0;0;0;0;0;0;0;1.437629622;327104.9948;0.3152857;0.321372;0.17282236
56;0;0;1;0;0;0;0;0;0;0;0;0;0;0;1;-0.495645414;327064.1088;0.3150275;0.322221;0.17269407
57;1;0;0;0;0;0;0;0;0;0;0;0;0;1;0;-6.056014476;327024.7437;0.3218915;0.317391;0.17538749
58;0;0;0;0;0;0;1;2;0;0;0;0;0;0;0;-6600.488118;326986.4282;0.3169237;0.309964;0.17246547
59;1;0;0;0;0;0;1;1;0;0;0;0;0;0;0;-407.5703579;326823.039;0.3080044;0.302133;0.1829822
60;0;0;1;0;0;0;0;2;0;0;0;0;0;0;0;3378.821029;326787.1036;0.3059923;0.300561;0.18312963
61;1;0;1;0;0;0;0;1;0;0;0;0;0;0;0;205.2774272;326732.5213;0.3038349;0.299148;0.18338547
62;0;1;0;0;0;0;0;0;1;0;0;0;0;0;0;-18.73416971;326699.6144;0.3063613;0.298664;0.18199868
63;0;0;1;0;0;1;0;1;0;0;0;0;0;0;0;175.3918028;326667.632;0.3042724;0.295646;0.18168407
64;0;0;0;0;0;0;0;0;0;0;0;1;0;0;1;-0.199523235;326638.3511;0.3039436;0.298171;0.18143757
65;0;1;0;0;0;0;0;1;0;0;0;0;0;0;1;2.453504236;326610.2439;0.3014272;0.296173;0.18307836
66;1;1;0;0;0;0;0;0;0;0;0;0;0;0;1;0.105437423;326572.1693;0.2968831;0.298762;0.17989452
67;2;0;0;0;0;1;0;1;0;0;0;0;0;0;0;-13.02145296;326545.0668;0.2916089;0.286224;0.16873258
68;1;1;0;0;0;0;0;1;0;0;0;0;0;0;0;93.69420043;326518.8532;0.2922817;0.287477;0.17180456
69;0;1;0;0;0;0;0;2;0;0;0;0;0;0;0;891.579186;326477.8636;0.2937254;0.282324;0.17346009
70;0;0;0;0;0;0;1;1;0;0;0;0;0;0;1;-6.214280718;326452.7548;0.2912463;0.281253;0.17507827
71;0;0;0;0;0;0;0;1;0;0;1;0;0;0;0;-112.5575417;326428.4006;0.2889505;0.281426;0.17557284
72;1;0;0;0;0;0;0;0;0;0;1;0;0;0;0;-5.268777967;326398.3201;0.2844463;0.281641;0.17252043
73;1;0;0;0;0;0;0;3;0;0;0;0;0;0;0;1129.770078;326374.036;0.2761004;0.26364;0.16223546
74;1;0;0;0;0;0;1;0;0;0;0;0;0;0;1;-0.291340222;326351.8364;0.2715004;0.266499;0.15765243
75;1;0;0;0;0;0;0;1;1;0;0;0;0;0;0;-56.53512998;326330.9222;0.269209;0.265842;0.15681253
\end{filecontents*}
\pgfplotstabletypeset[
        multicolumn names, 
        col sep=semicolon, 
        display columns/0/.style={column name=$\bm{k}$,dec sep align},
        display columns/1/.style={column type=C,column name=$\bm{r_{k}^{1}}$},
        display columns/2/.style={column type=C,column name=$\bm{r_{k}^{2}}$},
        display columns/3/.style={column type=C,column name=$\bm{r_{k}^{3}}$},
        display columns/4/.style={column type=C,column name=$\bm{r_{k}^{4}}$},
        display columns/5/.style={column type=C,column name=$\bm{r_{k}^{5}}$},
        display columns/6/.style={column type=C,column name=$\bm{r_{k}^{6}}$},
        display columns/7/.style={column type=C,column name=$\bm{r_{k}^{7}}$},
        display columns/8/.style={column type=C,column name=$\bm{r_{k}^{8}}$},
        display columns/9/.style={column type=C,column name=$\bm{r_{k}^{9}}$},
        display columns/10/.style={column type=C,column name=$\bm{r_{k}^{10}}$},
        display columns/11/.style={column type=C,column name=$\bm{r_{k}^{11}}$},
        display columns/12/.style={column type=C,column name=$\bm{r_{k}^{12}}$},
        display columns/13/.style={column type=C,column name=$\bm{r_{k}^{13}}$},
        display columns/14/.style={column type=C,column name=$\bm{r_{k}^{14}}$},
        display columns/15/.style={column type=C,column name=$\bm{r_{k}^{15}}$},
        display columns/16/.style={column name=$\bm{\widehat{\beta}_{\textbf{\text{OLS}},k}}$,fixed,fixed zerofill,precision=2,set thousands separator={,},dec sep align},
        display columns/17/.style={column name={\textbf{AIC}},fixed,fixed zerofill,precision=0,set thousands separator={,},dec sep align},
        display columns/18/.style={column name={\textbf{v.mae}},fixed,fixed zerofill,precision=3,set thousands separator={,},dec sep align},
        display columns/19/.style={column name={\textbf{ns.mae}},fixed,fixed zerofill,precision=3,set thousands separator={,},dec sep align},
        display columns/20/.style={column name={\textbf{cr.mae}},fixed,fixed zerofill,precision=3,set thousands separator={,},dec sep align},
	    	every head row/.style={before row=\toprule,after row=\midrule},
	    	every row no 1/.style={before row=\midrule},
	    	every nth row={10[+1]}{before row=\midrule},
      	every last row/.style={after row=\bottomrule},
      ]{BGROOLSadalgwre1.csv} 
    \caption{OLS proxy function of BEL derived under $150$-$443$ in the adaptive algorithm with the final coefficients. Furthermore, AIC scores and out-of-sample MAEs in $\%$ after each iteration.}
    \label{tab:BGROOLS1}
  \end{center}
\end{table}

\begin{table}[htb]
\ContinuedFloat
  \tiny
  \begin{center}
  \tabcolsep=0.106cm
  \renewcommand{\arraystretch}{0.90}
      \begin{filecontents*}{BGROOLSadalgwre2.csv}
k;r1;r2;r3;r4;r5;r6;r7;r8;r9;r10;r11;r12;r13;r14;r15;beta;AIC;v.wre.y;ns.wre.y;cr.wre.y
76;2;0;0;0;0;0;0;0;1;0;0;0;0;0;0;-3.018803731;326312.806;0.2707765;0.266302;0.15513565
77;1;0;0;0;0;1;1;0;0;0;0;0;0;0;0;-10.58857351;326294.6974;0.2644754;0.270402;0.15125574
78;0;1;0;0;0;1;0;0;0;0;0;0;0;0;0;-6.990571848;326277.9291;0.2643033;0.274868;0.15250997
79;1;0;0;0;0;2;0;0;0;0;0;0;0;0;0;-2.250120718;326261.2692;0.2522373;0.284877;0.1537699
80;0;0;0;0;0;0;0;0;2;0;0;0;0;0;0;-14.7679623;326245.0608;0.2626108;0.308787;0.15723232
81;2;1;0;0;0;0;0;0;0;0;0;0;0;0;0;1.94942285;326229.1837;0.2671515;0.305669;0.15496196
82;0;1;0;1;0;0;0;0;0;0;0;0;0;0;0;2248.537128;326214.1354;0.2663075;0.3066;0.15587825
83;0;0;0;0;0;0;0;3;0;0;0;0;0;0;1;-111.7661887;326200.5001;0.2631118;0.301523;0.15820325
84;1;0;0;0;0;0;0;0;1;0;0;0;0;0;1;-0.113240581;326187.0709;0.2622106;0.302094;0.15680625
85;0;0;0;0;0;0;0;0;0;0;0;0;1;0;1;-0.17975695;326173.6812;0.2631535;0.304627;0.1557477
86;0;1;0;0;0;1;0;1;0;0;0;0;0;0;0;45.57524828;326160.9026;0.2649574;0.303215;0.1567363
87;0;0;0;1;0;0;0;2;0;0;0;0;0;0;0;-83291.88928;326148.6568;0.2672798;0.307622;0.15606296
88;0;0;1;0;0;0;1;0;0;0;0;0;0;0;0;-56.19707818;326136.5849;0.2672232;0.30754;0.15577187
89;1;0;0;0;0;0;0;0;0;0;0;0;1;0;0;-5.316657257;326126.0939;0.2666623;0.309883;0.15571433
90;0;0;0;0;0;2;1;0;0;0;0;0;0;0;0;-10.87486001;326115.6586;0.2673225;0.313173;0.15836889
91;0;0;0;1;0;0;0;0;0;0;0;0;0;0;1;-32.75072803;326105.8157;0.2651858;0.316666;0.15804356
92;0;0;0;0;0;0;0;2;0;0;0;0;0;0;2;-0.08752715;326097.3265;0.2653569;0.308276;0.15055448
93;0;1;0;0;0;0;0;0;0;0;0;0;0;1;0;10.86529147;326088.9526;0.2647319;0.308212;0.15111561
94;1;0;0;0;0;1;1;1;0;0;0;0;0;0;0;-48.93337884;326080.7093;0.2635902;0.306109;0.14777798
95;0;0;0;0;0;0;2;0;0;0;0;0;0;0;0;69.57419419;326072.8235;0.2559173;0.288382;0.14142352
96;0;0;0;1;0;0;0;3;0;0;0;0;0;0;0;-542688.1867;326065.5322;0.2561173;0.288959;0.14110021
97;0;0;0;0;0;0;0;0;0;0;0;0;0;2;0;10.43510251;326058.3098;0.24765;0.275445;0.13606405
98;0;0;0;0;0;0;0;1;1;0;0;0;0;0;1;-1.076034855;326051.1906;0.2483201;0.275665;0.1358892
99;0;0;1;0;0;0;1;1;0;0;0;0;0;0;0;419.0492315;326044.6673;0.2487559;0.275445;0.13609492
100;0;1;1;0;0;0;0;0;0;0;0;0;0;0;0;12.79709265;326038.3955;0.2495997;0.275657;0.13576231
101;0;0;0;0;0;1;0;0;0;0;1;0;0;0;0;-3.940438245;326032.7551;0.2502623;0.276236;0.13560358
102;1;0;0;0;0;0;0;2;0;0;0;0;0;0;1;-10.1211971;326027.1998;0.2477768;0.280788;0.13776766
103;2;0;0;0;0;0;0;1;0;0;0;0;0;0;1;-0.363770808;326017.4618;0.2444871;0.283073;0.13498913
104;0;0;1;0;0;0;0;1;0;0;0;0;0;0;1;1.738476714;326011.5086;0.2440426;0.282417;0.13557825
105;0;0;0;0;0;0;0;0;0;0;0;0;0;0;3;-3.23E-05;326006.0614;0.2417151;0.268432;0.13215556
106;2;0;0;0;0;0;1;0;0;0;0;0;0;0;0;-7.085845528;326000.6255;0.2383197;0.265458;0.13122807
107;2;0;0;0;0;0;1;1;0;0;0;0;0;0;0;-109.4560786;325982.3515;0.237758;0.263302;0.12893856
108;0;0;0;0;0;0;0;0;0;0;1;0;0;0;1;-0.100999584;325977.363;0.2374842;0.263333;0.12839481
109;0;1;0;0;0;0;0;0;0;0;0;1;0;0;0;5.763991912;325972.4519;0.2351102;0.263389;0.1287256
110;1;0;0;0;0;0;0;1;0;0;1;0;0;0;0;54.5088643;325967.8195;0.2367017;0.263586;0.1291863
111;1;0;0;0;0;0;1;2;0;0;0;0;0;0;0;-1386.729174;325963.217;0.2354816;0.264381;0.12938195
112;0;0;0;0;0;0;0;0;1;0;0;0;0;0;2;-0.002204779;325958.8503;0.2368094;0.264654;0.13017395
113;0;1;0;0;0;0;0;0;1;0;0;0;0;0;1;0.108361944;325955.0744;0.2352776;0.264752;0.1299097
114;0;1;0;0;0;1;0;0;0;0;0;0;0;0;1;0.052071773;325951.4298;0.2341433;0.26642;0.13004019
115;1;0;1;0;0;1;0;0;0;0;0;0;0;0;0;4.295165287;325947.7678;0.2359483;0.265182;0.12713471
116;1;0;0;0;0;2;0;1;0;0;0;0;0;0;0;-19.81062009;325944.4243;0.2372127;0.262176;0.12599667
117;2;0;0;0;0;2;0;0;0;0;0;0;0;0;0;-0.865676891;325937.8194;0.2411466;0.266637;0.12379891
118;0;1;0;0;0;1;0;1;0;0;0;0;0;0;1;-0.361377589;325934.6622;0.2409495;0.266729;0.12449166
119;0;1;1;0;0;0;0;1;0;0;0;0;0;0;0;-80.29186716;325931.3702;0.2412053;0.266606;0.12452601
120;0;0;0;0;0;0;0;0;1;0;0;0;1;0;0;-6.953758741;325928.4071;0.2407819;0.266658;0.1242983
121;0;0;0;0;0;1;0;0;0;0;0;0;0;0;2;-0.000883843;325925.4889;0.2426842;0.25874;0.12056789
122;0;0;0;0;0;0;0;2;0;0;1;0;0;0;0;436.5596471;325922.7801;0.2409452;0.2586;0.12120498
123;0;0;0;0;0;2;0;0;0;0;0;0;0;0;1;-0.031510356;325920.2659;0.2426163;0.263262;0.12107269
124;0;0;0;0;0;1;0;0;1;0;0;0;0;0;0;2.988998888;325917.935;0.2419631;0.262764;0.1203935
125;1;0;0;0;0;1;0;1;0;0;0;0;0;0;1;-0.593812872;325915.7393;0.2412079;0.260618;0.11931734
126;2;0;0;0;0;1;0;0;0;0;0;0;0;0;1;-0.023734047;325908.4727;0.2471174;0.264745;0.12397021
127;0;0;0;0;0;1;0;2;0;0;0;0;0;0;1;-4.65702713;325902.0655;0.2493741;0.279167;0.12325458
128;0;0;0;0;0;0;1;3;0;0;0;0;0;0;0;-8179.681842;325900.0814;0.2494525;0.280192;0.12416067
129;0;0;0;0;0;1;0;3;0;0;0;0;0;0;0;691.4032004;325898.1302;0.2491477;0.280208;0.12301807
130;1;0;0;0;0;0;0;0;0;0;0;0;1;0;1;0.036355635;325896.2645;0.2500524;0.281126;0.12249777
131;0;0;0;0;0;0;0;0;0;1;0;0;0;0;0;7.04324373;325894.4565;0.2459405;0.264353;0.12039654
132;0;0;1;0;0;0;0;0;0;1;0;0;0;0;0;-27.71753805;325892.274;0.2470996;0.264184;0.11898739
133;2;0;0;0;0;0;0;0;0;0;1;0;0;0;0;1.255732704;325890.6451;0.2472813;0.263913;0.11876762
134;0;0;0;0;0;1;0;0;0;0;0;1;0;0;0;-2.673229914;325889.0274;0.2491554;0.265135;0.11793071
135;1;0;0;0;0;1;0;0;0;0;0;1;0;0;0;1.526037133;325887.0104;0.2503738;0.266094;0.11857341
136;0;0;0;0;0;0;0;0;0;0;0;0;0;1;1;-0.071440938;325885.4369;0.249942;0.265196;0.11960413
137;1;0;0;0;0;0;0;1;0;0;0;1;0;0;0;40.44133305;325883.8708;0.2505697;0.265491;0.11905005
138;0;0;0;0;0;0;0;2;0;0;0;1;0;0;0;434.5031927;325878.084;0.2494132;0.264126;0.11851469
139;0;0;0;0;0;0;0;0;1;0;0;1;0;0;0;-5.992535973;325876.5637;0.248154;0.264097;0.1188042
140;0;0;0;0;0;0;0;0;2;0;0;1;0;0;0;14.63657629;325872.6555;0.2462088;0.262717;0.1196701
141;0;0;0;0;0;2;0;2;0;0;0;0;0;0;0;-119.421544;325871.1949;0.2469842;0.269636;0.1210609
142;0;0;0;0;0;0;0;1;0;0;0;0;0;0;3;0.000175158;325869.7476;0.2476105;0.270571;0.1209917
143;1;0;0;0;0;0;0;0;0;0;0;1;0;0;1;0.073275202;325868.3132;0.2475406;0.271206;0.12065139
144;0;0;0;0;0;0;0;1;0;0;0;1;0;0;1;1.061053216;325860.7124;0.2461969;0.271434;0.12080655
145;1;0;0;0;0;0;1;1;0;0;0;0;0;0;1;-0.738876476;325859.2523;0.2468653;0.271008;0.12075376
146;0;0;0;0;0;0;0;0;1;0;0;0;0;1;0;-5.614581822;325857.9621;0.2461839;0.270591;0.12117554
147;0;1;0;0;0;0;0;0;0;0;0;0;0;1;1;-0.081989373;325856.6707;0.2465515;0.270461;0.1214354
148;0;0;0;0;0;0;1;0;0;0;0;0;1;0;0;-37.16486937;325855.39;0.2466194;0.270768;0.12167671
149;0;0;0;0;0;0;1;0;0;0;0;0;1;0;1;0.412288862;325851.0706;0.247278;0.270834;0.12198138
150;0;1;0;1;0;0;0;1;0;0;0;0;0;0;0;-7290.991386;325849.7177;0.247487;0.271045;0.12219633
\end{filecontents*}
\pgfplotstabletypeset[
        multicolumn names, 
        col sep=semicolon, 
        display columns/0/.style={column name=$\bm{k}$,dec sep align},
        display columns/1/.style={column type=C,column name=$\bm{r_{k}^{1}}$},
        display columns/2/.style={column type=C,column name=$\bm{r_{k}^{2}}$},
        display columns/3/.style={column type=C,column name=$\bm{r_{k}^{3}}$},
        display columns/4/.style={column type=C,column name=$\bm{r_{k}^{4}}$},
        display columns/5/.style={column type=C,column name=$\bm{r_{k}^{5}}$},
        display columns/6/.style={column type=C,column name=$\bm{r_{k}^{6}}$},
        display columns/7/.style={column type=C,column name=$\bm{r_{k}^{7}}$},
        display columns/8/.style={column type=C,column name=$\bm{r_{k}^{8}}$},
        display columns/9/.style={column type=C,column name=$\bm{r_{k}^{9}}$},
        display columns/10/.style={column type=C,column name=$\bm{r_{k}^{10}}$},
        display columns/11/.style={column type=C,column name=$\bm{r_{k}^{11}}$},
        display columns/12/.style={column type=C,column name=$\bm{r_{k}^{12}}$},
        display columns/13/.style={column type=C,column name=$\bm{r_{k}^{13}}$},
        display columns/14/.style={column type=C,column name=$\bm{r_{k}^{14}}$},
        display columns/15/.style={column type=C,column name=$\bm{r_{k}^{15}}$},
        display columns/16/.style={column name=$\bm{\widehat{\beta}_{\textbf{\text{OLS}},k}}$,fixed,fixed zerofill,precision=2,set thousands separator={,},dec sep align},
        display columns/17/.style={column name={\textbf{AIC}},fixed,fixed zerofill,precision=0,set thousands separator={,},dec sep align},
        display columns/18/.style={column name={\textbf{v.mae}},fixed,fixed zerofill,precision=3,set thousands separator={,},dec sep align},
        display columns/19/.style={column name={\textbf{ns.mae}},fixed,fixed zerofill,precision=3,set thousands separator={,},dec sep align},
        display columns/20/.style={column name={\textbf{cr.mae}},fixed,fixed zerofill,precision=3,set thousands separator={,},dec sep align},
	    	every head row/.style={before row=\toprule,after row=\midrule},
	    	every nth row={10[+5]}{before row=\midrule},
      	every last row/.style={after row=\bottomrule},
      ]{BGROOLSadalgwre2.csv} 
    \caption[]{Cont.}
    \label{tab:BGROOLS2}
  \end{center}
\end{table}

\begin{table}[htb]
  \tiny
  \begin{center}
  \tabcolsep=0.106cm
  \renewcommand{\arraystretch}{0.90}
      \begin{filecontents*}{PGROOLSadalgwre1.csv}
k;r1;r2;r3;r4;r5;r6;r7;r8;r9;r10;r11;r12;r13;r14;r15;beta;AIC;v.wre.y;ns.wre.y;cr.wre.y
0;0;0;0;0;0;0;0;0;0;0;0;0;0;0;0;745.3487958;391374.6779;60.619885;97.518451;257.761878
1;0;0;0;0;0;0;0;1;0;0;0;0;0;0;0;5766.613319;382610.1336;50.401757;99.306094;256.7893996
2;1;0;0;0;0;0;0;0;0;0;0;0;0;0;0;272.7475969;367667.0873;35.285037;38.123512;99.9021971
3;0;0;0;0;0;0;0;0;0;0;0;0;0;0;1;5.457477253;359997.2406;30.739238;18.209629;72.7189605
4;0;0;0;0;0;1;0;0;0;0;0;0;0;0;0;128.4148007;356705.1315;30.118867;25.087674;29.3571584
5;1;0;0;0;0;0;0;1;0;0;0;0;0;0;0;-1750.717586;355353.8996;30.867075;28.172812;21.8700427
6;0;0;0;0;0;0;0;2;0;0;0;0;0;0;0;-19127.26556;351002.4498;22.942162;14.947737;44.6684657
7;2;0;0;0;0;0;0;0;0;0;0;0;0;0;0;-33.25061478;349146.9508;19.029697;12.141773;42.5348616
8;0;0;0;0;0;0;1;0;0;0;0;0;0;0;0;307.3195404;347777.2487;18.221074;10.928423;35.4196768
9;0;0;0;0;0;1;0;1;0;0;0;0;0;0;0;-868.0491318;346423.1068;16.662386;11.526572;35.9406719
10;0;1;0;0;0;0;0;0;0;0;0;0;0;0;0;-87.54250027;345025.3299;15.986842;10.264416;31.4613306
11;0;0;0;0;0;0;0;1;0;0;0;0;0;0;1;-30.51258504;343569.7614;14.858333;11.186831;34.5019961
12;1;0;0;0;0;0;0;0;0;0;0;0;0;0;1;-1.656599426;339282.0543;13.09232;12.669138;23.1739686
13;1;0;0;0;0;1;0;0;0;0;0;0;0;0;0;-33.33339154;337648.4279;10.427032;20.975668;30.4017974
14;0;0;0;0;0;0;0;0;0;0;0;1;0;0;0;-70.63150879;336840.1475;11.086689;21.598269;29.9722019
15;0;0;0;0;0;0;0;0;1;0;0;0;0;0;0;-41.3717284;336120.0608;11.436395;21.764195;30.408183
16;0;0;0;0;0;0;1;1;0;0;0;0;0;0;0;-2666.438628;335494.9023;11.088095;21.543147;29.8900689
17;1;0;0;0;0;0;1;0;0;0;0;0;0;0;0;-96.47956122;335021.5869;10.545152;22.478934;32.3344069
18;1;0;0;0;0;0;0;1;0;0;0;0;0;0;1;6.298721897;334562.8449;10.804362;23.09474;31.5185518
19;0;0;0;0;0;0;0;2;0;0;0;0;0;0;1;47.02400313;334058.1378;10.232254;19.912622;28.1279222
20;0;0;0;0;0;0;0;0;0;0;0;0;0;1;0;-48.77007051;333609.6282;10.291748;19.163175;26.9953565
21;1;0;0;0;0;0;0;2;0;0;0;0;0;0;0;3412.540718;333280.8203;10.083364;17.437529;24.1897285
22;0;0;0;0;0;0;0;0;0;0;0;0;0;0;2;-0.018127923;332970.3556;10.246117;15.328378;21.3258872
23;2;0;0;0;0;0;0;0;0;0;0;0;0;0;1;0.117408683;332713.8549;10.020177;14.435682;22.6706471
24;0;0;1;0;0;0;0;0;0;0;0;0;0;0;0;-120.6819517;332457.374;9.833627;14.282922;21.6081564
25;0;0;1;0;0;0;0;1;0;0;0;0;0;0;0;1287.627362;332108.2031;9.724615;13.969374;21.2726896
26;1;0;1;0;0;0;0;0;0;0;0;0;0;0;0;44.70549696;331831.9618;9.754871;13.660895;20.5008787
27;0;0;0;0;0;0;0;3;0;0;0;0;0;0;0;24899.66005;331569.4923;9.275013;12.461801;19.873182
28;2;0;0;0;0;0;0;1;0;0;0;0;0;0;0;87.04123745;331003.7281;9.292295;10.757103;17.021517
29;0;0;0;0;0;0;0;0;0;0;0;0;1;0;0;-43.38242341;330742.4406;9.171068;11.183257;16.0228382
30;0;0;0;0;0;1;0;0;0;0;0;0;0;0;1;-0.55189418;330542.6927;9.443616;13.408763;15.7662238
31;0;0;0;0;0;1;1;0;0;0;0;0;0;0;0;-77.35337285;330345.4395;9.324491;14.207183;16.1919909
32;1;0;0;0;0;0;0;0;1;0;0;0;0;0;0;-25.20087659;330160.7497;9.245746;14.202642;15.6918209
33;0;0;0;0;0;2;0;0;0;0;0;0;0;0;0;-14.3679492;330007.4669;8.67228;15.764175;16.9635025
34;0;1;0;0;0;0;0;0;0;0;0;0;0;0;1;0.39466766;329858.9354;8.682458;16.031191;17.2226957
35;0;0;0;0;0;0;0;0;0;0;1;0;0;0;0;-27.79648761;329728.2174;8.665241;16.109505;17.2637553
36;0;0;0;1;0;0;0;0;0;0;0;0;0;0;0;-8757.491412;329618.5713;8.870773;16.530087;17.0045356
37;0;0;0;0;0;1;0;1;0;0;0;0;0;0;1;2.169846125;329513.0433;8.936681;16.276449;16.7902106
38;0;1;0;0;0;0;0;1;0;0;0;0;0;0;0;369.1646727;329407.7061;8.842072;16.168864;16.7383549
39;1;1;0;0;0;0;0;0;0;0;0;0;0;0;0;17.96757493;329108.9152;8.636962;16.387355;17.5265177
40;0;0;0;0;0;0;0;1;1;0;0;0;0;0;0;-222.5492742;329008.0507;8.656012;16.35901;17.2712312
41;0;0;0;0;0;1;0;2;0;0;0;0;0;0;0;1791.695434;328909.936;8.297143;14.282343;14.7477689
42;1;0;0;0;0;1;0;1;0;0;0;0;0;0;0;101.2322046;328110.6529;6.783188;11.112373;14.1436998
43;0;0;0;0;0;0;1;0;0;0;0;0;0;0;1;-0.698647806;328041.3999;6.713247;11.355389;14.0134841
44;0;0;0;0;0;0;0;0;1;0;0;0;0;0;1;-0.56703732;327972.0514;6.683461;11.324723;13.8671715
45;1;0;0;1;0;0;0;0;0;0;0;0;0;0;0;3083.046565;327905.1639;6.654448;11.456348;13.594687
46;0;0;0;1;0;0;0;1;0;0;0;0;0;0;0;12863.79387;327836.7381;6.700217;11.720604;13.5001027
47;1;0;0;0;0;0;0;0;0;0;0;1;0;0;0;17.77758633;327779.9126;6.710456;11.777369;13.4497985
48;0;0;0;0;0;0;0;1;0;0;0;1;0;0;0;190.4633002;327710.6205;6.824068;11.817715;13.4680595
49;0;0;0;0;0;1;1;1;0;0;0;0;0;0;0;300.7649745;327656.7963;6.723924;11.792827;13.7157867
50;1;0;0;0;0;1;0;0;0;0;0;0;0;0;1;0.088871535;327606.859;6.717981;12.565016;13.1823811
51;0;0;0;0;0;2;0;1;0;0;0;0;0;0;0;60.82627963;327556.7386;6.543407;12.533277;13.5578415
52;0;0;1;0;0;1;0;0;0;0;0;0;0;0;0;20.90545994;327506.6202;6.415192;12.529993;13.3939289
53;1;0;0;0;0;0;0;0;0;0;0;0;0;0;2;0.003724268;327463.2777;6.313557;12.117699;12.2519305
54;0;0;0;0;0;0;0;1;0;0;0;0;0;0;2;0.084756394;327327.1393;6.175599;11.485663;11.0489599
55;2;0;0;0;0;1;0;0;0;0;0;0;0;0;0;-1.457815173;327283.8039;5.751357;10.339064;10.294833
56;0;0;1;0;0;0;0;0;0;0;0;0;0;0;1;0.502782823;327242.0873;5.746232;10.366768;10.2868152
57;1;0;0;0;0;0;0;0;0;0;0;0;0;1;0;6.077269608;327202.7313;5.87051;10.211417;10.449654
58;0;0;0;0;0;0;1;2;0;0;0;0;0;0;0;6593.976146;327164.6166;5.780472;9.972774;10.2735631
59;1;0;0;0;0;0;1;1;0;0;0;0;0;0;0;406.7263379;327003.0817;5.617524;9.722103;10.8969986
60;0;0;1;0;0;0;0;2;0;0;0;0;0;0;0;-3364.020245;326967.5784;5.580913;9.671341;10.9039555
61;1;0;1;0;0;0;0;1;0;0;0;0;0;0;0;-204.1235183;326914.0422;5.541798;9.62631;10.9205093
62;0;1;0;0;0;0;0;0;1;0;0;0;0;0;0;18.89770286;326880.7736;5.588272;9.610712;10.8374604
63;0;0;1;0;0;1;0;1;0;0;0;0;0;0;0;-175.1723912;326849.1727;5.5462;9.51412;10.8167834
64;0;0;0;0;0;0;0;0;0;0;0;1;0;0;1;0.205490224;326818.3718;5.540039;9.59654;10.7989357
65;0;1;0;0;0;0;0;1;0;0;0;0;0;0;1;-2.443231853;326790.6295;5.494267;9.53246;10.8960165
66;1;1;0;0;0;0;0;0;0;0;0;0;0;0;1;-0.105492836;326752.8253;5.41275;9.615933;10.708144
67;2;0;0;0;0;1;0;1;0;0;0;0;0;0;0;12.99022347;326726.1225;5.316847;9.215011;10.0459342
68;1;1;0;0;0;0;0;1;0;0;0;0;0;0;0;-93.57091434;326700.1715;5.329103;9.255211;10.2312187
69;0;1;0;0;0;0;0;2;0;0;0;0;0;0;0;-890.6181786;326659.5343;5.355415;9.089716;10.3260144
70;0;0;0;0;0;0;0;1;0;0;1;0;0;0;0;113.0402758;326634.7648;5.313013;9.095344;10.3572533
71;1;0;0;0;0;0;0;0;0;0;1;0;0;0;0;5.231741611;326605.246;5.231262;9.101282;10.1639859
72;0;0;0;0;0;0;1;1;0;0;0;0;0;0;1;6.199237346;326580.552;5.186329;9.06787;10.2647291
73;1;0;0;0;0;0;0;3;0;0;0;0;0;0;0;-1133.834649;326556.2706;5.033599;8.487933;9.6473013
74;1;0;0;0;0;0;1;0;0;0;0;0;0;0;1;0.291725838;326534.1794;4.949603;8.57979;9.3744341
75;1;0;0;0;0;0;0;1;1;0;0;0;0;0;0;56.56028953;326513.4404;4.907829;8.558644;9.3225041
\end{filecontents*}
\pgfplotstabletypeset[
        multicolumn names, 
        col sep=semicolon, 
        display columns/0/.style={column name=$\bm{k}$,dec sep align},
        display columns/1/.style={column type=C,column name=$\bm{r_{k}^{1}}$},
        display columns/2/.style={column type=C,column name=$\bm{r_{k}^{2}}$},
        display columns/3/.style={column type=C,column name=$\bm{r_{k}^{3}}$},
        display columns/4/.style={column type=C,column name=$\bm{r_{k}^{4}}$},
        display columns/5/.style={column type=C,column name=$\bm{r_{k}^{5}}$},
        display columns/6/.style={column type=C,column name=$\bm{r_{k}^{6}}$},
        display columns/7/.style={column type=C,column name=$\bm{r_{k}^{7}}$},
        display columns/8/.style={column type=C,column name=$\bm{r_{k}^{8}}$},
        display columns/9/.style={column type=C,column name=$\bm{r_{k}^{9}}$},
        display columns/10/.style={column type=C,column name=$\bm{r_{k}^{10}}$},
        display columns/11/.style={column type=C,column name=$\bm{r_{k}^{11}}$},
        display columns/12/.style={column type=C,column name=$\bm{r_{k}^{12}}$},
        display columns/13/.style={column type=C,column name=$\bm{r_{k}^{13}}$},
        display columns/14/.style={column type=C,column name=$\bm{r_{k}^{14}}$},
        display columns/15/.style={column type=C,column name=$\bm{r_{k}^{15}}$},
        display columns/16/.style={column name=$\bm{\widehat{\beta}_{\textbf{\text{OLS}},k}}$,fixed,fixed zerofill,precision=2,set thousands separator={,},dec sep align},
        display columns/17/.style={column name={\textbf{AIC}},fixed,fixed zerofill,precision=0,set thousands separator={,},dec sep align},
        display columns/18/.style={column name={\textbf{v.mae}},fixed,fixed zerofill,precision=3,set thousands separator={,},dec sep align},
        display columns/19/.style={column name={\textbf{ns.mae}},fixed,fixed zerofill,precision=3,set thousands separator={,},dec sep align},
        display columns/20/.style={column name={\textbf{cr.mae}},fixed,fixed zerofill,precision=3,set thousands separator={,},dec sep align},
	    	every head row/.style={before row=\toprule,after row=\midrule},
	    	every row no 1/.style={before row=\midrule},
	    	every nth row={10[+1]}{before row=\midrule},
      	every last row/.style={after row=\bottomrule},
      ]{PGROOLSadalgwre1.csv} 
    \caption{OLS proxy function of AC derived under $150$-$443$ in the adaptive algorithm with the final coefficients. Furthermore, AIC scores and out-of-sample MAEs in $\%$ after each iteration.}
    \label{tab:PGROOLS1}
  \end{center}
\end{table}

\begin{table}[htb]
\ContinuedFloat
  \tiny
  \begin{center}
  \tabcolsep=0.106cm
  \renewcommand{\arraystretch}{0.90}
      \begin{filecontents*}{PGROOLSadalgwre2.csv}
k;r1;r2;r3;r4;r5;r6;r7;r8;r9;r10;r11;r12;r13;r14;r15;beta;AIC;v.wre.y;ns.wre.y;cr.wre.y
76;2;0;0;0;0;0;0;0;1;0;0;0;0;0;0;3.024375656;326495.3895;4.936469;8.573485;9.2234705
77;1;0;0;0;0;1;1;0;0;0;0;0;0;0;0;10.60960554;326477.336;4.823532;8.705251;8.9955276
78;0;1;0;0;0;1;0;0;0;0;0;0;0;0;0;6.9683905;326460.7394;4.820951;8.848602;9.0714645
79;1;0;0;0;0;2;0;0;0;0;0;0;0;0;0;2.250586713;326444.3088;4.602269;9.170227;9.1616515
80;2;1;0;0;0;0;0;0;0;0;0;0;0;0;0;-1.944941202;326428.687;4.688216;9.068788;8.997388
81;0;1;0;1;0;0;0;0;0;0;0;0;0;0;0;-2257.396102;326413.6086;4.676335;9.098727;9.0695505
82;0;0;0;0;0;0;0;0;2;0;0;0;0;0;0;14.06273029;326399.1546;4.853057;9.830771;9.2781129
83;1;0;0;0;0;0;0;0;1;0;0;0;0;0;1;0.113873967;326385.4081;4.844055;9.850621;9.2030107
84;0;0;0;0;0;0;0;0;0;0;0;0;1;0;1;0.18238584;326371.6936;4.861265;9.9346;9.1735709
85;0;0;0;0;0;0;0;3;0;0;0;0;0;0;1;111.5848176;326358.199;4.795817;9.769368;9.2697357
86;0;1;0;0;0;1;0;1;0;0;0;0;0;0;0;-45.10563699;326345.7074;4.825558;9.724391;9.3300744
87;0;0;0;1;0;0;0;2;0;0;0;0;0;0;0;82935.6642;326333.6769;4.870714;9.86475;9.2839722
88;0;0;1;0;0;0;1;0;0;0;0;0;0;0;0;56.00019591;326321.8113;4.866619;9.862104;9.2667562
89;1;0;0;0;0;0;0;0;0;0;0;0;1;0;0;5.351775991;326311.1475;4.857064;9.937869;9.2576134
90;0;0;0;0;0;2;1;0;0;0;0;0;0;0;0;10.8806552;326300.7907;4.870078;10.04327;9.41375
91;0;0;0;1;0;0;0;0;0;0;0;0;0;0;1;32.81482003;326290.9872;4.832626;10.155545;9.394396
92;1;0;0;0;0;1;1;1;0;0;0;0;0;0;0;48.96478769;326282.6002;4.811797;10.084985;9.1850142
93;0;1;0;0;0;0;0;0;0;0;0;0;0;1;0;-10.90406621;326274.3346;4.801496;10.083297;9.2096331
94;0;0;0;0;0;0;0;2;0;0;0;0;0;0;2;0.087203419;326266.1426;4.803084;9.817663;8.7865947
95;0;0;0;0;0;0;2;0;0;0;0;0;0;0;0;-69.45082349;326258.3655;4.659343;9.249992;8.4128625
96;0;0;0;1;0;0;0;3;0;0;0;0;0;0;0;543840.2643;326251.0844;4.663003;9.268656;8.3934751
97;0;0;0;0;0;0;0;0;0;0;0;0;0;2;0;-10.3066516;326244.1526;4.510492;8.840642;8.1013404
98;0;0;0;0;0;0;0;1;1;0;0;0;0;0;1;1.066704922;326237.2633;4.522602;8.84749;8.0908607
99;0;0;1;0;0;0;1;1;0;0;0;0;0;0;0;-417.8768574;326230.8493;4.530525;8.84043;8.1014626
100;0;1;1;0;0;0;0;0;0;0;0;0;0;0;0;-12.91990522;326224.4893;4.546051;8.847298;8.0812325
101;0;0;0;0;0;1;0;0;0;0;1;0;0;0;0;3.942422004;326218.8963;4.558141;8.866091;8.071804
102;1;0;0;0;0;0;0;2;0;0;0;0;0;0;1;10.0998535;326213.4207;4.512891;9.012004;8.2031155
103;2;0;0;0;0;0;0;1;0;0;0;0;0;0;1;0.362986678;326203.8289;4.453058;9.084421;8.0351163
104;0;0;1;0;0;0;0;1;0;0;0;0;0;0;1;-1.73560981;326197.9684;4.444969;9.063425;8.0700125
105;2;0;0;0;0;0;1;0;0;0;0;0;0;0;0;7.089043992;326192.5931;4.383103;8.967039;8.0083315
106;2;0;0;0;0;0;1;1;0;0;0;0;0;0;0;109.4967614;326174.346;4.37143;8.898877;7.8891044
107;0;0;0;0;0;0;0;0;0;0;0;0;0;0;3;3.20E-05;326169.0682;4.332261;8.45394;7.6692586
108;0;1;0;0;0;0;0;0;0;0;0;1;0;0;0;-5.851761412;326163.99;4.289836;8.455764;7.6891924
109;0;0;0;0;0;0;0;0;0;0;1;0;0;0;1;0.099781862;326159.2312;4.281774;8.456802;7.6573251
110;1;0;0;0;0;0;0;1;0;0;1;0;0;0;0;-54.88227329;326154.465;4.312884;8.463434;7.6887483
111;1;0;0;0;0;0;1;2;0;0;0;0;0;0;0;1380.739627;326149.8555;4.290542;8.48903;7.7004118
112;0;0;0;0;0;0;0;0;1;0;0;0;0;0;2;0.002207218;326145.5227;4.314574;8.49797;7.7514741
113;0;1;0;0;0;0;0;0;1;0;0;0;0;0;1;-0.109265883;326141.6985;4.28664;8.501158;7.7356627
114;1;0;1;0;0;1;0;0;0;0;0;0;0;0;0;-4.298323895;326138.1278;4.319518;8.460823;7.5579502
115;0;1;0;0;0;1;0;0;0;0;0;0;0;0;1;-0.051713899;326134.5499;4.299051;8.514302;7.5659943
116;1;0;0;0;0;2;0;1;0;0;0;0;0;0;0;20.08752936;326131.1258;4.31994;8.416688;7.4977704
117;2;0;0;0;0;2;0;0;0;0;0;0;0;0;0;0.871812799;326124.5085;4.392687;8.560709;7.3705191
118;0;1;0;0;0;1;0;1;0;0;0;0;0;0;1;0.356041347;326121.5446;4.389148;8.563563;7.4085941
119;0;1;1;0;0;0;0;1;0;0;0;0;0;0;0;79.51240587;326118.3863;4.39377;8.559616;7.410615
120;0;0;0;0;0;1;0;0;0;0;0;0;0;0;2;0.000890821;326115.4274;4.429833;8.303748;7.1869156
121;0;0;0;0;0;0;0;0;1;0;0;0;1;0;0;6.911511003;326112.5597;4.420337;8.305336;7.1755993
122;0;0;0;0;0;0;0;2;0;0;1;0;0;0;0;-435.8079998;326109.9524;4.389585;8.300789;7.2119115
123;0;0;0;0;0;2;0;0;0;0;0;0;0;0;1;0.031494679;326107.4763;4.419311;8.450256;7.2056875
124;0;0;0;0;0;1;0;0;1;0;0;0;0;0;0;-2.989255102;326105.1751;4.407398;8.434123;7.1627216
125;1;0;0;0;0;1;0;1;0;0;0;0;0;0;1;0.593403335;326103.0747;4.393733;8.36563;7.095485
126;2;0;0;0;0;1;0;0;0;0;0;0;0;0;1;0.023780774;326095.8382;4.501718;8.499146;7.3820513
127;0;0;0;0;0;1;0;2;0;0;0;0;0;0;1;4.663043487;326089.4729;4.542918;8.962205;7.3399355
128;0;0;0;0;0;1;0;3;0;0;0;0;0;0;0;-692.5937702;326087.5293;4.537347;8.961166;7.2484225
129;0;0;0;0;0;0;1;3;0;0;0;0;0;0;0;8097.699039;326085.6626;4.538766;8.994996;7.3160425
130;1;0;0;0;0;0;0;0;0;0;0;0;1;0;1;-0.036362211;326083.8276;4.55526;9.024475;7.2850606
131;0;0;0;0;0;1;0;0;0;0;0;1;0;0;0;2.734081285;326082.0754;4.590259;9.065444;7.2458806
132;1;0;0;0;0;1;0;0;0;0;0;1;0;0;0;-1.526376373;326080.1112;4.612454;9.096845;7.2803928
133;2;0;0;0;0;0;0;0;0;0;1;0;0;0;0;-1.277948452;326078.3517;4.616023;9.086214;7.2514953
134;0;0;0;0;0;0;0;0;0;0;0;0;0;1;1;0.07332622;326076.6214;4.607206;9.055027;7.2873683
135;0;0;0;0;0;0;0;0;0;1;0;0;0;0;0;-6.955924458;326074.941;4.533229;8.526703;7.2299387
136;0;0;1;0;0;0;0;0;0;1;0;0;0;0;0;27.74365682;326072.7676;4.555711;8.519751;7.1149945
137;0;0;0;0;0;2;0;2;0;0;0;0;0;0;0;122.0793947;326071.1877;4.571255;8.745717;7.1707302
138;0;0;0;0;0;0;0;0;1;0;0;1;0;0;0;6.000051741;326069.689;4.556373;8.745059;7.1902207
139;0;0;0;0;0;0;0;0;2;0;0;1;0;0;0;-14.49778946;326065.9184;4.53301;8.698849;7.1991655
140;1;0;0;0;0;0;0;0;0;0;0;1;0;0;1;-0.073492093;326064.4493;4.531634;8.722103;7.2271308
141;0;0;0;0;0;0;0;1;0;0;0;1;0;0;1;-1.054237619;326057.0299;4.507457;8.733271;7.2503197
142;1;0;0;0;0;0;1;1;0;0;0;0;0;0;1;0.738273372;326055.6276;4.514722;8.71917;7.2383316
143;0;0;0;0;0;0;0;0;1;0;0;0;0;1;0;5.706824565;326054.2615;4.502989;8.705601;7.2626682
144;1;0;0;0;0;0;0;1;0;0;0;1;0;0;0;-39.86705573;326052.9013;4.499292;8.714817;7.2440942
145;0;0;0;0;0;0;0;2;0;0;0;1;0;0;0;-431.7100898;326047.2341;4.469725;8.668689;7.2145745
146;0;0;0;0;0;0;0;1;0;0;0;0;0;0;3;-0.000172795;326045.9211;4.487662;8.697807;7.2070043
147;0;1;0;0;0;0;0;0;0;0;0;0;0;1;1;0.082549755;326044.6096;4.49441;8.693614;7.222536
148;0;0;0;0;0;0;1;0;0;0;0;0;1;0;0;37.3306937;326043.3839;4.49569;8.703464;7.2364759
149;0;0;0;0;0;0;1;0;0;0;0;0;1;0;1;-0.416635421;326038.9788;4.507826;8.705527;7.2525581
150;0;1;0;1;0;0;0;1;0;0;0;0;0;0;0;7224.251636;326037.7116;4.511601;8.712216;7.2652011
\end{filecontents*}
\pgfplotstabletypeset[
        multicolumn names, 
        col sep=semicolon, 
        display columns/0/.style={column name=$\bm{k}$,dec sep align},
        display columns/1/.style={column type=C,column name=$\bm{r_{k}^{1}}$},
        display columns/2/.style={column type=C,column name=$\bm{r_{k}^{2}}$},
        display columns/3/.style={column type=C,column name=$\bm{r_{k}^{3}}$},
        display columns/4/.style={column type=C,column name=$\bm{r_{k}^{4}}$},
        display columns/5/.style={column type=C,column name=$\bm{r_{k}^{5}}$},
        display columns/6/.style={column type=C,column name=$\bm{r_{k}^{6}}$},
        display columns/7/.style={column type=C,column name=$\bm{r_{k}^{7}}$},
        display columns/8/.style={column type=C,column name=$\bm{r_{k}^{8}}$},
        display columns/9/.style={column type=C,column name=$\bm{r_{k}^{9}}$},
        display columns/10/.style={column type=C,column name=$\bm{r_{k}^{10}}$},
        display columns/11/.style={column type=C,column name=$\bm{r_{k}^{11}}$},
        display columns/12/.style={column type=C,column name=$\bm{r_{k}^{12}}$},
        display columns/13/.style={column type=C,column name=$\bm{r_{k}^{13}}$},
        display columns/14/.style={column type=C,column name=$\bm{r_{k}^{14}}$},
        display columns/15/.style={column type=C,column name=$\bm{r_{k}^{15}}$},
        display columns/16/.style={column name=$\bm{\widehat{\beta}_{\textbf{\text{OLS}},k}}$,fixed,fixed zerofill,precision=2,set thousands separator={,},dec sep align},
        display columns/17/.style={column name={\textbf{AIC}},fixed,fixed zerofill,precision=0,set thousands separator={,},dec sep align},
        display columns/18/.style={column name={\textbf{v.mae}},fixed,fixed zerofill,precision=3,set thousands separator={,},dec sep align},
        display columns/19/.style={column name={\textbf{ns.mae}},fixed,fixed zerofill,precision=3,set thousands separator={,},dec sep align},
        display columns/20/.style={column name={\textbf{cr.mae}},fixed,fixed zerofill,precision=3,set thousands separator={,},dec sep align},
	    	every head row/.style={before row=\toprule,after row=\midrule},
	    	every nth row={10[+5]}{before row=\midrule},
      	every last row/.style={after row=\bottomrule},
      ]{PGROOLSadalgwre2.csv} 
    \caption{Cont.}
    \label{tab:PGROOLS2}
  \end{center}
\end{table}

\begin{table}[htb]
  \tiny
  \begin{center}
  \tabcolsep=0.106cm
  \renewcommand{\arraystretch}{0.90}
      \begin{filecontents*}{BGROOLS300886adalgwre1.csv}
k;r1;r2;r3;r4;r5;r6;r7;r8;r9;r10;r11;r12;r13;r14;r15;beta;AIC;v.wre.y;ns.wre.y;cr.wre.y
0;0;0;0;0;0;0;0;0;0;0;0;0;0;0;0;14689.75189;437251.0284;4.556602;3.23053;4.02699629
1;0;0;0;0;0;0;0;1;0;0;0;0;0;0;0;7990.979119;386721.8775;2.4738689;0.845084;0.91309099
2;1;0;0;0;0;0;0;0;0;0;0;0;0;0;0;-274.2372099;375143.8396;2.0650148;2.139015;1.83147434
3;0;0;0;0;0;1;0;0;0;0;0;0;0;0;0;145.734269;366567.289;1.6562468;0.443792;0.49630791
4;0;0;0;0;0;0;0;0;0;0;0;0;0;0;1;-5.109429974;358893.518;1.6467626;1.006377;0.55626101
5;0;0;0;0;0;0;1;0;0;0;0;0;0;0;0;416.7903676;355731.9286;1.6347756;0.853357;0.46869164
6;1;0;0;0;0;0;0;1;0;0;0;0;0;0;0;2332.908669;354317.7979;1.6790248;0.956304;0.37400957
7;0;0;0;0;0;0;0;2;0;0;0;0;0;0;0;24914.35587;349758.5786;1.2340493;0.491057;0.62820691
8;2;0;0;0;0;0;0;0;0;0;0;0;0;0;0;49.41974102;347795.9634;0.9985358;0.339591;0.59410883
9;0;0;0;0;0;1;0;1;0;0;0;0;0;0;0;859.4921834;346443.9203;0.9119805;0.357314;0.60191958
10;0;0;0;0;0;0;0;1;0;0;0;0;0;0;1;29.50213516;345045.2474;0.8386126;0.388718;0.64994998
11;1;0;0;0;0;0;0;0;0;0;0;0;0;0;1;1.707231457;341082.8124;0.7587438;0.397868;0.46541559
12;0;1;0;0;0;0;0;0;0;0;0;0;0;0;0;91.65096366;339360.374;0.7178702;0.394231;0.38998224
13;1;0;0;0;0;1;0;0;0;0;0;0;0;0;0;36.34407091;337730.9855;0.5735332;0.653032;0.51187199
14;0;0;0;0;0;0;0;0;1;0;0;0;0;0;0;51.78308001;336842.8235;0.5889452;0.658094;0.51788433
15;0;0;0;0;0;0;0;0;0;0;0;1;0;0;0;68.02450531;335980.0391;0.6276231;0.678399;0.51220045
16;0;0;0;0;0;0;1;1;0;0;0;0;0;0;0;2661.470358;335351.1441;0.6085193;0.671499;0.50343614
17;1;0;0;0;0;0;1;0;0;0;0;0;0;0;0;109.1420626;334875.5029;0.5791571;0.700678;0.54495249
18;1;0;0;0;0;0;0;1;0;0;0;0;0;0;1;-12.63466788;334412.7937;0.5933914;0.719882;0.53111425
19;0;0;0;0;0;0;0;2;0;0;0;0;0;0;1;-114.4767507;333904.2185;0.5619575;0.620595;0.47362037
20;0;0;0;0;0;0;0;0;0;0;0;0;0;1;0;35.40147412;333447.4362;0.5648041;0.59708;0.45439558
21;1;0;0;0;0;0;0;2;0;0;0;0;0;0;0;-4570.146567;333116.2226;0.5534635;0.543166;0.40661664
22;0;0;0;0;0;0;0;0;0;0;0;0;0;0;2;0.018189602;332805.7679;0.5621878;0.477606;0.35839493
23;2;0;0;0;0;0;0;0;0;0;0;0;0;0;1;-0.257590507;332547.2224;0.5497891;0.449694;0.38124004
24;0;0;0;0;0;0;0;0;0;0;0;0;1;0;0;47.16705688;332293.8077;0.5445717;0.46758;0.37847101
25;0;0;1;0;0;0;0;0;0;0;0;0;0;0;0;123.4721039;332042.1939;0.5300925;0.46364;0.36198325
26;0;0;1;0;0;0;0;1;0;0;0;0;0;0;0;-1240.441912;331686.6184;0.5217526;0.453302;0.35476415
27;1;0;1;0;0;0;0;0;0;0;0;0;0;0;0;-43.81963379;331404.843;0.5247251;0.443834;0.34290691
28;0;0;0;0;0;0;0;3;0;0;0;0;0;0;0;-32661.61191;331136.4147;0.4985978;0.404711;0.32693856
29;2;0;0;0;0;0;0;1;0;0;0;0;0;0;0;-140.9037144;330562.055;0.5037983;0.348409;0.26837872
30;0;0;0;0;0;1;0;0;0;0;0;0;0;0;1;0.560822216;330361.1506;0.5184664;0.41797;0.26442846
31;0;0;0;0;0;1;1;0;0;0;0;0;0;0;0;87.33234885;330163.0398;0.5122091;0.442838;0.27185191
32;1;0;0;0;0;0;0;0;1;0;0;0;0;0;0;25.30989553;329988.2028;0.5082752;0.442784;0.26411044
33;0;0;0;0;0;2;0;0;0;0;0;0;0;0;0;14.22126628;329834.2063;0.4766377;0.491447;0.2856491
34;0;1;0;0;0;0;0;0;0;0;0;0;0;0;1;-0.439186882;329687.797;0.4769757;0.499703;0.29008603
35;0;0;0;0;0;0;0;0;0;0;1;0;0;0;0;26.88413;329550.3981;0.4759476;0.502168;0.29069359
36;0;1;0;0;0;0;0;1;0;0;0;0;0;0;0;-391.805173;329441.9896;0.4721628;0.498611;0.28832454
37;1;1;0;0;0;0;0;0;0;0;0;0;0;0;0;-18.57692872;329146.5443;0.4624657;0.505377;0.30108613
38;0;0;0;1;0;0;0;0;0;0;0;0;0;0;0;11959.3191;329042.672;0.4722863;0.518349;0.29975566
39;0;0;0;0;0;1;0;1;0;0;0;0;0;0;1;-2.150856512;328934.797;0.4740133;0.510304;0.29532844
40;0;0;0;0;0;0;0;1;1;0;0;0;0;0;0;228.3176427;328832.2073;0.4749207;0.509415;0.29103128
41;0;0;0;0;0;1;0;2;0;0;0;0;0;0;0;-1938.367038;328732.9494;0.4551975;0.444528;0.24823832
42;1;0;0;0;0;1;0;1;0;0;0;0;0;0;0;-112.8261834;327927.0187;0.3719387;0.34547;0.23732419
43;0;0;0;0;0;0;1;0;0;0;0;0;0;0;1;0.711359024;327857.5891;0.3681123;0.353053;0.23519227
44;0;0;0;0;0;0;0;0;1;0;0;0;0;0;1;0.717991678;327792.2809;0.3664425;0.352132;0.23280702
45;1;0;0;1;0;0;0;0;0;0;0;0;0;0;0;-4230.29296;327728.9214;0.365056;0.356124;0.2283351
46;0;0;0;1;0;0;0;1;0;0;0;0;0;0;0;-10720.3048;327659.061;0.3675825;0.36443;0.22679792
47;1;0;0;0;0;0;0;0;0;0;0;1;0;0;0;-18.39217271;327603.2303;0.3682433;0.366201;0.2259735
48;0;0;0;0;0;0;0;1;0;0;0;1;0;0;0;-212.7774529;327537.0309;0.374316;0.36744;0.22628398
49;0;0;0;0;0;1;1;1;0;0;0;0;0;0;0;-177.6382629;327482.8636;0.3688262;0.366659;0.23044851
50;1;0;0;0;0;1;0;0;0;0;0;0;0;0;1;-0.085371181;327432.0902;0.3684989;0.390834;0.22138584
51;0;0;0;0;0;2;0;1;0;0;0;0;0;0;0;-57.39510605;327381.5018;0.3589157;0.389841;0.22771688
52;0;0;1;0;0;1;0;0;0;0;0;0;0;0;0;-23.54660417;327331.0188;0.3518843;0.389732;0.22502614
53;1;0;0;0;0;0;0;0;0;0;0;0;0;0;2;-0.003724868;327286.7993;0.3462766;0.376815;0.20560833
54;0;0;0;0;0;0;0;1;0;0;0;0;0;0;2;-0.082577923;327148.6298;0.338558;0.357072;0.18545478
55;2;0;0;0;0;1;0;0;0;0;0;0;0;0;0;1.153067627;327104.9948;0.3152857;0.321372;0.17282236
56;0;0;1;0;0;0;0;0;0;0;0;0;0;0;1;-0.654823107;327064.1088;0.3150275;0.322221;0.17269407
57;1;0;0;0;0;0;0;0;0;0;0;0;0;1;0;-4.41437225;327024.7437;0.3218915;0.317391;0.17538749
58;0;0;0;0;0;0;1;2;0;0;0;0;0;0;0;-6095.967325;326986.4282;0.3169237;0.309964;0.17246547
59;1;0;0;0;0;0;1;1;0;0;0;0;0;0;0;-332.8837738;326823.039;0.3080044;0.302133;0.1829822
60;0;0;1;0;0;0;0;2;0;0;0;0;0;0;0;3624.772101;326787.1036;0.3059923;0.300561;0.18312963
61;1;0;1;0;0;0;0;1;0;0;0;0;0;0;0;191.4615557;326732.5213;0.3038349;0.299148;0.18338547
62;0;1;0;0;0;0;0;0;1;0;0;0;0;0;0;-17.4859327;326699.6144;0.3063613;0.298664;0.18199868
63;0;0;1;0;0;1;0;1;0;0;0;0;0;0;0;183.6765183;326667.632;0.3042724;0.295646;0.18168407
64;0;0;0;0;0;0;0;0;0;0;0;1;0;0;1;-0.201175973;326638.3511;0.3039436;0.298171;0.18143757
65;0;1;0;0;0;0;0;1;0;0;0;0;0;0;1;2.549036478;326610.2439;0.3014272;0.296173;0.18307836
66;1;1;0;0;0;0;0;0;0;0;0;0;0;0;1;0.130537716;326572.1693;0.2968831;0.298762;0.17989452
67;2;0;0;0;0;1;0;1;0;0;0;0;0;0;0;-29.57119719;326545.0668;0.2916089;0.286224;0.16873258
68;1;1;0;0;0;0;0;1;0;0;0;0;0;0;0;95.54871537;326518.8532;0.2922817;0.287477;0.17180456
69;0;1;0;0;0;0;0;2;0;0;0;0;0;0;0;922.4841086;326477.8636;0.2937254;0.282324;0.17346009
70;0;0;0;0;0;0;1;1;0;0;0;0;0;0;1;-6.221311024;326452.7548;0.2912463;0.281253;0.17507827
71;0;0;0;0;0;0;0;1;0;0;1;0;0;0;0;-134.954514;326428.4006;0.2889505;0.281426;0.17557284
72;1;0;0;0;0;0;0;0;0;0;1;0;0;0;0;-4.471670823;326398.3201;0.2844463;0.281641;0.17252043
73;1;0;0;0;0;0;0;3;0;0;0;0;0;0;0;-26186.71931;326374.036;0.2761004;0.26364;0.16223546
74;1;0;0;0;0;0;1;0;0;0;0;0;0;0;1;-0.287330657;326351.8364;0.2715004;0.266499;0.15765243
75;1;0;0;0;0;0;0;1;1;0;0;0;0;0;0;-58.01220629;326330.9222;0.269209;0.265842;0.15681253
\end{filecontents*}
\pgfplotstabletypeset[
        multicolumn names, 
        col sep=semicolon, 
        display columns/0/.style={column name=$\bm{k}$,dec sep align},
        display columns/1/.style={column type=C,column name=$\bm{r_{k}^{1}}$},
        display columns/2/.style={column type=C,column name=$\bm{r_{k}^{2}}$},
        display columns/3/.style={column type=C,column name=$\bm{r_{k}^{3}}$},
        display columns/4/.style={column type=C,column name=$\bm{r_{k}^{4}}$},
        display columns/5/.style={column type=C,column name=$\bm{r_{k}^{5}}$},
        display columns/6/.style={column type=C,column name=$\bm{r_{k}^{6}}$},
        display columns/7/.style={column type=C,column name=$\bm{r_{k}^{7}}$},
        display columns/8/.style={column type=C,column name=$\bm{r_{k}^{8}}$},
        display columns/9/.style={column type=C,column name=$\bm{r_{k}^{9}}$},
        display columns/10/.style={column type=C,column name=$\bm{r_{k}^{10}}$},
        display columns/11/.style={column type=C,column name=$\bm{r_{k}^{11}}$},
        display columns/12/.style={column type=C,column name=$\bm{r_{k}^{12}}$},
        display columns/13/.style={column type=C,column name=$\bm{r_{k}^{13}}$},
        display columns/14/.style={column type=C,column name=$\bm{r_{k}^{14}}$},
        display columns/15/.style={column type=C,column name=$\bm{r_{k}^{15}}$},
        display columns/16/.style={column name=$\bm{\widehat{\beta}_{\textbf{\text{OLS}},k}}$,fixed,fixed zerofill,precision=2,set thousands separator={,},dec sep align},
        display columns/17/.style={column name={\textbf{AIC}},fixed,fixed zerofill,precision=0,set thousands separator={,},dec sep align},
        display columns/18/.style={column name={\textbf{v.mae}},fixed,fixed zerofill,precision=3,set thousands separator={,},dec sep align},
        display columns/19/.style={column name={\textbf{ns.mae}},fixed,fixed zerofill,precision=3,set thousands separator={,},dec sep align},
        display columns/20/.style={column name={\textbf{cr.mae}},fixed,fixed zerofill,precision=3,set thousands separator={,},dec sep align},
	    	every head row/.style={before row=\toprule,after row=\midrule},
	    	every row no 1/.style={before row=\midrule},
	    	every nth row={10[+1]}{before row=\midrule},
      	every last row/.style={after row=\bottomrule},
      ]{BGROOLS300886adalgwre1.csv} 
    \caption{OLS proxy function of BEL derived under $300$-$886$ in the adaptive algorithm with the final coefficients. Furthermore, AIC scores and out-of-sample MAEs in $\%$ after each iteration.}
    \label{tab:BGROOLS3008861}
  \end{center}
\end{table}

\begin{table}[htb]
\ContinuedFloat
  \tiny
  \begin{center}
  \tabcolsep=0.104cm
  \renewcommand{\arraystretch}{0.90}
      \begin{filecontents*}{BGROOLS300886adalgwre2.csv}
k;r1;r2;r3;r4;r5;r6;r7;r8;r9;r10;r11;r12;r13;r14;r15;beta;AIC;v.wre.y;ns.wre.y;cr.wre.y
76;2;0;0;0;0;0;0;0;1;0;0;0;0;0;0;-3.107626242;326312.806;0.2707765;0.266302;0.15513565
77;1;0;0;0;0;1;1;0;0;0;0;0;0;0;0;-2.099797805;326294.6974;0.2644754;0.270402;0.15125574
78;0;1;0;0;0;1;0;0;0;0;0;0;0;0;0;-8.72582392;326277.9291;0.2643033;0.274868;0.15250997
79;1;0;0;0;0;2;0;0;0;0;0;0;0;0;0;-1.932617465;326261.2692;0.2522373;0.284877;0.1537699
80;0;0;0;0;0;0;0;0;2;0;0;0;0;0;0;-14.89619165;326245.0608;0.2626108;0.308787;0.15723232
81;2;1;0;0;0;0;0;0;0;0;0;0;0;0;0;-1.216103373;326229.1837;0.2671515;0.305669;0.15496196
82;0;1;0;1;0;0;0;0;0;0;0;0;0;0;0;3341.286146;326214.1354;0.2663075;0.3066;0.15587825
83;0;0;0;0;0;0;0;3;0;0;0;0;0;0;1;-43.83973514;326200.5001;0.2631118;0.301523;0.15820325
84;1;0;0;0;0;0;0;0;1;0;0;0;0;0;1;-0.118507265;326187.0709;0.2622106;0.302094;0.15680625
85;0;0;0;0;0;0;0;0;0;0;0;0;1;0;1;-0.182998917;326173.6812;0.2631535;0.304627;0.1557477
86;0;1;0;0;0;1;0;1;0;0;0;0;0;0;0;67.18879088;326160.9026;0.2649574;0.303215;0.1567363
87;0;0;0;1;0;0;0;2;0;0;0;0;0;0;0;-432954.9759;326148.6568;0.2672798;0.307622;0.15606296
88;0;0;1;0;0;0;1;0;0;0;0;0;0;0;0;-34.57737915;326136.5849;0.2672232;0.30754;0.15577187
89;1;0;0;0;0;0;0;0;0;0;0;0;1;0;0;-5.101231075;326126.0939;0.2666623;0.309883;0.15571433
90;0;0;0;0;0;2;1;0;0;0;0;0;0;0;0;-10.78422232;326115.6586;0.2673225;0.313173;0.15836889
91;0;0;0;1;0;0;0;0;0;0;0;0;0;0;1;-66.9890856;326105.8157;0.2651858;0.316666;0.15804356
92;0;0;0;0;0;0;0;2;0;0;0;0;0;0;2;-0.086171533;326097.3265;0.2653569;0.308276;0.15055448
93;0;1;0;0;0;0;0;0;0;0;0;0;0;1;0;0.352607426;326088.9526;0.2647319;0.308212;0.15111561
94;1;0;0;0;0;1;1;1;0;0;0;0;0;0;0;-93.83311594;326080.7093;0.2635902;0.306109;0.14777798
95;0;0;0;0;0;0;2;0;0;0;0;0;0;0;0;70.4456741;326072.8235;0.2559173;0.288382;0.14142352
96;0;0;0;1;0;0;0;3;0;0;0;0;0;0;0;-1073454.044;326065.5322;0.2561173;0.288959;0.14110021
97;0;0;0;0;0;0;0;0;0;0;0;0;0;2;0;-21.5875168;326058.3098;0.24765;0.275445;0.13606405
98;0;0;0;0;0;0;0;1;1;0;0;0;0;0;1;-1.09520347;326051.1906;0.2483201;0.275665;0.1358892
99;0;0;1;0;0;0;1;1;0;0;0;0;0;0;0;398.9428514;326044.6673;0.2487559;0.275445;0.13609492
100;0;1;1;0;0;0;0;0;0;0;0;0;0;0;0;22.02928063;326038.3955;0.2495997;0.275657;0.13576231
101;0;0;0;0;0;1;0;0;0;0;1;0;0;0;0;-4.119592506;326032.7551;0.2502623;0.276236;0.13560358
102;1;0;0;0;0;0;0;2;0;0;0;0;0;0;1;1.297862672;326027.1998;0.2477768;0.280788;0.13776766
103;2;0;0;0;0;0;0;1;0;0;0;0;0;0;1;0.197807648;326017.4618;0.2444871;0.283073;0.13498913
104;1;0;0;0;0;0;0;3;0;0;0;0;0;0;1;351.1092188;326008.9468;0.244694;0.288861;0.13816264
105;0;0;1;0;0;0;0;1;0;0;0;0;0;0;1;1.092033786;326002.5352;0.2442413;0.28833;0.1389673
106;0;0;0;0;0;0;0;0;0;0;0;0;0;0;3;-3.26E-05;325997.1269;0.24192;0.274323;0.1359288
107;2;0;0;0;0;0;1;0;0;0;0;0;0;0;0;-7.781782101;325991.7513;0.2385386;0.271232;0.13449288
108;2;0;0;0;0;0;1;1;0;0;0;0;0;0;0;-126.2797715;325973.4549;0.2379093;0.269056;0.13191575
109;0;0;0;0;0;0;0;0;0;0;1;0;0;0;1;-0.095289232;325968.4199;0.2377028;0.269156;0.13137902
110;1;0;0;0;0;0;0;1;0;0;1;0;0;0;0;57.61000535;325963.4171;0.2391895;0.269454;0.1319992
111;0;1;0;0;0;0;0;0;0;0;0;1;0;0;0;9.907133996;325958.5478;0.2369059;0.269482;0.13234775
112;1;0;0;0;0;0;1;2;0;0;0;0;0;0;0;-1698.916237;325953.9548;0.2356868;0.270205;0.1320058
113;0;0;0;0;0;0;0;0;1;0;0;0;0;0;2;-0.007654173;325949.5329;0.2369125;0.270448;0.13281753
114;0;1;0;0;0;0;0;0;1;0;0;0;0;0;1;0.104155151;325945.8643;0.2355009;0.270504;0.13252191
115;0;1;0;0;0;1;0;0;0;0;0;0;0;0;1;0.053937705;325942.2261;0.2343674;0.272176;0.1322872
116;1;0;1;0;0;1;0;0;0;0;0;0;0;0;0;4.998108568;325938.5559;0.2361734;0.2709;0.12937915
117;1;0;0;0;0;2;0;1;0;0;0;0;0;0;0;-17.59504976;325935.2016;0.2377253;0.267784;0.12726982
118;2;0;0;0;0;2;0;0;0;0;0;0;0;0;0;-0.786909563;325928.5907;0.2415519;0.27263;0.12780055
119;0;1;0;0;0;1;0;1;0;0;0;0;0;0;1;-0.552283606;325925.4934;0.2413667;0.272735;0.12849247
120;0;1;1;0;0;0;0;1;0;0;0;0;0;0;0;-119.8086677;325922.293;0.2416919;0.272591;0.12854567
121;0;0;0;0;0;0;0;0;1;0;0;0;1;0;0;-7.161268755;325919.2174;0.2412877;0.272679;0.12836937
122;0;0;0;0;0;1;0;0;0;0;0;0;0;0;2;-0.000943652;325916.3105;0.2431871;0.264786;0.12432049
123;0;0;0;0;0;0;0;2;0;0;1;0;0;0;0;497.0162886;325913.631;0.2414384;0.264657;0.12512927
124;0;0;0;0;0;2;0;0;0;0;0;0;0;0;1;-0.033629972;325911.1559;0.2431156;0.26923;0.1251744
125;1;0;0;0;0;1;0;1;0;0;0;0;0;0;1;-0.577848932;325908.8106;0.2423518;0.26701;0.12290663
126;2;0;0;0;0;1;0;0;0;0;0;0;0;0;1;-0.023251788;325901.4819;0.2482813;0.271399;0.12943094
127;0;0;0;0;0;1;0;2;0;0;0;0;0;0;1;-4.482536407;325895.3233;0.2505054;0.285538;0.12871053
128;0;0;0;0;0;1;0;0;1;0;0;0;0;0;0;2.931680223;325892.8997;0.2498536;0.285073;0.12775118
129;0;0;0;0;0;0;1;3;0;0;0;0;0;0;0;-5069.148482;325890.8582;0.2499337;0.286101;0.12844833
130;1;0;0;0;0;0;0;0;0;0;0;0;1;0;1;0.034014356;325888.8676;0.2508547;0.287005;0.12747051
131;0;0;0;0;0;1;0;3;0;0;0;0;0;0;0;2631.072937;325886.9306;0.2505491;0.286856;0.12468626
132;0;0;0;0;0;0;0;0;0;1;0;0;0;0;0;30.02854841;325885.1479;0.2464511;0.270277;0.12400054
133;0;0;1;0;0;0;0;0;0;1;0;0;0;0;0;-27.79172765;325882.9787;0.2475883;0.270072;0.12250971
134;0;0;0;0;0;1;0;0;0;0;0;1;0;0;0;-2.68243641;325881.3696;0.2494651;0.271351;0.12235544
135;1;0;0;0;0;1;0;0;0;0;0;1;0;0;0;2.183254242;325879.3927;0.2506763;0.272289;0.12272023
136;0;0;0;0;0;0;0;0;0;0;0;0;0;1;1;-0.073586224;325877.8241;0.2502043;0.271408;0.12383185
137;1;0;0;0;0;0;0;1;0;0;0;1;0;0;0;52.059328;325876.2897;0.2508686;0.271694;0.12346175
138;0;0;0;0;0;0;0;2;0;0;0;1;0;0;0;507.7922452;325870.4422;0.2497091;0.270256;0.12251003
139;0;0;0;0;0;0;0;0;1;0;0;1;0;0;0;0.089363091;325868.9358;0.2481791;0.270237;0.122734
140;0;0;0;0;0;0;0;0;2;0;0;1;0;0;0;14.53435232;325865.0117;0.246235;0.268809;0.12283709
141;0;0;0;0;0;0;0;1;0;0;0;0;0;0;3;0.000161009;325863.5001;0.2470163;0.269662;0.12194748
142;2;0;0;0;0;0;0;0;0;0;1;0;0;0;0;1.481113338;325861.9891;0.2471944;0.269335;0.12141583
143;0;0;0;0;0;2;0;2;0;0;0;0;0;0;0;-98.05900138;325860.5439;0.2478664;0.276177;0.12245487
144;1;0;0;0;0;0;1;1;0;0;0;0;0;0;1;-0.683840241;325859.1762;0.24826;0.275706;0.12203938
145;1;0;0;0;0;0;0;0;0;0;0;1;0;0;1;0.075930162;325857.7959;0.2481927;0.276325;0.12191352
146;0;0;0;0;0;0;0;1;0;0;0;1;0;0;1;1.096923297;325850.0589;0.2474318;0.276615;0.12208283
147;0;0;0;0;0;0;0;0;1;0;0;0;0;1;0;-5.635148122;325848.7705;0.2467507;0.276205;0.12259155
148;0;1;0;0;0;0;0;0;0;0;0;0;0;1;1;-0.080279891;325847.4797;0.247118;0.27607;0.12280183
149;1;0;0;1;0;0;0;0;0;0;0;0;0;0;1;20.58322625;325846.1979;0.2455706;0.27698;0.12281638
150;0;0;0;1;0;0;0;1;0;0;0;0;0;0;1;-60.88995512;325840.9428;0.2421701;0.274369;0.12281914
\end{filecontents*}
\pgfplotstabletypeset[
        multicolumn names, 
        col sep=semicolon, 
        display columns/0/.style={column name=$\bm{k}$,dec sep align},
        display columns/1/.style={column type=C,column name=$\bm{r_{k}^{1}}$},
        display columns/2/.style={column type=C,column name=$\bm{r_{k}^{2}}$},
        display columns/3/.style={column type=C,column name=$\bm{r_{k}^{3}}$},
        display columns/4/.style={column type=C,column name=$\bm{r_{k}^{4}}$},
        display columns/5/.style={column type=C,column name=$\bm{r_{k}^{5}}$},
        display columns/6/.style={column type=C,column name=$\bm{r_{k}^{6}}$},
        display columns/7/.style={column type=C,column name=$\bm{r_{k}^{7}}$},
        display columns/8/.style={column type=C,column name=$\bm{r_{k}^{8}}$},
        display columns/9/.style={column type=C,column name=$\bm{r_{k}^{9}}$},
        display columns/10/.style={column type=C,column name=$\bm{r_{k}^{10}}$},
        display columns/11/.style={column type=C,column name=$\bm{r_{k}^{11}}$},
        display columns/12/.style={column type=C,column name=$\bm{r_{k}^{12}}$},
        display columns/13/.style={column type=C,column name=$\bm{r_{k}^{13}}$},
        display columns/14/.style={column type=C,column name=$\bm{r_{k}^{14}}$},
        display columns/15/.style={column type=C,column name=$\bm{r_{k}^{15}}$},
        display columns/16/.style={column name=$\bm{\widehat{\beta}_{\textbf{\text{OLS}},k}}$,fixed,fixed zerofill,precision=2,set thousands separator={,},dec sep align},
        display columns/17/.style={column name={\textbf{AIC}},fixed,fixed zerofill,precision=0,set thousands separator={,},dec sep align},
        display columns/18/.style={column name={\textbf{v.mae}},fixed,fixed zerofill,precision=3,set thousands separator={,},dec sep align},
        display columns/19/.style={column name={\textbf{ns.mae}},fixed,fixed zerofill,precision=3,set thousands separator={,},dec sep align},
        display columns/20/.style={column name={\textbf{cr.mae}},fixed,fixed zerofill,precision=3,set thousands separator={,},dec sep align},
	    	every head row/.style={before row=\toprule,after row=\midrule},
	    	every nth row={10[+5]}{before row=\midrule},
      	every last row/.style={after row=\bottomrule},
      ]{BGROOLS300886adalgwre2.csv} 
    \caption[]{Cont.}
    \label{tab:BGROOLS3008862}
  \end{center}
\end{table}

\begin{table}[htb]
\ContinuedFloat
  \tiny
  \begin{center}
  \tabcolsep=0.106cm
  \renewcommand{\arraystretch}{0.90}
      \begin{filecontents*}{BGROOLS300886adalgwre3.csv}
k;r1;r2;r3;r4;r5;r6;r7;r8;r9;r10;r11;r12;r13;r14;r15;beta;AIC;v.wre.y;ns.wre.y;cr.wre.y
151;0;0;0;0;0;0;1;0;0;0;0;0;1;0;0;-26.95050828;325839.677;0.2424315;0.274666;0.12307002
152;0;0;0;0;0;0;1;0;0;0;0;0;1;0;1;0.418942458;325835.2374;0.2430724;0.274706;0.12330498
153;0;1;0;1;0;0;0;1;0;0;0;0;0;0;0;-10592.62097;325833.9639;0.2432839;0.274879;0.12345501
154;2;0;0;0;0;0;0;0;0;0;0;0;0;1;0;0.934865038;325832.7121;0.2427718;0.275066;0.12463033
155;1;0;0;0;0;0;0;0;1;0;0;1;0;0;0;2.958947667;325831.5496;0.2438579;0.275197;0.12431869
156;0;0;0;0;0;1;0;0;1;0;0;1;0;0;0;-3.867686314;325830.3919;0.2443411;0.275489;0.12515464
157;0;0;0;0;0;0;2;0;0;0;0;0;1;0;0;-68.28611422;325829.312;0.2430157;0.27737;0.1250373
158;0;0;0;1;0;0;1;0;0;0;0;0;0;0;0;-9773.541506;325828.2681;0.2434643;0.277856;0.12494914
159;0;0;0;1;0;0;1;0;0;0;0;0;0;0;1;120.5119972;325821.8395;0.2420457;0.27849;0.12519151
160;1;0;0;0;0;0;0;0;0;0;0;0;0;1;1;0.030257554;325820.8272;0.24268;0.277675;0.12676389
161;0;1;0;1;0;0;0;0;0;0;0;0;0;0;1;-19.67990911;325819.8278;0.2428634;0.277813;0.12693441
162;0;0;0;0;0;0;0;0;0;2;0;0;0;0;0;-24.62311069;325818.8352;0.2399069;0.261214;0.12710096
163;0;0;0;0;0;0;0;0;1;0;0;0;0;0;3;3.31E-05;325817.9265;0.2387851;0.261458;0.12774815
164;0;0;0;0;0;0;0;0;0;0;1;0;0;1;0;-5.280990164;325817.0621;0.2387954;0.261541;0.127629
165;1;1;0;0;0;1;0;0;0;0;0;0;0;0;0;2.362827401;325816.2002;0.2398272;0.262222;0.12852521
166;1;1;0;0;0;1;0;0;0;0;0;0;0;0;1;-0.023495931;325814.1396;0.2381109;0.263541;0.12862548
167;1;1;1;0;0;0;0;0;0;0;0;0;0;0;0;-5.056948284;325813.1653;0.2379774;0.263872;0.12873677
168;1;0;1;0;0;1;0;1;0;0;0;0;0;0;0;20.1784746;325812.315;0.2375515;0.263433;0.12880462
169;1;1;0;1;0;0;0;0;0;0;0;0;0;0;0;-461.0485399;325811.5205;0.2389723;0.263684;0.12956465
170;0;1;0;0;0;0;0;0;1;0;0;1;0;0;0;6.135659463;325810.9067;0.2383048;0.26463;0.12981998
171;0;0;0;1;0;0;0;2;0;0;0;0;0;0;1;2708.640285;325810.206;0.2372498;0.265443;0.12974755
172;0;0;0;1;0;0;0;3;0;0;0;0;0;0;1;9307.24875;325805.4156;0.2386137;0.264958;0.12861419
173;0;1;1;0;0;0;0;0;0;0;0;0;0;0;1;-0.1672332;325804.8103;0.2383906;0.265095;0.12867542
174;0;1;0;0;0;0;0;0;0;0;0;0;0;2;0;5.943918252;325804.2484;0.2383869;0.263663;0.12766657
175;0;1;0;0;0;0;0;0;0;0;0;1;0;0;1;-0.072163855;325803.6853;0.238024;0.263779;0.12749108
176;0;0;1;0;0;0;1;2;0;0;0;0;0;0;0;-1367.326828;325803.1312;0.2375614;0.263727;0.12755105
177;0;0;0;1;0;0;0;0;0;0;0;1;0;0;0;1133.780081;325802.6001;0.2374175;0.263837;0.12776936
178;1;1;0;0;0;0;0;0;0;0;0;1;0;0;0;-1.855597384;325802.0875;0.2370188;0.263923;0.12781271
179;3;0;0;0;0;0;0;0;0;0;0;0;0;0;0;0.991529761;325801.6876;0.24067;0.273655;0.13077892
180;3;0;0;0;0;0;0;0;0;0;0;0;0;0;1;-0.009156728;325766.2164;0.2406211;0.299769;0.14873691
181;3;0;0;0;0;1;0;0;0;0;0;0;0;0;0;-0.682413181;325744.0482;0.2475599;0.334602;0.17218421
182;3;0;0;0;0;0;0;1;0;0;0;0;0;0;0;-70.02490993;325726.5848;0.2453103;0.325847;0.15743926
183;2;0;0;0;0;0;0;2;0;0;0;0;0;0;0;-1883.772771;325699.8968;0.2381308;0.31303;0.1435803
184;4;0;0;0;0;0;0;0;0;0;0;0;0;0;0;-1.212542296;325671.562;0.2308995;0.327288;0.17307531
185;0;0;0;0;0;0;0;4;0;0;0;0;0;0;0;-157391.7646;325654.7567;0.2254437;0.308922;0.17453655
186;0;0;0;0;0;0;0;4;0;0;0;0;0;0;1;2127.737257;325643.7084;0.2207518;0.303092;0.17608894
187;2;0;0;0;0;0;0;2;0;0;0;0;0;0;1;21.16879879;325583.4508;0.205778;0.296198;0.19033057
188;3;0;0;0;0;0;0;1;0;0;0;0;0;0;1;0.624868933;325524.3552;0.1976252;0.268038;0.1643352
189;0;0;0;1;0;0;0;4;0;0;0;0;0;0;0;5216336.052;325514.8487;0.1991956;0.269855;0.16602096
190;3;0;0;0;0;0;1;0;0;0;0;0;0;0;0;-0.542970772;325506.0811;0.2009273;0.27544;0.17264937
191;4;0;0;0;0;0;0;0;0;0;0;0;0;0;1;0.007916445;325499.9902;0.1947409;0.280733;0.18377235
192;2;0;0;0;0;0;1;2;0;0;0;0;0;0;0;136.6845264;325498.8615;0.193133;0.279334;0.18212607
193;0;0;0;0;0;0;2;1;0;0;0;0;0;0;0;-526.8334518;325498.0046;0.1937328;0.279634;0.18167831
194;1;0;0;0;0;0;2;0;0;0;0;0;0;0;0;-32.63483108;325493.6289;0.1916108;0.269776;0.17848672
195;0;0;0;0;0;0;2;2;0;0;0;0;0;0;0;-2791.140869;325491.8213;0.1895837;0.261278;0.17601907
196;2;0;0;0;0;0;2;0;0;0;0;0;0;0;0;11.06010599;325491.0221;0.1910154;0.265157;0.17843712
197;0;0;1;0;0;1;0;0;0;0;0;0;0;0;1;0.094664073;325490.6453;0.190253;0.265218;0.17866998
198;0;0;2;0;0;0;0;0;0;0;0;0;0;0;0;13.2304011;325490.2738;0.1859563;0.258189;0.17767511
199;0;0;2;0;0;0;1;0;0;0;0;0;0;0;0;143.4834048;325488.0896;0.187463;0.260984;0.17887833
200;2;1;0;0;0;1;0;0;0;0;0;0;0;0;0;0.45956277;325487.7602;0.1860621;0.262403;0.18100537
201;2;0;0;0;0;0;0;0;0;0;0;1;0;0;0;0.98273283;325487.4373;0.185246;0.261712;0.18077937
202;0;0;0;0;0;0;0;0;0;1;0;0;0;1;0;8.974087847;325487.2226;0.1845218;0.263356;0.17953985
203;0;0;0;1;0;0;0;4;0;0;0;0;0;0;1;-33222.0971;325487.0684;0.1835812;0.263221;0.17909482
204;2;1;0;0;0;0;0;0;0;0;0;0;0;0;1;0.012137508;325486.8958;0.1839957;0.263827;0.18039757
205;3;1;0;0;0;0;0;0;0;0;0;0;0;0;0;-0.318826617;325486.6907;0.1843614;0.262811;0.17849477
206;4;1;0;0;0;0;0;0;0;0;0;0;0;0;0;0.195596431;325486.4366;0.1833236;0.26374;0.17733683
207;2;0;0;0;0;1;1;0;0;0;0;0;0;0;0;-2.436791657;325486.3445;0.1851654;0.265088;0.17880219
208;3;0;0;0;0;1;1;0;0;0;0;0;0;0;0;-1.755410935;325484.5212;0.1844763;0.26116;0.17337714
209;2;0;0;0;0;1;1;1;0;0;0;0;0;0;0;-12.47514148;325481.949;0.1832466;0.260019;0.17272587
210;2;0;0;0;0;2;0;1;0;0;0;0;0;0;0;3.932470328;325481.863;0.1839071;0.2581;0.17014148
211;0;0;0;0;0;2;0;3;0;0;0;0;0;0;0;-495.9167986;325480.8838;0.1835542;0.257265;0.1678172
212;0;0;0;0;0;1;1;2;0;0;0;0;0;0;0;-434.1231286;325480.792;0.184875;0.259865;0.16870778
213;0;0;0;0;0;1;1;3;0;0;0;0;0;0;0;-2854.583838;325479.033;0.1845491;0.260486;0.16749834
214;2;0;0;0;0;0;0;1;0;0;1;0;0;0;0;6.580541187;325478.8696;0.1841692;0.260902;0.1674463
215;1;0;0;0;0;0;0;0;0;0;0;0;0;2;0;7.078397233;325478.8268;0.18322;0.257304;0.16662795
216;0;0;1;0;0;0;0;0;0;0;0;0;0;1;0;-20.05861145;325478.7988;0.1841903;0.257349;0.16651203
217;1;0;1;0;0;0;0;0;0;0;0;0;0;1;0;11.89989763;325468.4025;0.1858017;0.257315;0.16581756
218;0;0;1;0;0;0;0;0;0;0;0;0;0;1;1;0.19824925;325468.3369;0.1858248;0.257163;0.16570658
219;0;0;0;0;0;1;0;1;0;0;0;1;0;0;0;18.33198083;325468.3183;0.1859786;0.257423;0.16549634
220;0;0;0;0;0;0;0;0;0;0;0;0;0;3;0;9.558269041;325468.3161;0.1846578;0.25833;0.16478878
221;0;0;0;0;0;0;0;0;0;0;0;0;0;4;0;37.24029314;325462.5752;0.1939784;0.264534;0.16779433
222;0;1;0;0;0;0;0;0;0;0;0;0;0;3;0;17.45973232;325459.9876;0.1957646;0.264641;0.16767096
223;1;0;0;0;0;0;0;0;0;0;0;0;0;3;0;-5.46827678;325459.5477;0.1939481;0.265565;0.16639035
224;1;0;0;0;0;0;0;0;0;0;0;0;0;4;0;-11.20825453;325458.9411;0.1943361;0.267728;0.16834735
\end{filecontents*}
\pgfplotstabletypeset[
        multicolumn names, 
        col sep=semicolon, 
        display columns/0/.style={column name=$\bm{k}$,dec sep align},
        display columns/1/.style={column type=C,column name=$\bm{r_{k}^{1}}$},
        display columns/2/.style={column type=C,column name=$\bm{r_{k}^{2}}$},
        display columns/3/.style={column type=C,column name=$\bm{r_{k}^{3}}$},
        display columns/4/.style={column type=C,column name=$\bm{r_{k}^{4}}$},
        display columns/5/.style={column type=C,column name=$\bm{r_{k}^{5}}$},
        display columns/6/.style={column type=C,column name=$\bm{r_{k}^{6}}$},
        display columns/7/.style={column type=C,column name=$\bm{r_{k}^{7}}$},
        display columns/8/.style={column type=C,column name=$\bm{r_{k}^{8}}$},
        display columns/9/.style={column type=C,column name=$\bm{r_{k}^{9}}$},
        display columns/10/.style={column type=C,column name=$\bm{r_{k}^{10}}$},
        display columns/11/.style={column type=C,column name=$\bm{r_{k}^{11}}$},
        display columns/12/.style={column type=C,column name=$\bm{r_{k}^{12}}$},
        display columns/13/.style={column type=C,column name=$\bm{r_{k}^{13}}$},
        display columns/14/.style={column type=C,column name=$\bm{r_{k}^{14}}$},
        display columns/15/.style={column type=C,column name=$\bm{r_{k}^{15}}$},
        display columns/16/.style={column name=$\bm{\widehat{\beta}_{\textbf{\text{OLS}},k}}$,fixed,fixed zerofill,precision=2,set thousands separator={,},dec sep align},
        display columns/17/.style={column name={\textbf{AIC}},fixed,fixed zerofill,precision=0,set thousands separator={,},dec sep align},
        display columns/18/.style={column name={\textbf{v.mae}},fixed,fixed zerofill,precision=3,set thousands separator={,},dec sep align},
        display columns/19/.style={column name={\textbf{ns.mae}},fixed,fixed zerofill,precision=3,set thousands separator={,},dec sep align},
        display columns/20/.style={column name={\textbf{cr.mae}},fixed,fixed zerofill,precision=3,set thousands separator={,},dec sep align},
	    	every head row/.style={before row=\toprule,after row=\midrule},
	    	every nth row={10}{before row=\midrule},
      	every last row/.style={after row=\bottomrule},
      ]{BGROOLS300886adalgwre3.csv} 
    \caption[]{Cont.}
    \label{tab:BGROOLS3008863}
  \end{center}
\end{table}

\begin{table}[htb]
  \tiny
  \begin{center}
  \tabcolsep=0.083cm
  \renewcommand{\arraystretch}{0.90}
      \begin{filecontents*}{BGROOLSvalfig.csv}
k;v.wre.y;v.a.wre.y;v.res.avg.y;v.wre.delta.y;v.res.avg.delta.y;ns.wre.y;ns.a.wre.y;ns.res.avg.y;ns.wre.delta.y;ns.res.avg.delta.y;cr.wre.y;cr.a.wre.y;cr.res.avg.y;cr.wre.delta.y;cr.res.avg.delta.y
0;4.556602;4.3565067;-237.5325499;100;38.252875;3.23053;3.121333;-0.294727;100;261.099235;4.02699629;3.94212147;105.619188;100;367.0131501
10;0.8386126;0.8017863;-0.1621256;21.4677825;104.479234;0.388718;0.375579;23.084391;21.65881;113.334288;0.64994998;0.63625134;88.50389457;27.11154378;178.7537913
20;0.5648041;0.5400017;-9.6425216;16.7795342;81.9455509;0.59708;0.576898;-75.11821;8.273502;2.0784;0.45439558;0.44481853;-39.67540492;10.08259277;37.52120479
30;0.5184664;0.4956989;1.2706449;17.501119;100.0249494;0.41797;0.403842;-47.27442;7.969525;37.088422;0.26442846;0.25885524;1.06815141;13.37811619;85.43099318
40;0.4749207;0.4540654;-10.0802125;16.8877617;98.1555422;0.509415;0.492196;-66.483768;6.233981;27.360524;0.29103128;0.28489737;-25.69595959;10.49700567;68.14833229
50;0.3684989;0.3523169;-15.2033082;13.2677566;77.6943374;0.390834;0.377623;-49.997272;6.059911;28.508911;0.22138584;0.21671981;-9.21674612;10.67401636;69.28943673
60;0.3059923;0.2925552;-17.1711148;10.7604672;61.9613566;0.300561;0.290402;-35.79946;5.863104;28.941549;0.18312963;0.1792699;4.60399207;10.65126644;69.34500059
70;0.2912463;0.2784568;-17.8768297;10.4511338;59.9776421;0.281253;0.271746;-33.130104;6.060042;30.332905;0.17507827;0.17138824;8.12076288;10.95844816;71.58377196
80;0.2626108;0.2510787;-23.1963879;9.389059;54.072869;0.308787;0.29835;-40.649427;4.83747;22.228367;0.15723232;0.15391842;-3.79955745;8.9449389;59.07823656
90;0.2673225;0.2555835;-24.4509864;9.1959709;53.5277726;0.313173;0.302587;-41.85074;4.689074;21.736556;0.15836889;0.15503103;-6.87641503;8.58650149;56.71088109
100;0.2495997;0.2386389;-18.3789974;9.152235;53.235207;0.275657;0.266339;-35.283024;4.63726;21.939718;0.13576231;0.13290092;-0.38504662;8.60570218;56.837695
110;0.2367017;0.2263073;-18.1080475;8.4944072;48.4026547;0.263586;0.254677;-34.080586;4.144096;18.038653;0.1291863;0.12646351;-1.69920289;7.63401503;50.4200365
120;0.2407819;0.2302083;-16.2653117;8.8964383;50.4176341;0.266658;0.257644;-34.452899;4.152891;17.838584;0.1242983;0.12167853;-1.57403742;7.67904525;50.7174456
130;0.2500524;0.2390718;-17.6873916;9.8388934;56.9116006;0.281126;0.271624;-36.645724;4.809753;23.561805;0.12249777;0.11991595;-1.42844209;8.8996452;58.77908729
140;0.2462088;0.235397;-14.6673748;9.8546538;56.7448605;0.262717;0.253836;-33.348242;4.809249;23.67253;0.1196701;0.11714788;1.24569801;8.82203073;58.26647047
150;0.247487;0.236619;-14.4860724;9.9236248;57.011081;0.271045;0.261883;-34.843727;4.612476;22.261963;0.12219633;0.11962086;-0.71987917;8.53728323;56.38581146
\end{filecontents*}
\pgfplotstabletypeset[
        multicolumn names, 
        col sep=semicolon, 
        display columns/0/.style={column name=$\bm{k}$,dec sep align},
        display columns/1/.style={column name={\textbf{v.mae}},fixed,fixed zerofill,precision=3,set thousands separator={,},dec sep align},
        display columns/2/.style={column name={$\bm{\textbf{\text{v.mae}}^a}$},fixed,fixed zerofill,precision=3,set thousands separator={,},dec sep align},
        display columns/3/.style={column name={\textbf{v.res}},fixed,fixed zerofill,precision=0,set thousands separator={,},dec sep align},
        display columns/4/.style={column name={$\bm{\textbf{\text{v.mae}}^0}$},fixed,fixed zerofill,precision=3,set thousands separator={,},dec sep align},
        display columns/5/.style={column name={$\bm{\textbf{\text{v.res}}^0}$},fixed,fixed zerofill,precision=0,set thousands separator={,},dec sep align},
        display columns/6/.style={column name={\textbf{ns.mae}},fixed,fixed zerofill,precision=3,set thousands separator={,},dec sep align},
        display columns/7/.style={column name={$\bm{\textbf{\text{ns.mae}}^a}$},fixed,fixed zerofill,precision=3,set thousands separator={,},dec sep align},
        display columns/8/.style={column name={\textbf{ns.res}},fixed,fixed zerofill,precision=0,set thousands separator={,},dec sep align},
        display columns/9/.style={column name={$\bm{\textbf{\text{ns.mae}}^0}$},fixed,fixed zerofill,precision=3,set thousands separator={,},dec sep align},
        display columns/10/.style={column name={$\bm{\textbf{\text{ns.res}}^0}$},fixed,fixed zerofill,precision=0,set thousands separator={,},dec sep align},
        display columns/11/.style={column name={\textbf{cr.mae}},fixed,fixed zerofill,precision=3,set thousands separator={,},dec sep align},
        display columns/12/.style={column name={$\bm{\textbf{\text{cr.mae}}^a}$},fixed,fixed zerofill,precision=3,set thousands separator={,},dec sep align},
        display columns/13/.style={column name={\textbf{cr.res}},fixed,fixed zerofill,precision=0,set thousands separator={,},dec sep align},
        display columns/14/.style={column name={$\bm{\textbf{\text{cr.mae}}^0}$},fixed,fixed zerofill,precision=3,set thousands separator={,},dec sep align},
        display columns/15/.style={column name={$\bm{\textbf{\text{cr.res}}^0}$},fixed,fixed zerofill,precision=0,set thousands separator={,},dec sep align},
	    	every head row/.style={before row=\toprule,after row=\midrule},
      	every last row/.style={after row=\bottomrule},
      ]{BGROOLSvalfig.csv} 
    \caption{Out-of-sample validation figures of the OLS proxy function of BEL under $150$-$443$ after each tenth iteration.}
    \label{tab:BGROOLSvalfig}
  \end{center}
\end{table}

\begin{table}[htb]
  \tiny
  \begin{center}
  \tabcolsep=0.083cm
  \renewcommand{\arraystretch}{0.90}
      \begin{filecontents*}{PGROOLSvalfig.csv}
k;v.wre.y;v.a.wre.y;v.res.avg.y;v.wre.delta.y;v.res.avg.delta.y;ns.wre.y;ns.a.wre.y;ns.res.avg.y;ns.wre.delta.y;ns.res.avg.delta.y;cr.wre.y;cr.a.wre.y;cr.res.avg.y;cr.wre.delta.y;cr.res.avg.delta.y
0;60.619885;3.178492;-295.892759;100;-207.047418;97.518451;2.935691;-453.396059;100;-369.007473;257.761878;4.2506704;-652.5503704;100;-568.1617845
10;15.986842;0.838241;-1.17785;29.161044;-109.704512;10.264416;0.309;-6.120753;32.49153;-119.10417;31.4613306;0.518819;-67.1443746;31.7036091;-180.1277911
20;10.291748;0.539629;9.765477;21.029112;-81.797891;19.163175;0.576887;75.041201;12.24002;-20.978923;26.9953565;0.445172;39.4387155;13.3241425;-56.5814084
30;9.443616;0.495159;-1.219392;21.970837;-100.021297;13.408763;0.403657;47.136774;15.582708;-56.121887;15.7662238;0.2599959;-1.4001689;18.7591967;-104.6588295
40;8.656012;0.453862;10.199859;21.196583;-98.127221;16.35901;0.492471;66.524621;12.739668;-46.259214;17.2712312;0.2848145;25.6280346;15.4338272;-87.1558004
50;6.717981;0.352245;15.330405;16.654805;-77.687456;12.565016;0.378257;50.102662;12.938354;-47.371954;13.1823811;0.2173865;9.266304;15.6662024;-88.2083123
60;5.580913;0.292625;17.279931;13.505817;-61.972229;9.671341;0.291146;35.930315;12.985385;-47.778601;10.9039555;0.1798137;-4.5030552;15.6395095;-88.2119714
70;5.313013;0.278578;18.765965;13.025911;-59.37515;9.095344;0.273806;33.623371;13.289453;-48.974499;10.3572533;0.1707982;-7.7468061;15.9748821;-90.3446766
80;4.688216;0.245818;20.912901;11.326063;-51.466468;9.068788;0.273006;35.887078;11.130734;-40.949047;8.997388;0.1483731;-0.0586056;13.5904063;-76.8947307
90;4.870078;0.255354;24.414876;11.524553;-53.434849;10.04327;0.302342;41.799557;10.995057;-40.506924;9.41375;0.1552392;6.8235004;13.285473;-75.4829803
100;4.546051;0.238364;18.379721;11.471235;-53.151016;8.847298;0.266339;35.27463;11.041018;-40.712862;8.0812325;0.1332651;0.3766168;13.3079834;-75.6108759
110;4.312884;0.226138;18.115867;10.650134;-48.330093;8.463434;0.254783;34.088605;9.998694;-36.814111;7.6887483;0.1267927;1.6954024;12.1809166;-69.2073132
120;4.429833;0.23227;16.151557;11.349828;-51.318755;8.303748;0.249976;32.875352;10.595779;-39.051715;7.1869156;0.1185172;-0.5875724;12.7630267;-72.5146401
130;4.55526;0.238847;17.68118;12.345439;-56.88275;9.024475;0.271672;36.637976;11.491029;-42.38271;7.2850606;0.1201357;1.3913371;13.6632471;-77.6293488
140;4.531634;0.237608;14.649892;12.470025;-57.399493;8.722103;0.26257;34.895189;11.281964;-41.610951;7.2271308;0.1191804;0.1015485;13.4476822;-76.4045912
150;4.511601;0.236558;14.489929;12.458836;-57.026459;8.712216;0.262272;34.899712;11.136043;-41.073432;7.2652011;0.1198082;0.7371207;13.242007;-75.2360232
\end{filecontents*}
\pgfplotstabletypeset[
        multicolumn names, 
        col sep=semicolon, 
        display columns/0/.style={column name=$\bm{k}$,dec sep align},
        display columns/1/.style={column name={\textbf{v.mae}},fixed,fixed zerofill,precision=3,set thousands separator={,},dec sep align},
        display columns/2/.style={column name={$\bm{\textbf{\text{v.mae}}^a}$},fixed,fixed zerofill,precision=3,set thousands separator={,},dec sep align},
        display columns/3/.style={column name={\textbf{v.res}},fixed,fixed zerofill,precision=0,set thousands separator={,},dec sep align},
        display columns/4/.style={column name={$\bm{\textbf{\text{v.mae}}^0}$},fixed,fixed zerofill,precision=3,set thousands separator={,},dec sep align},
        display columns/5/.style={column name={$\bm{\textbf{\text{v.res}}^0}$},fixed,fixed zerofill,precision=0,set thousands separator={,},dec sep align},
        display columns/6/.style={column name={\textbf{ns.mae}},fixed,fixed zerofill,precision=3,set thousands separator={,},dec sep align},
        display columns/7/.style={column name={$\bm{\textbf{\text{ns.mae}}^a}$},fixed,fixed zerofill,precision=3,set thousands separator={,},dec sep align},
        display columns/8/.style={column name={\textbf{ns.res}},fixed,fixed zerofill,precision=0,set thousands separator={,},dec sep align},
        display columns/9/.style={column name={$\bm{\textbf{\text{ns.mae}}^0}$},fixed,fixed zerofill,precision=3,set thousands separator={,},dec sep align},
        display columns/10/.style={column name={$\bm{\textbf{\text{ns.res}}^0}$},fixed,fixed zerofill,precision=0,set thousands separator={,},dec sep align},
        display columns/11/.style={column name={\textbf{cr.mae}},fixed,fixed zerofill,precision=3,set thousands separator={,},dec sep align},
        display columns/12/.style={column name={$\bm{\textbf{\text{cr.mae}}^a}$},fixed,fixed zerofill,precision=3,set thousands separator={,},dec sep align},
        display columns/13/.style={column name={\textbf{cr.res}},fixed,fixed zerofill,precision=0,set thousands separator={,},dec sep align},
        display columns/14/.style={column name={$\bm{\textbf{\text{cr.mae}}^0}$},fixed,fixed zerofill,precision=3,set thousands separator={,},dec sep align},
        display columns/15/.style={column name={$\bm{\textbf{\text{cr.res}}^0}$},fixed,fixed zerofill,precision=0,set thousands separator={,},dec sep align},
	    	every head row/.style={before row=\toprule,after row=\midrule},
      	every last row/.style={after row=\bottomrule},
      ]{PGROOLSvalfig.csv} 
    \caption{Out-of-sample validation figures of the OLS proxy function of AC under $150$-$443$ after each tenth iteration.}
    \label{tab:PGROOLSvalfig}
  \end{center}
\end{table}

\begin{table}[htb]
  \tiny
  \begin{center}
  \tabcolsep=0.083cm
  \renewcommand{\arraystretch}{0.90}
      \begin{filecontents*}{BGROOLS300886valfig.csv}
k;v.wre.y;v.a.wre.y;v.res.avg.y;v.wre.delta.y;v.res.avg.delta.y;ns.wre.y;ns.a.wre.y;ns.res.avg.y;ns.wre.delta.y;ns.res.avg.delta.y;cr.wre.y;cr.a.wre.y;cr.res.avg.y;cr.wre.delta.y;cr.res.avg.delta.y
0;4.556602;4.3565067;-237.5325499;100;38.252875;3.23053;3.121333;-0.294727;100;261.099235;4.02699629;3.94212147;105.619188;100;367.0131501
10;0.8386126;0.8017863;-0.1621256;21.4677825;104.479234;0.388718;0.375579;23.084391;21.65881;113.334288;0.64994998;0.63625134;88.50389457;27.11154378;178.7537913
20;0.5648041;0.5400017;-9.6425216;16.7795342;81.9455509;0.59708;0.576898;-75.11821;8.273502;2.0784;0.45439558;0.44481853;-39.67540492;10.08259277;37.52120479
30;0.5184664;0.4956989;1.2706449;17.501119;100.0249494;0.41797;0.403842;-47.27442;7.969525;37.088422;0.26442846;0.25885524;1.06815141;13.37811619;85.43099318
40;0.4749207;0.4540654;-10.0802125;16.8877617;98.1555422;0.509415;0.492196;-66.483768;6.233981;27.360524;0.29103128;0.28489737;-25.69595959;10.49700567;68.14833229
50;0.3684989;0.3523169;-15.2033082;13.2677566;77.6943374;0.390834;0.377623;-49.997272;6.059911;28.508911;0.22138584;0.21671981;-9.21674612;10.67401636;69.28943673
60;0.3059923;0.2925552;-17.1711148;10.7604672;61.9613566;0.300561;0.290402;-35.79946;5.863104;28.941549;0.18312963;0.1792699;4.60399207;10.65126644;69.34500059
70;0.2912463;0.2784568;-17.8768297;10.4511338;59.9776421;0.281253;0.271746;-33.130104;6.060042;30.332905;0.17507827;0.17138824;8.12076288;10.95844816;71.58377196
80;0.2626108;0.2510787;-23.1963879;9.389059;54.072869;0.308787;0.29835;-40.649427;4.83747;22.228367;0.15723232;0.15391842;-3.79955745;8.9449389;59.07823656
90;0.2673225;0.2555835;-24.4509864;9.1959709;53.5277726;0.313173;0.302587;-41.85074;4.689074;21.736556;0.15836889;0.15503103;-6.87641503;8.58650149;56.71088109
100;0.2495997;0.2386389;-18.3789974;9.152235;53.235207;0.275657;0.266339;-35.283024;4.63726;21.939718;0.13576231;0.13290092;-0.38504662;8.60570218;56.837695
110;0.2391895;0.2286859;-18.2697562;9.1318849;52.4401679;0.269454;0.260346;-34.817088;4.576795;21.501374;0.1319992;0.12921712;-1.11359284;8.35847858;55.20486842
120;0.2416919;0.2310784;-16.2447359;9.5189854;54.4798139;0.272591;0.263377;-35.185585;4.568789;21.147502;0.12854567;0.12583638;-0.98573617;8.3800516;55.34735076
130;0.2508547;0.2398389;-17.9936575;10.5057972;60.9790765;0.287005;0.277304;-37.239566;5.421029;27.341705;0.12747051;0.12478388;-0.36020699;9.723606;64.22106419
140;0.246235;0.235422;-14.7233683;10.5295697;60.7143138;0.268809;0.259723;-34.109933;5.328821;26.936286;0.12283709;0.12024812;1.86835267;9.52579215;62.91457192
150;0.2421701;0.2315356;-14.3337955;10.555538;61.0133723;0.274369;0.265095;-35.233973;5.119336;25.721732;0.12281914;0.12023054;0.20945758;9.26091694;61.16516256
160;0.24268;0.2320232;-14.7932579;10.4833875;60.3997938;0.277675;0.268289;-35.819964;5.017658;24.981625;0.12676389;0.12409215;-0.40640356;9.14433596;60.39518535
170;0.2383048;0.22784;-13.0482087;10.1398687;58.1514403;0.26463;0.255686;-32.601165;4.967993;24.207021;0.12981998;0.12708383;1.86675397;8.88387646;58.67494019
180;0.2406211;0.2300546;-12.1162162;10.128107;57.4920735;0.299769;0.289636;-36.805884;4.552101;18.410943;0.14873691;0.14560206;2.31307619;8.71601141;57.52990312
190;0.2009273;0.1921039;-12.5994577;6.4578412;32.3504691;0.27544;0.26613;-32.913782;4.124379;-2.355318;0.17264937;0.16901053;-3.65510729;4.72105174;26.90335671
200;0.1860621;0.1778915;-9.4924849;6.1108683;29.4113408;0.262403;0.253533;-28.678883;4.460319;-4.16652;0.18100537;0.17719042;2.59409644;4.91978969;27.10645939
210;0.1839071;0.1758312;-9.0048296;6.2100666;30.3658668;0.2581;0.249376;-28.117908;4.336919;-3.138675;0.17014148;0.1665555;2.94971024;4.84606976;27.92894379
220;0.1846578;0.1765489;-7.622243;6.4330316;32.22885;0.25833;0.249598;-28.373627;4.285722;-2.913997;0.16478878;0.16131562;2.52621979;4.8495868;27.98584994
224;0.1943361;0.1858021;-8.7292706;6.658858;34.2822165;0.267728;0.258678;-30.207234;4.199764;-1.58721;0.16834735;0.16479918;0.75346231;5.00742678;29.37348663
\end{filecontents*}
\pgfplotstabletypeset[
        multicolumn names, 
        col sep=semicolon, 
        display columns/0/.style={column name=$\bm{k}$,dec sep align},
        display columns/1/.style={column name={\textbf{v.mae}},fixed,fixed zerofill,precision=3,set thousands separator={,},dec sep align},
        display columns/2/.style={column name={$\bm{\textbf{\text{v.mae}}^a}$},fixed,fixed zerofill,precision=3,set thousands separator={,},dec sep align},
        display columns/3/.style={column name={\textbf{v.res}},fixed,fixed zerofill,precision=0,set thousands separator={,},dec sep align},
        display columns/4/.style={column name={$\bm{\textbf{\text{v.mae}}^0}$},fixed,fixed zerofill,precision=3,set thousands separator={,},dec sep align},
        display columns/5/.style={column name={$\bm{\textbf{\text{v.res}}^0}$},fixed,fixed zerofill,precision=0,set thousands separator={,},dec sep align},
        display columns/6/.style={column name={\textbf{ns.mae}},fixed,fixed zerofill,precision=3,set thousands separator={,},dec sep align},
        display columns/7/.style={column name={$\bm{\textbf{\text{ns.mae}}^a}$},fixed,fixed zerofill,precision=3,set thousands separator={,},dec sep align},
        display columns/8/.style={column name={\textbf{ns.res}},fixed,fixed zerofill,precision=0,set thousands separator={,},dec sep align},
        display columns/9/.style={column name={$\bm{\textbf{\text{ns.mae}}^0}$},fixed,fixed zerofill,precision=3,set thousands separator={,},dec sep align},
        display columns/10/.style={column name={$\bm{\textbf{\text{ns.res}}^0}$},fixed,fixed zerofill,precision=0,set thousands separator={,},dec sep align},
        display columns/11/.style={column name={\textbf{cr.mae}},fixed,fixed zerofill,precision=3,set thousands separator={,},dec sep align},
        display columns/12/.style={column name={$\bm{\textbf{\text{cr.mae}}^a}$},fixed,fixed zerofill,precision=3,set thousands separator={,},dec sep align},
        display columns/13/.style={column name={\textbf{cr.res}},fixed,fixed zerofill,precision=0,set thousands separator={,},dec sep align},
        display columns/14/.style={column name={$\bm{\textbf{\text{cr.mae}}^0}$},fixed,fixed zerofill,precision=3,set thousands separator={,},dec sep align},
        display columns/15/.style={column name={$\bm{\textbf{\text{cr.res}}^0}$},fixed,fixed zerofill,precision=0,set thousands separator={,},dec sep align},
	    	every head row/.style={before row=\toprule,after row=\midrule},
      	every last row/.style={after row=\bottomrule},
      ]{BGROOLS300886valfig.csv} 
    \caption{Out-of-sample validation figures of the OLS proxy function of BEL under $300$-$886$ after each tenth and the final iteration.}
    \label{tab:BGROOLS300886valfig}
  \end{center}
\end{table}

\begin{table}[htb]
  \tiny
  \begin{center}
  \tabcolsep=0.083cm
  \renewcommand{\arraystretch}{0.90}
      \begin{filecontents*}{BGROOLSstderrvalfig.csv}
k;v.wre.y;v.a.wre.y;v.res.avg.y;v.wre.delta.y;v.res.avg.delta.y;ns.wre.y;ns.a.wre.y;ns.res.avg.y;ns.wre.delta.y;ns.res.avg.delta.y;cr.wre.y;cr.a.wre.y;cr.res.avg.y;cr.wre.delta.y;cr.res.avg.delta.y
150;0.285507435;0.272684585;-29.83352489;9.8777783;56.58983704;0.329980147;0.318600181;-45.63966297;3.9153352;16.36016537;0.151481886;0.14813849;-12.83740417;7.473377868;49.16242417
150;0.247487;0.236619;-14.4860724;9.9236248;57.011081;0.271045;0.261883;-34.843727;4.612476;22.261963;0.12219633;0.11962086;-0.71987917;8.53728323;56.38581146
150;0.231048576;0.221133344;0.861380095;9.976781264;57.43232502;0.219376562;0.212118048;-24.02072181;5.472628517;28.19083058;0.130151797;0.127484647;11.39764521;9.591186501;63.6091976
224;0.236055607;0.225453762;-24.07674405;6.756650065;33.86106238;0.325430598;0.314207533;-41.01768977;4.609756634;-7.503416927;0.190749196;0.18653912;-11.3641268;4.307088421;22.15014604
224;0.1943361;0.1858021;-8.7292706;6.658858;34.2822165;0.267728;0.258678;-30.207234;4.199764;-1.58721;0.16834735;0.16479918;0.75346231;5.00742678;29.37348663
224;0.184470908;0.176554513;6.618160932;6.624720016;34.70355036;0.21782081;0.210613771;-19.39874861;3.981844377;4.327248281;0.172608433;0.169071237;12.87092258;5.812867253;36.59691947
\end{filecontents*}
\pgfplotstabletypeset[
        multicolumn names, 
        col sep=semicolon, 
        display columns/0/.style={column name=$\bm{k}$},
        display columns/1/.style={column name={\textbf{v.mae}},column type={r},fixed,fixed zerofill,precision=3},
        display columns/2/.style={column name={$\bm{\textbf{\text{v.mae}}^a}$},column type={r},fixed,fixed zerofill,precision=3},
        display columns/3/.style={column name={\textbf{v.res}},column type={r},fixed,fixed zerofill,precision=0},
        display columns/4/.style={column name={$\bm{\textbf{\text{v.mae}}^0}$},column type={r},fixed,fixed zerofill,precision=3},
        display columns/5/.style={column name={$\bm{\textbf{\text{v.res}}^0}$},column type={r},fixed,fixed zerofill,precision=0},
        display columns/6/.style={column name={\textbf{ns.mae}},column type={r},fixed,fixed zerofill,precision=3},
        display columns/7/.style={column name={$\bm{\textbf{\text{ns.mae}}^a}$},column type={r},fixed,fixed zerofill,precision=3},
        display columns/8/.style={column name={\textbf{ns.res}},column type={r},fixed,fixed zerofill,precision=0},
        display columns/9/.style={column name={$\bm{\textbf{\text{ns.mae}}^0}$},column type={r},fixed,fixed zerofill,precision=3},
        display columns/10/.style={column name={$\bm{\textbf{\text{ns.res}}^0}$},column type={r},fixed,fixed zerofill,precision=0},
        display columns/11/.style={column name={\textbf{cr.mae}},column type={r},fixed,fixed zerofill,precision=3},
        display columns/12/.style={column name={$\bm{\textbf{\text{cr.mae}}^a}$},column type={r},fixed,fixed zerofill,precision=3},
        display columns/13/.style={column name={\textbf{cr.res}},column type={r},fixed,fixed zerofill,precision=0},
        display columns/14/.style={column name={$\bm{\textbf{\text{cr.mae}}^0}$},column type={r},fixed,fixed zerofill,precision=3},
        display columns/15/.style={column name={$\bm{\textbf{\text{cr.res}}^0}$},column type={r},fixed,fixed zerofill,precision=0},
	    	every head row/.style={before row=\toprule,after row=\midrule},
	    	every row no 0/.style={before row={\multicolumn{16}{l}{\textbf{150-443 figures based on validation values minus $\bm{1.96}$ times standard errors}}\\ \\}},
	    	every row no 1/.style={before row={\midrule \multicolumn{16}{l}{\textbf{150-443 figures based on validation values}}\\ \\}},
	    	every row no 2/.style={before row={\midrule \multicolumn{16}{l}{\textbf{150-443 figures based on validation values plus $\bm{1.96}$ times standard errors}}\\ \\}},
	    	every row no 3/.style={before row={\midrule \multicolumn{16}{l}{\textbf{300-886 figures based on validation values minus $\bm{1.96}$ times standard errors}}\\ \\}},
	    	every row no 4/.style={before row={\midrule \multicolumn{16}{l}{\textbf{300-886 figures based on validation values}}\\ \\}},
	    	every row no 5/.style={before row={\midrule \multicolumn{16}{l}{\textbf{300-886 figures based on validation values plus $\bm{1.96}$ times standard errors}}\\ \\}},
      	every last row/.style={after row=\bottomrule},
      ]{BGROOLSstderrvalfig.csv} 
    \caption{Out-of-sample validation figures of the derived OLS proxy functions of BEL under $150$-$443$ and $300$-$886$ after the final iteration based on three different sets of validation value estimates. Thereby emerges the first set of validation value estimates from pointwise subtraction of $1.96$ times the standard errors from the original set of validation values. The second set is the original set. The third set is the addition counterpart of the first set.}
    \label{tab:BGROOLSstderrvalfig}
  \end{center}
\end{table}

\begin{table}[htb]
  \tiny
  \begin{center}
  \tabcolsep=0.052cm
  \renewcommand{\arraystretch}{0.90}
      \begin{filecontents*}{BGROGLMgaussianidinvlogvalfig.csv}
k;AIC;v.wre.y;v.a.wre.y;v.res.avg.y;v.wre.delta.y;v.res.avg.delta.y;ns.wre.y;ns.a.wre.y;ns.res.avg.y;ns.wre.delta.y;ns.res.avg.delta.y;cr.wre.y;cr.a.wre.y;cr.res.avg.y;cr.wre.delta.y;cr.res.avg.delta.y
0;437251.0284;4.556602;4.3565067;-237.5325499;100;38.252875;3.23053;3.121333;-0.294727;100;261.099235;4.02699629;3.94212147;105.619188;100;367.0131501
10;345045.2474;0.8386126;0.8017863;-0.1621256;21.4677825;104.479234;0.388718;0.375579;23.084391;21.65881;113.334288;0.64994998;0.63625134;88.50389457;27.11154378;178.7537913
20;333447.4362;0.5648041;0.5400017;-9.6425216;16.7795342;81.9455509;0.59708;0.576898;-75.11821;8.273502;2.0784;0.45439558;0.44481853;-39.67540492;10.08259277;37.52120479
30;330361.1506;0.5184664;0.4956989;1.2706449;17.501119;100.0249494;0.41797;0.403842;-47.27442;7.969525;37.088422;0.26442846;0.25885524;1.06815141;13.37811619;85.43099318
40;328832.2073;0.4749207;0.4540654;-10.0802125;16.8877617;98.1555422;0.509415;0.492196;-66.483768;6.233981;27.360524;0.29103128;0.28489737;-25.69595959;10.49700567;68.14833229
50;327432.0902;0.3684989;0.3523169;-15.2033082;13.2677566;77.6943374;0.390834;0.377623;-49.997272;6.059911;28.508911;0.22138584;0.21671981;-9.21674612;10.67401636;69.28943673
60;326787.1036;0.3059923;0.2925552;-17.1711148;10.7604672;61.9613566;0.300561;0.290402;-35.79946;5.863104;28.941549;0.18312963;0.1792699;4.60399207;10.65126644;69.34500059
70;326452.7548;0.2912463;0.2784568;-17.8768297;10.4511338;59.9776421;0.281253;0.271746;-33.130104;6.060042;30.332905;0.17507827;0.17138824;8.12076288;10.95844816;71.58377196
80;326245.0608;0.2626108;0.2510787;-23.1963879;9.389059;54.072869;0.308787;0.29835;-40.649427;4.83747;22.228367;0.15723232;0.15391842;-3.79955745;8.9449389;59.07823656
90;326115.6586;0.2673225;0.2555835;-24.4509864;9.1959709;53.5277726;0.313173;0.302587;-41.85074;4.689074;21.736556;0.15836889;0.15503103;-6.87641503;8.58650149;56.71088109
100;326038.3955;0.2495997;0.2386389;-18.3789974;9.152235;53.235207;0.275657;0.266339;-35.283024;4.63726;21.939718;0.13576231;0.13290092;-0.38504662;8.60570218;56.837695
110;325967.8195;0.2367017;0.2263073;-18.1080475;8.4944072;48.4026547;0.263586;0.254677;-34.080586;4.144096;18.038653;0.1291863;0.12646351;-1.69920289;7.63401503;50.4200365
120;325928.4071;0.2407819;0.2302083;-16.2653117;8.8964383;50.4176341;0.266658;0.257644;-34.452899;4.152891;17.838584;0.1242983;0.12167853;-1.57403742;7.67904525;50.7174456
130;325896.2645;0.2500524;0.2390718;-17.6873916;9.8388934;56.9116006;0.281126;0.271624;-36.645724;4.809753;23.561805;0.12249777;0.11991595;-1.42844209;8.8996452;58.77908729
140;325872.6555;0.2462088;0.235397;-14.6673748;9.8546538;56.7448605;0.262717;0.253836;-33.348242;4.809249;23.67253;0.1196701;0.11714788;1.24569801;8.82203073;58.26647047
150;325849.7177;0.247487;0.236619;-14.4860724;9.9236248;57.011081;0.271045;0.261883;-34.843727;4.612476;22.261963;0.12219633;0.11962086;-0.71987917;8.53728323;56.38581146
0;437251.0284;4.556602;4.356507;-237.53255;100;38.252875;3.23053;3.121333;-0.294727;100;261.099235;4.0269963;3.9421215;105.6191879;100;367.0131501
10;343425.573;1.035822;0.990336;0.95747;33.705291;192.399545;0.650216;0.628238;-63.149782;21.480893;113.90083;0.3906022;0.3823697;44.0862646;33.4819718;221.136877
20;334984.9692;0.689454;0.659177;-6.452451;21.312708;118.472926;0.515445;0.498022;-61.623747;10.319182;48.910168;0.3236438;0.3168226;-4.0056833;16.4927571;106.5282311
30;331426.1329;0.512223;0.48973;-15.559177;18.835933;108.678125;0.393484;0.380184;-44.977252;12.276885;64.868587;0.2481155;0.2428861;14.8013397;18.9598812;124.6471792
40;328875.4835;0.433163;0.414141;-5.072545;14.353638;82.448451;0.31665;0.305946;-25.923256;9.312083;47.206278;0.2938246;0.2876318;25.996328;15.1876994;99.1258612
50;327876.6139;0.382976;0.366159;-8.463925;12.958692;75.684936;0.28544;0.275792;-23.511261;8.961099;46.246138;0.2712127;0.2654965;25.2246805;14.5915226;94.9820791
60;327273.5134;0.337446;0.322628;-16.271187;12.572085;72.987357;0.327504;0.316433;-36.647847;7.635989;38.219234;0.219136;0.2145173;10.1392807;13.0874836;85.0063612
70;326874.6213;0.28963;0.276911;-13.855417;11.247847;64.340537;0.270597;0.26145;-32.353253;6.232988;31.451239;0.1561271;0.1528365;5.6966873;10.5880666;69.5011792
80;326602.8124;0.259478;0.248084;-16.297278;9.975877;57.788291;0.287232;0.277524;-37.738076;5.041717;21.956031;0.1580564;0.1547251;-7.6188188;8.0138476;52.0752878
90;326390.4865;0.254097;0.242938;-19.842845;8.46189;46.824473;0.391958;0.378709;-51.45543;4.450825;0.820426;0.2198171;0.2151841;-16.5024189;5.6759814;35.7734364
100;326224.9856;0.269534;0.257698;-21.080107;8.883787;49.156682;0.392563;0.379294;-50.544321;4.454096;5.301005;0.2192287;0.2146081;-12.1182572;6.7318149;43.7270686
110;326152.3287;0.271758;0.259824;-19.704609;8.557664;46.517714;0.375469;0.362778;-47.626582;4.440712;4.204278;0.2083432;0.2039521;-9.6634159;6.5454701;42.1674443
120;326093.6707;0.266555;0.25485;-18.654649;8.418231;46.995058;0.379986;0.367142;-48.625081;4.413727;2.633163;0.2091272;0.2047196;-11.6645351;6.1944145;39.5937084
130;326058.3696;0.26572;0.254052;-18.722508;8.637887;48.20904;0.379378;0.366554;-48.574848;4.329111;3.965238;0.2033879;0.1991012;-11.4615072;6.3624565;41.0785785
140;325981.5025;0.258153;0.246817;-17.146624;8.353216;45.276086;0.363197;0.35092;-45.799537;4.380084;2.231709;0.1966771;0.1925318;-9.6171569;6.05929;38.4140897
150;325951.9558;0.2579;0.246574;-16.111193;8.467954;45.477528;0.353335;0.341392;-44.239747;4.282015;2.957511;0.1922;0.1881491;-8.1613171;6.0876558;39.0359406
0;437251.0284;4.556602;4.356507;-237.53255;100;38.252875;3.23053;3.121333;-0.294727;100;261.099235;4.0269963;3.9421215;105.619188;100;367.0131501
10;342324.9882;0.879074;0.840471;26.095933;25.171472;131.716526;0.422018;0.407753;-16.731005;15.628478;74.498125;0.5297507;0.5185854;51.6355244;22.0337172;142.864655
20;334417.0764;0.661225;0.632188;-4.620473;22.474495;124.925204;0.532397;0.514401;-63.887798;10.763558;51.266416;0.3301542;0.3231957;-3.2142854;17.3165597;111.9399288
30;330901.2308;0.560213;0.535612;-2.653987;21.779706;125.680724;0.473879;0.457862;-55.158295;11.198974;58.784953;0.2661307;0.2605216;3.013064;17.8020813;116.9563119
40;328443.953;0.410597;0.392567;-9.682479;13.638798;77.764333;0.314745;0.304106;-29.184759;8.610211;43.87059;0.2637863;0.2582266;19.1644384;14.1622734;92.2197871
50;327574.0967;0.340919;0.325948;-16.329659;12.935904;75.10565;0.333964;0.322676;-35.149285;8.293502;41.894561;0.2622818;0.2567538;12.166462;13.6416065;89.2103079
60;327028.9244;0.315376;0.301527;-16.622307;11.991017;69.085291;0.311732;0.301195;-35.618557;7.023723;35.697577;0.1918097;0.187767;10.3224743;12.4647403;81.638609
70;326637.0038;0.279457;0.267185;-15.760574;10.620443;60.719324;0.266469;0.257462;-31.038027;6.142487;31.050408;0.1615128;0.1581086;8.8371179;10.7970817;70.9255528
80;326448.5991;0.265553;0.253892;-20.507529;10.06936;58.668835;0.304396;0.294107;-39.977211;5.195307;24.80769;0.1525042;0.14929;-3.7985857;9.233838;60.9863155
90;326287.036;0.272635;0.260663;-22.447165;9.741739;56.631101;0.299805;0.289671;-39.783521;5.082161;24.903283;0.1409103;0.1379404;-5.3088683;8.9903158;59.3779353
100;326081.6052;0.269021;0.257208;-23.019765;8.052251;44.990192;0.370289;0.357772;-47.980698;4.093714;5.637796;0.2098516;0.2054287;-12.7924931;6.3141246;40.826001
110;326020.817;0.258196;0.246857;-18.716463;8.043415;43.915942;0.342761;0.331175;-42.877499;4.102127;5.363443;0.197523;0.1933599;-7.0678993;6.3813769;41.1730432
120;325950.3586;0.251834;0.240775;-16.592844;7.891166;41.834459;0.329237;0.318108;-40.803717;4.086244;3.232123;0.1909989;0.1869733;-6.5926251;5.8827406;37.4432152
130;325880.7713;0.250598;0.239593;-17.717416;8.048625;44.567083;0.35937;0.347222;-46.012893;4.237631;1.880144;0.1940341;0.1899445;-9.8598666;5.9244404;38.03317
140;325849.4731;0.245224;0.234455;-16.999587;7.978471;44.188944;0.339513;0.328037;-42.70796;4.044587;4.089108;0.1832864;0.1794234;-7.153829;6.1312218;39.6432393
150;325823.254;0.239702;0.229176;-14.615343;7.979548;43.868304;0.315585;0.304918;-38.326049;4.013879;5.766135;0.1704513;0.1668588;-2.2305725;6.4343916;41.8616117
\end{filecontents*}
\pgfplotstabletypeset[
        multicolumn names, 
        col sep=semicolon, 
        display columns/0/.style={column name=$\bm{k}$,dec sep align},
        display columns/1/.style={column name={\textbf{AIC}},fixed,fixed zerofill,precision=0,set thousands separator={,},dec sep align},
        display columns/2/.style={column name={\textbf{v.mae}},fixed,fixed zerofill,precision=3,set thousands separator={,},dec sep align},
        display columns/3/.style={column name={$\bm{\textbf{\text{v.mae}}^a}$},fixed,fixed zerofill,precision=3,set thousands separator={,},dec sep align},
        display columns/4/.style={column name={\textbf{v.res}},fixed,fixed zerofill,precision=0,set thousands separator={,},dec sep align},
        display columns/5/.style={column name={$\bm{\textbf{\text{v.mae}}^0}$},fixed,fixed zerofill,precision=3,set thousands separator={,},dec sep align},
        display columns/6/.style={column name={$\bm{\textbf{\text{v.res}}^0}$},fixed,fixed zerofill,precision=0,set thousands separator={,},dec sep align},
        display columns/7/.style={column name={\textbf{ns.mae}},fixed,fixed zerofill,precision=3,set thousands separator={,},dec sep align},
        display columns/8/.style={column name={$\bm{\textbf{\text{ns.mae}}^a}$},fixed,fixed zerofill,precision=3,set thousands separator={,},dec sep align},
        display columns/9/.style={column name={\textbf{ns.res}},fixed,fixed zerofill,precision=0,set thousands separator={,},dec sep align},
        display columns/10/.style={column name={$\bm{\textbf{\text{ns.mae}}^0}$},fixed,fixed zerofill,precision=3,set thousands separator={,},dec sep align},
        display columns/11/.style={column name={$\bm{\textbf{\text{ns.res}}^0}$},fixed,fixed zerofill,precision=0,set thousands separator={,},dec sep align},
        display columns/12/.style={column name={\textbf{cr.mae}},fixed,fixed zerofill,precision=3,set thousands separator={,},dec sep align},
        display columns/13/.style={column name={$\bm{\textbf{\text{cr.mae}}^a}$},fixed,fixed zerofill,precision=3,set thousands separator={,},dec sep align},
        display columns/14/.style={column name={\textbf{cr.res}},fixed,fixed zerofill,precision=0,set thousands separator={,},dec sep align},
        display columns/15/.style={column name={$\bm{\textbf{\text{cr.mae}}^0}$},fixed,fixed zerofill,precision=3,set thousands separator={,},dec sep align},
        display columns/16/.style={column name={$\bm{\textbf{\text{cr.res}}^0}$},fixed,fixed zerofill,precision=0,set thousands separator={,},dec sep align},
	    	every head row/.style={before row=\toprule,after row=\midrule},
	    	every row no 0/.style={before row={\multicolumn{18}{l}{\textbf{Gaussian with identity link}}\\ \\}},
	    	every row no 16/.style={before row={\midrule \multicolumn{18}{l}{\textbf{Gaussian with inverse link}}\\ \\}},
	    	every row no 32/.style={before row={\midrule \multicolumn{18}{l}{\textbf{Gaussian with log link}}\\ \\}},
      	every last row/.style={after row=\bottomrule},
      ]{BGROGLMgaussianidinvlogvalfig.csv} 
    \caption{AIC scores and out-of-sample validation figures of the gaussian GLMs of BEL with identity, inverse and log link functions under $150$-$443$ after each tenth iteration.}
    \label{tab:BGROGLMgaussian}
  \end{center}
\end{table}

\begin{table}[htb]
  \tiny
  \begin{center}
  \tabcolsep=0.052cm
  \renewcommand{\arraystretch}{0.90}
      \begin{filecontents*}{BGROGLMgammaidinvlogvalfig.csv}
k;AIC;v.wre.y;v.a.wre.y;v.res.avg.y;v.wre.delta.y;v.res.avg.delta.y;ns.wre.y;ns.a.wre.y;ns.res.avg.y;ns.wre.delta.y;ns.res.avg.delta.y;cr.wre.y;cr.a.wre.y;cr.res.avg.y;cr.wre.delta.y;cr.res.avg.delta.y
0;437243.2175;4.556602;4.3565067;-237.5325499;100;38.252875;3.23053;3.121333;-0.294727;100;261.099235;4.0269963;3.9421215;105.619188;100;367.0131501
10;345605.0672;0.8724865;0.8341728;1.0212847;23.4849764;113.6540732;0.315086;0.304436;6.478426;19.860868;104.719752;0.5303067;0.5191297;68.4198828;25.2657885;166.6612084
20;333911.2248;0.553423;0.5291204;-12.1459861;16.2651817;78.6658911;0.599058;0.578809;-76.315337;8.267944;0.105078;0.4637109;0.4539375;-42.5560411;9.8948027;33.8643734
30;330707.0488;0.5025829;0.4805128;0.306677;17.4037553;99.443565;0.425124;0.410754;-49.374944;7.753764;35.370481;0.2674069;0.2617709;-2.4427733;12.9590229;82.3026519
40;328589.1292;0.3756894;0.3591916;-12.9080179;13.3165305;75.8352181;0.341423;0.329882;-39.018541;7.187466;35.333232;0.237549;0.2325423;5.7904439;12.3405912;80.1422171
50;327668.4039;0.3483876;0.3330887;-15.2198613;13.17271;76.8060464;0.356147;0.344109;-44.122741;6.656171;33.511704;0.2266366;0.2218599;-3.7836417;11.3478881;73.8508032
60;327135.1604;0.3053299;0.2919219;-16.1178396;11.1897254;65.4438824;0.303971;0.293696;-37.026897;6.059143;30.143362;0.1754068;0.1717098;3.4424615;10.8427282;70.6127208
70;326685.7668;0.272734;0.2607574;-14.5160898;9.729664;55.2933948;0.257308;0.24861;-29.650018;5.363749;25.768004;0.1645075;0.1610403;9.2940546;9.9276306;64.7120763
80;326460.6304;0.2682971;0.2565153;-20.8223956;9.4705941;54.3811796;0.28683;0.277135;-36.227448;5.150638;24.584664;0.148865;0.1457275;2.2549697;9.5488834;63.067082
90;326328.165;0.2590969;0.2477191;-23.0339498;8.8894928;51.685104;0.303766;0.293499;-40.495118;4.372883;19.832473;0.1484838;0.1453543;-5.8052608;8.2551366;54.5223302
100;326246.3935;0.2379352;0.2274867;-19.5104974;8.3213242;47.76369;0.26209;0.253231;-33.635876;4.278683;19.246848;0.1369685;0.1340817;-1.0694746;7.8449592;51.81325
110;326184.3893;0.2327343;0.2225141;-17.5724647;8.0452883;45.3224355;0.254748;0.246137;-32.672222;3.906659;15.831216;0.1300657;0.1273243;-1.0693282;7.1819207;47.4341092
120;326135.3607;0.2282077;0.2181864;-15.7085391;8.1913455;46.3121086;0.253149;0.244592;-32.639787;3.69642;14.989398;0.1287639;0.12605;-2.2845649;6.8695226;45.3446199
130;326093.3412;0.2440161;0.2333005;-16.6633713;9.5304806;54.8738353;0.271794;0.262607;-34.998732;4.628417;22.147011;0.1241786;0.1215614;-0.3745156;8.5956385;56.7712281
140;326067.9443;0.2382437;0.2277817;-16.9151217;9.4158742;54.0395166;0.270543;0.261399;-35.004808;4.522682;21.558367;0.125293;0.1226522;-1.278742;8.3705254;55.2844335
150;326041.0456;0.236068;0.2257015;-14.202776;9.3290335;53.3460621;0.25986;0.251076;-32.953058;4.321119;20.204318;0.1210146;0.118464;1.0416692;8.2061884;54.1990446
0;437243.2175;4.556602;4.356507;-237.53255;100;38.252875;3.23052978;3.12133335;-0.29472736;100;261.0992348;4.026996;3.942121;105.619188;100;367.01315
10;343969.0271;1.036898;0.991365;-0.252404;33.817959;193.027524;0.6610411;0.63869698;-64.35481308;21.60142004;114.5336524;0.397261;0.388888;44.034328;33.752375;222.922794
20;335495.3297;0.679245;0.649417;-6.933254;20.887957;115.353105;0.53033463;0.51240858;-64.72226444;9.63707587;43.17263261;0.335193;0.328128;-9.281621;15.409948;98.613276
30;332646.4609;0.627374;0.599824;-9.09313;26.097704;151.519447;0.6211058;0.60011155;-81.72846967;12.36066255;64.49264398;0.346478;0.339175;-24.230572;18.470388;121.990541
40;329192.4066;0.408737;0.390788;-10.279964;14.061336;81.238395;0.31717285;0.30645197;-26.64395773;9.71863496;50.48293883;0.289213;0.283117;23.459862;15.40537;100.586759
50;328114.0887;0.339018;0.32413;-11.648162;12.598642;72.968445;0.31272015;0.30214977;-30.21532237;8.08371008;40.00982156;0.271212;0.265496;15.090279;13.145706;85.315423
60;327513.2436;0.327857;0.31346;-15.650642;12.247212;71.330096;0.29444908;0.28449629;-29.25305101;8.34070918;43.33622341;0.239725;0.234673;18.013854;13.902317;90.603128
70;327115.4068;0.284821;0.272313;-11.565975;11.127175;63.9563;0.25144482;0.24294563;-28.17094637;6.4632199;32.95986605;0.165677;0.162185;10.544517;10.914985;71.67533
80;326794.538;0.252072;0.241003;-17.256824;8.375999;45.344911;0.31524755;0.30459175;-39.2549171;4.06872418;8.95535481;0.195716;0.191591;-7.766589;6.415921;40.443683
90;326614.7936;0.24987;0.238897;-19.803608;8.11348;44.738162;0.38435105;0.37135944;-50.54380586;4.41427224;-0.39349903;0.217925;0.213332;-16.172058;5.477788;33.978249
100;326444.8417;0.263102;0.251549;-20.206236;8.724098;47.856758;0.38163441;0.36873463;-49.07036388;4.40963317;4.60116719;0.210945;0.206499;-10.835206;6.594848;42.836325
110;326369.6602;0.266481;0.254779;-18.997816;8.250947;44.51094;0.3693151;0.35683173;-46.8363075;4.49441405;2.28098589;0.204884;0.200566;-8.896347;6.288426;40.220946
120;326310.3326;0.258428;0.24708;-17.351576;8.00347;43.601121;0.35717964;0.34510647;-44.88184393;4.43515769;1.67939071;0.195631;0.191508;-7.793925;6.087271;38.767309
130;326277.17;0.259363;0.247974;-16.765572;8.330687;46.545854;0.35654313;0.34449147;-44.67880464;4.35578066;4.24115854;0.186903;0.182964;-6.828503;6.50924;42.09146
140;326245.9942;0.261844;0.250345;-16.856737;8.583068;47.926429;0.35733867;0.34526012;-44.93909725;4.30394721;5.45260581;0.182576;0.178728;-7.200259;6.619818;43.191444
150;326221.9441;0.254004;0.242849;-14.609355;8.409522;46.165413;0.32658471;0.31554569;-39.76753288;4.11104413;6.61577254;0.170609;0.167013;-2.514794;6.721812;43.868511
0;437243.2175;4.556602;4.356507;-237.53255;100;38.252875;3.23053;3.121333;-0.294727;100;261.099235;4.026996;3.942121;105.619188;100;367.01315
1;388233.765;2.364856;2.261007;-3.66361;67.494106;276.646501;0.773154;0.74702;21.535526;54.213673;287.454174;1.19268;1.167543;169.538323;65.931826;435.456971
10;342942.0281;0.870093;0.831884;21.262721;24.998151;130.932935;0.439924;0.425054;-24.375103;15.144975;70.903648;0.504537;0.493903;42.901829;21.395886;138.18058
20;334880.7277;0.649385;0.620868;-4.957791;19.898755;110.311123;0.518668;0.501136;-64.875152;8.282936;36.0023;0.312454;0.305869;-11.051452;14.104785;89.825999
30;331226.9081;0.544384;0.520478;-3.823593;21.752147;125.576481;0.479357;0.463154;-57.188251;11.010403;57.820359;0.262425;0.256894;-0.273088;17.457715;114.735523
40;328726.6754;0.373801;0.357386;-9.573659;14.008847;80.929633;0.329499;0.318362;-32.810344;8.553123;43.301484;0.268226;0.262573;15.031891;13.989908;91.14372
50;327805.9938;0.32782;0.313424;-16.17095;12.750126;74.048326;0.327019;0.315965;-33.44612;8.325053;42.381693;0.271709;0.265982;13.89953;13.779393;89.727344
60;327270.3997;0.301756;0.288505;-15.276165;11.825008;68.333726;0.297154;0.28711;-32.623446;7.147325;36.594982;0.197258;0.193101;13.536342;12.636714;82.75477
70;326866.0963;0.264456;0.252843;-14.970107;10.158632;57.894607;0.248982;0.240566;-27.786434;6.070654;30.686818;0.165032;0.161554;11.718734;10.692937;70.191985
80;326669.1452;0.254859;0.243667;-18.987205;9.818549;57.062259;0.288264;0.27852;-37.338229;5.084742;24.319772;0.145842;0.142768;-1.621975;9.089956;60.036027
90;326433.4819;0.265646;0.253981;-23.435584;8.890852;51.246814;0.326614;0.315574;-44.879561;4.078671;15.411374;0.170768;0.167169;-11.843472;7.353004;48.447463
100;326301.9194;0.265029;0.253391;-22.711517;7.83933;43.811805;0.361422;0.349205;-46.774459;4.030193;5.3574;0.204922;0.200603;-11.701113;6.24566;40.430746
110;326224.4641;0.255632;0.244406;-18.076979;8.139202;45.191217;0.334983;0.32366;-41.024515;4.211252;7.852217;0.191143;0.187114;-2.775086;7.043486;46.101647
120;326147.3633;0.249881;0.238908;-17.743674;7.817431;43.13548;0.33996;0.328469;-42.659043;4.121829;3.828649;0.187838;0.183879;-5.979494;6.24708;40.508198
130;326111.2447;0.246885;0.236043;-17.276266;7.750011;43.002212;0.340712;0.329195;-42.971152;4.11525;2.915863;0.186443;0.182514;-6.725423;6.060194;39.161592
140;326049.6004;0.247287;0.236428;-17.376224;7.730267;42.66456;0.335794;0.324444;-42.079195;4.073376;3.570127;0.179466;0.175683;-5.911514;6.116937;39.737807
150;326022.3145;0.243052;0.232379;-14.794787;7.820028;42.5755;0.323209;0.312284;-39.967415;4.039941;3.011409;0.173713;0.170052;-4.231749;6.009838;38.747075
\end{filecontents*}
\pgfplotstabletypeset[
        multicolumn names, 
        col sep=semicolon, 
        display columns/0/.style={column name=$\bm{k}$,dec sep align},
        display columns/1/.style={column name={\textbf{AIC}},fixed,fixed zerofill,precision=0,set thousands separator={,},dec sep align},
        display columns/2/.style={column name={\textbf{v.mae}},fixed,fixed zerofill,precision=3,set thousands separator={,},dec sep align},
        display columns/3/.style={column name={$\bm{\textbf{\text{v.mae}}^a}$},fixed,fixed zerofill,precision=3,set thousands separator={,},dec sep align},
        display columns/4/.style={column name={\textbf{v.res}},fixed,fixed zerofill,precision=0,set thousands separator={,},dec sep align},
        display columns/5/.style={column name={$\bm{\textbf{\text{v.mae}}^0}$},fixed,fixed zerofill,precision=3,set thousands separator={,},dec sep align},
        display columns/6/.style={column name={$\bm{\textbf{\text{v.res}}^0}$},fixed,fixed zerofill,precision=0,set thousands separator={,},dec sep align},
        display columns/7/.style={column name={\textbf{ns.mae}},fixed,fixed zerofill,precision=3,set thousands separator={,},dec sep align},
        display columns/8/.style={column name={$\bm{\textbf{\text{ns.mae}}^a}$},fixed,fixed zerofill,precision=3,set thousands separator={,},dec sep align},
        display columns/9/.style={column name={\textbf{ns.res}},fixed,fixed zerofill,precision=0,set thousands separator={,},dec sep align},
        display columns/10/.style={column name={$\bm{\textbf{\text{ns.mae}}^0}$},fixed,fixed zerofill,precision=3,set thousands separator={,},dec sep align},
        display columns/11/.style={column name={$\bm{\textbf{\text{ns.res}}^0}$},fixed,fixed zerofill,precision=0,set thousands separator={,},dec sep align},
        display columns/12/.style={column name={\textbf{cr.mae}},fixed,fixed zerofill,precision=3,set thousands separator={,},dec sep align},
        display columns/13/.style={column name={$\bm{\textbf{\text{cr.mae}}^a}$},fixed,fixed zerofill,precision=3,set thousands separator={,},dec sep align},
        display columns/14/.style={column name={\textbf{cr.res}},fixed,fixed zerofill,precision=0,set thousands separator={,},dec sep align},
        display columns/15/.style={column name={$\bm{\textbf{\text{cr.mae}}^0}$},fixed,fixed zerofill,precision=3,set thousands separator={,},dec sep align},
        display columns/16/.style={column name={$\bm{\textbf{\text{cr.res}}^0}$},fixed,fixed zerofill,precision=0,set thousands separator={,},dec sep align},
	    	every head row/.style={before row=\toprule,after row=\midrule},
	    	every row no 0/.style={before row={\multicolumn{18}{l}{\textbf{Gamma with identity link}}\\ \\}},
	    	every row no 16/.style={before row={\midrule \multicolumn{18}{l}{\textbf{Gamma with inverse link}}\\ \\}},
	    	every row no 32/.style={before row={\midrule \multicolumn{18}{l}{\textbf{Gamma with log link}}\\ \\}},
      	every last row/.style={after row=\bottomrule},
      ]{BGROGLMgammaidinvlogvalfig.csv} 
    \caption{AIC scores and out-of-sample validation figures of the gamma GLMs of BEL with identity, inverse and log link functions under $150$-$443$ after each tenth iteration.}
    \label{tab:BGROGLMgamma}
  \end{center}
\end{table}

\begin{table}[htb]
  \tiny
  \begin{center}
  \tabcolsep=0.052cm
  \renewcommand{\arraystretch}{0.90}
      \begin{filecontents*}{BGROGLMinvgaussidinvlogdivmusqvalfig.csv}
k;AIC;v.wre.y;v.a.wre.y;v.res.avg.y;v.wre.delta.y;v.res.avg.delta.y;ns.wre.y;ns.a.wre.y;ns.res.avg.y;ns.wre.delta.y;ns.res.avg.delta.y;cr.wre.y;cr.a.wre.y;cr.res.avg.y;cr.wre.delta.y;cr.res.avg.delta.y
0;437337.824;4.556602;4.356507;-237.53255;100;38.252875;3.23053;3.121333;-0.294727;100;261.099235;4.0269963;3.9421215;105.619188;100;367.0131501
10;346131.9225;0.871108;0.832855;0.713793;23.559167;115.058412;0.314178;0.303559;7.078512;20.268605;107.031668;0.5339306;0.5226773;69.5053102;25.6731254;169.458466
20;334430.1768;0.548513;0.524426;-13.43352;15.995781;76.964459;0.599478;0.579215;-76.80317;8.272797;-0.796654;0.4683173;0.4584468;-43.8485341;9.8093322;32.1579812
30;331452.9427;0.488026;0.466595;-3.855963;15.939114;88.522158;0.516886;0.499414;-66.742392;6.531527;11.244266;0.4133809;0.4046683;-40.432431;9.2798444;37.5542276
40;328985.278;0.369987;0.35374;-12.741961;13.27921;75.556357;0.338108;0.32668;-38.528659;7.192921;35.378196;0.2379894;0.2329734;6.0085474;12.3007425;79.9154031
50;328064.426;0.331745;0.317177;-15.003395;12.727137;74.296265;0.338327;0.326891;-40.278972;6.871132;34.629225;0.2317861;0.2269009;1.0312772;11.6641499;75.9394737
60;327533.1937;0.297839;0.28476;-16.793441;10.99407;64.249496;0.304005;0.293729;-37.464368;5.868003;29.187106;0.1720514;0.1684251;2.8322592;10.6458388;69.4837337
70;327081.9222;0.27385;0.261825;-15.495551;9.3874;53.189243;0.243242;0.23502;-26.842832;5.535433;27.450499;0.1708122;0.167212;12.6007881;10.2529205;66.8941191
80;326849.2943;0.267203;0.255469;-19.622478;9.425509;54.373854;0.277883;0.26849;-34.211392;5.271381;25.393477;0.1515451;0.148351;5.0111314;9.7834027;64.6160007
90;326714.8779;0.246552;0.235725;-20.736193;8.545667;49.153838;0.275055;0.265758;-35.4139;4.399296;20.084668;0.1399033;0.1369546;-0.6641388;8.3023911;54.8344297
100;326630.3311;0.2356;0.225254;-19.734263;7.878792;44.564748;0.262264;0.253399;-33.87616;3.978613;16.031388;0.1396022;0.1366598;-2.0297652;7.2490966;47.8777828
110;326564.2915;0.225079;0.215195;-16.906014;7.728123;43.260311;0.242875;0.234665;-30.660979;3.850164;15.113883;0.1289153;0.1261982;0.1360861;6.9576886;45.9109483
120;326506.9173;0.236504;0.226118;-17.958947;8.775582;50.299602;0.269523;0.260412;-35.31307;4.120097;18.554017;0.1298038;0.127068;-2.9449331;7.7100397;50.9221535
130;326474.8862;0.240284;0.229732;-16.505958;9.225363;53.131231;0.264505;0.255564;-33.810221;4.516483;21.435505;0.1229317;0.1203407;0.2354327;8.4003112;55.4811587
140;326447.4142;0.240905;0.230326;-16.329103;9.41487;54.277541;0.270469;0.261326;-34.776177;4.542605;21.439004;0.1244378;0.1218151;-0.5652029;8.4258719;55.6499783
150;326352.1317;0.249149;0.238208;-16.775489;9.375421;53.551643;0.337168;0.325771;-43.9068;4.223941;12.028869;0.1495211;0.1463698;-3.6004468;7.9304563;52.3352221
0;437337.824;4.556602;4.356507;-237.53255;100;38.252875;3.23053;3.121333;-0.294727;100;261.099235;4.026996;3.942121;105.619188;100;367.01315
10;344458.4028;1.128647;1.079084;-24.812685;35.685428;202.225205;1.137887;1.099425;-149.827914;14.423476;62.818513;0.639191;0.625719;-63.488961;22.713184;149.157466
20;336003.5098;0.682051;0.6521;-4.761222;21.011309;117.387045;0.534168;0.516112;-67.030553;8.865545;40.726251;0.32119;0.31442;-12.306896;14.894979;95.449908
30;333059.5068;0.625724;0.598246;-9.99764;24.463198;141.803373;0.622649;0.601602;-82.659857;10.858507;54.749694;0.376467;0.368532;-30.85212;16.233247;106.55743
40;329632.0078;0.411752;0.393671;-13.721669;15.91162;93.418915;0.344855;0.333199;-28.82722;12.095934;63.921901;0.318142;0.311436;28.226176;18.445587;120.975297
50;328514.7513;0.335021;0.320309;-11.647239;12.38651;71.477703;0.305342;0.295021;-28.508127;8.121905;40.225352;0.275644;0.269834;17.702028;13.33318;86.435507
60;327915.5826;0.321206;0.307101;-15.404543;11.9697;69.601687;0.285955;0.276289;-27.05341;8.385318;43.561357;0.246612;0.241414;20.481474;13.97291;91.096242
70;327542.9472;0.278313;0.266092;-12.144585;10.488295;60.46237;0.24596;0.237646;-27.68964;6.105928;30.525852;0.164213;0.160752;8.88946;10.331249;67.104953
80;327195.5635;0.249294;0.238347;-17.07464;8.22664;44.508527;0.307691;0.297291;-38.150254;4.037213;9.04145;0.19296;0.188893;-6.976912;6.381318;40.214793
90;327012.4129;0.247238;0.236381;-18.864803;8.016337;43.977166;0.376019;0.363309;-49.33525;4.390038;-0.884744;0.211752;0.207289;-15.098537;5.407165;33.35197
100;326836.7536;0.261201;0.249731;-20.488397;8.469311;46.257025;0.375421;0.362731;-48.127506;4.427732;4.226453;0.208482;0.204088;-9.838112;6.569092;42.515846
110;326761.7026;0.261684;0.250193;-18.43058;8.089756;43.977361;0.365343;0.352994;-46.233813;4.504978;1.782666;0.20139;0.197145;-8.13121;6.241666;39.885268
120;326698.9583;0.259305;0.247918;-17.963216;8.105632;44.858151;0.367018;0.354613;-46.791474;4.402232;1.63843;0.192451;0.188395;-9.365444;6.082035;39.064461
130;326667.2634;0.258694;0.247334;-16.845788;7.987114;43.660935;0.352404;0.340493;-44.464205;4.303295;1.651055;0.187091;0.183147;-8.094849;5.957936;38.020411
140;326642.0975;0.257728;0.24641;-16.130455;8.242927;45.791767;0.339791;0.328305;-42.019313;4.227545;5.511446;0.173061;0.169414;-4.514039;6.602192;43.01672
150;326617.1348;0.253324;0.242199;-14.571099;8.15202;44.370231;0.323912;0.312964;-39.376865;4.147759;5.173003;0.172357;0.168725;-2.551035;6.476346;41.998833
0;437337.824;4.556602;4.356507;-237.53255;100;38.252875;3.23053;3.121333;-0.294727;100;261.099235;4.026996;3.942121;105.619188;100;367.01315
10;343530.4892;0.866193;0.828155;18.799286;24.925012;130.649954;0.449963;0.434754;-28.083858;14.939768;69.375347;0.494273;0.483855;38.734477;21.12156;136.193682
20;335354.8054;0.644254;0.615963;-5.029689;19.653097;108.854341;0.526343;0.508552;-66.586947;7.946656;32.905619;0.317869;0.311169;-14.103594;13.489567;85.388973
30;331675.4225;0.535872;0.51234;-4.364227;21.697061;125.269777;0.481537;0.46526;-58.088018;10.885011;57.154524;0.261787;0.25627;-1.894086;17.244563;113.348456
40;329139.5419;0.366118;0.350041;-9.759329;13.912628;80.279582;0.32473;0.313754;-31.966068;8.604199;43.68138;0.269195;0.263521;15.670977;14.010675;91.318425
50;328189.5002;0.324062;0.309832;-16.04414;12.640048;73.410434;0.319169;0.30838;-31.629069;8.481909;43.434042;0.274015;0.26824;15.96218;13.965641;91.025291
60;327665.9438;0.295842;0.28285;-15.184014;11.626056;67.088658;0.289833;0.280036;-31.071019;7.180515;36.81019;0.20111;0.196871;15.24763;12.694847;83.128838
70;327262.5332;0.261298;0.249823;-14.572764;9.94776;56.57654;0.243752;0.235513;-26.644994;6.042153;30.112848;0.171727;0.168108;12.305423;10.531302;69.063265
80;327061.4271;0.250668;0.23966;-18.493849;9.746092;56.499597;0.284301;0.274691;-36.817095;4.98778;23.784889;0.145374;0.14231;-1.4004;8.963615;59.201584
90;326824.5402;0.262579;0.251048;-23.306741;8.768996;50.578253;0.320785;0.309942;-44.020719;4.05923;15.472812;0.168192;0.164647;-11.285628;7.315749;48.207903
100;326694.8884;0.260938;0.249479;-22.18394;7.727419;43.190348;0.351961;0.340064;-45.137586;4.047648;5.845239;0.202807;0.198532;-9.848216;6.340962;41.13461
110;326597.5188;0.239286;0.228778;-16.811892;7.408203;40.036285;0.343118;0.33152;-43.301838;4.443587;-0.845123;0.1851;0.181198;-7.270049;5.571502;35.186665
120;326530.1609;0.248929;0.237998;-18.136524;7.51961;40.91005;0.342844;0.331255;-43.226038;4.246755;1.429074;0.191494;0.187458;-6.869825;5.927585;37.785287
130;326494.4164;0.245907;0.235109;-17.286973;7.601794;42.03965;0.337303;0.325902;-42.500038;4.108171;2.435122;0.183053;0.179195;-6.40178;5.96407;38.53338
140;326471.0264;0.246202;0.235391;-16.872566;7.771793;42.687048;0.331878;0.32066;-41.560306;4.068169;3.607846;0.177043;0.173311;-5.669258;6.092296;39.498893
150;326412.6525;0.247431;0.236565;-14.766914;7.716267;41.907191;0.323553;0.312617;-39.989455;4.095487;2.293187;0.171862;0.16824;-4.44379;5.891975;37.838852
0;437337.824;4.556602;4.356507;-237.53255;100;38.252875;3.23053;3.121333;-0.294727;100;261.099235;4.026996;3.942121;105.619188;100;367.01315
10;344467.1505;0.984597;0.94136;-14.471219;31.473092;175.749876;0.992861;0.959301;-130.19041;12.573059;45.639223;0.561024;0.5492;-52.073943;18.986311;123.755689
20;336815.4432;0.667854;0.638527;-6.627196;21.404117;121.540766;0.590588;0.570625;-75.313317;9.506364;38.463183;0.372172;0.364328;-22.316968;14.521288;91.459531
30;331791.9199;0.47753;0.456561;-5.426245;15.821218;89.882398;0.366522;0.354133;-28.009002;10.573225;52.908178;0.373308;0.36544;33.416127;17.495939;114.333307
40;330089.1104;0.421373;0.402869;-0.829876;15.183183;88.667596;0.294643;0.284684;-19.335645;10.659964;55.770365;0.315948;0.309289;33.959663;16.657004;109.065672
50;329020.2601;0.376002;0.35949;-10.244725;14.44279;85.145902;0.299994;0.289854;-20.715849;11.438972;60.283315;0.320249;0.313499;34.073511;17.552575;115.072676
60;328451.5922;0.330259;0.315756;-11.590698;12.904577;75.151328;0.289542;0.279756;-24.455882;9.196308;47.894681;0.273024;0.26727;25.318209;14.951954;97.668772
70;327925.2138;0.316202;0.302316;-16.341403;11.733491;68.611742;0.300967;0.290794;-35.300707;7.089979;35.260975;0.199591;0.195384;5.594425;11.700869;76.156107
80;327639.3861;0.261697;0.250205;-17.850745;8.127888;42.557403;0.29777;0.287704;-34.953136;4.425188;11.063549;0.207664;0.203287;-1.174324;7.204954;44.842361
90;327264.7457;0.27811;0.265898;-21.517399;8.310569;45.733254;0.354818;0.342825;-44.177672;4.38254;8.681518;0.201502;0.197255;-6.689055;7.090184;46.170135
100;327148.0483;0.28806;0.275411;-22.439591;8.166407;43.85558;0.356746;0.344688;-44.309473;4.408436;7.594235;0.207045;0.202681;-6.361797;7.039261;45.54191
110;327077.5608;0.273521;0.26151;-19.640597;7.942726;43.031379;0.35415;0.342179;-44.14153;4.450605;4.138983;0.196075;0.191942;-7.221705;6.43373;41.058808
120;326919.6763;0.268727;0.256927;-17.800875;8.350237;46.273653;0.373777;0.361142;-47.109526;4.579304;2.57354;0.197607;0.193442;-8.587401;6.419249;41.095665
130;326886.5583;0.270094;0.258233;-18.043424;8.437041;46.990385;0.360342;0.348162;-44.441746;4.543673;6.2006;0.195697;0.191573;-4.348795;7.150683;46.293552
140;326807.0689;0.266569;0.254863;-17.763861;8.192664;44.593266;0.344927;0.333268;-42.563613;4.317519;5.402052;0.187665;0.18371;-4.975707;6.661027;42.989958
150;326778.1994;0.261519;0.250035;-15.691022;8.257955;44.219915;0.332116;0.32089;-40.519593;4.237989;4.999882;0.177478;0.173737;-3.475248;6.517505;42.044226
\end{filecontents*}
\pgfplotstabletypeset[
        multicolumn names, 
        col sep=semicolon, 
        display columns/0/.style={column name=$\bm{k}$,dec sep align},
        display columns/1/.style={column name={\textbf{AIC}},fixed,fixed zerofill,precision=0,set thousands separator={,},dec sep align},
        display columns/2/.style={column name={\textbf{v.mae}},fixed,fixed zerofill,precision=3,set thousands separator={,},dec sep align},
        display columns/3/.style={column name={$\bm{\textbf{\text{v.mae}}^a}$},fixed,fixed zerofill,precision=3,set thousands separator={,},dec sep align},
        display columns/4/.style={column name={\textbf{v.res}},fixed,fixed zerofill,precision=0,set thousands separator={,},dec sep align},
        display columns/5/.style={column name={$\bm{\textbf{\text{v.mae}}^0}$},fixed,fixed zerofill,precision=3,set thousands separator={,},dec sep align},
        display columns/6/.style={column name={$\bm{\textbf{\text{v.res}}^0}$},fixed,fixed zerofill,precision=0,set thousands separator={,},dec sep align},
        display columns/7/.style={column name={\textbf{ns.mae}},fixed,fixed zerofill,precision=3,set thousands separator={,},dec sep align},
        display columns/8/.style={column name={$\bm{\textbf{\text{ns.mae}}^a}$},fixed,fixed zerofill,precision=3,set thousands separator={,},dec sep align},
        display columns/9/.style={column name={\textbf{ns.res}},fixed,fixed zerofill,precision=0,set thousands separator={,},dec sep align},
        display columns/10/.style={column name={$\bm{\textbf{\text{ns.mae}}^0}$},fixed,fixed zerofill,precision=3,set thousands separator={,},dec sep align},
        display columns/11/.style={column name={$\bm{\textbf{\text{ns.res}}^0}$},fixed,fixed zerofill,precision=0,set thousands separator={,},dec sep align},
        display columns/12/.style={column name={\textbf{cr.mae}},fixed,fixed zerofill,precision=3,set thousands separator={,},dec sep align},
        display columns/13/.style={column name={$\bm{\textbf{\text{cr.mae}}^a}$},fixed,fixed zerofill,precision=3,set thousands separator={,},dec sep align},
        display columns/14/.style={column name={\textbf{cr.res}},fixed,fixed zerofill,precision=0,set thousands separator={,},dec sep align},
        display columns/15/.style={column name={$\bm{\textbf{\text{cr.mae}}^0}$},fixed,fixed zerofill,precision=3,set thousands separator={,},dec sep align},
        display columns/16/.style={column name={$\bm{\textbf{\text{cr.res}}^0}$},fixed,fixed zerofill,precision=0,set thousands separator={,},dec sep align},
	    	every head row/.style={before row=\toprule,after row=\midrule},
	    	every row no 0/.style={before row={\multicolumn{18}{l}{\textbf{inverse gaussian with identity link}}\\ \\}},
	    	every row no 16/.style={before row={\midrule \multicolumn{18}{l}{\textbf{Inverse gaussian with inverse link}}\\ \\}},
	    	every row no 32/.style={before row={\midrule \multicolumn{18}{l}{\textbf{Inverse gaussian with log link}}\\ \\}},
	    	every row no 48/.style={before row={\midrule \multicolumn{18}{l}{\textbf{Inverse gaussian with $\frac{1}{\mu^2}$ link}}\\ \\}},
      	every last row/.style={after row=\bottomrule},
      ]{BGROGLMinvgaussidinvlogdivmusqvalfig.csv} 
    \caption{AIC scores and out-of-sample validation figures of the inverse gaussian GLMs of BEL with identity, inverse, log and $\frac{1}{\mu^2}$ link functions under $150$-$443$ after each tenth iteration.}
    \label{tab:BGROGLMinvgauss}
  \end{center}
\end{table}

\begin{table}[htb]
  \tiny
  \begin{center}
  \tabcolsep=0.052cm
  \renewcommand{\arraystretch}{0.90}
      \begin{filecontents*}{BGROGLMgaussianidinvlog300886valfig.csv}
k;AIC;v.wre.y;v.a.wre.y;v.res.avg.y;v.wre.delta.y;v.res.avg.delta.y;ns.wre.y;ns.a.wre.y;ns.res.avg.y;ns.wre.delta.y;ns.res.avg.delta.y;cr.wre.y;cr.a.wre.y;cr.res.avg.y;cr.wre.delta.y;cr.res.avg.delta.y
0;437251.0284;4.556602;4.3565067;-237.5325499;100;38.252875;3.23053;3.121333;-0.294727;100;261.099235;4.02699629;3.94212147;105.619188;100;367.0131501
10;345045.2474;0.8386126;0.8017863;-0.1621256;21.4677825;104.479234;0.388718;0.375579;23.084391;21.65881;113.334288;0.64994998;0.63625134;88.50389457;27.11154378;178.7537913
20;333447.4362;0.5648041;0.5400017;-9.6425216;16.7795342;81.9455509;0.59708;0.576898;-75.11821;8.273502;2.0784;0.45439558;0.44481853;-39.67540492;10.08259277;37.52120479
30;330361.1506;0.5184664;0.4956989;1.2706449;17.501119;100.0249494;0.41797;0.403842;-47.27442;7.969525;37.088422;0.26442846;0.25885524;1.06815141;13.37811619;85.43099318
40;328832.2073;0.4749207;0.4540654;-10.0802125;16.8877617;98.1555422;0.509415;0.492196;-66.483768;6.233981;27.360524;0.29103128;0.28489737;-25.69595959;10.49700567;68.14833229
50;327432.0902;0.3684989;0.3523169;-15.2033082;13.2677566;77.6943374;0.390834;0.377623;-49.997272;6.059911;28.508911;0.22138584;0.21671981;-9.21674612;10.67401636;69.28943673
60;326787.1036;0.3059923;0.2925552;-17.1711148;10.7604672;61.9613566;0.300561;0.290402;-35.79946;5.863104;28.941549;0.18312963;0.1792699;4.60399207;10.65126644;69.34500059
70;326452.7548;0.2912463;0.2784568;-17.8768297;10.4511338;59.9776421;0.281253;0.271746;-33.130104;6.060042;30.332905;0.17507827;0.17138824;8.12076288;10.95844816;71.58377196
80;326245.0608;0.2626108;0.2510787;-23.1963879;9.389059;54.072869;0.308787;0.29835;-40.649427;4.83747;22.228367;0.15723232;0.15391842;-3.79955745;8.9449389;59.07823656
90;326115.6586;0.2673225;0.2555835;-24.4509864;9.1959709;53.5277726;0.313173;0.302587;-41.85074;4.689074;21.736556;0.15836889;0.15503103;-6.87641503;8.58650149;56.71088109
100;326038.3955;0.2495997;0.2386389;-18.3789974;9.152235;53.235207;0.275657;0.266339;-35.283024;4.63726;21.939718;0.13576231;0.13290092;-0.38504662;8.60570218;56.837695
110;325963.4171;0.2391895;0.2286859;-18.2697562;9.1318849;52.4401679;0.269454;0.260346;-34.817088;4.576795;21.501374;0.1319992;0.12921712;-1.11359284;8.35847858;55.20486842
120;325922.293;0.2416919;0.2310784;-16.2447359;9.5189854;54.4798139;0.272591;0.263377;-35.185585;4.568789;21.147502;0.12854567;0.12583638;-0.98573617;8.3800516;55.34735076
130;325888.8676;0.2508547;0.2398389;-17.9936575;10.5057972;60.9790765;0.287005;0.277304;-37.239566;5.421029;27.341705;0.12747051;0.12478388;-0.36020699;9.723606;64.22106419
140;325865.0117;0.246235;0.235422;-14.7233683;10.5295697;60.7143138;0.268809;0.259723;-34.109933;5.328821;26.936286;0.12283709;0.12024812;1.86835267;9.52579215;62.91457192
150;325840.9428;0.2421701;0.2315356;-14.3337955;10.555538;61.0133723;0.274369;0.265095;-35.233973;5.119336;25.721732;0.12281914;0.12023054;0.20945758;9.26091694;61.16516256
160;325820.8272;0.24268;0.2320232;-14.7932579;10.4833875;60.3997938;0.277675;0.268289;-35.819964;5.017658;24.981625;0.12676389;0.12409215;-0.40640356;9.14433596;60.39518535
170;325810.9067;0.2383048;0.22784;-13.0482087;10.1398687;58.1514403;0.26463;0.255686;-32.601165;4.967993;24.207021;0.12981998;0.12708383;1.86675397;8.88387646;58.67494019
180;325766.2164;0.2406211;0.2300546;-12.1162162;10.128107;57.4920735;0.299769;0.289636;-36.805884;4.552101;18.410943;0.14873691;0.14560206;2.31307619;8.71601141;57.52990312
190;325506.0811;0.2009273;0.1921039;-12.5994577;6.4578412;32.3504691;0.27544;0.26613;-32.913782;4.124379;-2.355318;0.17264937;0.16901053;-3.65510729;4.72105174;26.90335671
200;325487.7602;0.1860621;0.1778915;-9.4924849;6.1108683;29.4113408;0.262403;0.253533;-28.678883;4.460319;-4.16652;0.18100537;0.17719042;2.59409644;4.91978969;27.10645939
210;325481.863;0.1839071;0.1758312;-9.0048296;6.2100666;30.3658668;0.2581;0.249376;-28.117908;4.336919;-3.138675;0.17014148;0.1665555;2.94971024;4.84606976;27.92894379
220;325468.3161;0.1846578;0.1765489;-7.622243;6.4330316;32.22885;0.25833;0.249598;-28.373627;4.285722;-2.913997;0.16478878;0.16131562;2.52621979;4.8495868;27.98584994
224;325458.9411;0.1943361;0.1858021;-8.7292706;6.658858;34.2822165;0.267728;0.258678;-30.207234;4.199764;-1.58721;0.16834735;0.16479918;0.75346231;5.00742678;29.37348663
0;437251.0284;4.556602;4.356507;-237.53255;100;38.252875;3.23053;3.121333;-0.294727;100;261.099235;4.0269963;3.9421215;105.6191879;100;367.0131501
10;343425.573;1.035822;0.990336;0.95747;33.705291;192.399545;0.650216;0.628238;-63.149782;21.480893;113.90083;0.3906022;0.3823697;44.0862646;33.4819718;221.136877
20;334984.9692;0.689454;0.659177;-6.452451;21.312708;118.472926;0.515445;0.498022;-61.623747;10.319182;48.910168;0.3236438;0.3168226;-4.0056833;16.4927571;106.5282311
30;331426.1329;0.512223;0.48973;-15.559177;18.835933;108.678125;0.393484;0.380184;-44.977252;12.276885;64.868587;0.2481155;0.2428861;14.8013397;18.9598812;124.6471792
40;328875.4835;0.433163;0.414141;-5.072545;14.353638;82.448451;0.31665;0.305946;-25.923256;9.312083;47.206278;0.2938246;0.2876318;25.996328;15.1876994;99.1258612
50;327876.6139;0.382976;0.366159;-8.463925;12.958692;75.684936;0.28544;0.275792;-23.511261;8.961099;46.246138;0.2712127;0.2654965;25.2246805;14.5915226;94.9820791
60;327273.5134;0.337446;0.322628;-16.271187;12.572085;72.987357;0.327504;0.316433;-36.647847;7.635989;38.219234;0.219136;0.2145173;10.1392807;13.0874836;85.0063612
70;326874.6213;0.28963;0.276911;-13.855417;11.247847;64.340537;0.270597;0.26145;-32.353253;6.232988;31.451239;0.1561271;0.1528365;5.6966873;10.5880666;69.5011792
80;326602.8124;0.259478;0.248084;-16.297278;9.975877;57.788291;0.287232;0.277524;-37.738076;5.041717;21.956031;0.1580564;0.1547251;-7.6188188;8.0138476;52.0752878
90;326390.4865;0.254097;0.242938;-19.842845;8.46189;46.824473;0.391958;0.378709;-51.45543;4.450825;0.820426;0.2198171;0.2151841;-16.5024189;5.6759814;35.7734364
100;326224.4958;0.268574;0.25678;-20.966928;9.365447;52.639451;0.402647;0.389037;-52.002338;4.500126;7.212578;0.2250965;0.2203523;-12.3480205;7.1740658;46.8668951
110;326134.5796;0.265871;0.254196;-19.109182;8.894023;49.355736;0.377161;0.364413;-48.790491;4.333835;5.282964;0.2054984;0.2011673;-12.0284571;6.4972455;42.0449981
120;326068.8477;0.26576;0.25409;-19.395519;8.563999;48.451685;0.381065;0.368184;-49.729147;4.270542;3.726594;0.203977;0.1996779;-14.1868688;6.1017397;39.2688723
130;326032.7829;0.265118;0.253476;-18.639132;8.498212;47.428035;0.38617;0.373117;-49.991394;4.445132;1.68431;0.2116627;0.2072016;-14.1508331;5.9169155;37.5248709
140;325950.1982;0.253261;0.242139;-16.834406;8.151488;43.787346;0.357979;0.345879;-45.59971;4.345327;0.630579;0.1886353;0.1846595;-10.8898302;5.5976468;35.3404589
150;325924.2385;0.254798;0.243609;-17.415469;8.485362;45.956907;0.3641;0.351793;-46.474812;4.287788;2.506101;0.192257;0.1882049;-11.0925637;5.8939586;37.8883497
160;325885.6849;0.258087;0.246753;-15.139712;8.842134;48.388155;0.349241;0.337436;-43.813365;4.198897;5.323039;0.1781205;0.1743663;-7.7862742;6.3594086;41.3501297
170;325868.7359;0.248777;0.237852;-13.670406;8.502911;45.541859;0.33114;0.319947;-40.080498;4.253725;4.740304;0.1744554;0.1707785;-4.8891939;6.1816965;39.9316081
180;325850.3219;0.24828;0.237377;-12.450476;8.50543;44.982016;0.31243;0.30187;-37.262112;4.098954;5.778917;0.1644182;0.1609528;-3.4778191;6.094915;39.5632102
190;325820.2931;0.238281;0.227817;-11.900787;8.240278;43.310021;0.313405;0.302811;-37.05294;4.137336;3.766405;0.1693703;0.1658005;-3.2654807;5.8252846;37.5538643
200;325803.0271;0.244356;0.233626;-12.855401;8.457786;45.227362;0.319754;0.308946;-38.182787;4.072941;5.508514;0.1706277;0.1670315;-3.9659345;6.1322575;39.7253657
210;325800.4334;0.241498;0.230893;-12.592697;8.375565;44.515078;0.312647;0.302079;-36.49105;4.058878;6.225262;0.170947;0.167344;-2.0899909;6.2484999;40.6263215
213;325797.3295;0.240981;0.230399;-12.305274;8.324796;44.039468;0.309614;0.299148;-35.697951;4.063452;6.255329;0.1709604;0.1673572;-1.0692338;6.2839868;40.8840457
0;437251.0284;4.556602;4.356507;-237.53255;100;38.252875;3.23053;3.121333;-0.294727;100;261.099235;4.0269963;3.9421215;105.619188;100;367.0131501
10;342324.9882;0.879074;0.840471;26.095933;25.171472;131.716526;0.422018;0.407753;-16.731005;15.628478;74.498125;0.5297507;0.5185854;51.6355244;22.0337172;142.864655
20;334417.0764;0.661225;0.632188;-4.620473;22.474495;124.925204;0.532397;0.514401;-63.887798;10.763558;51.266416;0.3301542;0.3231957;-3.2142854;17.3165597;111.9399288
30;330901.2308;0.560213;0.535612;-2.653987;21.779706;125.680724;0.473879;0.457862;-55.158295;11.198974;58.784953;0.2661307;0.2605216;3.013064;17.8020813;116.9563119
40;328443.953;0.410597;0.392567;-9.682479;13.638798;77.764333;0.314745;0.304106;-29.184759;8.610211;43.87059;0.2637863;0.2582266;19.1644384;14.1622734;92.2197871
50;327574.0967;0.340919;0.325948;-16.329659;12.935904;75.10565;0.333964;0.322676;-35.149285;8.293502;41.894561;0.2622818;0.2567538;12.166462;13.6416065;89.2103079
60;327028.9244;0.315376;0.301527;-16.622307;11.991017;69.085291;0.311732;0.301195;-35.618557;7.023723;35.697577;0.1918097;0.187767;10.3224743;12.4647403;81.638609
70;326637.0038;0.279457;0.267185;-15.760574;10.620443;60.719324;0.266469;0.257462;-31.038027;6.142487;31.050408;0.1615128;0.1581086;8.8371179;10.7970817;70.9255528
80;326448.5991;0.265553;0.253892;-20.507529;10.06936;58.668835;0.304396;0.294107;-39.977211;5.195307;24.80769;0.1525042;0.14929;-3.7985857;9.233838;60.9863155
90;326287.036;0.272635;0.260663;-22.447165;9.741739;56.631101;0.299805;0.289671;-39.783521;5.082161;24.903283;0.1409103;0.1379404;-5.3088683;8.9903158;59.3779353
100;326081.6052;0.269021;0.257208;-23.019765;8.052251;44.990192;0.370289;0.357772;-47.980698;4.093714;5.637796;0.2098516;0.2054287;-12.7924931;6.3141246;40.826001
110;326020.817;0.258196;0.246857;-18.716463;8.043415;43.915942;0.342761;0.331175;-42.877499;4.102127;5.363443;0.197523;0.1933599;-7.0678993;6.3813769;41.1730432
120;325950.3586;0.251834;0.240775;-16.592844;7.891166;41.834459;0.329237;0.318108;-40.803717;4.086244;3.232123;0.1909989;0.1869733;-6.5926251;5.8827406;37.4432152
130;325743.2949;0.208305;0.199157;-12.738024;6.208349;30.018153;0.309935;0.299458;-38.329067;4.993533;-9.964352;0.1909054;0.1868818;-7.5343821;4.2729737;20.8303325
140;325693.1431;0.210842;0.201583;-13.088989;6.62042;34.050174;0.301707;0.291509;-36.233875;4.522227;-3.486175;0.1856211;0.1817089;-3.1219514;5.0367512;29.6257488
150;325664.767;0.209541;0.200339;-13.021882;6.729259;34.792582;0.298198;0.288118;-35.677808;4.384978;-2.254807;0.1796845;0.1758973;-2.556242;5.1681534;30.8667595
160;325626.2395;0.214219;0.204812;-14.435645;6.549138;33.248496;0.302055;0.291846;-36.471716;4.409883;-3.179038;0.1830694;0.1792109;-3.5427421;5.0756886;29.7499362
170;325610.1638;0.213572;0.204193;-13.536779;6.590491;33.18956;0.290949;0.281114;-34.848396;4.272798;-2.51352;0.1729976;0.1693515;-2.3757046;5.0279955;29.9591711
180;325584.2008;0.213611;0.204231;-13.059597;6.586856;33.020309;0.295555;0.285565;-35.319734;4.386192;-3.63129;0.1757953;0.1720902;-2.2559076;4.9729055;29.4325365
190;325574.6894;0.212379;0.203052;-11.869009;6.501678;31.986464;0.282506;0.272957;-33.063045;4.363336;-3.599035;0.1731563;0.1695068;-0.4010116;4.9497039;29.0629978
200;325566.9295;0.200797;0.191979;-8.996988;6.271873;29.71704;0.263514;0.254607;-28.746331;4.490675;-4.423766;0.1714761;0.167862;3.0566617;4.8627349;27.3792266
210;325553.4008;0.20458;0.195596;-9.216891;6.65484;32.34497;0.266613;0.257601;-29.173357;4.398281;-2.002959;0.176271;0.1725558;2.7719501;5.1650826;29.942348
214;325552.0671;0.20602;0.196973;-9.601323;6.640115;32.059351;0.267493;0.258452;-29.265488;4.401831;-1.996277;0.1766801;0.1729563;2.8082184;5.1799491;30.0774301
\end{filecontents*}
\pgfplotstabletypeset[
        multicolumn names, 
        col sep=semicolon, 
        display columns/0/.style={column name=$\bm{k}$,dec sep align},
        display columns/1/.style={column name={\textbf{AIC}},fixed,fixed zerofill,precision=0,set thousands separator={,},dec sep align},
        display columns/2/.style={column name={\textbf{v.mae}},fixed,fixed zerofill,precision=3,set thousands separator={,},dec sep align},
        display columns/3/.style={column name={$\bm{\textbf{\text{v.mae}}^a}$},fixed,fixed zerofill,precision=3,set thousands separator={,},dec sep align},
        display columns/4/.style={column name={\textbf{v.res}},fixed,fixed zerofill,precision=0,set thousands separator={,},dec sep align},
        display columns/5/.style={column name={$\bm{\textbf{\text{v.mae}}^0}$},fixed,fixed zerofill,precision=3,set thousands separator={,},dec sep align},
        display columns/6/.style={column name={$\bm{\textbf{\text{v.res}}^0}$},fixed,fixed zerofill,precision=0,set thousands separator={,},dec sep align},
        display columns/7/.style={column name={\textbf{ns.mae}},fixed,fixed zerofill,precision=3,set thousands separator={,},dec sep align},
        display columns/8/.style={column name={$\bm{\textbf{\text{ns.mae}}^a}$},fixed,fixed zerofill,precision=3,set thousands separator={,},dec sep align},
        display columns/9/.style={column name={\textbf{ns.res}},fixed,fixed zerofill,precision=0,set thousands separator={,},dec sep align},
        display columns/10/.style={column name={$\bm{\textbf{\text{ns.mae}}^0}$},fixed,fixed zerofill,precision=3,set thousands separator={,},dec sep align},
        display columns/11/.style={column name={$\bm{\textbf{\text{ns.res}}^0}$},fixed,fixed zerofill,precision=0,set thousands separator={,},dec sep align},
        display columns/12/.style={column name={\textbf{cr.mae}},fixed,fixed zerofill,precision=3,set thousands separator={,},dec sep align},
        display columns/13/.style={column name={$\bm{\textbf{\text{cr.mae}}^a}$},fixed,fixed zerofill,precision=3,set thousands separator={,},dec sep align},
        display columns/14/.style={column name={\textbf{cr.res}},fixed,fixed zerofill,precision=0,set thousands separator={,},dec sep align},
        display columns/15/.style={column name={$\bm{\textbf{\text{cr.mae}}^0}$},fixed,fixed zerofill,precision=3,set thousands separator={,},dec sep align},
        display columns/16/.style={column name={$\bm{\textbf{\text{cr.res}}^0}$},fixed,fixed zerofill,precision=0,set thousands separator={,},dec sep align},
	    	every head row/.style={before row=\toprule,after row=\midrule},
	    	every row no 0/.style={before row={\multicolumn{18}{l}{\textbf{Gaussian with identity link}}\\ \\}},
	    	every row no 24/.style={before row={\midrule \multicolumn{18}{l}{\textbf{Gaussian with inverse link}}\\ \\}},
	    	every row no 47/.style={before row={\midrule \multicolumn{18}{l}{\textbf{Gaussian with log link}}\\ \\}},
      	every last row/.style={after row=\bottomrule},
      ]{BGROGLMgaussianidinvlog300886valfig.csv} 
    \caption{AIC scores and out-of-sample validation figures of the gaussian GLMs of BEL with identity, inverse and log link functions under $300$-$886$ after each tenth and the final iteration.}
    \label{tab:BGROGLMgaussian300886}
  \end{center}
\end{table}

\begin{table}[htb]
  \tiny
  \begin{center}
  \tabcolsep=0.052cm
  \renewcommand{\arraystretch}{0.90}
      \begin{filecontents*}{BGROGLMgammaidinvlog300886valfig.csv}
k;AIC;v.wre.y;v.a.wre.y;v.res.avg.y;v.wre.delta.y;v.res.avg.delta.y;ns.wre.y;ns.a.wre.y;ns.res.avg.y;ns.wre.delta.y;ns.res.avg.delta.y;cr.wre.y;cr.a.wre.y;cr.res.avg.y;cr.wre.delta.y;cr.res.avg.delta.y
0;437243.2175;4.556602;4.3565067;-237.5325499;100;38.252875;3.23053;3.121333;-0.294727;100;261.099235;4.02699629;3.94212147;105.619188;100;367.0131501
10;345605.0672;0.8724865;0.8341728;1.0212847;23.4849764;113.6540732;0.315086;0.304436;6.478426;19.860868;104.719752;0.5303067;0.51912971;68.41988277;25.26578851;166.6612084
20;333911.2248;0.553423;0.5291204;-12.1459861;16.2651817;78.6658911;0.599058;0.578809;-76.315337;8.267944;0.105078;0.4637109;0.45393751;-42.55604108;9.89480266;33.86437339
30;330707.0488;0.5025829;0.4805128;0.306677;17.4037553;99.443565;0.425124;0.410754;-49.374944;7.753764;35.370481;0.26740693;0.26177094;-2.44277333;12.9590229;82.30265186
40;328589.1292;0.3756894;0.3591916;-12.9080179;13.3165305;75.8352181;0.341423;0.329882;-39.018541;7.187466;35.333232;0.23754899;0.2325423;5.79044389;12.34059118;80.14221713
50;327668.4039;0.3483876;0.3330887;-15.2198613;13.17271;76.8060464;0.356147;0.344109;-44.122741;6.656171;33.511704;0.2266366;0.2218599;-3.78364174;11.34788805;73.85080316
60;327135.1604;0.3053299;0.2919219;-16.1178396;11.1897254;65.4438824;0.303971;0.293696;-37.026897;6.059143;30.143362;0.17540678;0.17170983;3.44246153;10.84272822;70.61272076
70;326685.7668;0.272734;0.2607574;-14.5160898;9.729664;55.2933948;0.257308;0.24861;-29.650018;5.363749;25.768004;0.16450749;0.16104025;9.29405457;9.92763059;64.71207632
80;326460.6304;0.2682971;0.2565153;-20.8223956;9.4705941;54.3811796;0.28683;0.277135;-36.227448;5.150638;24.584664;0.148865;0.14572746;2.25496969;9.54888345;63.06708202
90;326328.165;0.2590969;0.2477191;-23.0339498;8.8894928;51.685104;0.303766;0.293499;-40.495118;4.372883;19.832473;0.14848384;0.14535433;-5.80526077;8.2551366;54.52233024
100;326243.5099;0.2400135;0.2294737;-19.8689771;9.2733261;53.7751618;0.28219;0.272652;-36.771041;4.759491;22.481635;0.14371406;0.14068508;-2.04596888;8.66157371;57.20670719
110;326178.315;0.2355689;0.2252243;-17.6982262;8.8368399;50.6231373;0.262395;0.253525;-33.573045;4.454055;20.356856;0.13499437;0.13214917;-0.17453031;8.13901245;53.75537029
120;326116.6949;0.2368476;0.2264469;-18.0777331;9.6679522;55.7955822;0.275447;0.266136;-35.942937;4.844588;23.538916;0.12918013;0.12645747;-1.37024136;8.79858372;58.11161105
130;326084.0301;0.2452898;0.2345183;-16.506733;10.1475194;58.6633881;0.269567;0.260455;-34.523127;5.235841;26.255531;0.12234762;0.11976897;1.14267092;9.37540689;61.92132918
140;326057.9249;0.2429765;0.2323066;-16.7425691;10.1525934;58.4994595;0.273335;0.264096;-35.430469;5.091543;25.420097;0.12452257;0.12189808;-0.60027344;9.12239794;60.25029235
150;326031.4132;0.2390473;0.22855;-13.8573636;10.1300794;58.2644256;0.263081;0.254189;-33.429384;4.91365;24.300943;0.12075136;0.11820635;1.80230844;9.01373861;59.53263489
160;325871.0241;0.2322071;0.2220101;-14.8620926;7.8981608;44.1316522;0.317463;0.306733;-39.498718;3.917622;5.103564;0.17415375;0.17048321;-4.15968036;6.23698132;40.44260167
170;325729.0274;0.1989703;0.1902329;-12.7928674;6.2347698;30.3123191;0.280132;0.270664;-33.568203;4.287956;-4.854479;0.17561875;0.17191733;-2.07984597;4.68399997;26.63387781
180;325717.749;0.2013145;0.1924742;-13.1577354;6.1707998;30.0588203;0.279312;0.269871;-33.539564;4.252579;-4.714472;0.17230036;0.16866888;-2.31974871;4.62277096;26.50534422
190;325703.1477;0.1973522;0.1886858;-12.4164636;6.1584196;30.0455363;0.277845;0.268453;-33.388171;4.268862;-5.317634;0.17136284;0.16775112;-2.50225487;4.52123281;25.56828217
200;325697.1132;0.193903;0.1853881;-11.0549566;5.9425884;27.8284505;0.264124;0.255196;-29.971015;4.416022;-5.479071;0.16867678;0.16512167;0.3721912;4.46962495;24.86413547
210;325689.3069;0.1896161;0.1812895;-10.4810984;5.991792;28.2968644;0.261209;0.25238;-29.472796;4.380678;-5.086296;0.16883131;0.16527295;1.02603743;4.53385064;25.41253741
212;325689.0763;0.1886277;0.1803444;-10.5899504;5.974842;28.18509;0.261123;0.252297;-29.425764;4.383878;-5.042186;0.16901819;0.16545589;1.11342257;4.5452945;25.4970002
0;437243.2175;4.556602;4.356507;-237.53255;100;38.252875;3.23052978;3.12133335;-0.29472736;100;261.0992348;4.026996;3.942121;105.619188;100;367.01315
10;343969.0271;1.036898;0.991365;-0.252404;33.817959;193.027524;0.6610411;0.63869698;-64.35481308;21.60142004;114.5336524;0.397261;0.388888;44.034328;33.752375;222.922794
20;335495.3297;0.679245;0.649417;-6.933254;20.887957;115.353105;0.53033463;0.51240858;-64.72226444;9.63707587;43.17263261;0.335193;0.328128;-9.281621;15.409948;98.613276
30;332646.4609;0.627374;0.599824;-9.09313;26.097704;151.519447;0.6211058;0.60011155;-81.72846967;12.36066255;64.49264398;0.346478;0.339175;-24.230572;18.470388;121.990541
40;329192.4066;0.408737;0.390788;-10.279964;14.061336;81.238395;0.31717285;0.30645197;-26.64395773;9.71863496;50.48293883;0.289213;0.283117;23.459862;15.40537;100.586759
50;328114.0887;0.339018;0.32413;-11.648162;12.598642;72.968445;0.31272015;0.30214977;-30.21532237;8.08371008;40.00982156;0.271212;0.265496;15.090279;13.145706;85.315423
60;327513.2436;0.327857;0.31346;-15.650642;12.247212;71.330096;0.29444908;0.28449629;-29.25305101;8.34070918;43.33622341;0.239725;0.234673;18.013854;13.902317;90.603128
70;327115.4068;0.284821;0.272313;-11.565975;11.127175;63.9563;0.25144482;0.24294563;-28.17094637;6.4632199;32.95986605;0.165677;0.162185;10.544517;10.914985;71.67533
80;326794.538;0.252072;0.241003;-17.256824;8.375999;45.344911;0.31524755;0.30459175;-39.2549171;4.06872418;8.95535481;0.195716;0.191591;-7.766589;6.415921;40.443683
90;326614.7936;0.24987;0.238897;-19.803608;8.11348;44.738162;0.38435105;0.37135944;-50.54380586;4.41427224;-0.39349903;0.217925;0.213332;-16.172058;5.477788;33.978249
100;326444.6768;0.263393;0.251827;-20.316528;9.213137;51.634774;0.38708694;0.37400286;-49.57853354;4.46932455;7.98130582;0.218726;0.214116;-9.652454;7.316346;47.907386
110;326354.9975;0.272242;0.260287;-21.21694;8.811541;48.947698;0.38446084;0.37146552;-50.34417548;4.31278593;5.42899988;0.209002;0.204597;-13.599343;6.489236;42.173832
120;326296.9103;0.266918;0.255197;-19.502863;8.378355;45.595537;0.37730471;0.36455128;-48.49607114;4.47023722;2.2108664;0.206427;0.202076;-11.438941;6.139618;39.267996
130;326247.7852;0.259491;0.248096;-17.168348;8.20957;45.053152;0.36452094;0.35219962;-46.42345751;4.43720108;1.40657917;0.199771;0.195561;-10.304901;5.932832;37.525136
140;326214.1542;0.258467;0.247117;-16.760881;8.211973;45.109893;0.35538165;0.34336925;-44.80482776;4.40401008;2.67448319;0.191915;0.18787;-8.6511;6.076731;38.828211
150;326190.2139;0.259774;0.248367;-16.890156;8.701183;48.589211;0.34886443;0.33707232;-43.76954312;4.21709384;7.31836095;0.17967;0.175883;-6.706632;6.780527;44.381272
160;326146.8867;0.247314;0.236454;-14.770968;8.555744;47.403526;0.32852956;0.3174248;-40.45757017;4.09113681;7.32546099;0.173902;0.170237;-4.435592;6.642747;43.34744
170;326069.7521;0.247352;0.23649;-14.766136;8.355005;45.709286;0.33173288;0.32051985;-41.16543699;4.07747317;4.91852207;0.17281;0.169168;-6.082434;6.182054;40.001525
180;326045.4442;0.243344;0.232658;-13.573676;8.142953;43.35002;0.3072852;0.29689853;-36.75629727;4.00097082;5.775936;0.163676;0.160227;-2.936006;6.107293;39.596227
190;326026.2291;0.235534;0.225191;-13.024538;7.995859;42.430666;0.30507136;0.29475952;-35.99596329;4.03935491;5.06777783;0.16464;0.16117;-2.228234;5.972524;38.835507
200;325979.2185;0.23919;0.228686;-11.964947;8.320144;44.849886;0.28394589;0.27434812;-31.0027624;4.16168114;11.42060802;0.154204;0.150954;4.535895;7.110025;46.959266
208;325968.5791;0.233665;0.223404;-11.013182;8.162493;43.503043;0.28803631;0.27830028;-31.13897214;4.1849728;8.98578952;0.157738;0.154414;4.98877;6.831676;45.113531
0;437243.2175;4.556602;4.356507;-237.53255;100;38.252875;3.23053;3.121333;-0.294727;100;261.099235;4.026996;3.942121;105.619188;100;367.01315
10;342942.0281;0.870093;0.831884;21.262721;24.998151;130.932935;0.439924;0.425054;-24.375103;15.144975;70.903648;0.504537;0.493903;42.901829;21.395886;138.18058
20;334880.7277;0.649385;0.620868;-4.957791;19.898755;110.311123;0.518668;0.501136;-64.875152;8.282936;36.0023;0.312454;0.305869;-11.051452;14.104785;89.825999
30;331226.9081;0.544384;0.520478;-3.823593;21.752147;125.576481;0.479357;0.463154;-57.188251;11.010403;57.820359;0.262425;0.256894;-0.273088;17.457715;114.735523
40;328726.6754;0.373801;0.357386;-9.573659;14.008847;80.929633;0.329499;0.318362;-32.810344;8.553123;43.301484;0.268226;0.262573;15.031891;13.989908;91.14372
50;327805.9938;0.32782;0.313424;-16.17095;12.750126;74.048326;0.327019;0.315965;-33.44612;8.325053;42.381693;0.271709;0.265982;13.89953;13.779393;89.727344
60;327270.3997;0.301756;0.288505;-15.276165;11.825008;68.333726;0.297154;0.28711;-32.623446;7.147325;36.594982;0.197258;0.193101;13.536342;12.636714;82.75477
70;326866.0963;0.264456;0.252843;-14.970107;10.158632;57.894607;0.248982;0.240566;-27.786434;6.070654;30.686818;0.165032;0.161554;11.718734;10.692937;70.191985
80;326669.1452;0.254859;0.243667;-18.987205;9.818549;57.062259;0.288264;0.27852;-37.338229;5.084742;24.319772;0.145842;0.142768;-1.621975;9.089956;60.036027
90;326433.4819;0.265646;0.253981;-23.435584;8.890852;51.246814;0.326614;0.315574;-44.879561;4.078671;15.411374;0.170768;0.167169;-11.843472;7.353004;48.447463
100;326301.9194;0.265029;0.253391;-22.711517;7.83933;43.811805;0.361422;0.349205;-46.774459;4.030193;5.3574;0.204922;0.200603;-11.701113;6.24566;40.430746
110;326224.4641;0.255632;0.244406;-18.076979;8.139202;45.191217;0.334983;0.32366;-41.024515;4.211252;7.852217;0.191143;0.187114;-2.775086;7.043486;46.101647
120;326015.0751;0.219613;0.209969;-17.243193;6.898218;36.177714;0.316994;0.306279;-39.895922;4.4113;-0.866477;0.194493;0.190394;-6.513396;5.364239;32.516049
130;325972.9521;0.216224;0.206729;-14.559588;6.654128;33.168501;0.306746;0.296378;-37.088568;4.544155;-3.751942;0.196469;0.192328;-3.536803;5.114463;29.799823
140;325919.3056;0.21243;0.203102;-14.786631;6.333965;31.389194;0.302366;0.292146;-36.701909;4.555545;-4.917547;0.19128;0.187249;-3.94544;4.88331;27.838922
150;325877.5803;0.214557;0.205135;-14.335708;6.485959;32.664383;0.297256;0.287209;-35.970665;4.374716;-3.362037;0.181103;0.177286;-3.127813;4.968243;29.480816
160;325858.1925;0.215837;0.206359;-13.570398;6.619103;33.810627;0.2992;0.289087;-35.466264;4.442377;-2.476702;0.181278;0.177457;-1.437485;5.275169;31.552077
170;325825.8798;0.212556;0.203222;-14.327737;6.485281;32.647393;0.302189;0.291974;-36.177624;4.464087;-3.593957;0.183432;0.179566;-2.597453;5.108994;29.986214
180;325816.18;0.213388;0.204017;-13.968889;6.504535;33.004156;0.300072;0.289929;-35.517439;4.468349;-2.935857;0.179388;0.175608;-1.406989;5.238364;31.174593
190;325797.2806;0.210383;0.201144;-13.729679;6.580288;33.02102;0.295402;0.285417;-34.999406;4.40566;-2.64017;0.179421;0.175639;-1.76838;5.156851;30.590857
200;325783.1123;0.208452;0.199298;-13.349895;6.495725;32.279872;0.289862;0.280065;-33.974699;4.420671;-2.736394;0.178022;0.17427;-1.099142;5.140478;30.139163
210;325776.661;0.200245;0.191452;-9.968606;6.259879;29.700802;0.262827;0.253944;-28.159758;4.471347;-2.881813;0.176417;0.172698;4.372819;5.106658;29.650764
220;325774.1964;0.198746;0.190018;-9.798841;6.247938;29.751171;0.264348;0.255413;-28.116869;4.541319;-2.95832;0.17905;0.175276;4.278595;5.084829;29.437144
226;325767.4126;0.198038;0.189342;-8.442633;6.256046;29.064311;0.249273;0.240847;-24.483423;4.532473;-1.367942;0.184232;0.180349;8.447233;5.417085;31.562714
\end{filecontents*}
\pgfplotstabletypeset[
        multicolumn names, 
        col sep=semicolon, 
        display columns/0/.style={column name=$\bm{k}$,dec sep align},
        display columns/1/.style={column name={\textbf{AIC}},fixed,fixed zerofill,precision=0,set thousands separator={,},dec sep align},
        display columns/2/.style={column name={\textbf{v.mae}},fixed,fixed zerofill,precision=3,set thousands separator={,},dec sep align},
        display columns/3/.style={column name={$\bm{\textbf{\text{v.mae}}^a}$},fixed,fixed zerofill,precision=3,set thousands separator={,},dec sep align},
        display columns/4/.style={column name={\textbf{v.res}},fixed,fixed zerofill,precision=0,set thousands separator={,},dec sep align},
        display columns/5/.style={column name={$\bm{\textbf{\text{v.mae}}^0}$},fixed,fixed zerofill,precision=3,set thousands separator={,},dec sep align},
        display columns/6/.style={column name={$\bm{\textbf{\text{v.res}}^0}$},fixed,fixed zerofill,precision=0,set thousands separator={,},dec sep align},
        display columns/7/.style={column name={\textbf{ns.mae}},fixed,fixed zerofill,precision=3,set thousands separator={,},dec sep align},
        display columns/8/.style={column name={$\bm{\textbf{\text{ns.mae}}^a}$},fixed,fixed zerofill,precision=3,set thousands separator={,},dec sep align},
        display columns/9/.style={column name={\textbf{ns.res}},fixed,fixed zerofill,precision=0,set thousands separator={,},dec sep align},
        display columns/10/.style={column name={$\bm{\textbf{\text{ns.mae}}^0}$},fixed,fixed zerofill,precision=3,set thousands separator={,},dec sep align},
        display columns/11/.style={column name={$\bm{\textbf{\text{ns.res}}^0}$},fixed,fixed zerofill,precision=0,set thousands separator={,},dec sep align},
        display columns/12/.style={column name={\textbf{cr.mae}},fixed,fixed zerofill,precision=3,set thousands separator={,},dec sep align},
        display columns/13/.style={column name={$\bm{\textbf{\text{cr.mae}}^a}$},fixed,fixed zerofill,precision=3,set thousands separator={,},dec sep align},
        display columns/14/.style={column name={\textbf{cr.res}},fixed,fixed zerofill,precision=0,set thousands separator={,},dec sep align},
        display columns/15/.style={column name={$\bm{\textbf{\text{cr.mae}}^0}$},fixed,fixed zerofill,precision=3,set thousands separator={,},dec sep align},
        display columns/16/.style={column name={$\bm{\textbf{\text{cr.res}}^0}$},fixed,fixed zerofill,precision=0,set thousands separator={,},dec sep align},
	    	every head row/.style={before row=\toprule,after row=\midrule},
	    	every row no 0/.style={before row={\multicolumn{18}{l}{\textbf{Gamma with identity link}}\\ \\}},
	    	every row no 23/.style={before row={\midrule \multicolumn{18}{l}{\textbf{Gamma with inverse link}}\\ \\}},
	    	every row no 45/.style={before row={\midrule \multicolumn{18}{l}{\textbf{Gamma with log link}}\\ \\}},
      	every last row/.style={after row=\bottomrule},
      ]{BGROGLMgammaidinvlog300886valfig.csv} 
    \caption{AIC scores and out-of-sample validation figures of the gamma GLMs of BEL with identity, inverse and log link functions under $300$-$886$ after each tenth and the final iteration.}
    \label{tab:BGROGLMgamma300886}
  \end{center}
\end{table}

\begin{table}[htb]
  \tiny
  \begin{center}
  \tabcolsep=0.052cm
  \renewcommand{\arraystretch}{0.90}
      \begin{filecontents*}{BGROGLMinvgaussidinv300886valfig.csv}
k;AIC;v.wre.y;v.a.wre.y;v.res.avg.y;v.wre.delta.y;v.res.avg.delta.y;ns.wre.y;ns.a.wre.y;ns.res.avg.y;ns.wre.delta.y;ns.res.avg.delta.y;cr.wre.y;cr.a.wre.y;cr.res.avg.y;cr.wre.delta.y;cr.res.avg.delta.y
0;437337.824;4.556602;4.356507;-237.53255;100;38.252875;3.23053;3.121333;-0.294727;100;261.099235;4.0269963;3.9421215;105.619188;100;367.0131501
10;346131.9225;0.871108;0.832855;0.713793;23.559167;115.058412;0.314178;0.303559;7.078512;20.268605;107.031668;0.5339306;0.5226773;69.5053102;25.6731254;169.458466
20;334430.1768;0.548513;0.524426;-13.43352;15.995781;76.964459;0.599478;0.579215;-76.80317;8.272797;-0.796654;0.4683173;0.4584468;-43.8485341;9.8093322;32.1579812
30;331452.9427;0.488026;0.466595;-3.855963;15.939114;88.522158;0.516886;0.499414;-66.742392;6.531527;11.244266;0.4133809;0.4046683;-40.432431;9.2798444;37.5542276
40;328985.278;0.369987;0.35374;-12.741961;13.27921;75.556357;0.338108;0.32668;-38.528659;7.192921;35.378196;0.2379894;0.2329734;6.0085474;12.3007425;79.9154031
50;328064.426;0.331745;0.317177;-15.003395;12.727137;74.296265;0.338327;0.326891;-40.278972;6.871132;34.629225;0.2317861;0.2269009;1.0312772;11.6641499;75.9394737
60;327533.1937;0.297839;0.28476;-16.793441;10.99407;64.249496;0.304005;0.293729;-37.464368;5.868003;29.187106;0.1720514;0.1684251;2.8322592;10.6458388;69.4837337
70;327081.9222;0.27385;0.261825;-15.495551;9.3874;53.189243;0.243242;0.23502;-26.842832;5.535433;27.450499;0.1708122;0.167212;12.6007881;10.2529205;66.8941191
80;326849.2943;0.267203;0.255469;-19.622478;9.425509;54.373854;0.277883;0.26849;-34.211392;5.271381;25.393477;0.1515451;0.148351;5.0111314;9.7834027;64.6160007
90;326714.8779;0.246552;0.235725;-20.736193;8.545667;49.153838;0.275055;0.265758;-35.4139;4.399296;20.084668;0.1399033;0.1369546;-0.6641388;8.3023911;54.8344297
100;326627.0221;0.234399;0.224105;-19.594301;8.454385;48.860104;0.266064;0.25707;-34.215449;4.413945;19.847494;0.143776;0.1407458;-1.0757839;8.0226988;52.9871587
110;326557.3278;0.225211;0.215321;-17.026916;8.350406;47.471059;0.246219;0.237896;-30.817134;4.336546;19.289379;0.1322554;0.1294679;1.6816411;7.8411594;51.7881536
120;326504.5114;0.232968;0.222737;-16.94445;8.897205;51.002471;0.25572;0.247076;-32.79817;4.4275;20.757289;0.1254586;0.1228144;-0.0168069;8.1061994;53.5386515
130;326464.5046;0.242668;0.232012;-16.244818;9.965435;57.543748;0.264952;0.255997;-33.817159;5.12624;25.579945;0.1222887;0.1197113;1.4707993;9.2159093;60.8679028
140;326441.8227;0.244137;0.233416;-16.038909;10.175323;58.757634;0.272821;0.263599;-35.230877;5.07854;25.174203;0.1249306;0.1222975;-0.3174803;9.0977649;60.0875997
150;326357.1133;0.252438;0.241353;-15.931101;10.132579;58.031846;0.351584;0.3397;-44.664081;4.600606;14.907403;0.1694847;0.1659126;-1.2441489;8.8312461;58.3273351
160;326130.1983;0.206043;0.196995;-14.599592;6.293761;30.542948;0.293068;0.283162;-35.576932;4.359992;-4.825855;0.1872365;0.1832902;-4.4410626;4.710593;26.3100141
170;326112.4571;0.204102;0.195139;-14.79924;6.173412;30.077159;0.288695;0.278936;-35.033744;4.284117;-4.548808;0.1791494;0.1753735;-3.6677561;4.6882284;26.8171798
180;326099.324;0.203062;0.194145;-13.785796;6.12992;29.803523;0.282512;0.272963;-34.042114;4.277377;-4.844257;0.1767657;0.1730401;-2.8951165;4.654362;26.3027401
190;326088.3254;0.204161;0.195196;-13.896405;6.143359;29.831603;0.28174;0.272217;-33.91865;4.279541;-4.582105;0.1780826;0.1743293;-2.7384418;4.6992237;26.5981038
200;326075.7263;0.203723;0.194777;-14.030225;6.17206;30.227764;0.286112;0.276441;-34.351091;4.347482;-4.484565;0.1835518;0.1796832;-2.6306311;4.8226749;27.235895
210;326070.9761;0.19919;0.190443;-12.367161;6.140239;29.83835;0.272903;0.263679;-32.050517;4.276803;-4.236469;0.1829455;0.1790897;-0.2538811;4.8676617;27.5601672
217;326068.5881;0.191294;0.182894;-11.112816;5.966507;27.846281;0.261027;0.252204;-29.314979;4.363665;-4.747346;0.1784737;0.1747121;2.1968028;4.7786287;26.7644366
0;437337.824;4.556602;4.356507;-237.53255;100;38.252875;3.23053;3.121333;-0.294727;100;261.099235;4.026996;3.942121;105.619188;100;367.01315
10;344458.4028;1.128647;1.079084;-24.812685;35.685428;202.225205;1.137887;1.099425;-149.827914;14.423476;62.818513;0.639191;0.625719;-63.488961;22.713184;149.157466
20;336003.5098;0.682051;0.6521;-4.761222;21.011309;117.387045;0.534168;0.516112;-67.030553;8.865545;40.726251;0.32119;0.31442;-12.306896;14.894979;95.449908
30;333059.5068;0.625724;0.598246;-9.99764;24.463198;141.803373;0.622649;0.601602;-82.659857;10.858507;54.749694;0.376467;0.368532;-30.85212;16.233247;106.55743
40;329632.0078;0.411752;0.393671;-13.721669;15.91162;93.418915;0.344855;0.333199;-28.82722;12.095934;63.921901;0.318142;0.311436;28.226176;18.445587;120.975297
50;328514.7513;0.335021;0.320309;-11.647239;12.38651;71.477703;0.305342;0.295021;-28.508127;8.121905;40.225352;0.275644;0.269834;17.702028;13.33318;86.435507
60;327915.5826;0.321206;0.307101;-15.404543;11.9697;69.601687;0.285955;0.276289;-27.05341;8.385318;43.561357;0.246612;0.241414;20.481474;13.97291;91.096242
70;327542.9472;0.278313;0.266092;-12.144585;10.488295;60.46237;0.24596;0.237646;-27.68964;6.105928;30.525852;0.164213;0.160752;8.88946;10.331249;67.104953
80;327195.5635;0.249294;0.238347;-17.07464;8.22664;44.508527;0.307691;0.297291;-38.150254;4.037213;9.04145;0.19296;0.188893;-6.976912;6.381318;40.214793
90;327012.4129;0.247238;0.236381;-18.864803;8.016337;43.977166;0.376019;0.363309;-49.33525;4.390038;-0.884744;0.211752;0.207289;-15.098537;5.407165;33.35197
100;326835.9256;0.261482;0.25;-20.257166;9.073455;50.67821;0.382309;0.369386;-48.927025;4.437523;7.616887;0.215478;0.210937;-9.162559;7.237328;47.381353
110;326750.0999;0.268403;0.256616;-20.779363;8.678547;47.363223;0.386293;0.373236;-49.958028;4.509715;3.793095;0.21674;0.212172;-11.827048;6.49033;41.924075
120;326673.9957;0.2625;0.250973;-19.142843;8.190539;45.301746;0.377963;0.365187;-48.738782;4.498717;1.314345;0.207273;0.202904;-11.839864;6.010573;38.213263
130;326636.4001;0.260959;0.2495;-18.017078;8.380258;46.32605;0.372743;0.360144;-48.014466;4.401854;1.9372;0.197547;0.193383;-11.628438;5.985193;38.323227
140;326606.6863;0.258196;0.246858;-16.748488;8.253393;45.715429;0.34927;0.337464;-43.897828;4.28935;4.174626;0.185052;0.181151;-7.575869;6.277145;40.496585
150;326580.7191;0.257662;0.246347;-17.203519;8.437141;46.892942;0.349635;0.337817;-44.03133;4.228088;5.673668;0.182681;0.178831;-7.322017;6.505007;42.38298
160;326537.8383;0.24616;0.235351;-14.699856;8.444583;46.752125;0.325824;0.314811;-40.107641;4.076698;6.952876;0.172669;0.169029;-4.193981;6.571675;42.866537
170;326521.952;0.248595;0.237678;-15.065836;8.147624;44.640751;0.322253;0.31136;-38.83059;4.119281;6.484534;0.175424;0.171727;-2.088982;6.603334;43.226143
180;326468.1318;0.244778;0.234029;-13.968847;8.582756;47.353137;0.297604;0.287544;-33.584668;4.302595;13.345853;0.16211;0.158693;4.086036;7.724333;51.016557
190;326454.644;0.243397;0.232709;-13.898362;8.505697;46.964096;0.299393;0.289273;-33.777587;4.290294;12.693408;0.163434;0.15999;3.98902;7.641405;50.460015
200;326398.8245;0.230757;0.220623;-11.772829;7.918046;42.141456;0.286274;0.276598;-30.76574;4.208392;8.757083;0.157859;0.154532;5.693691;6.855867;45.216514
210;326364.8276;0.232884;0.222657;-11.552713;7.982945;42.554124;0.288293;0.278549;-31.1994;4.208397;8.515974;0.158643;0.155299;4.891399;6.765396;44.606773
219;326362.8883;0.232976;0.222745;-11.196841;8.040331;42.889211;0.283241;0.273667;-30.793801;4.129863;8.900789;0.152804;0.149583;5.059882;6.786417;44.754471
\end{filecontents*}
\pgfplotstabletypeset[
        multicolumn names, 
        col sep=semicolon, 
        display columns/0/.style={column name=$\bm{k}$,dec sep align},
        display columns/1/.style={column name={\textbf{AIC}},fixed,fixed zerofill,precision=0,set thousands separator={,},dec sep align},
        display columns/2/.style={column name={\textbf{v.mae}},fixed,fixed zerofill,precision=3,set thousands separator={,},dec sep align},
        display columns/3/.style={column name={$\bm{\textbf{\text{v.mae}}^a}$},fixed,fixed zerofill,precision=3,set thousands separator={,},dec sep align},
        display columns/4/.style={column name={\textbf{v.res}},fixed,fixed zerofill,precision=0,set thousands separator={,},dec sep align},
        display columns/5/.style={column name={$\bm{\textbf{\text{v.mae}}^0}$},fixed,fixed zerofill,precision=3,set thousands separator={,},dec sep align},
        display columns/6/.style={column name={$\bm{\textbf{\text{v.res}}^0}$},fixed,fixed zerofill,precision=0,set thousands separator={,},dec sep align},
        display columns/7/.style={column name={\textbf{ns.mae}},fixed,fixed zerofill,precision=3,set thousands separator={,},dec sep align},
        display columns/8/.style={column name={$\bm{\textbf{\text{ns.mae}}^a}$},fixed,fixed zerofill,precision=3,set thousands separator={,},dec sep align},
        display columns/9/.style={column name={\textbf{ns.res}},fixed,fixed zerofill,precision=0,set thousands separator={,},dec sep align},
        display columns/10/.style={column name={$\bm{\textbf{\text{ns.mae}}^0}$},fixed,fixed zerofill,precision=3,set thousands separator={,},dec sep align},
        display columns/11/.style={column name={$\bm{\textbf{\text{ns.res}}^0}$},fixed,fixed zerofill,precision=0,set thousands separator={,},dec sep align},
        display columns/12/.style={column name={\textbf{cr.mae}},fixed,fixed zerofill,precision=3,set thousands separator={,},dec sep align},
        display columns/13/.style={column name={$\bm{\textbf{\text{cr.mae}}^a}$},fixed,fixed zerofill,precision=3,set thousands separator={,},dec sep align},
        display columns/14/.style={column name={\textbf{cr.res}},fixed,fixed zerofill,precision=0,set thousands separator={,},dec sep align},
        display columns/15/.style={column name={$\bm{\textbf{\text{cr.mae}}^0}$},fixed,fixed zerofill,precision=3,set thousands separator={,},dec sep align},
        display columns/16/.style={column name={$\bm{\textbf{\text{cr.res}}^0}$},fixed,fixed zerofill,precision=0,set thousands separator={,},dec sep align},
	    	every head row/.style={before row=\toprule,after row=\midrule},
	    	every row no 0/.style={before row={\multicolumn{18}{l}{\textbf{Inverse gaussian with identity link}}\\ \\}},
	    	every row no 23/.style={before row={\midrule \multicolumn{18}{l}{\textbf{Inverse gaussian with inverse link}}\\ \\}},
      	every last row/.style={after row=\bottomrule},
      ]{BGROGLMinvgaussidinv300886valfig.csv} 
    \caption{AIC scores and out-of-sample validation figures of the inverse gaussian GLMs of BEL with identity, inverse, log and $\frac{1}{\mu^2}$ link functions under $300$-$886$ after each tenth and the final iteration.}
    \label{tab:BGROGLMinvgauss3008861}
  \end{center}
\end{table}

\begin{table}[htb]
\ContinuedFloat
  \tiny
  \begin{center}
  \tabcolsep=0.052cm
  \renewcommand{\arraystretch}{0.90}
      \begin{filecontents*}{BGROGLMinvgausslogdivmusq300886valfig.csv}
k;AIC;v.wre.y;v.a.wre.y;v.res.avg.y;v.wre.delta.y;v.res.avg.delta.y;ns.wre.y;ns.a.wre.y;ns.res.avg.y;ns.wre.delta.y;ns.res.avg.delta.y;cr.wre.y;cr.a.wre.y;cr.res.avg.y;cr.wre.delta.y;cr.res.avg.delta.y
0;437337.824;4.556602;4.356507;-237.53255;100;38.252875;3.23053;3.121333;-0.294727;100;261.099235;4.0269963;3.9421215;105.619188;100;367.0131501
10;343530.4892;0.866193;0.828155;18.799286;24.925012;130.649954;0.449963;0.434754;-28.083858;14.939768;69.375347;0.4942728;0.4838553;38.7344769;21.1215603;136.1936819
20;335354.8054;0.644254;0.615963;-5.029689;19.653097;108.854341;0.526343;0.508552;-66.586947;7.946656;32.905619;0.3178686;0.311169;-14.1035936;13.4895667;85.3889731
30;331675.4225;0.535872;0.51234;-4.364227;21.697061;125.269777;0.481537;0.46526;-58.088018;10.885011;57.154524;0.2617875;0.2562699;-1.8940862;17.2445625;113.3484557
40;329139.5419;0.366118;0.350041;-9.759329;13.912628;80.279582;0.32473;0.313754;-31.966068;8.604199;43.68138;0.2691949;0.2635212;15.6709769;14.0106752;91.3184247
50;328189.5002;0.324062;0.309832;-16.04414;12.640048;73.410434;0.319169;0.30838;-31.629069;8.481909;43.434042;0.2740155;0.2682402;15.96218;13.9656413;91.025291
60;327665.9438;0.295842;0.28285;-15.184014;11.626056;67.088658;0.289833;0.280036;-31.071019;7.180515;36.81019;0.2011096;0.196871;15.2476297;12.6948471;83.1288385
70;327262.5332;0.261298;0.249823;-14.572764;9.94776;56.57654;0.243752;0.235513;-26.644994;6.042153;30.112848;0.171727;0.1681076;12.3054233;10.5313024;69.0632654
80;327061.4271;0.250668;0.23966;-18.493849;9.746092;56.499597;0.284301;0.274691;-36.817095;4.98778;23.784889;0.1453745;0.1423105;-1.4003999;8.9636147;59.2015837
90;326824.5402;0.262579;0.251048;-23.306741;8.768996;50.578253;0.320785;0.309942;-44.020719;4.05923;15.472812;0.168192;0.1646471;-11.2856283;7.3157494;48.2079027
100;326694.8884;0.260938;0.249479;-22.18394;7.727419;43.190348;0.351961;0.340064;-45.137586;4.047648;5.845239;0.2028069;0.1985324;-9.8482156;6.3409619;41.1346098
110;326588.5731;0.24048;0.22992;-18.94303;7.483766;41.245616;0.341766;0.330214;-44.29758;4.124484;1.499603;0.1918098;0.1877672;-10.8759516;5.4838485;34.9212314
120;326408.7578;0.216462;0.206957;-16.484903;6.396654;32.044529;0.299134;0.289023;-36.620918;4.533505;-2.482948;0.195078;0.1909664;-4.1453293;5.1701862;29.9926403
130;326363.17;0.215992;0.206507;-15.446273;6.313692;30.828215;0.308265;0.297845;-37.403317;4.693106;-5.520292;0.200518;0.1962917;-4.2375544;4.9573393;27.6454706
140;326331.4671;0.217651;0.208093;-14.550974;6.537156;32.949602;0.302628;0.292398;-35.830995;4.505445;-2.721881;0.1954502;0.1913308;-1.0927201;5.3619142;32.0163935
150;326269.7984;0.216042;0.206555;-14.452406;6.456526;31.992942;0.30157;0.291377;-36.111816;4.52449;-4.057932;0.1886873;0.1847105;-2.4871239;5.0492836;29.5667606
160;326249.2084;0.217281;0.207739;-13.949918;6.595693;33.520434;0.298335;0.288251;-35.512493;4.417642;-2.433604;0.1817659;0.1779349;-1.321437;5.2914558;31.7574523
170;326231.1781;0.216805;0.207284;-14.602638;6.49232;32.40481;0.296339;0.286322;-35.205648;4.39141;-2.589664;0.1788456;0.1750762;-1.5874399;5.1889052;31.0285449
180;326206.4923;0.214164;0.204759;-14.593771;6.42577;31.868309;0.301582;0.291388;-36.325738;4.466032;-4.255121;0.1789196;0.1751487;-3.1642229;4.9503596;28.9063941
190;326190.9876;0.206446;0.19738;-12.599234;6.471893;32.523181;0.288414;0.278665;-33.685024;4.421951;-2.954072;0.1733838;0.1697295;-0.0306195;5.1487558;30.7003329
200;326176.1362;0.207639;0.198521;-12.622313;6.545447;32.814517;0.286092;0.276422;-32.970686;4.430027;-1.925319;0.1791347;0.1753592;0.4082195;5.2884393;31.4535865
210;326160.8794;0.207964;0.198832;-12.890463;6.500538;32.510371;0.285712;0.276055;-32.798956;4.439277;-1.789585;0.1837522;0.1798793;0.5504986;5.318452;31.5598694
220;326153.1533;0.201561;0.19271;-9.809059;6.280467;29.719613;0.259708;0.25093;-27.368478;4.455171;-2.231269;0.1782123;0.1744562;5.3955328;5.1901842;30.5327418
222;326152.973;0.200578;0.19177;-9.727104;6.291246;29.767289;0.261319;0.252486;-27.65102;4.494332;-2.548089;0.1804175;0.1766149;5.1590833;5.1761217;30.2620139
0;437337.824;4.556602;4.356507;-237.53255;100;38.252875;3.2305298;3.1213333;-0.2947274;100;261.0992347;4.0269963;3.9421215;105.619188;100;367.0131501
10;344467.1505;0.984597;0.94136;-14.471219;31.473092;175.749876;0.9928607;0.9593006;-130.1904101;12.5730595;45.6392228;0.5610239;0.5491995;-52.0739435;18.9863114;123.7556894
20;336815.4432;0.667854;0.638527;-6.627196;21.404117;121.540766;0.5905878;0.5706251;-75.3133167;9.5063641;38.4631828;0.3721724;0.3643283;-22.3169684;14.521288;91.4595312
30;331791.9199;0.47753;0.456561;-5.426245;15.821218;89.882398;0.366522;0.354133;-28.0090024;10.5732252;52.9081775;0.3733079;0.3654399;33.4161271;17.4959385;114.333307
40;330089.1104;0.421373;0.402869;-0.829876;15.183183;88.667596;0.2946434;0.2846841;-19.3356446;10.6599641;55.7703652;0.3159476;0.3092885;33.9596626;16.6570041;109.0656724
50;329020.2601;0.376002;0.35949;-10.244725;14.44279;85.145902;0.2999941;0.2898539;-20.715849;11.4389724;60.2833154;0.3202489;0.3134992;34.0735115;17.5525751;115.0726759
60;328451.5922;0.330259;0.315756;-11.590698;12.904577;75.151328;0.2895425;0.2797556;-24.4558822;9.1963081;47.8946811;0.2730242;0.2672698;25.3182089;14.9519543;97.6687721
70;327925.2138;0.316202;0.302316;-16.341403;11.733491;68.611742;0.3009671;0.290794;-35.3007073;7.0899788;35.260975;0.1995911;0.1953844;5.594425;11.7008693;76.1561073
80;327639.3861;0.261697;0.250205;-17.850745;8.127888;42.557403;0.2977695;0.2877045;-34.9531357;4.4251884;11.0635495;0.2076643;0.2032874;-1.1743237;7.2049543;44.8423614
90;327264.7457;0.27811;0.265898;-21.517399;8.310569;45.733254;0.3548184;0.342825;-44.1776721;4.3825404;8.6815179;0.2015018;0.1972548;-6.6890548;7.0901835;46.1701352
100;327148.0483;0.28806;0.275411;-22.439591;8.166407;43.85558;0.3567463;0.3446878;-44.3094729;4.4084358;7.5942348;0.2070452;0.2026814;-6.3617974;7.0392614;45.5419104
110;327077.4139;0.27455;0.262494;-19.567037;7.965202;42.328158;0.3658588;0.3534922;-45.2211517;4.6761639;2.2825806;0.2066837;0.2023275;-7.0251456;6.4099792;40.4785867
120;326915.7222;0.274481;0.262428;-17.917259;8.312839;44.545793;0.392914;0.379633;-46.72368;5.1329366;1.3479096;0.2278577;0.2230553;-5.0143686;6.7899328;43.057221
130;326875.5866;0.269319;0.257492;-17.72813;8.132671;43.267013;0.3957694;0.3823918;-46.7972727;5.2174731;-0.193593;0.2335937;0.2286704;-5.000935;6.6252467;41.6027447
140;326788.8714;0.258904;0.247535;-17.846019;8.149241;43.951881;0.3948403;0.3814941;-46.5153225;5.074095;0.8911153;0.2492496;0.2439963;-5.5121437;6.6966979;41.8942941
150;326576.1292;0.226939;0.216973;-14.871504;6.896343;33.686501;0.340536;0.3290254;-39.164011;5.2911642;-4.9974686;0.2212022;0.21654;-3.1562045;5.5104955;31.0103379
160;326478.7249;0.214143;0.204739;-15.665205;6.273707;28.559472;0.2911301;0.2812895;-35.3376057;4.5714784;-5.5043916;0.2064538;0.2021025;-8.0318451;4.6173554;21.801369
170;326450.736;0.210069;0.200844;-14.974864;6.035067;25.973338;0.2846362;0.2750151;-34.3509241;4.6107724;-7.794185;0.2021355;0.1978752;-7.9878194;4.4407098;18.5689197
180;326426.3027;0.195776;0.187179;-12.513416;5.753402;24.514243;0.2503688;0.241906;-28.1861807;4.373368;-5.5499848;0.1869401;0.1830001;-2.0863717;4.4258127;20.5498242
190;326407.6083;0.195316;0.186739;-12.645673;5.681978;24.027262;0.2489979;0.2405815;-27.9019564;4.3603978;-5.6204836;0.1880953;0.1841309;-1.6929167;4.4642185;20.588556
200;326396.5363;0.192689;0.184228;-12.739612;5.686066;24.069253;0.2449151;0.2366366;-27.4379399;4.2516123;-5.020538;0.1860068;0.1820864;-2.8832835;4.3817897;19.5341184
210;326304.8853;0.186731;0.178531;-12.634824;5.720575;27.059223;0.2368857;0.2288787;-25.781121;3.8112223;-0.4785373;0.1619812;0.1585672;1.6553823;4.5096905;26.957966
220;326171.6156;0.175576;0.167866;-14.266407;5.109901;25.811052;0.1974224;0.1907492;-21.6768061;3.3462201;4.0091907;0.1461721;0.1430913;5.6525196;4.9185056;31.3385164
230;326160.1243;0.175252;0.167557;-13.781392;4.993739;24.929271;0.2063134;0.1993397;-21.4522144;3.5834496;2.8669866;0.158835;0.1554873;8.0057775;5.1141406;32.3249785
240;326140.5027;0.166371;0.159066;-10.836283;5.012066;24.269477;0.1968154;0.1901627;-16.1509526;3.9085987;4.5633441;0.1819747;0.1781393;14.1928232;5.5599857;34.9071199
250;326123.6999;0.173502;0.165883;-12.390515;5.05767;25.455753;0.1925144;0.1860071;-14.6231243;3.8331417;8.8316804;0.1880412;0.184078;17.302053;6.2657858;40.7568577
\end{filecontents*}
\pgfplotstabletypeset[
        multicolumn names, 
        col sep=semicolon, 
        display columns/0/.style={column name=$\bm{k}$,dec sep align},
        display columns/1/.style={column name={\textbf{AIC}},fixed,fixed zerofill,precision=0,set thousands separator={,},dec sep align},
        display columns/2/.style={column name={\textbf{v.mae}},fixed,fixed zerofill,precision=3,set thousands separator={,},dec sep align},
        display columns/3/.style={column name={$\bm{\textbf{\text{v.mae}}^a}$},fixed,fixed zerofill,precision=3,set thousands separator={,},dec sep align},
        display columns/4/.style={column name={\textbf{v.res}},fixed,fixed zerofill,precision=0,set thousands separator={,},dec sep align},
        display columns/5/.style={column name={$\bm{\textbf{\text{v.mae}}^0}$},fixed,fixed zerofill,precision=3,set thousands separator={,},dec sep align},
        display columns/6/.style={column name={$\bm{\textbf{\text{v.res}}^0}$},fixed,fixed zerofill,precision=0,set thousands separator={,},dec sep align},
        display columns/7/.style={column name={\textbf{ns.mae}},fixed,fixed zerofill,precision=3,set thousands separator={,},dec sep align},
        display columns/8/.style={column name={$\bm{\textbf{\text{ns.mae}}^a}$},fixed,fixed zerofill,precision=3,set thousands separator={,},dec sep align},
        display columns/9/.style={column name={\textbf{ns.res}},fixed,fixed zerofill,precision=0,set thousands separator={,},dec sep align},
        display columns/10/.style={column name={$\bm{\textbf{\text{ns.mae}}^0}$},fixed,fixed zerofill,precision=3,set thousands separator={,},dec sep align},
        display columns/11/.style={column name={$\bm{\textbf{\text{ns.res}}^0}$},fixed,fixed zerofill,precision=0,set thousands separator={,},dec sep align},
        display columns/12/.style={column name={\textbf{cr.mae}},fixed,fixed zerofill,precision=3,set thousands separator={,},dec sep align},
        display columns/13/.style={column name={$\bm{\textbf{\text{cr.mae}}^a}$},fixed,fixed zerofill,precision=3,set thousands separator={,},dec sep align},
        display columns/14/.style={column name={\textbf{cr.res}},fixed,fixed zerofill,precision=0,set thousands separator={,},dec sep align},
        display columns/15/.style={column name={$\bm{\textbf{\text{cr.mae}}^0}$},fixed,fixed zerofill,precision=3,set thousands separator={,},dec sep align},
        display columns/16/.style={column name={$\bm{\textbf{\text{cr.res}}^0}$},fixed,fixed zerofill,precision=0,set thousands separator={,},dec sep align},
	    	every head row/.style={before row=\toprule,after row=\midrule},
	    	every row no 0/.style={before row={\multicolumn{18}{l}{\textbf{Inverse gaussian with log link}}\\ \\}},
	    	every row no 24/.style={before row={\midrule \multicolumn{18}{l}{\textbf{Inverse gaussian with }$\bm{\frac{1}{\mu^2}}$\textbf{ link}}\\ \\}},
      	every last row/.style={after row=\bottomrule},
      ]{BGROGLMinvgausslogdivmusq300886valfig.csv} 
    \caption[]{Cont.}
    \label{tab:BGROGLMinvgauss3008862}
  \end{center}
\end{table}

\begin{table}[htb]
  \tiny
  \begin{center}
  \tabcolsep=0.052cm
  \renewcommand{\arraystretch}{0.90}
      \begin{filecontents*}{BGROGLMgaussiangammainvgauss.csv}
k;AIC;v.wre.y;v.a.wre.y;v.res.avg.y;v.wre.delta.y;v.res.avg.delta.y;ns.wre.y;ns.a.wre.y;ns.res.avg.y;ns.wre.delta.y;ns.res.avg.delta.y;cr.wre.y;cr.a.wre.y;cr.res.avg.y;cr.wre.delta.y;cr.res.avg.delta.y
150;325849.7177;0.247487;0.236619;-14.4860724;9.9236248;57.011081;0.271045;0.261883;-34.843727;4.612476;22.261963;0.12219633;0.11962086;-0.71987917;8.53728323;56.38581146
150;325951.9558;0.2579;0.246574;-16.111193;8.467954;45.477528;0.353335;0.341392;-44.239747;4.282015;2.957511;0.1922;0.1881491;-8.1613171;6.0876558;39.0359406
150;325823.254;0.239702;0.229176;-14.615343;7.979548;43.868304;0.315585;0.304918;-38.326049;4.013879;5.766135;0.1704513;0.1668588;-2.2305725;6.4343916;41.8616117
150;326041.0456;0.236068;0.2257015;-14.202776;9.3290335;53.3460621;0.25986;0.251076;-32.953058;4.321119;20.204318;0.1210146;0.118464;1.0416692;8.2061884;54.1990446
150;326221.9441;0.254004;0.242849;-14.609355;8.409522;46.165413;0.32658471;0.31554569;-39.76753288;4.11104413;6.61577254;0.170609;0.167013;-2.514794;6.721812;43.868511
150;326022.3145;0.243052;0.232379;-14.794787;7.820028;42.5755;0.323209;0.312284;-39.967415;4.039941;3.011409;0.173713;0.170052;-4.231749;6.009838;38.747075
150;326352.1317;0.249149;0.238208;-16.775489;9.375421;53.551643;0.337168;0.325771;-43.9068;4.223941;12.028869;0.1495211;0.1463698;-3.6004468;7.9304563;52.3352221
150;326617.1348;0.253324;0.242199;-14.571099;8.15202;44.370231;0.323912;0.312964;-39.376865;4.147759;5.173003;0.172357;0.168725;-2.551035;6.476346;41.998833
150;326412.6525;0.247431;0.236565;-14.766914;7.716267;41.907191;0.323553;0.312617;-39.989455;4.095487;2.293187;0.171862;0.16824;-4.44379;5.891975;37.838852
150;326778.1994;0.261519;0.250035;-15.691022;8.257955;44.219915;0.332116;0.32089;-40.519593;4.237989;4.999882;0.177478;0.173737;-3.475248;6.517505;42.044226
224;325458.9411;0.1943361;0.1858021;-8.7292706;6.658858;34.2822165;0.267728;0.258678;-30.207234;4.199764;-1.58721;0.16834735;0.16479918;0.75346231;5.00742678;29.37348663
213;325797.3295;0.240981;0.230399;-12.305274;8.324796;44.039468;0.309614;0.299148;-35.697951;4.063452;6.255329;0.1709604;0.1673572;-1.0692338;6.2839868;40.8840457
214;325552.0671;0.20602;0.196973;-9.601323;6.640115;32.059351;0.267493;0.258452;-29.265488;4.401831;-1.996277;0.1766801;0.1729563;2.8082184;5.1799491;30.0774301
212;325689.0763;0.1886277;0.1803444;-10.5899504;5.974842;28.18509;0.261123;0.252297;-29.425764;4.383878;-5.042186;0.16901819;0.16545589;1.11342257;4.5452945;25.4970002
208;325968.5791;0.233665;0.223404;-11.013182;8.162493;43.503043;0.28803631;0.27830028;-31.13897214;4.1849728;8.98578952;0.157738;0.154414;4.98877;6.831676;45.113531
226;325767.4126;0.198038;0.189342;-8.442633;6.256046;29.064311;0.249273;0.240847;-24.483423;4.532473;-1.367942;0.184232;0.180349;8.447233;5.417085;31.562714
217;326068.5881;0.191294;0.182894;-11.112816;5.966507;27.846281;0.261027;0.252204;-29.314979;4.363665;-4.747346;0.1784737;0.1747121;2.1968028;4.7786287;26.7644366
219;326362.8883;0.232976;0.222745;-11.196841;8.040331;42.889211;0.283241;0.273667;-30.793801;4.129863;8.900789;0.152804;0.149583;5.059882;6.786417;44.754471
222;326152.973;0.200578;0.19177;-9.727104;6.291246;29.767289;0.261319;0.252486;-27.65102;4.494332;-2.548089;0.1804175;0.1766149;5.1590833;5.1761217;30.2620139
250;326123.6999;0.173502;0.165883;-12.390515;5.05767;25.455753;0.1925144;0.1860071;-14.6231243;3.8331417;8.8316804;0.1880412;0.184078;17.302053;6.2657858;40.7568577
\end{filecontents*}
\pgfplotstabletypeset[
        multicolumn names, 
        col sep=semicolon, 
        display columns/0/.style={column name=$\bm{k}$,dec sep align},
        display columns/1/.style={column name={\textbf{AIC}},column type={r},fixed,fixed zerofill,precision=0,set thousands separator={,}},
        display columns/2/.style={column name={\textbf{v.mae}},column type={r},fixed,fixed zerofill,precision=3,set thousands separator={,}},
        display columns/3/.style={column name={$\bm{\textbf{\text{v.mae}}^a}$},column type={r},fixed,fixed zerofill,precision=3,set thousands separator={,}},
        display columns/4/.style={column name={\textbf{v.res}},column type={r},fixed,fixed zerofill,precision=0,set thousands separator={,}},
        display columns/5/.style={column name={$\bm{\textbf{\text{v.mae}}^0}$},column type={r},fixed,fixed zerofill,precision=3,set thousands separator={,}},
        display columns/6/.style={column name={$\bm{\textbf{\text{v.res}}^0}$},column type={r},fixed,fixed zerofill,precision=0,set thousands separator={,}},
        display columns/7/.style={column name={\textbf{ns.mae}},column type={r},fixed,fixed zerofill,precision=3,set thousands separator={,}},
        display columns/8/.style={column name={$\bm{\textbf{\text{ns.mae}}^a}$},column type={r},fixed,fixed zerofill,precision=3,set thousands separator={,}},
        display columns/9/.style={column name={\textbf{ns.res}},column type={r},fixed,fixed zerofill,precision=0,set thousands separator={,}},
        display columns/10/.style={column name={$\bm{\textbf{\text{ns.mae}}^0}$},column type={r},fixed,fixed zerofill,precision=3,set thousands separator={,}},
        display columns/11/.style={column name={$\bm{\textbf{\text{ns.res}}^0}$},column type={r},fixed,fixed zerofill,precision=0,set thousands separator={,}},
        display columns/12/.style={column name={\textbf{cr.mae}},column type={r},fixed,fixed zerofill,precision=3,set thousands separator={,}},
        display columns/13/.style={column name={$\bm{\textbf{\text{cr.mae}}^a}$},column type={r},fixed,fixed zerofill,precision=3,set thousands separator={,}},
        display columns/14/.style={column name={\textbf{cr.res}},column type={r},fixed,fixed zerofill,precision=0,set thousands separator={,}},
        display columns/15/.style={column name={$\bm{\textbf{\text{cr.mae}}^0}$},column type={r},fixed,fixed zerofill,precision=3,set thousands separator={,}},
        display columns/16/.style={column name={$\bm{\textbf{\text{cr.res}}^0}$},column type={r},fixed,fixed zerofill,precision=0,set thousands separator={,}},
	    	every head row/.style={before row=\toprule,after row=\midrule},
	    	every row no 0/.style={before row={\multicolumn{18}{l}{\textbf{Gaussian with identity link under 150-443}}\\ \\}},
	    	every row no 1/.style={before row={\midrule \multicolumn{18}{l}{\textbf{Gaussian with inverse link under 150-443}}\\ \\}},
	    	every row no 2/.style={before row={\midrule \multicolumn{18}{l}{\textbf{Gaussian with log link under 150-443}}\\ \\}},
	    	every row no 3/.style={before row={\midrule \multicolumn{18}{l}{\textbf{Gamma with identity link under 150-443}}\\ \\}},
	    	every row no 4/.style={before row={\midrule \multicolumn{18}{l}{\textbf{Gamma with inverse link under 150-443}}\\ \\}},
	    	every row no 5/.style={before row={\midrule \multicolumn{18}{l}{\textbf{Gamma with log link under 150-443}}\\ \\}},
	    	every row no 6/.style={before row={\midrule \multicolumn{18}{l}{\textbf{Inverse gaussian with identity link under 150-443}}\\ \\}},
	    	every row no 7/.style={before row={\midrule \multicolumn{18}{l}{\textbf{Inverse gaussian with inverse link under 150-443}}\\ \\}},
	    	every row no 8/.style={before row={\midrule \multicolumn{18}{l}{\textbf{Inverse gaussian with log link under 150-443}}\\ \\}},
	    	every row no 9/.style={before row={\midrule \multicolumn{18}{l}{\textbf{Inverse gaussian with }$\bm{\frac{1}{\mu^2}}$\textbf{ link under 150-443}}\\ \\}},
	    	every row no 10/.style={before row={\midrule \multicolumn{18}{l}{\textbf{Gaussian with identity link under 300-886}}\\ \\}},
	    	every row no 11/.style={before row={\midrule \multicolumn{18}{l}{\textbf{Gaussian with inverse link under 300-886}}\\ \\}},
	    	every row no 12/.style={before row={\midrule \multicolumn{18}{l}{\textbf{Gaussian with log link under 300-886}}\\ \\}},
	    	every row no 13/.style={before row={\midrule \multicolumn{18}{l}{\textbf{Gamma with identity link under 300-886}}\\ \\}},
	    	every row no 14/.style={before row={\midrule \multicolumn{18}{l}{\textbf{Gamma with inverse link under 300-886}}\\ \\}},
	    	every row no 15/.style={before row={\midrule \multicolumn{18}{l}{\textbf{Gamma with log link under 300-886}}\\ \\}},
	    	every row no 16/.style={before row={\midrule \multicolumn{18}{l}{\textbf{Inverse gaussian with identity link under 300-886}}\\ \\}},
	    	every row no 17/.style={before row={\midrule \multicolumn{18}{l}{\textbf{Inverse gaussian with inverse link under 300-886}}\\ \\}},
	    	every row no 18/.style={before row={\midrule \multicolumn{18}{l}{\textbf{Inverse gaussian with log link under 300-886}}\\ \\}},
	    	every row no 19/.style={before row={\midrule \multicolumn{18}{l}{\textbf{Inverse gaussian with }$\bm{\frac{1}{\mu^2}}$\textbf{ link under 300-886}}\\ \\}},
      	every last row/.style={after row=\bottomrule},
      	every row 10 column 1/.style={postproc cell content/.append style={/pgfplots/table/@cell content/.add={\cellcolor{green!26!white}}{},}},
      	every row 19 column 2/.style={postproc cell content/.append style={/pgfplots/table/@cell content/.add={\cellcolor{green!26!white}}{},}},
      	every row 19 column 3/.style={postproc cell content/.append style={/pgfplots/table/@cell content/.add={\cellcolor{green!26!white}}{},}},
      	every row 15 column 4/.style={postproc cell content/.append style={/pgfplots/table/@cell content/.add={\cellcolor{green!26!white}}{},}},
      	every row 19 column 5/.style={postproc cell content/.append style={/pgfplots/table/@cell content/.add={\cellcolor{green!26!white}}{},}},
      	every row 19 column 6/.style={postproc cell content/.append style={/pgfplots/table/@cell content/.add={\cellcolor{green!26!white}}{},}},
      	every row 19 column 7/.style={postproc cell content/.append style={/pgfplots/table/@cell content/.add={\cellcolor{green!26!white}}{},}},
      	every row 19 column 8/.style={postproc cell content/.append style={/pgfplots/table/@cell content/.add={\cellcolor{green!26!white}}{},}},
      	every row 19 column 9/.style={postproc cell content/.append style={/pgfplots/table/@cell content/.add={\cellcolor{green!26!white}}{},}},
      	every row 19 column 10/.style={postproc cell content/.append style={/pgfplots/table/@cell content/.add={\cellcolor{green!26!white}}{},}},
      	every row 15 column 11/.style={postproc cell content/.append style={/pgfplots/table/@cell content/.add={\cellcolor{green!26!white}}{},}},
      	every row 3 column 12/.style={postproc cell content/.append style={/pgfplots/table/@cell content/.add={\cellcolor{green!26!white}}{},}},
      	every row 3 column 13/.style={postproc cell content/.append style={/pgfplots/table/@cell content/.add={\cellcolor{green!26!white}}{},}},
      	every row 0 column 14/.style={postproc cell content/.append style={/pgfplots/table/@cell content/.add={\cellcolor{green!26!white}}{},}},
      	every row 13 column 15/.style={postproc cell content/.append style={/pgfplots/table/@cell content/.add={\cellcolor{green!26!white}}{},}},
      	every row 13 column 16/.style={postproc cell content/.append style={/pgfplots/table/@cell content/.add={\cellcolor{green!26!white}}{},}},
      	every row 9 column 1/.style={postproc cell content/.append style={/pgfplots/table/@cell content/.add={\cellcolor{red!15!white}}{},}},
      	every row 9 column 2/.style={postproc cell content/.append style={/pgfplots/table/@cell content/.add={\cellcolor{red!15!white}}{},}},
      	every row 9 column 3/.style={postproc cell content/.append style={/pgfplots/table/@cell content/.add={\cellcolor{red!15!white}}{},}},
      	every row 6 column 4/.style={postproc cell content/.append style={/pgfplots/table/@cell content/.add={\cellcolor{red!15!white}}{},}},
      	every row 0 column 5/.style={postproc cell content/.append style={/pgfplots/table/@cell content/.add={\cellcolor{red!15!white}}{},}},
      	every row 0 column 6/.style={postproc cell content/.append style={/pgfplots/table/@cell content/.add={\cellcolor{red!15!white}}{},}},
      	every row 1 column 7/.style={postproc cell content/.append style={/pgfplots/table/@cell content/.add={\cellcolor{red!15!white}}{},}},
      	every row 1 column 8/.style={postproc cell content/.append style={/pgfplots/table/@cell content/.add={\cellcolor{red!15!white}}{},}},
      	every row 1 column 9/.style={postproc cell content/.append style={/pgfplots/table/@cell content/.add={\cellcolor{red!15!white}}{},}},
      	every row 0 column 10/.style={postproc cell content/.append style={/pgfplots/table/@cell content/.add={\cellcolor{red!15!white}}{},}},
      	every row 0 column 11/.style={postproc cell content/.append style={/pgfplots/table/@cell content/.add={\cellcolor{red!15!white}}{},}},
      	every row 1 column 12/.style={postproc cell content/.append style={/pgfplots/table/@cell content/.add={\cellcolor{red!15!white}}{},}},
      	every row 1 column 13/.style={postproc cell content/.append style={/pgfplots/table/@cell content/.add={\cellcolor{red!15!white}}{},}},
      	every row 19 column 14/.style={postproc cell content/.append style={/pgfplots/table/@cell content/.add={\cellcolor{red!15!white}}{},}},
      	every row 0 column 15/.style={postproc cell content/.append style={/pgfplots/table/@cell content/.add={\cellcolor{red!15!white}}{},}},
      	every row 0 column 16/.style={postproc cell content/.append style={/pgfplots/table/@cell content/.add={\cellcolor{red!15!white}}{},}},
      ]{BGROGLMgaussiangammainvgauss.csv} 
    \caption{AIC scores and out-of-sample validation figures of the gaussian, gamma and inverse gaussian GLMs of BEL with identity, inverse, log and $\frac{1}{\mu^2}$ link functions under $150$-$443$ and $300$-$886$ after the final iteration. Highlighted in green and red respectively the best and worst AIC scores and validation figures.}
    \label{tab:BGROGLMgaussiangammainvgauss}
  \end{center}
\end{table}

\begin{table}[htb]
  \tiny
  \begin{center}
  \tabcolsep=0.052cm
  \renewcommand{\arraystretch}{0.90}
      \begin{filecontents*}{BGROGAMsplinedim.csv}
k;Kmax;v.wre.y;v.a.wre.y;v.res.avg.y;v.wre.delta.y;v.res.avg.delta.y;ns.wre.y;ns.a.wre.y;ns.res.avg.y;ns.wre.delta.y;ns.res.avg.delta.y;cr.wre.y;cr.a.wre.y;cr.res.avg.y;cr.wre.delta.y;cr.res.avg.delta.y
0;150;4.55660203;4.35650671;-237.5325499;100;38.252875;3.23053;3.121333;-0.294727;100;261.099235;4.026996;3.942121;105.619188;100;367.01315
10;150;0.63169691;0.60395703;28.28162141;22.01939211;115.9154439;0.345482;0.333804;-7.976598;13.246935;65.265762;0.479183;0.469083;65.931607;21.072102;139.173967
20;150;0.40603596;0.38820559;-0.00934966;11.33014865;44.28758971;0.375014;0.362338;-42.022458;7.254174;-12.116982;0.340834;0.333651;-6.26066;7.708887;23.644817
30;150;0.39946731;0.38192539;-10.59725392;12.2675228;59.13348244;0.46477;0.44906;-61.294279;5.744495;-5.955006;0.314069;0.30745;-26.063124;6.116269;29.276149
40;150;0.37139202;0.35508298;-8.31885279;11.41452271;53.34522743;0.479541;0.463332;-63.629762;6.379583;-16.357145;0.339561;0.332404;-33.806;5.282513;13.466617
50;150;0.39170582;0.37450474;-12.95082144;12.07872509;59.46619859;0.520136;0.502555;-69.878592;5.960567;-11.853035;0.365425;0.357723;-38.645217;5.367919;19.38034
60;150;0.30570547;0.29228094;-15.23565922;9.83348745;48.28336149;0.40473;0.39105;-51.440253;5.282647;-2.312695;0.272514;0.26677;-10.384153;6.484493;38.743405
70;150;0.27172673;0.25979432;-14.97706662;9.8956806;56.45822398;0.32056;0.309725;-35.20444;5.227053;21.839388;0.232452;0.227553;11.968091;10.45977;69.011919
80;150;0.24915737;0.23821605;-16.87323002;8.62665534;48.92089608;0.307682;0.297282;-35.507133;4.587725;15.895531;0.204934;0.200615;8.60929;9.099637;60.011953
90;150;0.26114977;0.24968183;-17.2420157;9.26184492;53.68348011;0.324662;0.313688;-38.86244;4.639398;17.671593;0.194805;0.190699;5.122119;9.339666;61.656152
100;150;0.25374317;0.24260048;-17.99409312;9.59277929;55.2797205;0.339712;0.328229;-41.54479;4.626252;17.337561;0.195762;0.191636;2.62069;9.312075;61.503041
110;150;0.25517378;0.24396826;-17.94116266;9.40712844;54.09887963;0.335851;0.324499;-39.747721;4.640196;17.900859;0.207024;0.202661;3.936792;9.32454;61.585372
120;150;0.24320856;0.23252848;-16.18038013;8.47398622;48.43189066;0.30685;0.296478;-37.533903;4.023211;12.686905;0.18621;0.182286;1.164855;7.819262;51.385663
130;150;0.24104732;0.23046214;-16.29424909;8.48093737;48.64910178;0.308062;0.297649;-37.317074;4.108099;13.234814;0.183343;0.179479;2.49908;8.075073;53.050968
140;150;0.2350851;0.22476174;-14.67498002;8.01815008;45.25298317;0.294538;0.284582;-35.404853;3.865487;10.131647;0.172696;0.169056;1.572209;7.181943;47.108709
150;150;0.23965953;0.22913529;-15.0377391;8.19158043;46.36306061;0.291186;0.281344;-34.50222;3.906892;12.507117;0.175562;0.171862;3.167473;7.640849;50.17681
0;100;4.556602;4.356507;-237.53255;100;38.252875;3.23053;3.121333;-0.294727;100;261.099235;4.0269963;3.9421215;105.619188;100;367.0131501
10;100;0.643283;0.615034;26.85882;23.277871;125.477058;0.343932;0.332306;-6.152116;15.237951;78.074659;0.4933311;0.4829335;68.6803177;23.1514122;152.9070932
20;100;0.387158;0.370156;0.507801;10.370666;35.186043;0.363814;0.351516;-40.423482;7.855084;-20.136703;0.3350484;0.3279868;-6.3704665;7.4539943;13.9163125
30;100;0.382454;0.365659;-9.947359;11.235024;50.473768;0.454228;0.438875;-59.644166;6.247011;-13.614502;0.3166488;0.309975;-28.4591164;5.6033817;17.5705481
40;100;0.368227;0.352057;-11.404389;10.931344;48.112175;0.462588;0.446952;-61.288508;6.265956;-16.163407;0.336524;0.3294313;-33.3109829;5.3434127;11.8141182
50;100;0.3546;0.339029;-11.333514;10.085561;39.535121;0.481168;0.464904;-64.13662;7.752063;-27.659449;0.3513998;0.3439936;-36.5181892;5.4809742;-0.0410176
60;100;0.344061;0.328952;-9.411007;10.015465;40.233908;0.49009;0.473524;-65.585352;8.152069;-30.331901;0.3639324;0.356262;-38.4861721;5.5933222;-3.2327205
70;100;0.339031;0.324143;-6.158755;10.0345;44.895641;0.476133;0.460039;-63.547346;7.577811;-26.884413;0.3445337;0.3372721;-36.9276225;5.0777335;-0.2646889
80;100;0.294997;0.282043;-11.064923;9.396941;48.510898;0.403537;0.389897;-51.430154;5.513296;-6.245795;0.2410093;0.2359297;-11.3648985;5.8202879;33.81946
90;100;0.296201;0.283194;-11.541786;9.69384;51.613315;0.392927;0.379646;-49.040112;5.154597;-0.276474;0.2063001;0.2019521;-7.3004243;6.6053161;41.4632139
100;100;0.28673;0.274139;-11.384574;9.430586;48.196286;0.396606;0.3832;-50.011256;5.40182;-4.821859;0.2019994;0.1977419;-9.0186537;5.9453404;36.1707432
0;150;4.556602;4.356507;-237.53255;100;38.252875;3.23053;3.121333;-0.294727;100;261.099235;4.026996;3.942121;105.619188;100;367.01315
10;150;0.638718;0.61067;27.046944;23.176015;124.881537;0.340015;0.328522;-3.437927;15.516636;80.005203;0.515968;0.505093;72.603594;23.626779;156.046724
20;150;0.375184;0.358709;3.079973;9.604019;26.094516;0.333719;0.322439;-32.928845;8.378276;-24.305765;0.340547;0.333369;1.1248;7.710613;9.74788
30;150;0.360578;0.344744;-6.754388;10.443977;40.840073;0.415431;0.401389;-52.088471;6.961107;-18.885473;0.303517;0.297119;-20.545817;5.870684;12.657182
40;150;0.356029;0.340394;-4.522897;10.098446;36.079496;0.424607;0.410255;-53.842117;7.920046;-27.631187;0.310992;0.304437;-27.003128;5.647187;-0.792198
50;150;0.338807;0.323929;-7.070305;9.711837;33.167706;0.418479;0.404334;-53.184495;7.745799;-27.337946;0.310929;0.304376;-25.738179;5.595924;0.10837
60;150;0.325472;0.311179;-6.1893;9.037418;26.118923;0.410788;0.396903;-52.068067;8.705569;-34.151307;0.310496;0.303952;-26.354386;5.850382;-8.437625
70;150;0.325199;0.310919;-4.041592;9.180181;30.995582;0.428743;0.414251;-54.74054;8.772847;-34.094829;0.325512;0.318651;-29.609565;5.911979;-8.963854
80;150;0.309103;0.295529;-5.252441;8.617743;28.781596;0.429693;0.415169;-54.583578;8.984159;-34.941004;0.336462;0.329371;-28.828992;6.382003;-9.186417
90;150;0.313244;0.299489;-4.502999;8.980724;31.872439;0.384423;0.371429;-48.468219;7.38981;-26.484244;0.29957;0.293256;-25.828669;5.429933;-3.844694
100;150;0.327722;0.313331;-6.419957;9.909584;46.663497;0.400094;0.38657;-50.955077;5.572283;-12.263085;0.290745;0.284617;-25.346592;5.064045;13.3454
110;150;0.255787;0.244555;-10.017321;7.984835;38.355113;0.32587;0.314855;-40.02053;4.654884;-6.039559;0.201344;0.197101;-5.569938;5.001981;28.411033
120;150;0.253125;0.24201;-9.394257;7.340128;29.815614;0.321221;0.310363;-39.091102;5.541779;-14.272694;0.208845;0.204443;-4.814277;4.541449;20.004131
130;150;0.252327;0.241246;-9.035242;7.766767;34.314464;0.326343;0.315312;-39.850737;5.196826;-10.892493;0.205035;0.200713;-4.620258;4.770377;24.337985
140;150;0.245224;0.234455;-8.267161;7.59181;32.683212;0.322074;0.311188;-41.19016;5.314639;-14.63125;0.196977;0.192825;-6.929586;4.316957;19.629324
150;150;0.217359;0.207814;-10.508446;6.477383;31.878887;0.239187;0.231102;-26.32208;3.652311;1.673791;0.178717;0.17495;6.298133;5.577649;34.294004
0;150;4.556602;4.356507;-237.53255;100;38.252875;3.23053;3.121333;-0.294727;100;261.099235;4.026996;3.942121;105.619188;100;367.01315
10;150;0.642367;0.614158;26.903347;23.353768;126.019546;0.343789;0.332169;-5.035036;15.462924;79.6897;0.509435;0.498698;71.499766;23.653695;156.224501
20;150;0.381679;0.364919;2.421519;10.101456;33.470898;0.34086;0.329339;-34.254563;7.77996;-17.596647;0.338115;0.330988;0.902823;7.727795;17.560739
30;150;0.370068;0.353817;-6.929981;10.92232;44.984722;0.416096;0.402031;-51.892609;6.496826;-14.369369;0.305019;0.29859;-19.73571;6.102609;17.78753
40;150;0.353617;0.338089;-6.882668;10.41224;38.581816;0.404209;0.390546;-50.6943;6.747216;-19.621278;0.307903;0.301413;-23.552727;5.599609;7.520295
50;150;0.346606;0.331386;-7.272392;10.119125;37.854485;0.426124;0.41172;-54.277478;7.258075;-23.542064;0.31009;0.303555;-26.59916;5.466679;4.136254
60;150;0.34201;0.326991;-4.09039;9.765715;34.054145;0.400445;0.38691;-49.751713;7.599844;-25.998641;0.298454;0.292164;-23.406885;5.614549;0.346187
70;150;0.333567;0.318919;-3.775789;9.600824;35.2937;0.428165;0.413693;-54.512778;8.15813;-29.834752;0.318061;0.311358;-29.431616;5.618172;-4.75359
80;150;0.315166;0.301326;-5.413924;9.092633;35.119904;0.432407;0.417791;-54.904459;8.112938;-28.762094;0.333836;0.3268;-28.928034;6.08679;-2.785669
90;150;0.323278;0.309082;-4.730688;9.435905;38.214804;0.388043;0.374927;-49.035221;6.55849;-20.481192;0.297324;0.291057;-26.268757;5.193904;2.285272
100;150;0.309223;0.295644;-5.542984;8.721503;26.65829;0.409056;0.395229;-54.102673;8.780207;-36.292862;0.260849;0.255351;-26.934863;4.993763;-9.125052
110;150;0.308701;0.295145;-6.28982;8.541742;25.770294;0.410965;0.397074;-54.281042;8.710562;-36.612391;0.283789;0.277808;-32.925015;4.768163;-15.256364
120;150;0.206038;0.19699;-8.841716;5.768462;25.180779;0.216464;0.209147;-23.241939;3.805759;-3.610907;0.164465;0.160999;4.79804;4.518531;24.429071
130;150;0.20478;0.195787;-9.743451;5.759172;24.360569;0.225687;0.218059;-24.369712;3.951593;-4.657156;0.175455;0.171757;4.025383;4.579448;23.737939
140;150;0.214247;0.204839;-10.404348;6.761438;34.332915;0.227903;0.2202;-24.961804;3.362739;5.383996;0.166733;0.163219;6.154112;5.761657;36.499911
150;150;0.211915;0.202609;-9.793723;7.069811;36.763217;0.230458;0.222668;-24.274291;3.575314;7.891186;0.173223;0.169572;8.25587;6.337026;40.421347
\end{filecontents*}
\pgfplotstabletypeset[
        multicolumn names, 
        col sep=semicolon, 
        display columns/0/.style={column name=$\bm{k}$,dec sep align},
        display columns/1/.style={column name={$\bm{K_\textbf{\text{max}}}$},column type={r},fixed,fixed zerofill,precision=0,set thousands separator={,}},
        display columns/2/.style={column name={\textbf{v.mae}},column type={r},fixed,fixed zerofill,precision=3,set thousands separator={,}},
        display columns/3/.style={column name={$\bm{\textbf{\text{v.mae}}^a}$},column type={r},fixed,fixed zerofill,precision=3,set thousands separator={,}},
        display columns/4/.style={column name={\textbf{v.res}},column type={r},fixed,fixed zerofill,precision=0,set thousands separator={,}},
        display columns/5/.style={column name={$\bm{\textbf{\text{v.mae}}^0}$},column type={r},fixed,fixed zerofill,precision=3,set thousands separator={,}},
        display columns/6/.style={column name={$\bm{\textbf{\text{v.res}}^0}$},column type={r},fixed,fixed zerofill,precision=0,set thousands separator={,}},
        display columns/7/.style={column name={\textbf{ns.mae}},column type={r},fixed,fixed zerofill,precision=3,set thousands separator={,}},
        display columns/8/.style={column name={$\bm{\textbf{\text{ns.mae}}^a}$},column type={r},fixed,fixed zerofill,precision=3,set thousands separator={,}},
        display columns/9/.style={column name={\textbf{ns.res}},column type={r},fixed,fixed zerofill,precision=0,set thousands separator={,}},
        display columns/10/.style={column name={$\bm{\textbf{\text{ns.mae}}^0}$},column type={r},fixed,fixed zerofill,precision=3,set thousands separator={,}},
        display columns/11/.style={column name={$\bm{\textbf{\text{ns.res}}^0}$},column type={r},fixed,fixed zerofill,precision=0,set thousands separator={,}},
        display columns/12/.style={column name={\textbf{cr.mae}},column type={r},fixed,fixed zerofill,precision=3,set thousands separator={,}},
        display columns/13/.style={column name={$\bm{\textbf{\text{cr.mae}}^a}$},column type={r},fixed,fixed zerofill,precision=3,set thousands separator={,}},
        display columns/14/.style={column name={\textbf{cr.res}},column type={r},fixed,fixed zerofill,precision=0,set thousands separator={,}},
        display columns/15/.style={column name={$\bm{\textbf{\text{cr.mae}}^0}$},column type={r},fixed,fixed zerofill,precision=3,set thousands separator={,}},
        display columns/16/.style={column name={$\bm{\textbf{\text{cr.res}}^0}$},column type={r},fixed,fixed zerofill,precision=0,set thousands separator={,}},
	    	every head row/.style={before row=\toprule,after row=\midrule},
	    	every row no 0/.style={before row={\multicolumn{18}{l}{\textbf{4 Thin plate regression splines under gaussian with identity link in stagewise selection of length $\bm{5}$}}\\ \\}},
	    	every row no 16/.style={before row={\midrule \multicolumn{18}{l}{\textbf{5 Thin plate regression splines under gaussian with identity link}}\\ \\}},
	    	every row no 27/.style={before row={\midrule \multicolumn{18}{l}{\textbf{8 Thin plate regression splines under gaussian with identity link}}\\ \\}},
	    	every row no 43/.style={before row={\midrule \multicolumn{18}{l}{\textbf{10 Thin plate regression splines under gaussian with identity link}}\\ \\}},
      	every last row/.style={after row=\bottomrule},
      ]{BGROGAMsplinedim.csv} 
    \caption{Out-of-sample validation figures of selected GAMs of BEL with varying spline function number per dimension and fixed spline function type under $150$-$443$ after each tenth and the finally selected smooth function.}
    \label{tab:BGROGAMsplinedim}
  \end{center}
\end{table}

\begin{table}[htb]
  \tiny
  \begin{center}
  \tabcolsep=0.100cm
  \renewcommand{\arraystretch}{0.80}
      \begin{filecontents*}{BGROGAMsplinedimdfk4k10part1.csv}
k,edf k4 50,p-value k4 50,* k4 50,edf k4 100,p-value k4 100,* k4 100,edf k4 150,p-value k4 150,* k4 150,edf k10 50,p-value k10 50,* k10 50,edf k10 100,p-value k10 100,*  k10 100,edf k10 150,p-value k10 150,* k10 150
1,2.858,2.00E-16,***,2.35,2.00E-16,***,1.948,2.00E-16,***,9,2.00E-16,***,8.941,2.00E-16,***,7.7242,2.00E-16,***
2,3,2.00E-16,***,2.104,2.00E-16,***,1,2.00E-16,***,7.857,2.00E-16,***,4.436,2.00E-16,***,1,2.00E-16,***
3,3,2.00E-16,***,2.901,2.00E-16,***,2.922,2.00E-16,***,5.6,2.00E-16,***,1,2.00E-16,***,1,2.00E-16,***
4,2.997,2.00E-16,***,2.962,2.00E-16,***,2.998,2.00E-16,***,7.073,2.00E-16,***,6.791,2.00E-16,***,7.2883,2.00E-16,***
5,2.729,2.00E-16,***,1,2.00E-16,***,1,2.00E-16,***,8.679,2.00E-16,***,8.87,2.00E-16,***,8.21,2.00E-16,***
6,3,2.00E-16,***,3,2.00E-16,***,1.043,2.00E-16,***,3.417,2.00E-16,***,1,2.00E-16,***,1,2.00E-16,***
7,3,2.00E-16,***,2.806,2.00E-16,***,2.841,2.00E-16,***,7.99,2.00E-16,***,8.608,2.00E-16,***,1,2.00E-16,***
8,3,2.00E-16,***,2.956,2.00E-16,***,2.961,2.00E-16,***,8.282,2.00E-16,***,8.292,2.00E-16,***,8.1221,2.00E-16,***
9,1,2.00E-16,***,1,2.00E-16,***,2.223,2.00E-16,***,7.71,2.00E-16,***,6.51,2.00E-16,***,6.5493,2.00E-16,***
10,2.991,2.00E-16,***,2.924,2.00E-16,***,3,2.00E-16,***,1,2.00E-16,***,1,2.00E-16,***,1,2.00E-16,***
11,2.587,2.00E-16,***,2.922,2.00E-16,***,2.889,2.00E-16,***,6.535,2.00E-16,***,7.014,2.00E-16,***,5.6717,2.00E-16,***
12,2.645,2.00E-16,***,1.874,2.00E-16,***,1,2.00E-16,***,7.235,2.00E-16,***,7.284,2.00E-16,***,8.3463,2.00E-16,***
13,2.244,2.00E-16,***,2.425,2.00E-16,***,1,2.00E-16,***,2.372,2.00E-16,***,2.531,2.00E-16,***,1,2.00E-16,***
14,1,2.00E-16,***,1,2.00E-16,***,1,2.00E-16,***,1,2.00E-16,***,1,2.00E-16,***,1,2.00E-16,***
15,3,2.00E-16,***,1,2.00E-16,***,2.285,2.00E-16,***,5.43,2.00E-16,***,5.64,2.00E-16,***,4.4374,2.00E-16,***
16,1,2.00E-16,***,1,2.00E-16,***,2.783,2.00E-16,***,1,2.00E-16,***,1,2.00E-16,***,1,2.00E-16,***
17,2.344,2.00E-16,***,1.67,2.00E-16,***,1.646,2.00E-16,***,3.886,2.00E-16,***,1.61,2.00E-16,***,1.624,2.00E-16,***
18,3,2.00E-16,***,3,2.00E-16,***,3,2.00E-16,***,8.751,2.00E-16,***,8.62,1.39E-05,***,5.3668,6.90E-05,***
19,1,2.00E-16,***,1,2.00E-16,***,1,2.00E-16,***,1,2.00E-16,***,1,2.00E-16,***,1,2.00E-16,***
20,1.497,2.00E-16,***,1.501,2.00E-16,***,2.148,2.00E-16,***,1.754,2.00E-16,***,1,2.00E-16,***,3.1407,8.05E-16,***
21,1.441,2.00E-16,***,1,2.00E-16,***,1,2.00E-16,***,1,2.00E-16,***,1,2.00E-16,***,1,2.00E-16,***
22,1.77,2.00E-16,***,2.192,2.00E-16,***,1.4,2.00E-16,***,1,2.00E-16,***,1,2.00E-16,***,3.985,1.92E-09,***
23,2.395,2.00E-16,***,2.746,2.00E-16,***,2.911,2.00E-16,***,2.057,2.00E-16,***,1.428,2.00E-16,***,2.6628,2.00E-16,***
24,1,2.00E-16,***,1,2.00E-16,***,1,2.00E-16,***,2.964,2.00E-16,***,1,3.32E-13,***,1,1.08E-13,***
25,1,2.00E-16,***,1,2.00E-16,***,1,2.00E-16,***,1,2.00E-16,***,1,2.00E-16,***,1,2.00E-16,***
26,1,2.00E-16,***,1.485,2.00E-16,***,1,2.00E-16,***,1,2.00E-16,***,1,2.00E-16,***,1,2.00E-16,***
27,1,2.00E-16,***,1,2.00E-16,***,1,2.16E-10,***,1,2.00E-16,***,1,2.00E-16,***,1,1.64E-10,***
28,1,2.00E-16,***,2.607,2.00E-16,***,1.839,2.00E-16,***,1,2.00E-16,***,2.78,2.00E-16,***,1.9136,2.00E-16,***
29,1,2.00E-16,***,1,2.00E-16,***,1.809,2.00E-16,***,1,2.00E-16,***,1,2.00E-16,***,1,2.00E-16,***
30,1,2.00E-16,***,1,2.00E-16,***,1,2.00E-16,***,6.74,2.00E-16,***,6.416,2.00E-16,***,6.5078,2.00E-16,***
31,1,2.00E-16,***,1,2.00E-16,***,1,2.36E-16,***,1,2.00E-16,***,1,2.00E-16,***,1,2.00E-16,***
32,1,2.00E-16,***,1,2.00E-16,***,1,2.00E-16,***,1,2.00E-16,***,1,2.00E-16,***,1,2.00E-16,***
33,1,2.00E-16,***,2.055,4.90E-15,***,1.893,2.15E-15,***,7.111,2.00E-16,***,7.175,6.27E-12,***,6.7277,2.00E-16,***
34,1,3.22E-16,***,1,2.90E-16,***,1,8.70E-11,***,1,2.00E-16,***,1.213,2.00E-16,***,1.635,4.92E-16,***
35,3,2.00E-16,***,1,2.00E-16,***,1,2.45E-16,***,4.78,2.00E-16,***,4.013,2.00E-16,***,4.2238,2.00E-16,***
36,1,2.00E-16,***,1,2.00E-16,***,1,2.00E-16,***,7.825,4.83E-16,***,7.867,1.10E-15,***,7.7378,0.002345,**
37,1,2.00E-16,***,1,2.00E-16,***,1,2.00E-16,***,1,4.56E-16,***,1,7.46E-16,***,1,2.00E-16,***
38,2.512,1.06E-14,***,2.303,2.00E-16,***,2.057,2.00E-16,***,1.233,2.00E-16,***,1,2.00E-16,***,1,0.000105,***
39,1,2.66E-12,***,1,1.16E-13,***,1,1.89E-13,***,1,1.09E-15,***,1,2.62E-16,***,1,1.16E-14,***
40,1.826,6.35E-11,***,1,2.00E-16,***,1.915,3.55E-15,***,1,1.17E-13,***,1.514,2.00E-16,***,1,2.00E-16,***
41,2.668,7.49E-16,***,2.701,5.30E-15,***,1.787,9.77E-07,***,1.823,8.11E-12,***,1.319,9.43E-15,***,1,2.00E-16,***
42,1,1.14E-15,***,1,2.00E-16,***,1,1.95E-15,***,1,2.92E-12,***,1,8.03E-12,***,5.2752,0.000375,***
43,1,3.78E-10,***,1,9.53E-10,***,1,1.97E-09,***,1,3.29E-10,***,1,7.65E-11,***,1,1.10E-10,***
44,1.713,1.29E-08,***,1.887,8.19E-09,***,1.892,6.18E-09,***,2.109,6.03E-08,***,1.779,5.30E-08,***,2.061,3.41E-08,***
45,1,5.69E-09,***,1,6.37E-09,***,1,1.87E-08,***,1,7.98E-09,***,1,2.14E-08,***,1,8.80E-09,***
46,1.917,3.47E-09,***,1,2.00E-16,***,1,1.31E-15,***,1.305,1.88E-06,***,1.61,1.12E-06,***,1,8.65E-08,***
47,1.451,1.16E-06,***,1.507,5.82E-07,***,1.234,1.04E-06,***,1,7.66E-13,***,1,5.48E-13,***,1,7.44E-12,***
48,2.753,3.23E-07,***,2.863,6.49E-08,***,2.804,2.14E-08,***,1,2.41E-08,***,1,7.82E-08,***,1,2.86E-06,***
49,1,5.45E-07,***,1,4.68E-14,***,1,1.57E-11,***,1,6.90E-07,***,1,9.63E-12,***,1,1.57E-12,***
50,1,9.16E-07,***,1.372,8.31E-11,***,1,1.05E-12,***,1,1.10E-06,***,1,1.95E-10,***,1,1.95E-11,***
51,,,,1.004,2.00E-16,***,1,2.00E-16,***,,,,1,1.06E-06,***,1,1.33E-06,***
52,,,,2.839,2.00E-16,***,1.334,2.00E-16,***,,,,1,4.27E-13,***,1,2.95E-13,***
53,,,,2.64,2.00E-16,***,2.421,2.00E-16,***,,,,1,4.69E-10,***,1,7.14E-11,***
54,,,,2.664,2.00E-16,***,1,2.00E-16,***,,,,3.237,2.75E-06,***,3.1676,4.94E-06,***
55,,,,1,9.23E-09,***,1,3.06E-06,***,,,,3.906,5.76E-08,***,3.4931,1.04E-09,***
56,,,,1,2.80E-09,***,2.376,2.34E-08,***,,,,1.098,3.49E-05,***,3.5132,2.00E-16,***
57,,,,1,3.25E-15,***,1,2.80E-13,***,,,,5.574,0.005055,**,5.0193,0.066613,.
58,,,,1,2.00E-16,***,1,2.00E-16,***,,,,1,7.32E-05,***,1,1.02E-05,***
59,,,,1,1.22E-11,***,1,2.04E-11,***,,,,1,1.75E-06,***,1,8.79E-08,***
60,,,,1,2.00E-16,***,1,2.00E-16,***,,,,3.717,5.16E-04,***,3.2864,0.00555,**
61,,,,1,7.49E-11,***,1,7.11E-11,***,,,,1,6.66E-05,***,1,1.54E-05,***
62,,,,2.613,4.24E-04,***,2.868,2.00E-16,***,,,,1,1.05E-05,***,1,4.61E-06,***
63,,,,1,7.92E-15,***,1.867,1.63E-14,***,,,,4.21,0.006637,**,3.543,0.000726,***
64,,,,1,2.38E-06,***,1,1.18E-06,***,,,,1,0.000173,***,1,0.000335,***
65,,,,2.96,2.25E-13,***,2.976,2.00E-16,***,,,,2.799,0.007089,**,2.8614,0.003004,**
66,,,,1.904,2.00E-16,***,2.115,2.00E-16,***,,,,3.054,1.74E-03,**,3.1587,8.78E-06,***
67,,,,2.859,9.09E-14,***,2.778,1.14E-13,***,,,,3.671,0.007586,**,3.7877,0.000842,***
68,,,,1,0.289593,,1,5.16E-11,***,,,,1,0.000398,***,1,0.000115,***
69,,,,2.797,2.84E-03,**,2.954,0.002169,**,,,,1,0.002819,**,1,0.003307,**
70,,,,1,2.44E-06,***,1,1.47E-06,***,,,,1,0.006651,**,1,0.001138,**
71,,,,2.957,6.04E-14,***,2.996,6.10E-15,***,,,,1,0.008608,**,1,0.004961,**
72,,,,2.612,1.36E-13,***,2.101,6.26E-11,***,,,,1,0.012069,*,1,0.008907,**
73,,,,1.196,2.00E-16,***,3,2.00E-16,***,,,,1,1.52E-02,*,1,6.07E-05,***
74,,,,2.994,3.82E-06,***,2.559,0.00175,**,,,,3.644,0.118344,,2.9877,0.137864,
75,,,,1,1.74E-14,***,1,3.00E-14,***,,,,1,0.016688,*,1,0.018361,*
\end{filecontents*}
\pgfplotstabletypeset[
        multicolumn names, 
        col sep=comma, 
        display columns/0/.style={column name={$\bm{k}$},dec sep align},
        display columns/1/.style={column name={\textbf{df}},column type={|r},fixed,fixed zerofill,precision=3,set thousands separator={,}},
        display columns/2/.style={column name={\textbf{p-val}},sci subscript,precision=1,dec sep align},
        display columns/3/.style={column name={\textbf{sign}},string type},
        display columns/4/.style={column name={\textbf{df}},column type={|r},fixed,fixed zerofill,precision=3,set thousands separator={,}},
        display columns/5/.style={column name={\textbf{p-val}},sci subscript,precision=1,dec sep align},
        display columns/6/.style={column name={\textbf{sign}},string type},
        display columns/7/.style={column name={\textbf{df}},column type={|r},fixed,fixed zerofill,precision=3,set thousands separator={,}},
        display columns/8/.style={column name={\textbf{p-val}},sci subscript,precision=1,dec sep align},
        display columns/9/.style={column name={\textbf{sign}},string type},
        display columns/10/.style={column name={\textbf{df}},column type={|r},fixed,fixed zerofill,precision=3,set thousands separator={,}},
        display columns/11/.style={column name={\textbf{p-val}},sci subscript,precision=1,dec sep align},
        display columns/12/.style={column name={\textbf{sign}},string type},
        display columns/13/.style={column name={\textbf{df}},column type={|r},fixed,fixed zerofill,precision=3,set thousands separator={,}},
        display columns/14/.style={column name={\textbf{p-val}},sci subscript,precision=1,dec sep align},
        display columns/15/.style={column name={\textbf{sign}},string type},
        display columns/16/.style={column name={\textbf{df}},column type={|r},fixed,fixed zerofill,precision=3,set thousands separator={,}},
        display columns/17/.style={column name={\textbf{p-val}},sci subscript,precision=1,dec sep align},
        display columns/18/.style={column name={\textbf{sign}},string type},
	    	every head row/.style={before row={\toprule
	    	& \multicolumn{5}{c}{$\bm{J=4}$, $\bm{k=50}$} & \multicolumn{4}{c}{$\bm{J=4}$, $\bm{k=100}$} & \multicolumn{4}{c}{$\bm{J=4}$, $\bm{k=150}$} & \multicolumn{4}{c}{$\bm{J=10}$, $\bm{k=50}$} & \multicolumn{4}{c}{$\bm{J=10}$, $\bm{k=100}$} & \multicolumn{4}{c}{$\bm{J=10}$, $\bm{k=150}$}\\},after row=\midrule},
	    	every nth row={10}{before row=\midrule},
      	every last row/.style={after row=\bottomrule},
      ]{BGROGAMsplinedimdfk4k10part1.csv} 
    \caption{Effective degrees of freedom, p-values and significance codes per dimension of GAMs of BEL built up of thin plate regression splines with gaussian random component and identity link function under $150$-$443$ for spline function numbers $J \in \left\{4, 10\right\}$ per dimension at stages $k \in \left\{50, 100, 150\right\}$. The confidence levels corresponding to the indicated significance codes are ‘***’ = 0.001, ‘**’ = 0.01, ‘*’ = 0.05, ‘.’ = 0.1, ‘ ’ = 1.}
    \label{tab:BGROGAMsplinedimdfk4k10part1}
  \end{center}
\end{table}

\begin{table}[htb]
  \ContinuedFloat
  \tiny
  \begin{center}
  \tabcolsep=0.106cm
  \renewcommand{\arraystretch}{0.90}
      \begin{filecontents*}{BGROGAMsplinedimdfk4k10part2.csv}
k,edf k4 50,p-value k4 50,* k4 50,edf k4 100,p-value k4 100,* k4 100,edf k4 150,p-value k4 150,* k4 150,edf k10 50,p-value k10 50,* k10 50,edf k10 100,p-value k10 100,*  k10 100,edf k10 150,p-value k10 150,* k10 150
76,,,,1,4.41E-13,***,2.334,3.84E-14,***,,,,2.469,0.102689,,2.0771,0.180605,
77,,,,1.353,4.03E-09,***,1.411,8.83E-09,***,,,,1,0.024821,*,1,0.011213,*
78,,,,1,1.47E-05,***,1,6.48E-06,***,,,,1,2.00E-16,***,1,0.000158,***
79,,,,1,2.99E-05,***,1,1.52E-05,***,,,,5.186,1.52E-06,***,1,2.00E-16,***
80,,,,1,1.03E-07,***,1,7.77E-08,***,,,,1.892,0.022366,*,1.7945,0.018934,*
81,,,,2.725,1.29E-04,***,2.739,7.07E-05,***,,,,1,5.24E-06,***,1,0.57627,
82,,,,1,7.62E-05,***,2.175,1.41E-05,***,,,,1,0.001752,**,1,0.511098,
83,,,,2.24,1.27E-03,**,2.075,0.000903,***,,,,7.02,2.00E-16,***,4.8089,0.002896,**
84,,,,1,6.82E-05,***,2.902,1.51E-05,***,,,,4.003,0.153832,,4.7221,0.009794,**
85,,,,1,7.48E-05,***,1,3.98E-06,***,,,,1,1.02E-09,***,1,0.00018,***
86,,,,1,0.000366,***,1,0.000771,***,,,,3.115,0.124774,,2.7476,0.115059,
87,,,,1,0.000341,***,1,9.10E-05,***,,,,5.294,0.138347,,5.5976,0.133323,
88,,,,1,0.00019,***,1,9.59E-05,***,,,,2.263,0.151162,,1.7882,0.254825,
89,,,,2.828,0.002069,**,3,6.04E-05,***,,,,1,0.000344,***,1,0.000334,***
90,,,,1,0.000779,***,1,0.000561,***,,,,1,0.037412,*,1,0.03761,*
91,,,,1,0.002514,**,1,0.002891,**,,,,1,0.001803,**,1,0.00123,**
92,,,,1,0.003751,**,1,0.003523,**,,,,1,0.017208,*,1,0.011693,*
93,,,,1,1.77E-03,**,1,0.00106,**,,,,1,0.037742,*,1,0.027832,*
94,,,,2.776,3.60E-05,***,1,1.79E-07,***,,,,5.921,0.004248,**,3.9624,2.00E-16,***
95,,,,2.103,0.048544,*,1.974,0.125369,,,,,8.154,2.00E-16,***,2.2898,2.00E-16,***
96,,,,2.023,0.000118,***,1,4.57E-10,***,,,,1,2.75E-12,***,1,1.58E-05,***
97,,,,2.811,0.014751,*,2.873,0.005851,**,,,,3.748,0.000713,***,1,1.20E-06,***
98,,,,1,0.007103,**,1,0.010669,*,,,,1,3.93E-06,***,7.3489,0.281552,
99,,,,1,0.014275,*,1,0.018899,*,,,,2.149,0.001244,**,1,2.82E-08,***
100,,,,2.764,0.028852,*,2.321,0.089987,.,,,,1,0.00312,**,1,0.207064,
101,,,,,,,1,0.000105,***,,,,,,,1,8.15E-10,***
102,,,,,,,1,0.076703,.,,,,,,,1,0.01599,*
103,,,,,,,1,0.002903,**,,,,,,,4.0844,0.000578,***
104,,,,,,,1,6.78E-05,***,,,,,,,1,0.031782,*
105,,,,,,,1,0.009292,**,,,,,,,1,0.068054,.
106,,,,,,,1,2.14E-09,***,,,,,,,1,0.005175,**
107,,,,,,,1,0.019439,*,,,,,,,3.397,0.103293,
108,,,,,,,2.187,0.095735,.,,,,,,,1.2478,0.343915,
109,,,,,,,1,0.002138,**,,,,,,,3.0786,0.38612,
110,,,,,,,1,0.046169,*,,,,,,,1,0.000385,***
111,,,,,,,1,2.00E-16,***,,,,,,,0.9794,4.25E-08,***
112,,,,,,,1,0.028833,*,,,,,,,8.5548,2.00E-16,***
113,,,,,,,1,0.954718,,,,,,,,8.9518,1.68E-12,***
114,,,,,,,1.644,0.096303,.,,,,,,,1,2.00E-16,***
115,,,,,,,1,0.019894,*,,,,,,,1,2.00E-16,***
116,,,,,,,1,0.017676,*,,,,,,,1,1.70E-13,***
117,,,,,,,1,0.004773,**,,,,,,,2.9876,3.36E-13,***
118,,,,,,,1,0.023942,*,,,,,,,8.4011,1.18E-10,***
119,,,,,,,2.704,0.08265,.,,,,,,,2.4931,4.72E-05,***
120,,,,,,,1,0.018333,*,,,,,,,1,4.10E-07,***
121,,,,,,,1.413,0.672923,,,,,,,,1,9.02E-05,***
122,,,,,,,1.886,0.61938,,,,,,,,2.7451,0.001205,**
123,,,,,,,1,1.36E-05,***,,,,,,,1,0.003404,**
124,,,,,,,2.499,0.177006,,,,,,,,1,0.015288,*
125,,,,,,,1,0.03607,*,,,,,,,1,0.014131,*
126,,,,,,,2.416,0.101792,,,,,,,,1,0.005795,**
127,,,,,,,1,5.08E-05,***,,,,,,,3.1198,0.057165,.
128,,,,,,,1,0.037597,*,,,,,,,1,0.000922,***
129,,,,,,,1,0.001295,**,,,,,,,1,0.003887,**
130,,,,,,,1,0.056859,.,,,,,,,3.7779,0.174255,
131,,,,,,,1,0.01343,*,,,,,,,2.7521,0.02727,*
132,,,,,,,1,0.011616,*,,,,,,,1,0.006946,**
133,,,,,,,1.97,0.252689,,,,,,,,1,0.004817,**
134,,,,,,,1,0.034867,*,,,,,,,1,0.054852,.
135,,,,,,,1,0.000594,***,,,,,,,1,0.038085,*
136,,,,,,,1.176,0.00707,**,,,,,,,5.289,0.135912,
137,,,,,,,2.357,0.336481,,,,,,,,1,0.036515,*
138,,,,,,,1,0.066552,.,,,,,,,1,0.000195,***
139,,,,,,,1,0.079096,.,,,,,,,1,0.005148,**
140,,,,,,,1,0.069227,.,,,,,,,1,0.163609,
141,,,,,,,1,0.047234,*,,,,,,,8.4525,0.002539,**
142,,,,,,,1,0.001252,**,,,,,,,1,0.039511,*
143,,,,,,,2.602,0.04052,*,,,,,,,3.9745,0.14114,
144,,,,,,,1.631,0.45806,,,,,,,,1,0.00042,***
145,,,,,,,1,0.082707,.,,,,,,,1,0.003744,**
146,,,,,,,1,0.010385,*,,,,,,,2.1468,0.192844,
147,,,,,,,1,0.035508,*,,,,,,,1,0.050466,.
148,,,,,,,1.251,0.162248,,,,,,,,1,0.041413,*
149,,,,,,,2.376,0.208535,,,,,,,,1,0.053564,.
150,,,,,,,1.482,0.204216,,,,,,,,1,0.063083,.
\end{filecontents*}
\pgfplotstabletypeset[
        multicolumn names, 
        col sep=comma, 
        display columns/0/.style={column name={$\bm{k}$},dec sep align},
        display columns/1/.style={column name={\textbf{df}},column type={|r},fixed,fixed zerofill,precision=3,set thousands separator={,}},
        display columns/2/.style={column name={\textbf{p-val}},sci subscript,precision=1,dec sep align},
        display columns/3/.style={column name={\textbf{sign}},string type},
        display columns/4/.style={column name={\textbf{df}},column type={|r},fixed,fixed zerofill,precision=3,set thousands separator={,}},
        display columns/5/.style={column name={\textbf{p-val}},sci subscript,precision=1,dec sep align},
        display columns/6/.style={column name={\textbf{sign}},string type},
        display columns/7/.style={column name={\textbf{df}},column type={|r},fixed,fixed zerofill,precision=3,set thousands separator={,}},
        display columns/8/.style={column name={\textbf{p-val}},sci subscript,precision=1,dec sep align},
        display columns/9/.style={column name={\textbf{sign}},string type},
        display columns/10/.style={column name={\textbf{df}},column type={|r},fixed,fixed zerofill,precision=3,set thousands separator={,}},
        display columns/11/.style={column name={\textbf{p-val}},sci subscript,precision=1,dec sep align},
        display columns/12/.style={column name={\textbf{sign}},string type},
        display columns/13/.style={column name={\textbf{df}},column type={|r},fixed,fixed zerofill,precision=3,set thousands separator={,}},
        display columns/14/.style={column name={\textbf{p-val}},sci subscript,precision=1,dec sep align},
        display columns/15/.style={column name={\textbf{sign}},string type},
        display columns/16/.style={column name={\textbf{df}},column type={|r},fixed,fixed zerofill,precision=3,set thousands separator={,}},
        display columns/17/.style={column name={\textbf{p-val}},sci subscript,precision=1,dec sep align},
        display columns/18/.style={column name={\textbf{sign}},string type},
	    	every head row/.style={before row={\toprule
	    	& \multicolumn{5}{c}{$\bm{J=4}$, $\bm{k=50}$} & \multicolumn{4}{c}{$\bm{J=4}$, $\bm{k=100}$} & \multicolumn{4}{c}{$\bm{J=4}$, $\bm{k=150}$} & \multicolumn{4}{c}{$\bm{J=10}$, $\bm{k=50}$} & \multicolumn{4}{c}{$\bm{J=10}$, $\bm{k=100}$} & \multicolumn{4}{c}{$\bm{J=10}$, $\bm{k=150}$}\\},after row=\midrule},
	    	every nth row={10[+5]}{before row=\midrule},
      	every last row/.style={after row=\bottomrule},
      ]{BGROGAMsplinedimdfk4k10part2.csv} 
    \caption{Cont.}
    \label{tab:BGROGAMsplinedimdfk4k10part2}
  \end{center}
\end{table}

\begin{table}[htb]
  \tiny
  \begin{center}
  \tabcolsep=0.052cm
  \renewcommand{\arraystretch}{0.90}
      \begin{filecontents*}{BGROGAMbfdim5.csv}
k;Kmax;v.wre.y;v.a.wre.y;v.res.avg.y;v.wre.delta.y;v.res.avg.delta.y;ns.wre.y;ns.a.wre.y;ns.res.avg.y;ns.wre.delta.y;ns.res.avg.delta.y;cr.wre.y;cr.a.wre.y;cr.res.avg.y;cr.wre.delta.y;cr.res.avg.delta.y
0;100;4.556602;4.356507;-237.53255;100;38.252875;3.23053;3.121333;-0.294727;100;261.099235;4.0269963;3.9421215;105.619188;100;367.0131501
10;100;0.643283;0.615034;26.85882;23.277871;125.477058;0.343932;0.332306;-6.152116;15.237951;78.074659;0.4933311;0.4829335;68.6803177;23.1514122;152.9070932
20;100;0.387158;0.370156;0.507801;10.370666;35.186043;0.363814;0.351516;-40.423482;7.855084;-20.136703;0.3350484;0.3279868;-6.3704665;7.4539943;13.9163125
30;100;0.382454;0.365659;-9.947359;11.235024;50.473768;0.454228;0.438875;-59.644166;6.247011;-13.614502;0.3166488;0.309975;-28.4591164;5.6033817;17.5705481
40;100;0.368227;0.352057;-11.404389;10.931344;48.112175;0.462588;0.446952;-61.288508;6.265956;-16.163407;0.336524;0.3294313;-33.3109829;5.3434127;11.8141182
50;100;0.3546;0.339029;-11.333514;10.085561;39.535121;0.481168;0.464904;-64.13662;7.752063;-27.659449;0.3513998;0.3439936;-36.5181892;5.4809742;-0.0410176
60;100;0.344061;0.328952;-9.411007;10.015465;40.233908;0.49009;0.473524;-65.585352;8.152069;-30.331901;0.3639324;0.356262;-38.4861721;5.5933222;-3.2327205
70;100;0.339031;0.324143;-6.158755;10.0345;44.895641;0.476133;0.460039;-63.547346;7.577811;-26.884413;0.3445337;0.3372721;-36.9276225;5.0777335;-0.2646889
80;100;0.294997;0.282043;-11.064923;9.396941;48.510898;0.403537;0.389897;-51.430154;5.513296;-6.245795;0.2410093;0.2359297;-11.3648985;5.8202879;33.81946
90;100;0.296201;0.283194;-11.541786;9.69384;51.613315;0.392927;0.379646;-49.040112;5.154597;-0.276474;0.2063001;0.2019521;-7.3004243;6.6053161;41.4632139
100;100;0.28673;0.274139;-11.384574;9.430586;48.196286;0.396606;0.3832;-50.011256;5.40182;-4.821859;0.2019994;0.1977419;-9.0186537;5.9453404;36.1707432
0;100;4.556602;4.356507;-237.53255;100;38.252875;3.23053;3.121333;-0.294727;100;261.099235;4.026996;3.942121;105.619188;100;367.01315
10;100;0.637271;0.609286;27.79044;22.739492;121.534472;0.337056;0.325663;-4.089098;14.733033;75.263471;0.504583;0.493948;71.110321;22.781339;150.46289
20;100;0.387983;0.370946;2.426996;10.0941;31.59178;0.357942;0.345843;-39.557203;8.256441;-24.783881;0.319386;0.312654;-4.514877;7.161067;10.258444
30;100;0.389105;0.372019;-5.624536;11.425833;50.027422;0.436075;0.421335;-55.295823;6.65233;-14.035328;0.28894;0.282851;-19.065074;5.848792;22.195422
40;100;0.358852;0.343094;-8.948078;10.507601;40.915897;0.448283;0.433131;-58.845349;7.171112;-23.372837;0.309886;0.303355;-29.132862;5.17541;6.33965
50;100;0.345072;0.329918;-9.012745;9.906221;34.584549;0.476463;0.460358;-63.4398;8.736104;-34.233968;0.328113;0.321198;-34.167688;5.372762;-4.961856
60;100;0.338285;0.32343;-6.562099;9.817395;33.721179;0.474803;0.458754;-63.19898;9.191876;-37.307165;0.330479;0.323514;-34.147052;5.491241;-8.255237
70;100;0.307409;0.29391;-7.722337;9.341374;46.872947;0.430075;0.415537;-58.363189;6.081378;-18.159368;0.234173;0.229237;-25.544951;3.870825;14.65887
80;100;0.289333;0.276627;-13.0368;10.156932;54.936471;0.410067;0.396206;-53.108409;5.10628;0.473399;0.237216;0.232216;-10.757301;6.939183;42.824507
90;100;0.283176;0.270741;-12.786947;10.307037;55.611034;0.407384;0.393613;-52.906904;5.066606;1.099615;0.22903;0.224203;-10.418596;7.035101;43.587922
100;100;0.268186;0.256409;-11.563618;9.903409;51.69244;0.399161;0.385669;-50.788418;5.18194;-1.923822;0.226086;0.221321;-9.088135;6.533176;39.77646
0;100;4.556602;4.356507;-237.53255;100;38.252875;3.23053;3.121333;-0.294727;100;261.099235;4.026996;3.942121;105.619188;100;367.01315
10;100;0.752548;0.719502;-3.64126;20.569796;97.816035;0.427661;0.413206;-38.561632;11.805867;48.5042;0.407712;0.399119;6.085418;15.240602;93.15125
20;100;0.703713;0.672811;-22.188223;17.488216;73.877589;0.441324;0.426406;-50.935069;8.605845;30.73928;0.380202;0.372189;-16.154269;11.599705;65.520081
30;100;0.660921;0.631898;-32.499407;19.698729;95.255754;0.375959;0.363251;-39.984326;14.235262;73.379372;0.318609;0.311893;10.945552;19.16767;124.309251
40;100;0.663346;0.634217;-21.098073;18.426238;83.727638;0.291755;0.281893;-17.711731;14.138105;72.722516;0.377467;0.369512;33.057363;19.007156;123.49161
50;100;0.665656;0.636425;-17.403627;18.533987;85.630645;0.286511;0.276827;-12.463294;14.784575;76.179516;0.410373;0.401724;40.940165;19.896069;129.582975
56;100;0.665511;0.636286;-17.517526;18.531937;85.875639;0.288366;0.278618;-13.59088;14.642597;75.410822;0.40605;0.397492;39.540388;19.757039;128.542091
0;100;4.556602;4.356507;-237.53255;100;38.252875;3.23053;3.121333;-0.294727;100;261.099235;4.026996;3.942121;105.619188;100;367.01315
10;100;0.643132;0.61489;28.995381;22.835551;122.669442;0.343878;0.332255;-8.80537;13.950748;70.477229;0.470508;0.460592;65.056584;21.854159;144.339183
20;100;0.388816;0.371741;1.339965;10.495734;36.720597;0.365245;0.352899;-40.58435;7.778241;-19.59518;0.335707;0.328632;-8.256076;7.402197;12.733094
30;100;0.384068;0.367203;-9.247083;11.376575;52.731582;0.459086;0.443568;-60.172245;6.137879;-12.585042;0.319515;0.312781;-30.134558;5.511834;17.452645
40;100;0.370552;0.354279;-9.707288;10.977339;49.112507;0.454032;0.438685;-59.940417;6.094714;-15.512085;0.326954;0.320063;-33.590096;5.092497;10.838235
50;100;0.357073;0.341393;-9.056741;10.458551;45.070308;0.467245;0.451451;-61.569709;6.908833;-21.834123;0.334551;0.3275;-33.755335;5.05855;5.980251
60;100;0.339069;0.32418;-10.23922;9.932362;42.7815;0.492396;0.475752;-66.200839;7.639863;-27.571582;0.364709;0.357023;-40.3466;5.154847;-1.717343
70;100;0.343305;0.328229;-10.383052;10.522783;52.029067;0.545803;0.527354;-75.215914;7.681233;-27.195258;0.36562;0.357914;-45.737563;4.576357;2.283093
80;100;0.334159;0.319485;-6.856983;9.919764;44.761689;0.520224;0.502639;-66.559986;8.655174;-29.332777;0.346047;0.338754;-36.161786;5.036165;1.065422
90;100;0.227705;0.217706;-10.411818;6.973055;34.549063;0.278804;0.26938;-30.950998;4.299392;-0.38158;0.208318;0.203927;3.355189;5.809805;33.924607
100;100;0.225364;0.215468;-11.365175;6.89739;34.375712;0.256212;0.247552;-29.645537;3.715881;1.703887;0.164217;0.160756;0.710134;5.212007;32.059559
\end{filecontents*}
\pgfplotstabletypeset[
        multicolumn names, 
        col sep=semicolon, 
        display columns/0/.style={column name=$\bm{k}$,dec sep align},
        display columns/1/.style={column name={$\bm{K_\textbf{\text{max}}}$},column type={r},fixed,fixed zerofill,precision=0,set thousands separator={,}},
        display columns/2/.style={column name={\textbf{v.mae}},column type={r},fixed,fixed zerofill,precision=3,set thousands separator={,}},
        display columns/3/.style={column name={$\bm{\textbf{\text{v.mae}}^a}$},column type={r},fixed,fixed zerofill,precision=3,set thousands separator={,}},
        display columns/4/.style={column name={\textbf{v.res}},column type={r},fixed,fixed zerofill,precision=0,set thousands separator={,}},
        display columns/5/.style={column name={$\bm{\textbf{\text{v.mae}}^0}$},column type={r},fixed,fixed zerofill,precision=3,set thousands separator={,}},
        display columns/6/.style={column name={$\bm{\textbf{\text{v.res}}^0}$},column type={r},fixed,fixed zerofill,precision=0,set thousands separator={,}},
        display columns/7/.style={column name={\textbf{ns.mae}},column type={r},fixed,fixed zerofill,precision=3,set thousands separator={,}},
        display columns/8/.style={column name={$\bm{\textbf{\text{ns.mae}}^a}$},column type={r},fixed,fixed zerofill,precision=3,set thousands separator={,}},
        display columns/9/.style={column name={\textbf{ns.res}},column type={r},fixed,fixed zerofill,precision=0,set thousands separator={,}},
        display columns/10/.style={column name={$\bm{\textbf{\text{ns.mae}}^0}$},column type={r},fixed,fixed zerofill,precision=3,set thousands separator={,}},
        display columns/11/.style={column name={$\bm{\textbf{\text{ns.res}}^0}$},column type={r},fixed,fixed zerofill,precision=0,set thousands separator={,}},
        display columns/12/.style={column name={\textbf{cr.mae}},column type={r},fixed,fixed zerofill,precision=3,set thousands separator={,}},
        display columns/13/.style={column name={$\bm{\textbf{\text{cr.mae}}^a}$},column type={r},fixed,fixed zerofill,precision=3,set thousands separator={,}},
        display columns/14/.style={column name={\textbf{cr.res}},column type={r},fixed,fixed zerofill,precision=0,set thousands separator={,}},
        display columns/15/.style={column name={$\bm{\textbf{\text{cr.mae}}^0}$},column type={r},fixed,fixed zerofill,precision=3,set thousands separator={,}},
        display columns/16/.style={column name={$\bm{\textbf{\text{cr.res}}^0}$},column type={r},fixed,fixed zerofill,precision=0,set thousands separator={,}},
	    	every head row/.style={before row=\toprule,after row=\midrule},
	    	every row no 0/.style={before row={\multicolumn{18}{l}{\textbf{5 Thin plate regression splines under gaussian with identity link}}\\ \\}},
	    	every row no 11/.style={before row={\midrule \multicolumn{18}{l}{\textbf{5 Cubic regression splines under gaussian with identity link}}\\ \\}},
	    	every row no 22/.style={before row={\midrule \multicolumn{18}{l}{\textbf{5 Duchon splines under gaussian with identity link}}\\ \\}},
	    	every row no 29/.style={before row={\midrule \multicolumn{18}{l}{\textbf{5 Eilers and Marx style P-splines under gaussian with identity link}}\\ \\}},
      	every last row/.style={after row=\bottomrule},
      ]{BGROGAMbfdim5.csv} 
    \caption{Out-of-sample validation figures of selected GAMs of BEL with varying spline function type and fixed spline function number of $5$ per dimension under $100$-$443$ after each tenth and the finally selected smooth function.}
    \label{tab:BGROGAMbfdim5}
  \end{center}
\end{table}

\begin{table}[htb]
  \tiny
  \begin{center}
  \tabcolsep=0.052cm
  \renewcommand{\arraystretch}{0.90}
      \begin{filecontents*}{BGROGAMbfdim10.csv}
k;Kmax;v.wre.y;v.a.wre.y;v.res.avg.y;v.wre.delta.y;v.res.avg.delta.y;ns.wre.y;ns.a.wre.y;ns.res.avg.y;ns.wre.delta.y;ns.res.avg.delta.y;cr.wre.y;cr.a.wre.y;cr.res.avg.y;cr.wre.delta.y;cr.res.avg.delta.y
0;150;4.556602;4.356507;-237.53255;100;38.252875;3.23053;3.121333;-0.294727;100;261.099235;4.026996;3.942121;105.619188;100;367.01315
10;150;0.642367;0.614158;26.903347;23.353768;126.019546;0.343789;0.332169;-5.035036;15.462924;79.6897;0.509435;0.498698;71.499766;23.653695;156.224501
20;150;0.381679;0.364919;2.421519;10.101456;33.470898;0.34086;0.329339;-34.254563;7.77996;-17.596647;0.338115;0.330988;0.902823;7.727795;17.560739
30;150;0.370068;0.353817;-6.929981;10.92232;44.984722;0.416096;0.402031;-51.892609;6.496826;-14.369369;0.305019;0.29859;-19.73571;6.102609;17.78753
40;150;0.353617;0.338089;-6.882668;10.41224;38.581816;0.404209;0.390546;-50.6943;6.747216;-19.621278;0.307903;0.301413;-23.552727;5.599609;7.520295
50;150;0.346606;0.331386;-7.272392;10.119125;37.854485;0.426124;0.41172;-54.277478;7.258075;-23.542064;0.31009;0.303555;-26.59916;5.466679;4.136254
60;150;0.34201;0.326991;-4.09039;9.765715;34.054145;0.400445;0.38691;-49.751713;7.599844;-25.998641;0.298454;0.292164;-23.406885;5.614549;0.346187
70;150;0.333567;0.318919;-3.775789;9.600824;35.2937;0.428165;0.413693;-54.512778;8.15813;-29.834752;0.318061;0.311358;-29.431616;5.618172;-4.75359
80;150;0.315166;0.301326;-5.413924;9.092633;35.119904;0.432407;0.417791;-54.904459;8.112938;-28.762094;0.333836;0.3268;-28.928034;6.08679;-2.785669
90;150;0.323278;0.309082;-4.730688;9.435905;38.214804;0.388043;0.374927;-49.035221;6.55849;-20.481192;0.297324;0.291057;-26.268757;5.193904;2.285272
100;150;0.309223;0.295644;-5.542984;8.721503;26.65829;0.409056;0.395229;-54.102673;8.780207;-36.292862;0.260849;0.255351;-26.934863;4.993763;-9.125052
110;150;0.308701;0.295145;-6.28982;8.541742;25.770294;0.410965;0.397074;-54.281042;8.710562;-36.612391;0.283789;0.277808;-32.925015;4.768163;-15.256364
120;150;0.206038;0.19699;-8.841716;5.768462;25.180779;0.216464;0.209147;-23.241939;3.805759;-3.610907;0.164465;0.160999;4.79804;4.518531;24.429071
130;150;0.20478;0.195787;-9.743451;5.759172;24.360569;0.225687;0.218059;-24.369712;3.951593;-4.657156;0.175455;0.171757;4.025383;4.579448;23.737939
140;150;0.214247;0.204839;-10.404348;6.761438;34.332915;0.227903;0.2202;-24.961804;3.362739;5.383996;0.166733;0.163219;6.154112;5.761657;36.499911
150;150;0.211915;0.202609;-9.793723;7.069811;36.763217;0.230458;0.222668;-24.274291;3.575314;7.891186;0.173223;0.169572;8.25587;6.337026;40.421347
0;125;4.556602;4.356507;-237.53255;100;38.252875;3.2305298;3.1213333;-0.2947274;100;261.0992347;4.026996;3.942121;105.619188;100;367.01315
10;125;0.638337;0.610306;27.20966;23.396963;126.563286;0.340843;0.329322;-3.0965361;15.8286307;81.8656271;0.519484;0.508535;73.284157;23.959816;158.24632
20;125;0.380341;0.363639;1.610192;10.03806;33.935401;0.3392201;0.327754;-33.8289472;7.6496068;-15.8952007;0.345401;0.338121;0.349002;7.864656;18.282749
30;125;0.377023;0.360467;-5.996421;11.457871;53.249542;0.4112848;0.3973828;-49.6684008;6.0345853;-4.8139015;0.308756;0.302249;-14.44314;6.976057;30.411359
40;125;0.364194;0.348201;-10.151468;10.929417;47.445083;0.4212648;0.4070254;-53.3028783;5.791172;-10.0977903;0.314956;0.308318;-25.044653;5.824358;18.160435
50;125;0.348364;0.333067;-10.614521;10.437393;44.40972;0.4361173;0.4213759;-55.626489;6.2629247;-14.9937113;0.319051;0.312326;-27.208284;5.636057;13.424494
60;125;0.342418;0.327381;-4.653035;9.7914;36.396846;0.402725;0.3891123;-49.5185434;7.282407;-22.8601255;0.308246;0.301749;-22.652426;5.788777;4.005992
70;125;0.355301;0.339699;-2.84036;10.502074;47.503298;0.4420932;0.4271498;-56.1381101;7.0008846;-20.1859149;0.327094;0.3202;-29.75028;5.569611;6.201915
80;125;0.348937;0.333614;-1.888742;10.274957;45.640127;0.4341061;0.4194327;-55.0647983;7.1592025;-21.9273925;0.326258;0.319381;-29.480797;5.59164;3.656608
90;125;0.281609;0.269242;-4.506063;7.978239;36.97119;0.2749315;0.2656384;-30.1949148;4.4260283;-3.109124;0.214669;0.210145;-2.205998;5.088199;24.879793
100;125;0.262659;0.251125;-5.340183;7.109406;28.572556;0.30094;0.290768;-36.701485;5.637381;-17.180208;0.20007;0.195854;-7.945534;3.968777;11.575742
110;125;0.255494;0.244275;-6.52705;6.998959;29.577832;0.302546;0.29232;-36.565835;5.434764;-14.852416;0.202113;0.197853;-5.863093;4.22992;15.850327
120;125;0.257467;0.246161;-7.311734;7.052022;29.679076;0.30406;0.293782;-36.976327;5.371105;-14.37698;0.200487;0.196261;-6.018846;4.231678;16.580501
125;125;0.253706;0.242565;-6.633897;7.138864;30.739981;0.298756;0.288658;-36.376623;5.189094;-13.394209;0.196553;0.19241;-5.581915;4.227767;17.4005
0;100;4.556602;4.356507;-237.53255;100;38.252875;3.23053;3.121333;-0.294727;100;261.099235;4.026996;3.942121;105.619188;100;367.01315
10;100;0.786312;0.751783;-5.425687;22.142723;110.057018;0.445274;0.430223;-43.763055;12.588326;57.328186;0.405773;0.397221;0.645714;16.238038;101.736956
20;100;0.782922;0.748541;-32.386432;20.488643;101.397589;0.493932;0.477236;-61.711521;11.318554;57.681037;0.357155;0.349628;-21.181798;15.31557;98.21076
30;100;0.782105;0.747761;-39.36499;21.133609;98.388477;0.538028;0.519842;-58.867558;12.715242;64.494447;0.42233;0.413428;-2.722774;18.621289;120.63923
40;100;0.816073;0.780236;-44.816119;22.124777;98.140149;0.558919;0.540027;-63.416067;13.07122;65.148739;0.449909;0.440427;-9.504619;18.616145;119.060187
50;100;0.822885;0.786749;-45.110557;21.473217;95.538609;0.555095;0.536332;-63.4605;12.671808;62.797204;0.450974;0.441469;-10.353377;18.114261;115.904327
53;100;0.821486;0.785412;-44.001406;21.348029;94.195126;0.544573;0.526166;-61.416936;12.59311;62.388133;0.446303;0.436897;-8.073715;18.091325;115.731354
0;150;4.556602;4.356507;-237.53255;100;38.252875;3.23053;3.121333;-0.294727;100;261.099235;4.026996;3.942121;105.619188;100;367.01315
10;150;0.647685;0.619243;26.855012;23.688136;128.255022;0.349223;0.337419;-6.749366;15.565803;80.259181;0.50585;0.495189;70.771458;23.889212;157.780006
20;150;0.397972;0.380496;0.658465;10.946417;44.90155;0.358254;0.346144;-37.344168;7.063195;-7.492546;0.337697;0.33058;1.460373;8.101689;31.311996
30;150;0.393366;0.376092;-8.985849;11.982734;58.505909;0.435306;0.420592;-54.884493;5.575074;-1.784197;0.299271;0.292964;-16.617816;6.927692;36.48248
40;150;0.371404;0.355095;-7.562847;11.373781;55.304613;0.449463;0.434271;-57.322643;5.738476;-8.846646;0.31445;0.307822;-25.693621;5.769749;22.782377
50;150;0.363197;0.347248;-9.395417;10.955503;50.060718;0.459688;0.44415;-59.512013;6.248897;-14.44734;0.314606;0.307975;-27.818055;5.491816;17.246618
60;150;0.349363;0.334021;-7.82748;10.479421;46.207572;0.44258;0.42762;-56.376027;6.525833;-16.732437;0.304546;0.298127;-26.026349;5.426833;13.617241
70;150;0.34851;0.333206;-6.207092;10.629381;50.633;0.46427;0.448577;-59.512018;6.687342;-17.063389;0.325119;0.318267;-29.496838;5.501447;12.951791
80;150;0.350374;0.334988;-6.721094;10.465415;47.856672;0.467647;0.45184;-59.666627;7.036092;-19.480324;0.334994;0.327934;-29.356393;5.56277;10.829911
90;150;0.350264;0.334883;-7.22017;10.639336;50.728602;0.470013;0.454126;-60.17473;6.682598;-16.617421;0.33014;0.323182;-29.439029;5.452528;14.118281
100;150;0.333904;0.319241;-8.32217;9.960236;45.868632;0.467729;0.451919;-59.685193;7.170289;-19.885854;0.338723;0.331584;-28.502599;5.834585;11.296739
110;150;0.337323;0.32251;-8.97809;10.24938;48.491743;0.450291;0.43507;-58.145128;6.17101;-15.066758;0.329171;0.322233;-31.069604;5.266663;12.008767
120;150;0.339117;0.324226;-7.473826;10.283256;45.210609;0.433448;0.418796;-55.377449;6.419765;-17.084477;0.320146;0.313399;-28.26062;5.339616;10.032352
130;150;0.268541;0.256748;-12.639368;8.9118;43.218026;0.364808;0.352477;-45.663885;4.890716;-4.197955;0.243592;0.238458;-11.910768;5.502872;29.555163
140;150;0.255007;0.243809;-11.924913;8.157094;36.105611;0.355671;0.343649;-43.974159;5.415324;-10.335098;0.245813;0.240632;-9.670257;5.196415;23.968804
150;150;0.26131;0.249835;-12.454371;8.513934;38.95221;0.367874;0.35544;-46.36384;5.267498;-9.348722;0.245347;0.240176;-12.240557;5.16153;24.774561
\end{filecontents*}
\pgfplotstabletypeset[
        multicolumn names, 
        col sep=semicolon, 
        display columns/0/.style={column name=$\bm{k}$,dec sep align},
        display columns/1/.style={column name={$\bm{K_\textbf{\text{max}}}$},column type={r},fixed,fixed zerofill,precision=0,set thousands separator={,}},
        display columns/2/.style={column name={\textbf{v.mae}},column type={r},fixed,fixed zerofill,precision=3,set thousands separator={,}},
        display columns/3/.style={column name={$\bm{\textbf{\text{v.mae}}^a}$},column type={r},fixed,fixed zerofill,precision=3,set thousands separator={,}},
        display columns/4/.style={column name={\textbf{v.res}},column type={r},fixed,fixed zerofill,precision=0,set thousands separator={,}},
        display columns/5/.style={column name={$\bm{\textbf{\text{v.mae}}^0}$},column type={r},fixed,fixed zerofill,precision=3,set thousands separator={,}},
        display columns/6/.style={column name={$\bm{\textbf{\text{v.res}}^0}$},column type={r},fixed,fixed zerofill,precision=0,set thousands separator={,}},
        display columns/7/.style={column name={\textbf{ns.mae}},column type={r},fixed,fixed zerofill,precision=3,set thousands separator={,}},
        display columns/8/.style={column name={$\bm{\textbf{\text{ns.mae}}^a}$},column type={r},fixed,fixed zerofill,precision=3,set thousands separator={,}},
        display columns/9/.style={column name={\textbf{ns.res}},column type={r},fixed,fixed zerofill,precision=0,set thousands separator={,}},
        display columns/10/.style={column name={$\bm{\textbf{\text{ns.mae}}^0}$},column type={r},fixed,fixed zerofill,precision=3,set thousands separator={,}},
        display columns/11/.style={column name={$\bm{\textbf{\text{ns.res}}^0}$},column type={r},fixed,fixed zerofill,precision=0,set thousands separator={,}},
        display columns/12/.style={column name={\textbf{cr.mae}},column type={r},fixed,fixed zerofill,precision=3,set thousands separator={,}},
        display columns/13/.style={column name={$\bm{\textbf{\text{cr.mae}}^a}$},column type={r},fixed,fixed zerofill,precision=3,set thousands separator={,}},
        display columns/14/.style={column name={\textbf{cr.res}},column type={r},fixed,fixed zerofill,precision=0,set thousands separator={,}},
        display columns/15/.style={column name={$\bm{\textbf{\text{cr.mae}}^0}$},column type={r},fixed,fixed zerofill,precision=3,set thousands separator={,}},
        display columns/16/.style={column name={$\bm{\textbf{\text{cr.res}}^0}$},column type={r},fixed,fixed zerofill,precision=0,set thousands separator={,}},
	    	every head row/.style={before row=\toprule,after row=\midrule},
	    	every row no 0/.style={before row={\multicolumn{18}{l}{\textbf{10 Thin plate regression splines under gaussian with identity link}}\\ \\}},
	    	every row no 16/.style={before row={\midrule \multicolumn{18}{l}{\textbf{10 Cubic regression splines under gaussian with identity link}}\\ \\}},
	    	every row no 30/.style={before row={\midrule \multicolumn{18}{l}{\textbf{10 Duchon splines under gaussian with identity link}}\\ \\}},
	    	every row no 37/.style={before row={\midrule \multicolumn{18}{l}{\textbf{10 Eilers and Marx style P-splines under gaussian with identity link in stagewise selection of length $\bm{5}$}}\\ \\}},
      	every last row/.style={after row=\bottomrule},
      ]{BGROGAMbfdim10.csv} 
    \caption{Out-of-sample validation figures of selected GAMs of BEL with varying spline function type and fixed spline function number of $10$ per dimension under between $100$-$443$ and $150$-$443$ after each tenth and the finally selected smooth function.}
    \label{tab:BGROGAMbfdim10}
  \end{center}
\end{table}

\begin{table}[htb]
  \tiny
  \begin{center}
  \tabcolsep=0.052cm
  \renewcommand{\arraystretch}{0.8}
      \begin{filecontents*}{BGROGAMrclinkdim4.csv}
k;Kmax;v.wre.y;v.a.wre.y;v.res.avg.y;v.wre.delta.y;v.res.avg.delta.y;ns.wre.y;ns.a.wre.y;ns.res.avg.y;ns.wre.delta.y;ns.res.avg.delta.y;cr.wre.y;cr.a.wre.y;cr.res.avg.y;cr.wre.delta.y;cr.res.avg.delta.y
0;150;4.55660203;4.35650671;-237.5325499;100;38.252875;3.23053;3.121333;-0.294727;100;261.099235;4.026996;3.942121;105.619188;100;367.01315
10;150;0.63169691;0.60395703;28.28162141;22.01939211;115.9154439;0.345482;0.333804;-7.976598;13.246935;65.265762;0.479183;0.469083;65.931607;21.072102;139.173967
20;150;0.40603596;0.38820559;-0.00934966;11.33014865;44.28758971;0.375014;0.362338;-42.022458;7.254174;-12.116982;0.340834;0.333651;-6.26066;7.708887;23.644817
30;150;0.39946731;0.38192539;-10.59725392;12.2675228;59.13348244;0.46477;0.44906;-61.294279;5.744495;-5.955006;0.314069;0.30745;-26.063124;6.116269;29.276149
40;150;0.37139202;0.35508298;-8.31885279;11.41452271;53.34522743;0.479541;0.463332;-63.629762;6.379583;-16.357145;0.339561;0.332404;-33.806;5.282513;13.466617
50;150;0.39170582;0.37450474;-12.95082144;12.07872509;59.46619859;0.520136;0.502555;-69.878592;5.960567;-11.853035;0.365425;0.357723;-38.645217;5.367919;19.38034
60;150;0.30570547;0.29228094;-15.23565922;9.83348745;48.28336149;0.40473;0.39105;-51.440253;5.282647;-2.312695;0.272514;0.26677;-10.384153;6.484493;38.743405
70;150;0.27172673;0.25979432;-14.97706662;9.8956806;56.45822398;0.32056;0.309725;-35.20444;5.227053;21.839388;0.232452;0.227553;11.968091;10.45977;69.011919
80;150;0.24915737;0.23821605;-16.87323002;8.62665534;48.92089608;0.307682;0.297282;-35.507133;4.587725;15.895531;0.204934;0.200615;8.60929;9.099637;60.011953
90;150;0.26114977;0.24968183;-17.2420157;9.26184492;53.68348011;0.324662;0.313688;-38.86244;4.639398;17.671593;0.194805;0.190699;5.122119;9.339666;61.656152
100;150;0.25374317;0.24260048;-17.99409312;9.59277929;55.2797205;0.339712;0.328229;-41.54479;4.626252;17.337561;0.195762;0.191636;2.62069;9.312075;61.503041
110;150;0.25517378;0.24396826;-17.94116266;9.40712844;54.09887963;0.335851;0.324499;-39.747721;4.640196;17.900859;0.207024;0.202661;3.936792;9.32454;61.585372
120;150;0.24320856;0.23252848;-16.18038013;8.47398622;48.43189066;0.30685;0.296478;-37.533903;4.023211;12.686905;0.18621;0.182286;1.164855;7.819262;51.385663
130;150;0.24104732;0.23046214;-16.29424909;8.48093737;48.64910178;0.308062;0.297649;-37.317074;4.108099;13.234814;0.183343;0.179479;2.49908;8.075073;53.050968
140;150;0.2350851;0.22476174;-14.67498002;8.01815008;45.25298317;0.294538;0.284582;-35.404853;3.865487;10.131647;0.172696;0.169056;1.572209;7.181943;47.108709
150;150;0.23965953;0.22913529;-15.0377391;8.19158043;46.36306061;0.291186;0.281344;-34.50222;3.906892;12.507117;0.175562;0.171862;3.167473;7.640849;50.17681
0;40;4.556602;4.356507;-237.53255;100;38.252875;3.23053;3.121333;-0.294727;100;261.099235;4.026996;3.942121;105.619188;100;367.01315
10;40;0.788472;0.753848;7.836182;23.011227;114.010595;0.422711;0.408423;26.009458;22.470981;117.792408;0.700037;0.685283;94.043317;28.248406;185.826267
20;40;0.45174;0.431902;-3.652287;12.76124;50.293247;0.420512;0.406298;-48.133126;7.625691;-8.579055;0.360059;0.35247;-11.045351;8.165942;28.50872
30;40;0.462411;0.442105;-10.487908;14.180163;71.59448;0.52665;0.508848;-68.226818;6.208694;-0.535894;0.367874;0.36012;-32.139667;7.115633;35.551257
40;40;0.438488;0.419232;-6.94254;13.3818;66.231541;0.5235;0.505805;-69.067149;6.188814;-10.284532;0.37285;0.364992;-38.689395;5.913453;20.093222
0;70;4.556602029;4.356506712;-237.5325499;100;38.252875;3.230529781;3.121333346;-0.294727358;100;261.0992348;4.026996295;3.942121469;105.619188;100;367.0131501
10;70;0.625452354;0.597986693;31.28690594;21.06792671;110.2405364;0.332478074;0.321239849;-4.724372934;12.42121005;59.83779468;0.485569623;0.475335534;67.51236721;19.9971887;132.0745348
20;70;0.39400683;0.376704699;1.427585125;10.8872167;40.65805312;0.357382604;0.345302571;-39.34744157;7.282631281;-14.50843638;0.340414768;0.33324003;-5.513996974;7.641296046;19.32500822
30;70;0.383374107;0.366538894;-10.4049212;11.98529854;56.2825191;0.466787127;0.451009068;-61.79750815;5.852690479;-9.501530653;0.330646118;0.323677268;-30.37576924;5.742254107;21.92020826
40;70;0.289340634;0.276634739;-10.802329;9.447497991;45.18992095;0.346315267;0.334609326;-41.16480816;5.15947915;0.435978991;0.255723742;0.250333991;-2.480462251;6.681822943;39.1203249
50;70;0.306720465;0.293251364;-11.12724183;10.33888835;52.55919334;0.389145254;0.375991599;-49.50379067;4.922458745;-0.208818313;0.252064682;0.246752051;-11.03238652;6.293723984;38.26258583
60;70;0.308217497;0.294682658;-14.33956073;10.45457124;55.71651079;0.372292319;0.359708317;-49.03538627;4.377072736;6.629222451;0.222439592;0.217751353;-9.436544859;7.143165998;46.22806386
70;70;0.270395356;0.25852141;-16.11439127;9.999486193;57.14253253;0.325175557;0.314184167;-36.32428006;5.279810173;22.54118093;0.244771271;0.239612359;9.637176902;10.41603509;68.5026379
0;120;4.556602029;4.356506712;-237.5325499;100;38.252875;3.230529781;3.121333346;-0.294727358;100;261.0992348;4.026996295;3.942121469;105.619188;100;367.0131501
10;120;0.779736129;0.745495362;12.32538525;22.10364417;101.2533569;0.435784937;0.421054795;35.46197792;21.14992713;109.9984868;0.735587128;0.720083556;100.5111686;26.69153387;175.0476775
20;120;0.496909906;0.475088964;-1.075831923;14.72076859;70.8722049;0.457330385;0.441871977;-55.34646746;6.793822631;2.210106556;0.359678241;0.352097497;-16.34156517;8.605043292;41.21500885
30;120;0.436749009;0.417569929;-7.050561621;13.58058979;66.27523294;0.483411492;0.467071506;-61.4966759;6.041539619;-2.562344148;0.364184173;0.35650846;-27.61316366;7.017595309;31.32116809
40;120;0.418430287;0.400055643;-7.243787818;12.5749401;58.1701945;0.505025111;0.487954554;-66.80859191;6.529646561;-15.7860724;0.382055511;0.374003134;-39.99008074;5.844481875;11.03243877
50;120;0.415585222;0.397335514;-11.39249539;12.45569389;57.69313355;0.522273171;0.504619606;-70.04353273;6.310468164;-15.3493666;0.392340513;0.384071364;-42.4871466;5.536026124;12.20701953
60;120;0.407427835;0.389536345;-11.39564436;12.2005493;58.52312543;0.547331573;0.528830999;-74.20753675;6.706443292;-18.68022977;0.411392986;0.402722279;-47.08481942;5.47612472;8.442487561
70;120;0.407391294;0.389501408;-6.799794287;12.10370393;58.56374935;0.48023633;0.464003669;-64.26259688;5.740840437;-13.29051605;0.356445816;0.3489332;-38.68074538;5.172706748;12.29133545
80;120;0.27395537;0.261925093;-8.740554278;10.46129302;59.50081858;0.319401895;0.308605664;-31.12985646;5.4089738;22.72005359;0.256821834;0.251408939;16.16011624;10.63571072;70.01002629
90;120;0.251657664;0.240606552;-10.36688521;9.361588561;52.33469539;0.288655431;0.278898472;-31.49903117;4.594211986;16.81108662;0.195224715;0.191110069;9.350365556;8.753070977;57.66048335
100;120;0.239326615;0.228817;-12.90238336;8.40436398;45.77250106;0.253904825;0.245322486;-26.3825461;4.422771635;17.90087551;0.181787699;0.177956258;13.15262679;8.709995025;57.4360484
110;120;0.251249357;0.240216175;-14.81937649;8.3071474;46.26607717;0.256440648;0.247772594;-27.66111515;4.441739385;19.0328757;0.174222015;0.170550031;10.71470039;8.707791235;57.40869124
120;120;0.251733843;0.240679385;-16.27251343;8.367951838;47.04947654;0.262947101;0.254059121;-29.09295088;4.584844097;19.83757628;0.170549859;0.166955272;9.230636626;8.830194389;58.16116379
0;85;4.556601647;4.356506347;-237.5323815;100;38.252875;3.230529807;3.121333371;-0.294558962;100;261.0992347;4.026996417;3.942121589;105.6193564;100;367.0131501
10;85;0.622327233;0.594998806;32.68140255;20.6432502;107.5099114;0.32795473;0.3168694;-3.349334282;12.0336475;57.08771172;0.488193076;0.477903694;68.17789435;19.47337717;128.6149404
20;85;0.442816018;0.423370517;0.365445057;13.17588688;62.62261124;0.411942423;0.398018192;-48.7196886;6.644429389;-0.853985217;0.336375302;0.329285701;-10.61087625;8.149026358;37.25482714
30;85;0.390174497;0.373040657;-10.0263409;12.08730043;60.18292553;0.481343877;0.46507378;-64.5935227;5.77097245;-8.775719076;0.33390271;0.326865223;-32.53915307;5.776787469;23.27865055
40;85;0.279964709;0.267670542;-9.174118726;9.655379219;47.98885747;0.338596552;0.327151514;-39.16986439;5.078639064;3.601648998;0.255471806;0.250087365;1.12910454;7.153585694;43.90061793
50;85;0.296388699;0.2833733;-10.3010693;9.741655357;48.40453903;0.374228876;0.361579415;-47.65660001;4.932580744;-3.342454489;0.241623984;0.236531406;-10.18094342;5.767642911;34.1332021
60;85;0.310388636;0.296758454;-13.91816161;10.40530262;54.00234487;0.366793169;0.354395046;-47.869638;4.592407595;5.659405675;0.231671595;0.226788778;-7.551192493;7.165167404;45.97785118
70;85;0.271550137;0.259625481;-12.11441;10.27912504;57.97589786;0.31312438;0.302540337;-34.07056842;5.205367458;21.62827664;0.249312204;0.244057585;11.78106552;10.28561858;67.47991058
80;85;0.247167482;0.236313548;-14.29881954;8.582531261;47.77695451;0.29264182;0.282750117;-32.5011998;4.594288896;15.18311145;0.217233827;0.212655307;9.877413573;8.77647421;57.56172482
85;85;0.250241199;0.239252289;-17.25056055;8.739133468;49.72740057;0.324578111;0.313606915;-38.11043112;4.584873932;14.47606719;0.217732458;0.213143428;5.546969491;8.87137433;58.13346781
0;75;4.556602029;4.356506712;-237.5325499;100;38.252875;3.230529781;3.121333346;-0.294727359;100;261.0992347;4.026996295;3.942121469;105.619188;100;367.0131501
10;75;0.777693965;0.743542876;14.47588838;21.78049746;94.8889592;0.445923668;0.430850823;40.12067137;20.52048587;106.1422794;0.755993216;0.740059555;103.9869475;25.96887602;170.0085556
20;75;0.49139966;0.469820692;-0.63911426;14.54242377;68.91386811;0.4524392;0.437146121;-54.80789018;6.759225569;0.353629392;0.362301084;0.35466506;-17.02471456;8.422558496;38.13680501
30;75;0.425287972;0.406612185;-6.764115846;13.14216152;62.20414257;0.472294648;0.456330427;-59.94757641;6.123483916;-5.370780795;0.365786904;0.358077411;-27.46656765;6.854337228;27.11022797
40;75;0.405738359;0.387921059;-7.488135578;12.15118255;54.34134568;0.498739703;0.481881602;-66.14335158;6.757490657;-18.70533312;0.388964429;0.380766436;-40.85257395;5.920326272;6.585444512
50;75;0.412146238;0.394047547;-10.72361271;12.54260865;56.43489294;0.512756746;0.495424849;-68.73788192;6.309212266;-15.97083908;0.396483777;0.388127302;-42.30563498;5.655168972;10.46140786
60;75;0.297811019;0.284733162;-12.44268053;9.51860068;47.38950801;0.391931015;0.378683197;-49.61347967;5.298193093;-4.172753935;0.265130289;0.259542281;-9.610147158;6.172189526;35.83057858
70;75;0.262605345;0.251073484;-12.94349198;9.78879852;55.55494329;0.298249582;0.288168328;-31.3339747;5.406106394;22.77299776;0.226875624;0.222093889;16.38365047;10.67348919;70.49062293
75;75;0.257500428;0.246192741;-14.01431423;9.180624531;51.90660768;0.300259325;0.290110139;-32.51851573;5.04919484;19.01094337;0.223330723;0.218623702;13.39232238;9.837345949;64.92178149
0;55;4.556602029;4.356506712;-237.5325499;100;38.252875;3.230529781;3.121333346;-0.294727358;100;261.0992347;4.026996295;3.942121469;105.619188;100;367.0131501
10;55;0.802771072;0.767518765;2.219814888;23.42530288;116.8864214;0.383265329;0.370310424;-24.25003339;15.19724126;76.02511033;0.435231515;0.426058375;26.62181082;19.71326135;126.8969545
20;55;0.448115131;0.428436928;8.364969139;12.64508877;61.13415849;0.331481006;0.320276483;-28.55268296;7.087594221;9.825043581;0.330334765;0.323372478;18.08506465;9.983404851;56.46279119
30;55;0.387356387;0.370346299;1.218213295;12.45750735;63.87024801;0.331412797;0.32021058;-28.61201341;6.700806379;19.64855849;0.310799423;0.304248872;22.16585493;11.09937869;70.42642683
40;55;0.341366015;0.326375516;-4.923685581;11.66093395;61.38138865;0.339139132;0.327675754;-34.75521479;5.91995888;17.15839664;0.271343138;0.265624185;11.08036769;9.85129428;62.99397912
45;55;0.34273573;0.327685082;-9.050629471;10.92794464;55.33382888;0.360746298;0.348552568;-37.59874006;6.110667867;12.39425549;0.300402001;0.29407059;8.970594623;9.450781237;58.96359017
50;55;0.335730236;0.320987222;-7.251385622;10.64516493;54.71191132;0.355002568;0.343002983;-39.97767997;5.318862765;7.594154171;0.250233824;0.244959781;6.740750069;8.524598014;54.31258421
55;55;0.328126448;0.313717342;-8.801319732;10.59510718;55.86352526;0.328093762;0.317003733;-34.80672395;5.325277012;15.46665824;0.240971318;0.235892496;16.45615115;10.24902833;66.72953334
\end{filecontents*}
\pgfplotstabletypeset[
        multicolumn names, 
        col sep=semicolon, 
        display columns/0/.style={column name=$\bm{k}$,dec sep align},
        display columns/1/.style={column name={$\bm{K_\textbf{\text{max}}}$},column type={r},fixed,fixed zerofill,precision=0,set thousands separator={,}},
        display columns/2/.style={column name={\textbf{v.mae}},column type={r},fixed,fixed zerofill,precision=3,set thousands separator={,}},
        display columns/3/.style={column name={$\bm{\textbf{\text{v.mae}}^a}$},column type={r},fixed,fixed zerofill,precision=3,set thousands separator={,}},
        display columns/4/.style={column name={\textbf{v.res}},column type={r},fixed,fixed zerofill,precision=0,set thousands separator={,}},
        display columns/5/.style={column name={$\bm{\textbf{\text{v.mae}}^0}$},column type={r},fixed,fixed zerofill,precision=3,set thousands separator={,}},
        display columns/6/.style={column name={$\bm{\textbf{\text{v.res}}^0}$},column type={r},fixed,fixed zerofill,precision=0,set thousands separator={,}},
        display columns/7/.style={column name={\textbf{ns.mae}},column type={r},fixed,fixed zerofill,precision=3,set thousands separator={,}},
        display columns/8/.style={column name={$\bm{\textbf{\text{ns.mae}}^a}$},column type={r},fixed,fixed zerofill,precision=3,set thousands separator={,}},
        display columns/9/.style={column name={\textbf{ns.res}},column type={r},fixed,fixed zerofill,precision=0,set thousands separator={,}},
        display columns/10/.style={column name={$\bm{\textbf{\text{ns.mae}}^0}$},column type={r},fixed,fixed zerofill,precision=3,set thousands separator={,}},
        display columns/11/.style={column name={$\bm{\textbf{\text{ns.res}}^0}$},column type={r},fixed,fixed zerofill,precision=0,set thousands separator={,}},
        display columns/12/.style={column name={\textbf{cr.mae}},column type={r},fixed,fixed zerofill,precision=3,set thousands separator={,}},
        display columns/13/.style={column name={$\bm{\textbf{\text{cr.mae}}^a}$},column type={r},fixed,fixed zerofill,precision=3,set thousands separator={,}},
        display columns/14/.style={column name={\textbf{cr.res}},column type={r},fixed,fixed zerofill,precision=0,set thousands separator={,}},
        display columns/15/.style={column name={$\bm{\textbf{\text{cr.mae}}^0}$},column type={r},fixed,fixed zerofill,precision=3,set thousands separator={,}},
        display columns/16/.style={column name={$\bm{\textbf{\text{cr.res}}^0}$},column type={r},fixed,fixed zerofill,precision=0,set thousands separator={,}},
	    	every head row/.style={before row=\toprule,after row=\midrule},
	    	every row no 0/.style={before row={\multicolumn{18}{l}{\textbf{4 Thin plate regression splines under gaussian with identity link in stagewise selection of length $\bm{5}$}}\\ \\}},
	    	every row no 16/.style={before row={\midrule \multicolumn{18}{l}{\textbf{4 Thin plate regression splines under gaussian with log link in stagewise selection of length $\bm{5}$}}\\ \\}},
	    	every row no 21/.style={before row={\midrule \multicolumn{18}{l}{\textbf{4 Thin plate regression splines under gamma with identity link in stagewise selection of length $\bm{5}$}}\\ \\}},
	    	every row no 29/.style={before row={\midrule \multicolumn{18}{l}{\textbf{4 Thin plate regression splines under gamma with log link in stagewise selection of length $\bm{5}$}}\\ \\}},
	    	every row no 42/.style={before row={\midrule \multicolumn{18}{l}{\textbf{4 Thin plate regression splines under inverse gaussian with identity link in stagewise selection of length $\bm{5}$}}\\ \\}},
	    	every row no 52/.style={before row={\midrule \multicolumn{18}{l}{\textbf{4 Thin plate regression splines under inverse gaussian with log link in stagewise selection of length $\bm{5}$}}\\ \\}},
	    	every row no 61/.style={before row={\midrule \multicolumn{18}{l}{\textbf{4 Thin plate regression splines under inverse gaussian with }$\bm{\frac{1}{\mu^2}}$\textbf{ link in stagewise selection of length $\bm{5}$}}\\ \\}},
      	every last row/.style={after row=\bottomrule},
      ]{BGROGAMrclinkdim4.csv} 
    \caption{Out-of-sample validation figures of selected GAMs of BEL with varying random component link function combination and fixed spline function number of $4$ per dimension under between $40$-$443$ and $150$-$443$ after each tenth and the finally selected smooth function.}
    \label{tab:BGROGAMrclinkdim4}
  \end{center}
\end{table}

\begin{table}[htb]
  \tiny
  \begin{center}
  \tabcolsep=0.052cm
  \renewcommand{\arraystretch}{0.90}
      \begin{filecontents*}{BGROGAMrclinkdim8.csv}
k;Kmax;v.wre.y;v.a.wre.y;v.res.avg.y;v.wre.delta.y;v.res.avg.delta.y;ns.wre.y;ns.a.wre.y;ns.res.avg.y;ns.wre.delta.y;ns.res.avg.delta.y;cr.wre.y;cr.a.wre.y;cr.res.avg.y;cr.wre.delta.y;cr.res.avg.delta.y
0;150;4.556602;4.356507;-237.53255;100;38.252875;3.23053;3.121333;-0.294727;100;261.099235;4.026996;3.942121;105.619188;100;367.01315
10;150;0.638718;0.61067;27.046944;23.176015;124.881537;0.340015;0.328522;-3.437927;15.516636;80.005203;0.515968;0.505093;72.603594;23.626779;156.046724
20;150;0.375184;0.358709;3.079973;9.604019;26.094516;0.333719;0.322439;-32.928845;8.378276;-24.305765;0.340547;0.333369;1.1248;7.710613;9.74788
30;150;0.360578;0.344744;-6.754388;10.443977;40.840073;0.415431;0.401389;-52.088471;6.961107;-18.885473;0.303517;0.297119;-20.545817;5.870684;12.657182
40;150;0.356029;0.340394;-4.522897;10.098446;36.079496;0.424607;0.410255;-53.842117;7.920046;-27.631187;0.310992;0.304437;-27.003128;5.647187;-0.792198
50;150;0.338807;0.323929;-7.070305;9.711837;33.167706;0.418479;0.404334;-53.184495;7.745799;-27.337946;0.310929;0.304376;-25.738179;5.595924;0.10837
60;150;0.325472;0.311179;-6.1893;9.037418;26.118923;0.410788;0.396903;-52.068067;8.705569;-34.151307;0.310496;0.303952;-26.354386;5.850382;-8.437625
70;150;0.325199;0.310919;-4.041592;9.180181;30.995582;0.428743;0.414251;-54.74054;8.772847;-34.094829;0.325512;0.318651;-29.609565;5.911979;-8.963854
80;150;0.309103;0.295529;-5.252441;8.617743;28.781596;0.429693;0.415169;-54.583578;8.984159;-34.941004;0.336462;0.329371;-28.828992;6.382003;-9.186417
90;150;0.313244;0.299489;-4.502999;8.980724;31.872439;0.384423;0.371429;-48.468219;7.38981;-26.484244;0.29957;0.293256;-25.828669;5.429933;-3.844694
100;150;0.327722;0.313331;-6.419957;9.909584;46.663497;0.400094;0.38657;-50.955077;5.572283;-12.263085;0.290745;0.284617;-25.346592;5.064045;13.3454
110;150;0.255787;0.244555;-10.017321;7.984835;38.355113;0.32587;0.314855;-40.02053;4.654884;-6.039559;0.201344;0.197101;-5.569938;5.001981;28.411033
120;150;0.253125;0.24201;-9.394257;7.340128;29.815614;0.321221;0.310363;-39.091102;5.541779;-14.272694;0.208845;0.204443;-4.814277;4.541449;20.004131
130;150;0.252327;0.241246;-9.035242;7.766767;34.314464;0.326343;0.315312;-39.850737;5.196826;-10.892493;0.205035;0.200713;-4.620258;4.770377;24.337985
140;150;0.245224;0.234455;-8.267161;7.59181;32.683212;0.322074;0.311188;-41.19016;5.314639;-14.63125;0.196977;0.192825;-6.929586;4.316957;19.629324
150;150;0.217359;0.207814;-10.508446;6.477383;31.878887;0.239187;0.231102;-26.32208;3.652311;1.673791;0.178717;0.17495;6.298133;5.577649;34.294004
0;50;4.556602;4.356507;-237.53255;100;38.252875;3.23053;3.121333;-0.294727;100;261.099235;4.026996;3.942121;105.619188;100;367.01315
10;50;0.756803;0.723569;10.124959;21.570175;101.218843;0.443557;0.428564;39.022173;22.141268;115.724595;0.754589;0.738685;105.532475;27.693485;182.234897
20;50;0.401035;0.383424;0.693968;10.277535;22.938905;0.358803;0.346675;-35.466151;9.154363;-27.612677;0.36162;0.353998;-1.210321;8.110169;6.643153
30;50;0.396082;0.378689;-4.704206;11.249059;43.011209;0.438422;0.423603;-53.045656;7.692381;-19.721703;0.339141;0.331993;-19.290084;6.802694;14.033869
40;50;0.38203;0.365254;-5.033933;11.036189;44.794548;0.469928;0.454044;-60.100156;7.84577;-24.663138;0.351127;0.343727;-31.306616;6.233664;4.130402
50;50;0.369624;0.353392;-7.523863;10.486505;39.221131;0.463546;0.447878;-60.217971;8.000237;-27.86444;0.33976864;0.33260752;-32.34466144;5.90064533;0.00886945
0;100;4.556602029;4.356506712;-237.5325499;100;38.252875;3.230529781;3.121333346;-0.294727358;100;261.0992348;4.026996295;3.942121469;105.619188;100;367.0131501
10;100;0.637238356;0.609255133;29.07906298;22.74279357;122.7513388;0.334197488;0.322901144;-2.572773239;14.94135573;76.70803975;0.510403484;0.499645986;71.77190715;22.87064461;151.0527201
20;100;0.370118523;0.353865407;3.616016961;9.5374577;27.10897033;0.32355459;0.312617992;-31.38882958;8.075636987;-22.28733901;0.340115602;0.332947169;0.95628878;7.725067688;10.05777935
30;100;0.359280326;0.343503151;-7.822267053;10.55760364;44.1195774;0.41393723;0.399945572;-52.22636324;6.414981393;-14.67598159;0.304731449;0.298308789;-21.67972931;5.908647646;15.87065234
40;100;0.328685264;0.314251618;-8.569619439;9.642987305;36.61169703;0.40182203;0.388239882;-51.37107875;6.673404728;-20.58122509;0.320591779;0.31383484;-26.45278092;5.702297055;4.337072745
50;100;0.342312733;0.32728066;-7.099758509;9.631045389;32.50035671;0.408702613;0.394887892;-52.13482776;7.552631504;-26.92617534;0.326464498;0.319583782;-27.75964288;5.863476412;-2.550990464
60;100;0.323783252;0.309564869;-5.628543689;9.114477632;28.14517606;0.408619145;0.394807245;-51.80808015;8.420599774;-32.42582321;0.327022901;0.320130416;-28.28271589;6.067076228;-8.90045895
70;100;0.328367406;0.313947718;-6.065247664;9.616502892;40.56523692;0.450512625;0.435284667;-58.57933094;7.631422464;-26.34030916;0.349009564;0.341653678;-34.56303625;5.79566305;-2.324014471
80;100;0.269822226;0.257973448;-8.74916732;7.944356145;36.68267949;0.32412972;0.313173681;-38.14232249;5.067834553;-7.101938489;0.221174024;0.216512458;-2.047611914;5.460510448;28.99277209
90;100;0.279202295;0.266941608;-10.20941497;8.925939209;46.70204255;0.34053576;0.329025174;-40.43101827;4.59524757;2.088976446;0.224176643;0.219451793;-1.893565747;6.713094062;40.62642897
100;100;0.272381003;0.260419861;-10.79009063;8.653598121;44.01494334;0.335176553;0.323847116;-40.16676654;4.532136405;0.246804623;0.215536829;0.210994075;-2.128002922;6.396608525;38.28556824
0;110;4.556602029;4.356506712;-237.5325499;100;38.252875;3.230529781;3.121333346;-0.294727358;100;261.0992348;4.026996295;3.942121469;105.619188;100;367.0131501
10;110;0.762365163;0.728887212;13.25090608;21.35952078;94.90324395;0.458321639;0.442829726;44.8009082;21.52742555;112.0617833;0.77253312;0.756250857;108.3705463;26.7430381;175.6314214
20;110;0.441805362;0.422404241;2.352000974;12.41626765;48.83956294;0.395699314;0.382324123;-44.04734119;7.514585159;-11.95124202;0.349100478;0.341742676;-7.809163132;8.083054404;24.28693603
30;110;0.386966583;0.369973613;-3.012897023;11.14706877;45.11721374;0.413718401;0.399734139;-49.37961778;7.058091667;-15.64096982;0.337971488;0.330848245;-17.9508168;6.84660353;15.78783116
40;110;0.372241604;0.355895256;-6.39091542;10.82649903;42.77184122;0.457537419;0.442072013;-58.89149885;7.545690746;-24.12020501;0.35992286;0.35233696;-33.95046371;6.225135875;0.820830124
50;110;0.357327484;0.341636064;-9.158130388;10.2395297;36.34458894;0.458251839;0.442762284;-59.89163716;7.976733319;-28.78038063;0.356947868;0.34942467;-36.08646848;6.07284725;-4.975211955
60;110;0.351385911;0.335955405;-5.336414501;9.866233895;29.51619762;0.439166278;0.424321843;-56.35822186;9.066375053;-35.89707255;0.35343194;0.345982846;-35.03392755;6.537262491;-14.57277824
70;110;0.354441272;0.338876595;-5.249347136;10.12992241;37.11322728;0.457676012;0.442205921;-59.14961674;8.442002291;-31.17850513;0.36368272;0.356017576;-36.83229529;6.270791438;-8.861183681
80;110;0.359469969;0.343684465;-5.522859736;10.12211146;37.16331668;0.46289588;0.44724935;-59.92696275;8.528669265;-31.63224913;0.371286826;0.363461414;-37.45662487;6.41191332;-9.161911254
90;110;0.282071931;0.269685229;-10.11488915;9.016621672;46.81044868;0.364431618;0.352113318;-44.12870958;4.990868357;-1.594834549;0.2487468;0.243504098;-6.29188474;6.286396367;36.24199029
100;110;0.268017519;0.256247992;-10.6771106;7.806810114;36.78057577;0.31967223;0.30886686;-38.27682583;4.747741746;-5.210602264;0.208703813;0.204305075;-0.599485928;5.604202737;32.46673764
110;110;0.258772757;0.247409198;-11.13341718;7.372904963;33.9635833;0.312136135;0.301585496;-37.33726545;4.801026371;-6.631727768;0.201292858;0.197050317;-0.051133872;5.354434332;30.65440381
\end{filecontents*}
\pgfplotstabletypeset[
        multicolumn names, 
        col sep=semicolon, 
        display columns/0/.style={column name=$\bm{k}$,dec sep align},
        display columns/1/.style={column name={$\bm{K_\textbf{\text{max}}}$},column type={r},fixed,fixed zerofill,precision=0,set thousands separator={,}},
        display columns/2/.style={column name={\textbf{v.mae}},column type={r},fixed,fixed zerofill,precision=3,set thousands separator={,}},
        display columns/3/.style={column name={$\bm{\textbf{\text{v.mae}}^a}$},column type={r},fixed,fixed zerofill,precision=3,set thousands separator={,}},
        display columns/4/.style={column name={\textbf{v.res}},column type={r},fixed,fixed zerofill,precision=0,set thousands separator={,}},
        display columns/5/.style={column name={$\bm{\textbf{\text{v.mae}}^0}$},column type={r},fixed,fixed zerofill,precision=3,set thousands separator={,}},
        display columns/6/.style={column name={$\bm{\textbf{\text{v.res}}^0}$},column type={r},fixed,fixed zerofill,precision=0,set thousands separator={,}},
        display columns/7/.style={column name={\textbf{ns.mae}},column type={r},fixed,fixed zerofill,precision=3,set thousands separator={,}},
        display columns/8/.style={column name={$\bm{\textbf{\text{ns.mae}}^a}$},column type={r},fixed,fixed zerofill,precision=3,set thousands separator={,}},
        display columns/9/.style={column name={\textbf{ns.res}},column type={r},fixed,fixed zerofill,precision=0,set thousands separator={,}},
        display columns/10/.style={column name={$\bm{\textbf{\text{ns.mae}}^0}$},column type={r},fixed,fixed zerofill,precision=3,set thousands separator={,}},
        display columns/11/.style={column name={$\bm{\textbf{\text{ns.res}}^0}$},column type={r},fixed,fixed zerofill,precision=0,set thousands separator={,}},
        display columns/12/.style={column name={\textbf{cr.mae}},column type={r},fixed,fixed zerofill,precision=3,set thousands separator={,}},
        display columns/13/.style={column name={$\bm{\textbf{\text{cr.mae}}^a}$},column type={r},fixed,fixed zerofill,precision=3,set thousands separator={,}},
        display columns/14/.style={column name={\textbf{cr.res}},column type={r},fixed,fixed zerofill,precision=0,set thousands separator={,}},
        display columns/15/.style={column name={$\bm{\textbf{\text{cr.mae}}^0}$},column type={r},fixed,fixed zerofill,precision=3,set thousands separator={,}},
        display columns/16/.style={column name={$\bm{\textbf{\text{cr.res}}^0}$},column type={r},fixed,fixed zerofill,precision=0,set thousands separator={,}},
	    	every head row/.style={before row=\toprule,after row=\midrule},
	    	every row no 0/.style={before row={\multicolumn{18}{l}{\textbf{8 Thin plate regression splines under gaussian with identity link}}\\ \\}},
	    	every row no 16/.style={before row={\midrule \multicolumn{18}{l}{\textbf{8 Thin plate regression splines under gaussian with log link in stagewise selection of length $\bm{5}$}}\\ \\}},
	    	every row no 22/.style={before row={\midrule \multicolumn{18}{l}{\textbf{8 Thin plate regression splines under gamma with identity link in stagewise selection of length $\bm{5}$}}\\ \\}},
	    	every row no 33/.style={before row={\midrule \multicolumn{18}{l}{\textbf{8 Thin plate regression splines under gamma with log link in stagewise selection of length $\bm{5}$}}\\ \\}},
      	every last row/.style={after row=\bottomrule},
      ]{BGROGAMrclinkdim8.csv} 
    \caption{Out-of-sample validation figures of selected GAMs of BEL with varying random component link function combination and fixed spline function number of $8$ per dimension under between $50$-$443$ and $150$-$443$ after each tenth and the finally selected smooth function.}
    \label{tab:BGROGAMrclinkdim8}
  \end{center}
\end{table}

\begin{table}[htb]
  \tiny
  \begin{center}
  \tabcolsep=0.052cm
  \renewcommand{\arraystretch}{0.90}
      \begin{filecontents*}{BGROGAMadalg.csv}
k;Kmax;v.wre.y;v.a.wre.y;v.res.avg.y;v.wre.delta.y;v.res.avg.delta.y;ns.wre.y;ns.a.wre.y;ns.res.avg.y;ns.wre.delta.y;ns.res.avg.delta.y;cr.wre.y;cr.a.wre.y;cr.res.avg.y;cr.wre.delta.y;cr.res.avg.delta.y
0;25;4.556602;4.356507;-237.53255;100;38.252875;3.23053;3.121333;-0.294727;100;261.099235;4.026996;3.942121;105.619188;100;367.01315
10;25;0.662613;0.633516;25.987132;23.297732;123.481039;0.34133;0.329793;0.812986;16.217722;83.915431;0.547336;0.5358;77.851097;24.369712;160.953542
20;25;0.398303;0.380812;1.664212;10.221037;22.622534;0.360949;0.348748;-34.890797;9.380327;-28.323938;0.375339;0.367428;-1.038961;8.460294;5.527898
25;25;0.410588;0.392558;2.409065;11.892322;46.933293;0.410491;0.396616;-47.352415;7.709421;-17.21965;0.323703;0.316881;-11.052579;7.11955;19.080186
0;50;4.556602;4.356507;-237.53255;100;38.252875;3.23053;3.121333;-0.294727;100;261.099235;4.026996;3.942121;105.619188;100;367.01315
10;50;0.756803;0.723569;10.124959;21.570175;101.218843;0.443557;0.428564;39.022173;22.141268;115.724595;0.754589;0.738685;105.532475;27.693485;182.234897
20;50;0.401035;0.383424;0.693968;10.277535;22.938905;0.358803;0.346675;-35.466151;9.154363;-27.612677;0.36162;0.353998;-1.210321;8.110169;6.643153
30;50;0.396082;0.378689;-4.704206;11.249059;43.011209;0.438422;0.423603;-53.045656;7.692381;-19.721703;0.339141;0.331993;-19.290084;6.802694;14.033869
40;50;0.38203;0.365254;-5.033933;11.036189;44.794548;0.469928;0.454044;-60.100156;7.84577;-24.663138;0.351127;0.343727;-31.306616;6.233664;4.130402
50;50;0.369624;0.353392;-7.523863;10.486505;39.221131;0.463546;0.447878;-60.217971;8.000237;-27.86444;0.33976864;0.33260752;-32.34466144;5.90064533;0.00886945
0;71;4.556602029;4.356506712;-237.5325499;100;38.252875;3.230529781;3.121333346;-0.294727358;100;261.0992348;4.026996295;3.942121469;105.619188;100;367.0131501
10;71;0.637238356;0.609255133;29.07906298;22.74279357;122.7513388;0.334197488;0.322901144;-2.572773239;14.94135573;76.70803975;0.510403484;0.499645986;71.77190715;22.87064461;151.0527201
20;71;0.385576339;0.368644419;7.861480313;10.14087308;30.67042844;0.309612412;0.299147079;-26.13136583;7.903577062;-17.71388051;0.357565535;0.350029319;7.790794379;8.139816936;16.2082797
30;71;0.359280326;0.343503151;-7.822267049;10.55760364;44.11957741;0.41393723;0.399945572;-52.22636324;6.414981392;-14.67598158;0.304731449;0.298308789;-21.67972931;5.908647647;15.87065235
40;71;0.328685264;0.314251618;-8.569619439;9.642987305;36.61169703;0.40182203;0.388239882;-51.37107875;6.673404728;-20.58122509;0.320591779;0.31383484;-26.45278092;5.702297055;4.337072745
50;71;0.338394224;0.323534226;-6.654100497;9.542935083;32.22159252;0.41246684;0.398524883;-52.88451515;7.747883371;-28.40028494;0.324438611;0.317600593;-28.94691244;5.805353389;-4.462682227
60;71;0.323783252;0.309564869;-5.628543689;9.114477632;28.14517606;0.408619145;0.394807245;-51.80808015;8.420599774;-32.42582321;0.327022901;0.320130416;-28.28271589;6.067076228;-8.90045895
70;71;0.327072879;0.312710038;-5.235114542;9.417023952;35.71312134;0.433519852;0.418866274;-55.72230476;8.017426438;-29.16553168;0.341898352;0.334692345;-31.81675461;5.966802166;-5.259981533
71;71;0.290912316;0.278137403;-3.91749065;8.638995464;41.35077917;0.340832396;0.329311783;-43.2472217;5.204743024;-12.37041468;0.196266742;0.192130134;-16.5491835;3.897679611;14.32762352
0;100;4.556602029;4.356506712;-237.5325499;100;38.252875;3.230529781;3.121333346;-0.294727358;100;261.0992348;4.026996295;3.942121469;105.619188;100;367.0131501
10;100;0.637238356;0.609255133;29.07906298;22.74279357;122.7513388;0.334197488;0.322901144;-2.572773239;14.94135573;76.70803975;0.510403484;0.499645986;71.77190715;22.87064461;151.0527201
20;100;0.370118523;0.353865407;3.616016961;9.5374577;27.10897033;0.32355459;0.312617992;-31.38882958;8.075636987;-22.28733901;0.340115602;0.332947169;0.95628878;7.725067688;10.05777935
30;100;0.359280326;0.343503151;-7.822267053;10.55760364;44.1195774;0.41393723;0.399945572;-52.22636324;6.414981393;-14.67598159;0.304731449;0.298308789;-21.67972931;5.908647646;15.87065234
40;100;0.328685264;0.314251618;-8.569619439;9.642987305;36.61169703;0.40182203;0.388239882;-51.37107875;6.673404728;-20.58122509;0.320591779;0.31383484;-26.45278092;5.702297055;4.337072745
50;100;0.342312733;0.32728066;-7.099758509;9.631045389;32.50035671;0.408702613;0.394887892;-52.13482776;7.552631504;-26.92617534;0.326464498;0.319583782;-27.75964288;5.863476412;-2.550990464
60;100;0.323783252;0.309564869;-5.628543689;9.114477632;28.14517606;0.408619145;0.394807245;-51.80808015;8.420599774;-32.42582321;0.327022901;0.320130416;-28.28271589;6.067076228;-8.90045895
70;100;0.328367406;0.313947718;-6.065247664;9.616502892;40.56523692;0.450512625;0.435284667;-58.57933094;7.631422464;-26.34030916;0.349009564;0.341653678;-34.56303625;5.79566305;-2.324014471
80;100;0.269822226;0.257973448;-8.74916732;7.944356145;36.68267949;0.32412972;0.313173681;-38.14232249;5.067834553;-7.101938489;0.221174024;0.216512458;-2.047611914;5.460510448;28.99277209
90;100;0.279202295;0.266941608;-10.20941497;8.925939209;46.70204255;0.34053576;0.329025174;-40.43101827;4.59524757;2.088976446;0.224176643;0.219451793;-1.893565747;6.713094062;40.62642897
100;100;0.272381003;0.260419861;-10.79009063;8.653598121;44.01494334;0.335176553;0.323847116;-40.16676654;4.532136405;0.246804623;0.215536829;0.210994075;-2.128002922;6.396608525;38.28556824
\end{filecontents*}
\pgfplotstabletypeset[
        multicolumn names, 
        col sep=semicolon, 
        display columns/0/.style={column name=$\bm{k}$,dec sep align},
        display columns/1/.style={column name={$\bm{K_\textbf{\text{max}}}$},column type={r},fixed,fixed zerofill,precision=0,set thousands separator={,}},
        display columns/2/.style={column name={\textbf{v.mae}},column type={r},fixed,fixed zerofill,precision=3,set thousands separator={,}},
        display columns/3/.style={column name={$\bm{\textbf{\text{v.mae}}^a}$},column type={r},fixed,fixed zerofill,precision=3,set thousands separator={,}},
        display columns/4/.style={column name={\textbf{v.res}},column type={r},fixed,fixed zerofill,precision=0,set thousands separator={,}},
        display columns/5/.style={column name={$\bm{\textbf{\text{v.mae}}^0}$},column type={r},fixed,fixed zerofill,precision=3,set thousands separator={,}},
        display columns/6/.style={column name={$\bm{\textbf{\text{v.res}}^0}$},column type={r},fixed,fixed zerofill,precision=0,set thousands separator={,}},
        display columns/7/.style={column name={\textbf{ns.mae}},column type={r},fixed,fixed zerofill,precision=3,set thousands separator={,}},
        display columns/8/.style={column name={$\bm{\textbf{\text{ns.mae}}^a}$},column type={r},fixed,fixed zerofill,precision=3,set thousands separator={,}},
        display columns/9/.style={column name={\textbf{ns.res}},column type={r},fixed,fixed zerofill,precision=0,set thousands separator={,}},
        display columns/10/.style={column name={$\bm{\textbf{\text{ns.mae}}^0}$},column type={r},fixed,fixed zerofill,precision=3,set thousands separator={,}},
        display columns/11/.style={column name={$\bm{\textbf{\text{ns.res}}^0}$},column type={r},fixed,fixed zerofill,precision=0,set thousands separator={,}},
        display columns/12/.style={column name={\textbf{cr.mae}},column type={r},fixed,fixed zerofill,precision=3,set thousands separator={,}},
        display columns/13/.style={column name={$\bm{\textbf{\text{cr.mae}}^a}$},column type={r},fixed,fixed zerofill,precision=3,set thousands separator={,}},
        display columns/14/.style={column name={\textbf{cr.res}},column type={r},fixed,fixed zerofill,precision=0,set thousands separator={,}},
        display columns/15/.style={column name={$\bm{\textbf{\text{cr.mae}}^0}$},column type={r},fixed,fixed zerofill,precision=3,set thousands separator={,}},
        display columns/16/.style={column name={$\bm{\textbf{\text{cr.res}}^0}$},column type={r},fixed,fixed zerofill,precision=0,set thousands separator={,}},
	    	every head row/.style={before row=\toprule,after row=\midrule},
	    	every row no 0/.style={before row={\multicolumn{18}{l}{\textbf{8 Thin plate regression splines under gaussian with log link}}\\ \\}},
	    	every row no 4/.style={before row={\midrule \multicolumn{18}{l}{\textbf{8 Thin plate regression splines under gaussian with log link in stagewise selection of length $\bm{5}$}}\\ \\}},
	    	every row no 10/.style={before row={\midrule \multicolumn{18}{l}{\textbf{8 Thin plate regression splines under gamma with identity link}}\\ \\}},
	    	every row no 19/.style={before row={\midrule \multicolumn{18}{l}{\textbf{8 Thin plate regression splines under gamma with identity link in stagewise selection of length $\bm{5}$}}\\ \\}},
      	every last row/.style={after row=\bottomrule},
      ]{BGROGAMadalg.csv} 
    \caption{Out-of-sample validation figures of selected GAMs of BEL in adaptive forward stepwise and stagewise selection of length $5$ under between $25$-$443$ and $100$-$443$ after each tenth and the finally selected smooth function.}
    \label{tab:BGROGAMadalg}
  \end{center}
\end{table}

\begin{table}[htb]
  \tiny
  \begin{center}
  \tabcolsep=0.03cm
  \renewcommand{\arraystretch}{0.90}
      \begin{filecontents*}{BGROGAMchallengePS.csv}
k;Kmax;v.wre.y;v.a.wre.y;v.res.avg.y;v.wre.delta.y;v.res.avg.delta.y;ns.wre.y;ns.a.wre.y;ns.res.avg.y;ns.wre.delta.y;ns.res.avg.delta.y;cr.wre.y;cr.a.wre.y;cr.res.avg.y;cr.wre.delta.y;cr.res.avg.delta.y
0;100;4.556602;4.356507;-237.53255;100;38.252875;3.23053;3.121333;-0.294727;100;261.099235;4.026996;3.942121;105.619188;100;367.01315
10;100;0.643132;0.61489;28.995381;22.835551;122.669442;0.343878;0.332255;-8.80537;13.950748;70.477229;0.470508;0.460592;65.056584;21.854159;144.339183
20;100;0.388816;0.371741;1.339965;10.495734;36.720597;0.365245;0.352899;-40.58435;7.778241;-19.59518;0.335707;0.328632;-8.256076;7.402197;12.733094
30;100;0.384068;0.367203;-9.247083;11.376575;52.731582;0.459086;0.443568;-60.172245;6.137879;-12.585042;0.319515;0.312781;-30.134558;5.511834;17.452645
40;100;0.370552;0.354279;-9.707288;10.977339;49.112507;0.454032;0.438685;-59.940417;6.094714;-15.512085;0.326954;0.320063;-33.590096;5.092497;10.838235
50;100;0.357073;0.341393;-9.056741;10.458551;45.070308;0.467245;0.451451;-61.569709;6.908833;-21.834123;0.334551;0.3275;-33.755335;5.05855;5.980251
60;100;0.339069;0.32418;-10.23922;9.932362;42.7815;0.492396;0.475752;-66.200839;7.639863;-27.571582;0.364709;0.357023;-40.3466;5.154847;-1.717343
70;100;0.343305;0.328229;-10.383052;10.522783;52.029067;0.545803;0.527354;-75.215914;7.681233;-27.195258;0.36562;0.357914;-45.737563;4.576357;2.283093
80;100;0.334159;0.319485;-6.856983;9.919764;44.761689;0.520224;0.502639;-66.559986;8.655174;-29.332777;0.346047;0.338754;-36.161786;5.036165;1.065422
90;100;0.227705;0.217706;-10.411818;6.973055;34.549063;0.278804;0.26938;-30.950998;4.299392;-0.38158;0.208318;0.203927;3.355189;5.809805;33.924607
100;100;0.225364;0.215468;-11.365175;6.89739;34.375712;0.256212;0.247552;-29.645537;3.715881;1.703887;0.164217;0.160756;0.710134;5.212007;32.059559
0;91;4.556602029;4.356506712;-237.5325499;100;38.252875;3.230529781;3.121333346;-0.294727358;100;261.0992347;4.026996295;3.942121469;105.619188;100;367.0131501
5;91;1.57369604;1.504589893;-17.55024418;41.68801276;232.8566377;0.732341982;0.707587797;-75.46903525;30.20067001;160.5463838;0.384461368;0.376358284;42.26913947;42.13460858;278.2845585
11;91;0.816649135;0.780787397;-2.756626729;22.38075367;113.0543632;0.396434244;0.383034211;-33.51388529;13.47479366;67.90564178;0.412447508;0.403754575;22.58324972;19.32234838;124.0027768
21;91;0.679370471;0.649537089;-9.36817192;24.2031484;137.5552487;0.763307327;0.737506469;-101.5646663;8.222076741;30.96729154;0.424366018;0.415421884;-43.56148948;13.54802407;88.97046836
37;91;0.525345343;0.502275709;1.328321437;15.48549636;78.75784486;0.521120391;0.503505791;-62.88837922;6.153947905;0.149681401;0.396927006;0.38856119;-30.04589125;7.46083218;32.99216936
62;91;0.504520082;0.482364953;-1.499389085;14.20779912;63.91851659;0.506836446;0.489704664;-61.33111269;6.841621571;-10.30466981;0.41843211;0.409613043;-33.48446953;7.404816529;17.54197334
91;91;0.309327254;0.295743681;-10.63582917;9.688294018;45.31414995;0.335324106;0.323989681;-35.89694215;5.238665383;5.661574157;0.279219948;0.27333498;1.661189245;7.419666363;43.21970556
0;150;4.556602;4.356507;-237.53255;100;38.252875;3.23053;3.121333;-0.294727;100;261.099235;4.026996;3.942121;105.619188;100;367.01315
10;150;0.647685;0.619243;26.855012;23.688136;128.255022;0.349223;0.337419;-6.749366;15.565803;80.259181;0.50585;0.495189;70.771458;23.889212;157.780006
20;150;0.397972;0.380496;0.658465;10.946417;44.90155;0.358254;0.346144;-37.344168;7.063195;-7.492546;0.337697;0.33058;1.460373;8.101689;31.311996
30;150;0.393366;0.376092;-8.985849;11.982734;58.505909;0.435306;0.420592;-54.884493;5.575074;-1.784197;0.299271;0.292964;-16.617816;6.927692;36.48248
40;150;0.371404;0.355095;-7.562847;11.373781;55.304613;0.449463;0.434271;-57.322643;5.738476;-8.846646;0.31445;0.307822;-25.693621;5.769749;22.782377
50;150;0.363197;0.347248;-9.395417;10.955503;50.060718;0.459688;0.44415;-59.512013;6.248897;-14.44734;0.314606;0.307975;-27.818055;5.491816;17.246618
60;150;0.349363;0.334021;-7.82748;10.479421;46.207572;0.44258;0.42762;-56.376027;6.525833;-16.732437;0.304546;0.298127;-26.026349;5.426833;13.617241
70;150;0.34851;0.333206;-6.207092;10.629381;50.633;0.46427;0.448577;-59.512018;6.687342;-17.063389;0.325119;0.318267;-29.496838;5.501447;12.951791
80;150;0.350374;0.334988;-6.721094;10.465415;47.856672;0.467647;0.45184;-59.666627;7.036092;-19.480324;0.334994;0.327934;-29.356393;5.56277;10.829911
90;150;0.350264;0.334883;-7.22017;10.639336;50.728602;0.470013;0.454126;-60.17473;6.682598;-16.617421;0.33014;0.323182;-29.439029;5.452528;14.118281
100;150;0.333904;0.319241;-8.32217;9.960236;45.868632;0.467729;0.451919;-59.685193;7.170289;-19.885854;0.338723;0.331584;-28.502599;5.834585;11.296739
110;150;0.337323;0.32251;-8.97809;10.24938;48.491743;0.450291;0.43507;-58.145128;6.17101;-15.066758;0.329171;0.322233;-31.069604;5.266663;12.008767
120;150;0.339117;0.324226;-7.473826;10.283256;45.210609;0.433448;0.418796;-55.377449;6.419765;-17.084477;0.320146;0.313399;-28.26062;5.339616;10.032352
130;150;0.268541;0.256748;-12.639368;8.9118;43.218026;0.364808;0.352477;-45.663885;4.890716;-4.197955;0.243592;0.238458;-11.910768;5.502872;29.555163
140;150;0.255007;0.243809;-11.924913;8.157094;36.105611;0.355671;0.343649;-43.974159;5.415324;-10.335098;0.245813;0.240632;-9.670257;5.196415;23.968804
150;150;0.26131;0.249835;-12.454371;8.513934;38.95221;0.367874;0.35544;-46.36384;5.267498;-9.348722;0.245347;0.240176;-12.240557;5.16153;24.774561
\end{filecontents*}
\pgfplotstabletypeset[
        multicolumn names, 
        col sep=semicolon, 
        display columns/0/.style={column name=$\bm{k}$,dec sep align},
        display columns/1/.style={column name={$\bm{K_\textbf{\text{max}}}$},column type={r},fixed,fixed zerofill,precision=0,set thousands separator={,}},
        display columns/2/.style={column name={\textbf{v.mae}},column type={r},fixed,fixed zerofill,precision=3,set thousands separator={,}},
        display columns/3/.style={column name={$\bm{\textbf{\text{v.mae}}^a}$},column type={r},fixed,fixed zerofill,precision=3,set thousands separator={,}},
        display columns/4/.style={column name={\textbf{v.res}},column type={r},fixed,fixed zerofill,precision=0,set thousands separator={,}},
        display columns/5/.style={column name={$\bm{\textbf{\text{v.mae}}^0}$},column type={r},fixed,fixed zerofill,precision=3,set thousands separator={,}},
        display columns/6/.style={column name={$\bm{\textbf{\text{v.res}}^0}$},column type={r},fixed,fixed zerofill,precision=0,set thousands separator={,}},
        display columns/7/.style={column name={\textbf{ns.mae}},column type={r},fixed,fixed zerofill,precision=3,set thousands separator={,}},
        display columns/8/.style={column name={$\bm{\textbf{\text{ns.mae}}^a}$},column type={r},fixed,fixed zerofill,precision=3,set thousands separator={,}},
        display columns/9/.style={column name={\textbf{ns.res}},column type={r},fixed,fixed zerofill,precision=0,set thousands separator={,}},
        display columns/10/.style={column name={$\bm{\textbf{\text{ns.mae}}^0}$},column type={r},fixed,fixed zerofill,precision=3,set thousands separator={,}},
        display columns/11/.style={column name={$\bm{\textbf{\text{ns.res}}^0}$},column type={r},fixed,fixed zerofill,precision=0,set thousands separator={,}},
        display columns/12/.style={column name={\textbf{cr.mae}},column type={r},fixed,fixed zerofill,precision=3,set thousands separator={,}},
        display columns/13/.style={column name={$\bm{\textbf{\text{cr.mae}}^a}$},column type={r},fixed,fixed zerofill,precision=3,set thousands separator={,}},
        display columns/14/.style={column name={\textbf{cr.res}},column type={r},fixed,fixed zerofill,precision=0,set thousands separator={,}},
        display columns/15/.style={column name={$\bm{\textbf{\text{cr.mae}}^0}$},column type={r},fixed,fixed zerofill,precision=3,set thousands separator={,}},
        display columns/16/.style={column name={$\bm{\textbf{\text{cr.res}}^0}$},column type={r},fixed,fixed zerofill,precision=0,set thousands separator={,}},
	    	every head row/.style={before row=\toprule,after row=\midrule},
	    	every row no 0/.style={before row={\multicolumn{18}{l}{\textbf{5 Eilers and Marx style P-splines under gaussian with identity link}}\\ \\}},
	    	every row no 11/.style={before row={\midrule \multicolumn{18}{l}{\textbf{8 Eilers and Marx style P-splines under inverse gaussian with }$\bm{\frac{1}{\mu^2}}$\textbf{ link in dynamically stagewise selection of proportion $\bm{0.25}$}}\\ \\}},
	    	every row no 18/.style={before row={\midrule \multicolumn{18}{l}{\textbf{10 Eilers and Marx style P-splines under gaussian with identity link in stagewise selection of length $\bm{5}$}}\\ \\}},
      	every last row/.style={after row=\bottomrule},
      ]{BGROGAMchallengePS.csv} 
    \caption{Out-of-sample validation figures of selected GAMs of BEL with varying spline function number per dimension and fixed spline function type under between $91$-$443$ and $150$-$443$ after each tenth and the finally selected smooth function or after each dynamically stagewise selected smooth function block. Thereby furthermore a variation in the random component link function combination.}
    \label{tab:BGROGAMchallengePS}
  \end{center}
\end{table}

\begin{table}[htb]
  \tiny
  \begin{center}
  \tabcolsep=0.052cm
  \renewcommand{\arraystretch}{0.8}
      \begin{filecontents*}{BGROGAMall.csv}
k;Kmax;v.wre.y;v.a.wre.y;v.res.avg.y;v.wre.delta.y;v.res.avg.delta.y;ns.wre.y;ns.a.wre.y;ns.res.avg.y;ns.wre.delta.y;ns.res.avg.delta.y;cr.wre.y;cr.a.wre.y;cr.res.avg.y;cr.wre.delta.y;cr.res.avg.delta.y
150;150;0.23965953;0.22913529;-15.0377391;8.19158043;46.36306061;0.291186;0.281344;-34.50222;3.906892;12.507117;0.175562;0.171862;3.167473;7.640849;50.17681
100;100;0.28673;0.274139;-11.384574;9.430586;48.196286;0.396606;0.3832;-50.011256;5.40182;-4.821859;0.2019994;0.1977419;-9.0186537;5.9453404;36.1707432
150;150;0.217359;0.207814;-10.508446;6.477383;31.878887;0.239187;0.231102;-26.32208;3.652311;1.673791;0.178717;0.17495;6.298133;5.577649;34.294004
150;150;0.211915;0.202609;-9.793723;7.069811;36.763217;0.230458;0.222668;-24.274291;3.575314;7.891186;0.173223;0.169572;8.25587;6.337026;40.421347
100;100;0.268186;0.256409;-11.563618;9.903409;51.69244;0.399161;0.385669;-50.788418;5.18194;-1.923822;0.226086;0.221321;-9.088135;6.533176;39.77646
56;100;0.665511;0.636286;-17.517526;18.531937;85.875639;0.288366;0.278618;-13.59088;14.642597;75.410822;0.40605;0.397492;39.540388;19.757039;128.542091
100;100;0.225364;0.215468;-11.365175;6.89739;34.375712;0.256212;0.247552;-29.645537;3.715881;1.703887;0.164217;0.160756;0.710134;5.212007;32.059559
125;125;0.253706;0.242565;-6.633897;7.138864;30.739981;0.298756;0.288658;-36.376623;5.189094;-13.394209;0.196553;0.19241;-5.581915;4.227767;17.4005
53;100;0.821486;0.785412;-44.001406;21.348029;94.195126;0.544573;0.526166;-61.416936;12.59311;62.388133;0.446303;0.436897;-8.073715;18.091325;115.731354
150;150;0.26131;0.249835;-12.454371;8.513934;-38.95221;0.367874;0.35544;-46.36384;5.267498;9.348722;0.245347;0.240176;-12.240557;5.16153;-24.774561
25;25;0.410588;0.392558;2.409065;11.892322;46.933293;0.410491;0.396616;-47.352415;7.709421;-17.21965;0.323703;0.316881;-11.052579;7.11955;19.080186
50;50;0.369624;0.353392;-7.523863;10.486505;39.221131;0.463546;0.447878;-60.217971;8.000237;-27.86444;0.33976864;0.33260752;-32.34466144;5.90064533;0.00886945
71;71;0.290912316;0.278137403;-3.91749065;8.638995464;41.35077917;0.340832396;0.329311783;-43.2472217;5.204743024;-12.37041468;0.196266742;0.192130134;-16.5491835;3.897679611;14.32762352
100;100;0.272381003;0.260419861;-10.79009063;8.653598121;44.01494334;0.335176553;0.323847116;-40.16676654;4.532136405;0.246804623;0.215536829;0.210994075;-2.128002922;6.396608525;38.28556824
150;150;0.23965953;0.22913529;-15.0377391;8.19158043;46.36306061;0.291186;0.281344;-34.50222;3.906892;12.507117;0.175562;0.171862;3.167473;7.640849;50.17681
40;40;0.438488;0.419232;-6.94254;13.3818;66.231541;0.5235;0.505805;-69.067149;6.188814;-10.284532;0.37285;0.364992;-38.689395;5.913453;20.093222
70;70;0.270395356;0.25852141;-16.11439127;9.999486193;57.14253253;0.325175557;0.314184167;-36.32428006;5.279810173;22.54118093;0.244771271;0.239612359;9.637176902;10.41603509;68.5026379
120;120;0.251733843;0.240679385;-16.27251343;8.367951838;47.04947654;0.262947101;0.254059121;-29.09295088;4.584844097;19.83757628;0.170549859;0.166955272;9.230636626;8.830194389;58.16116379
85;85;0.250241199;0.239252289;-17.25056055;8.739133468;49.72740057;0.324578111;0.313606915;-38.11043112;4.584873932;14.47606719;0.217732458;0.213143428;5.546969491;8.87137433;58.13346781
75;75;0.257500428;0.246192741;-14.01431423;9.180624531;51.90660768;0.300259325;0.290110139;-32.51851573;5.04919484;19.01094337;0.223330723;0.218623702;13.39232238;9.837345949;64.92178149
55;55;0.328126448;0.313717342;-8.801319732;10.59510718;55.86352526;0.328093762;0.317003733;-34.80672395;5.325277012;15.46665824;0.240971318;0.235892496;16.45615115;10.24902833;66.72953334
110;110;0.258772757;0.247409198;-11.13341718;7.372904963;33.9635833;0.312136135;0.301585496;-37.33726545;4.801026371;-6.631727768;0.201292858;0.197050317;-0.051133872;5.354434332;30.65440381
91;91;0.309327254;0.295743681;-10.63582917;9.688294018;45.31414995;0.335324106;0.323989681;-35.89694215;5.238665383;5.661574157;0.279219948;0.27333498;1.661189245;7.419666363;43.21970556
\end{filecontents*}
\pgfplotstabletypeset[
        multicolumn names, 
        col sep=semicolon, 
        display columns/0/.style={column name=$\bm{k}$,dec sep align},
        display columns/1/.style={column name={$\bm{K_\textbf{\text{max}}}$},column type={r},fixed,fixed zerofill,precision=0,set thousands separator={,}},
        display columns/2/.style={column name={\textbf{v.mae}},column type={r},fixed,fixed zerofill,precision=3,set thousands separator={,}},
        display columns/3/.style={column name={$\bm{\textbf{\text{v.mae}}^a}$},column type={r},fixed,fixed zerofill,precision=3,set thousands separator={,}},
        display columns/4/.style={column name={\textbf{v.res}},column type={r},fixed,fixed zerofill,precision=0,set thousands separator={,}},
        display columns/5/.style={column name={$\bm{\textbf{\text{v.mae}}^0}$},column type={r},fixed,fixed zerofill,precision=3,set thousands separator={,}},
        display columns/6/.style={column name={$\bm{\textbf{\text{v.res}}^0}$},column type={r},fixed,fixed zerofill,precision=0,set thousands separator={,}},
        display columns/7/.style={column name={\textbf{ns.mae}},column type={r},fixed,fixed zerofill,precision=3,set thousands separator={,}},
        display columns/8/.style={column name={$\bm{\textbf{\text{ns.mae}}^a}$},column type={r},fixed,fixed zerofill,precision=3,set thousands separator={,}},
        display columns/9/.style={column name={\textbf{ns.res}},column type={r},fixed,fixed zerofill,precision=0,set thousands separator={,}},
        display columns/10/.style={column name={$\bm{\textbf{\text{ns.mae}}^0}$},column type={r},fixed,fixed zerofill,precision=3,set thousands separator={,}},
        display columns/11/.style={column name={$\bm{\textbf{\text{ns.res}}^0}$},column type={r},fixed,fixed zerofill,precision=0,set thousands separator={,}},
        display columns/12/.style={column name={\textbf{cr.mae}},column type={r},fixed,fixed zerofill,precision=3,set thousands separator={,}},
        display columns/13/.style={column name={$\bm{\textbf{\text{cr.mae}}^a}$},column type={r},fixed,fixed zerofill,precision=3,set thousands separator={,}},
        display columns/14/.style={column name={\textbf{cr.res}},column type={r},fixed,fixed zerofill,precision=0,set thousands separator={,}},
        display columns/15/.style={column name={$\bm{\textbf{\text{cr.mae}}^0}$},column type={r},fixed,fixed zerofill,precision=3,set thousands separator={,}},
        display columns/16/.style={column name={$\bm{\textbf{\text{cr.res}}^0}$},column type={r},fixed,fixed zerofill,precision=0,set thousands separator={,}},
	    	every head row/.style={before row=\toprule,after row=\midrule},
	    	every row no 0/.style={before row={\multicolumn{18}{l}{\textbf{4 Thin plate regression splines under gaussian with identity link}}\\ \\}},
	    	every row no 1/.style={before row={\midrule \multicolumn{18}{l}{\textbf{5 Thin plate regression splines under gaussian with identity link}}\\ \\}},
	    	every row no 2/.style={before row={\midrule \multicolumn{18}{l}{\textbf{8 Thin plate regression splines under gaussian with identity link}}\\ \\}},
	    	every row no 3/.style={before row={\midrule \multicolumn{18}{l}{\textbf{10 Thin plate regression splines under gaussian with identity link}}\\ \\}},
	    	every row no 4/.style={before row={\midrule \multicolumn{18}{l}{\textbf{5 Cubic regression splines under gaussian with identity link}}\\ \\}},
	    	every row no 5/.style={before row={\midrule \multicolumn{18}{l}{\textbf{5 Duchon splines under gaussian with identity link}}\\ \\}},
	    	every row no 6/.style={before row={\midrule \multicolumn{18}{l}{\textbf{5 Eilers and Marx style P-splines under gaussian with identity link}}\\ \\}},
	    	every row no 7/.style={before row={\midrule \multicolumn{18}{l}{\textbf{10 Cubic regression splines under gaussian with identity link}}\\ \\}},
	    	every row no 8/.style={before row={\midrule \multicolumn{18}{l}{\textbf{10 Duchon splines under gaussian with identity link}}\\ \\}},
	    	every row no 9/.style={before row={\midrule \multicolumn{18}{l}{\textbf{10 Eilers and Marx style P-splines under gaussian with identity link in stagewise selection of length $\bm{5}$}}\\ \\}},
	    	every row no 10/.style={before row={\midrule \multicolumn{18}{l}{\textbf{8 Thin plate regression splines under gaussian with log link}}\\ \\}},
	    	every row no 11/.style={before row={\midrule \multicolumn{18}{l}{\textbf{8 Thin plate regression splines under gaussian with log link in stagewise selection of length $\bm{5}$}}\\ \\}},
	    	every row no 12/.style={before row={\midrule \multicolumn{18}{l}{\textbf{8 Thin plate regression splines under gamma with identity link}}\\ \\}},
	    	every row no 13/.style={before row={\midrule \multicolumn{18}{l}{\textbf{8 Thin plate regression splines under gamma with identity link in stagewise selection of length $\bm{5}$}}\\ \\}},
	    	every row no 14/.style={before row={\midrule \multicolumn{18}{l}{\textbf{4 Thin plate regression splines under gaussian with identity link in stagewise selection of length $\bm{5}$}}\\ \\}},
	    	every row no 15/.style={before row={\midrule \multicolumn{18}{l}{\textbf{4 Thin plate regression splines under gaussian with log link in stagewise selection of length $\bm{5}$}}\\ \\}},
	    	every row no 16/.style={before row={\midrule \multicolumn{18}{l}{\textbf{4 Thin plate regression splines under gamma with identity link in stagewise selection of length $\bm{5}$}}\\ \\}},
	    	every row no 17/.style={before row={\midrule \multicolumn{18}{l}{\textbf{4 Thin plate regression splines under gaussian with log link in stagewise selection of length $\bm{5}$}}\\ \\}},
	    	every row no 18/.style={before row={\midrule \multicolumn{18}{l}{\textbf{4 Thin plate regression splines under inverse gaussian with identity link in stagewise selection of length $\bm{5}$}}\\ \\}},
	    	every row no 19/.style={before row={\midrule \multicolumn{18}{l}{\textbf{4 Thin plate regression splines under inverse gaussian with log link in stagewise selection of length $\bm{5}$}}\\ \\}},
	    	every row no 20/.style={before row={\midrule \multicolumn{18}{l}{\textbf{4 Thin plate regression splines under inverse gaussian with }$\bm{\frac{1}{\mu^2}}$\textbf{ link in stagewise selection of length $\bm{5}$}}\\ \\}},
	    	every row no 21/.style={before row={\midrule \multicolumn{18}{l}{\textbf{8 Thin plate regression splines under gamma with log link in stagewise selection of length $\bm{5}$}}\\ \\}},
	    	every row no 22/.style={before row={\midrule \multicolumn{18}{l}{\textbf{8 Eilers and Marx style P-splines under inverse gaussian with }$\bm{\frac{1}{\mu^2}}$\textbf{ link in dynamic stagewise selection of proportion $\bm{0.25}$}}\\ \\}},
      	every last row/.style={after row=\bottomrule},
      	every row 3 column 2/.style={postproc cell content/.append style={/pgfplots/table/@cell content/.add={\cellcolor{green!26!white}}{},}},
      	every row 3 column 3/.style={postproc cell content/.append style={/pgfplots/table/@cell content/.add={\cellcolor{green!26!white}}{},}},
      	every row 10 column 4/.style={postproc cell content/.append style={/pgfplots/table/@cell content/.add={\cellcolor{green!26!white}}{},}},
      	every row 2 column 5/.style={postproc cell content/.append style={/pgfplots/table/@cell content/.add={\cellcolor{green!26!white}}{},}},
      	every row 7 column 6/.style={postproc cell content/.append style={/pgfplots/table/@cell content/.add={\cellcolor{green!26!white}}{},}},
      	every row 3 column 7/.style={postproc cell content/.append style={/pgfplots/table/@cell content/.add={\cellcolor{green!26!white}}{},}},
      	every row 3 column 8/.style={postproc cell content/.append style={/pgfplots/table/@cell content/.add={\cellcolor{green!26!white}}{},}},
      	every row 5 column 9/.style={postproc cell content/.append style={/pgfplots/table/@cell content/.add={\cellcolor{green!26!white}}{},}},
      	every row 3 column 10/.style={postproc cell content/.append style={/pgfplots/table/@cell content/.add={\cellcolor{green!26!white}}{},}},
      	every row 13 column 11/.style={postproc cell content/.append style={/pgfplots/table/@cell content/.add={\cellcolor{green!26!white}}{},}},
      	every row 6 column 12/.style={postproc cell content/.append style={/pgfplots/table/@cell content/.add={\cellcolor{green!26!white}}{},}},
      	every row 6 column 13/.style={postproc cell content/.append style={/pgfplots/table/@cell content/.add={\cellcolor{green!26!white}}{},}},
      	every row 21 column 14/.style={postproc cell content/.append style={/pgfplots/table/@cell content/.add={\cellcolor{green!26!white}}{},}},
      	every row 12 column 15/.style={postproc cell content/.append style={/pgfplots/table/@cell content/.add={\cellcolor{green!26!white}}{},}},
      	every row 11 column 16/.style={postproc cell content/.append style={/pgfplots/table/@cell content/.add={\cellcolor{green!26!white}}{},}},
      	every row 8 column 2/.style={postproc cell content/.append style={/pgfplots/table/@cell content/.add={\cellcolor{red!15!white}}{},}},
      	every row 8 column 3/.style={postproc cell content/.append style={/pgfplots/table/@cell content/.add={\cellcolor{red!15!white}}{},}},
      	every row 8 column 4/.style={postproc cell content/.append style={/pgfplots/table/@cell content/.add={\cellcolor{red!15!white}}{},}},
      	every row 8 column 5/.style={postproc cell content/.append style={/pgfplots/table/@cell content/.add={\cellcolor{red!15!white}}{},}},
      	every row 8 column 6/.style={postproc cell content/.append style={/pgfplots/table/@cell content/.add={\cellcolor{red!15!white}}{},}},
      	every row 8 column 7/.style={postproc cell content/.append style={/pgfplots/table/@cell content/.add={\cellcolor{red!15!white}}{},}},
      	every row 8 column 8/.style={postproc cell content/.append style={/pgfplots/table/@cell content/.add={\cellcolor{red!15!white}}{},}},
      	every row 15 column 9/.style={postproc cell content/.append style={/pgfplots/table/@cell content/.add={\cellcolor{red!15!white}}{},}},
      	every row 5 column 10/.style={postproc cell content/.append style={/pgfplots/table/@cell content/.add={\cellcolor{red!15!white}}{},}},
      	every row 5 column 11/.style={postproc cell content/.append style={/pgfplots/table/@cell content/.add={\cellcolor{red!15!white}}{},}},
      	every row 8 column 12/.style={postproc cell content/.append style={/pgfplots/table/@cell content/.add={\cellcolor{red!15!white}}{},}},
      	every row 8 column 13/.style={postproc cell content/.append style={/pgfplots/table/@cell content/.add={\cellcolor{red!15!white}}{},}},
      	every row 5 column 14/.style={postproc cell content/.append style={/pgfplots/table/@cell content/.add={\cellcolor{red!15!white}}{},}},
      	every row 5 column 15/.style={postproc cell content/.append style={/pgfplots/table/@cell content/.add={\cellcolor{red!15!white}}{},}},
      	every row 5 column 16/.style={postproc cell content/.append style={/pgfplots/table/@cell content/.add={\cellcolor{red!15!white}}{},}},
      ]{BGROGAMall.csv} 
    \caption{Maximum allowed numbers of smooth functions and out-of-sample validation figures of all derived GAMs of BEL under between $25$-$443$ and $150$-$443$ after the final iteration. Highlighted in green and red respectively the best and worst validation figures.}
    \label{tab:BGROGAMall}
  \end{center}
\end{table}

\begin{table}[htb]
  \tiny
  \begin{center}
  \tabcolsep=0.10cm
  \renewcommand{\arraystretch}{0.90}
      \begin{filecontents*}{BGROGLSvarmod.csv}
m,r1,r2,r3,r4,r5,r6,r7,r8,r9,r10,r11,r12,r13,r14,r15,BP.p.val,AIC,v.wre.y,ns.wre.y,cr.wre.y
0,0,0,0,0,0,0,0,0,0,0,0,0,0,0,0,1.00E-20,325849.7177,0.238115981,0.251724221,0.153979127
1,1,0,0,0,0,0,0,0,0,0,0,0,0,0,0,1.00E-20,322451.8114,0.238179183,0.245824536,0.122371466
2,0,0,0,0,0,0,0,1,0,0,0,0,0,0,0,1.00E-20,315980.418,0.239464156,0.255095858,0.152702071
3,0,0,0,1,0,0,0,0,0,0,0,0,0,0,0,1.00E-20,314076.805,0.236720123,0.225553408,0.165284096
4,0,0,0,0,0,0,0,0,0,0,0,0,0,0,1,1.00E-20,312280.2523,0.230831959,0.206131472,0.183760339
5,0,0,0,0,0,0,0,0,1,0,0,0,0,0,0,1.00E-20,312113.7315,0.231210486,0.20467789,0.184505738
6,0,0,0,0,1,0,0,0,0,0,0,0,0,0,0,1.00E-20,311949.2027,0.230636006,0.203356331,0.186451574
7,0,1,0,0,0,0,0,0,0,0,0,0,0,0,0,1.00E-20,311793.6232,0.231974966,0.20178665,0.187364886
8,0,0,0,0,0,0,0,0,0,0,0,1,0,0,0,1.00E-20,311699.6818,0.23494409,0.200099268,0.190262502
9,1,0,0,0,0,0,0,1,0,0,0,0,0,0,0,1.00E-20,311609.918,0.233096374,0.198175226,0.190322239
10,0,0,0,0,0,0,0,2,0,0,0,0,0,0,0,1.00E-20,311363.2967,0.226880727,0.194085946,0.194804319
11,0,0,0,0,0,1,0,0,0,0,0,0,0,0,0,1.00E-20,311292.688,0.228536767,0.193806187,0.19733015
12,0,0,0,0,2,0,0,0,0,0,0,0,0,0,0,1.00E-20,311236.6621,0.227807415,0.192914034,0.198068142
13,0,0,0,0,0,1,0,1,0,0,0,0,0,0,0,1.00E-20,311196.2298,0.229804523,0.193261495,0.198268388
14,0,0,0,0,0,0,1,0,0,0,0,0,0,0,0,1.53E-20,311160.6866,0.230909895,0.192814,0.199553771
15,1,0,0,0,0,0,0,0,0,0,0,0,0,0,1,7.05E-19,311135.9822,0.230671461,0.191185753,0.201684677
16,0,0,0,0,0,0,0,1,0,0,0,0,0,0,1,4.99E-15,311090.838,0.228070362,0.18887839,0.201270591
17,0,0,1,0,0,0,0,0,0,0,0,0,0,0,0,5.77E-13,311067.3948,0.228042446,0.18835867,0.203305529
18,0,0,0,0,0,0,2,0,0,0,0,0,0,0,0,8.31E-13,311048.4659,0.228009502,0.187429515,0.204456319
19,0,0,0,0,1,0,0,1,0,0,0,0,0,0,0,3.19E-12,311030.0101,0.228475679,0.187983328,0.204424989
20,1,0,0,0,1,0,0,0,0,0,0,0,0,0,0,2.72E-12,311003.1306,0.229510273,0.188249774,0.205077589
21,0,0,0,0,0,1,1,0,0,0,0,0,0,0,0,1.28E-11,310988.005,0.230051034,0.187574565,0.205662443
22,0,0,0,1,0,0,0,1,0,0,0,0,0,0,0,9.37E-11,310974.0209,0.229833019,0.186753543,0.207178324
\end{filecontents*}
\pgfplotstabletypeset[
        multicolumn names, 
        col sep=comma, 
        display columns/0/.style={column name=$\bm{m}$,dec sep align},
        display columns/1/.style={column type=C,column name=$\bm{r_{m}^{1}}$},
        display columns/2/.style={column type=C,column name=$\bm{r_{m}^{2}}$},
        display columns/3/.style={column type=C,column name=$\bm{r_{m}^{3}}$},
        display columns/4/.style={column type=C,column name=$\bm{r_{m}^{4}}$},
        display columns/5/.style={column type=C,column name=$\bm{r_{m}^{5}}$},
        display columns/6/.style={column type=C,column name=$\bm{r_{m}^{6}}$},
        display columns/7/.style={column type=C,column name=$\bm{r_{m}^{7}}$},
        display columns/8/.style={column type=C,column name=$\bm{r_{m}^{8}}$},
        display columns/9/.style={column type=C,column name=$\bm{r_{m}^{9}}$},
        display columns/10/.style={column type=C,column name=$\bm{r_{m}^{10}}$},
        display columns/11/.style={column type=C,column name=$\bm{r_{m}^{11}}$},
        display columns/12/.style={column type=C,column name=$\bm{r_{m}^{12}}$},
        display columns/13/.style={column type=C,column name=$\bm{r_{m}^{13}}$},
        display columns/14/.style={column type=C,column name=$\bm{r_{m}^{14}}$},
        display columns/15/.style={column type=C,column name=$\bm{r_{m}^{15}}$},
        display columns/16/.style={column name=$\bm{\textbf{\text{BP.p-val}}}$,sci subscript,precision=1,dec sep align},
        display columns/17/.style={column name={\textbf{AIC}},fixed,fixed zerofill,precision=0,set thousands separator={,},dec sep align},
        display columns/18/.style={column name={\textbf{v.mae}},fixed,fixed zerofill,precision=3,set thousands separator={,},dec sep align},
        display columns/19/.style={column name={\textbf{ns.mae}},fixed,fixed zerofill,precision=3,set thousands separator={,},dec sep align},
        display columns/20/.style={column name={\textbf{cr.mae}},fixed,fixed zerofill,precision=3,set thousands separator={,},dec sep align},
	    	every head row/.style={before row=\toprule,after row=\midrule},
	    	every row no 1/.style={before row=\midrule},
	    	every nth row={4[+3]}{before row=\midrule},
      	every last row/.style={after row=\bottomrule},
      ]{BGROGLSvarmod.csv} 
    \caption{FGLS variance models of BEL corresponding to $M_{\text{max}} \in \left\{2, 6, 10, 14, 18, 22\right\}$ derived by adaptive selection from the set of basis functions of the $150$-$443$ OLS proxy function given in Table \ref{tab:BGROOLS1} with exponents summing up to at max two. Furthermore, p-values of Breusch-Pagan test, AIC scores and out-of-sample MAEs in $\%$ after each iteration.}
    \label{tab:BGROGLSvarmod}
  \end{center}
\end{table}

\begin{table}[htb]
  \tiny
  \begin{center}
  \tabcolsep=0.10cm
  \renewcommand{\arraystretch}{0.90}
      \begin{filecontents*}{BGROGLSvarmod300886.csv}
m,r1,r2,r3,r4,r5,r6,r7,r8,r9,r10,r11,r12,r13,r14,r15,BP.p.val,AIC,v.wre.y,ns.wre.y,cr.wre.y
0,0,0,0,0,0,0,0,0,0,0,0,0,0,0,0,1.00E-20,325458.9411,0.194940084,0.275281611,0.174923745
1,1,0,0,0,0,0,0,0,0,0,0,0,0,0,0,1.00E-20,322076.9654,0.198694341,0.272558278,0.165921202
2,0,0,0,0,0,0,0,1,0,0,0,0,0,0,0,1.00E-20,315614.7932,0.195546773,0.275337354,0.174680263
3,0,0,0,1,0,0,0,0,0,0,0,0,0,0,0,1.00E-20,313658.6914,0.19506455,0.255471486,0.174741196
4,0,0,0,0,0,0,0,0,0,0,0,0,0,0,1,1.00E-20,311864.1658,0.198038725,0.238916421,0.181904756
5,0,0,0,0,0,0,0,0,1,0,0,0,0,0,0,1.00E-20,311703.6556,0.198365322,0.235972149,0.182257928
6,0,1,0,0,0,0,0,0,0,0,0,0,0,0,0,1.00E-20,311554.4018,0.199702422,0.239872557,0.183007473
7,2,0,0,0,0,0,0,0,0,0,0,0,0,0,0,1.00E-20,311453.8823,0.19918418,0.240896043,0.183183243
8,0,0,0,0,0,0,0,0,0,0,0,1,0,0,0,1.00E-20,311359.9791,0.199181118,0.237895617,0.185697451
9,0,0,0,0,0,1,0,0,0,0,0,0,0,0,0,1.00E-20,311318.0467,0.20140655,0.235939293,0.187962046
10,0,0,0,0,0,0,1,0,0,0,0,0,0,0,0,1.00E-20,311287.0343,0.202849686,0.234119668,0.189419206
11,0,0,0,0,0,1,0,1,0,0,0,0,0,0,0,1.00E-20,311260.3461,0.203462002,0.233151278,0.189472055
12,0,0,0,0,0,0,0,2,0,0,0,0,0,0,0,1.00E-20,311236.7797,0.202934187,0.232196758,0.189497438
13,1,0,0,0,0,0,0,1,0,0,0,0,0,0,0,3.70E-17,311001.1848,0.199862388,0.223345973,0.192132155
14,1,0,0,0,0,0,0,0,0,0,0,0,0,0,1,1.69E-16,310980.0188,0.20024496,0.221802173,0.194365881
15,0,0,0,0,0,0,0,1,0,0,0,0,0,0,1,7.60E-13,310934.4842,0.200227168,0.220137747,0.195836286
16,0,0,1,0,0,0,0,0,0,0,0,0,0,0,0,4.16E-11,310912.3112,0.199962741,0.21813351,0.196801854
17,0,0,0,0,0,0,2,0,0,0,0,0,0,0,0,1.26E-10,310894.8594,0.200434665,0.218777265,0.197678169
18,0,0,0,0,0,1,1,0,0,0,0,0,0,0,0,2.31E-10,310880.8341,0.199935116,0.217240717,0.198094352
19,0,0,0,0,0,0,0,0,0,0,0,0,0,2,0,7.61E-10,310867.1281,0.200020698,0.217764798,0.196716757
20,0,0,0,0,0,0,0,1,0,0,1,0,0,0,0,3.35E-09,310854.3702,0.200002572,0.218232316,0.195732696
21,0,0,0,0,0,0,0,0,0,0,0,0,0,1,0,9.94E-09,310843.2581,0.200315906,0.21793089,0.195711708
22,1,0,0,0,0,0,0,0,0,0,1,0,0,0,0,3.13E-08,310832.2798,0.200430635,0.217220372,0.195868093
\end{filecontents*}
\pgfplotstabletypeset[
        multicolumn names, 
        col sep=comma, 
        display columns/0/.style={column name=$\bm{m}$,dec sep align},
        display columns/1/.style={column type=C,column name=$\bm{r_{m}^{1}}$},
        display columns/2/.style={column type=C,column name=$\bm{r_{m}^{2}}$},
        display columns/3/.style={column type=C,column name=$\bm{r_{m}^{3}}$},
        display columns/4/.style={column type=C,column name=$\bm{r_{m}^{4}}$},
        display columns/5/.style={column type=C,column name=$\bm{r_{m}^{5}}$},
        display columns/6/.style={column type=C,column name=$\bm{r_{m}^{6}}$},
        display columns/7/.style={column type=C,column name=$\bm{r_{m}^{7}}$},
        display columns/8/.style={column type=C,column name=$\bm{r_{m}^{8}}$},
        display columns/9/.style={column type=C,column name=$\bm{r_{m}^{9}}$},
        display columns/10/.style={column type=C,column name=$\bm{r_{m}^{10}}$},
        display columns/11/.style={column type=C,column name=$\bm{r_{m}^{11}}$},
        display columns/12/.style={column type=C,column name=$\bm{r_{m}^{12}}$},
        display columns/13/.style={column type=C,column name=$\bm{r_{m}^{13}}$},
        display columns/14/.style={column type=C,column name=$\bm{r_{m}^{14}}$},
        display columns/15/.style={column type=C,column name=$\bm{r_{m}^{15}}$},
        display columns/16/.style={column name=$\bm{\textbf{\text{BP.p-val}}}$,sci subscript,precision=1,dec sep align},
        display columns/17/.style={column name={\textbf{AIC}},fixed,fixed zerofill,precision=0,set thousands separator={,},dec sep align},
        display columns/18/.style={column name={\textbf{v.mae}},fixed,fixed zerofill,precision=3,set thousands separator={,},dec sep align},
        display columns/19/.style={column name={\textbf{ns.mae}},fixed,fixed zerofill,precision=3,set thousands separator={,},dec sep align},
        display columns/20/.style={column name={\textbf{cr.mae}},fixed,fixed zerofill,precision=3,set thousands separator={,},dec sep align},
	    	every head row/.style={before row=\toprule,after row=\midrule},
	    	every row no 1/.style={before row=\midrule},
	    	every nth row={4[+3]}{before row=\midrule},
      	every last row/.style={after row=\bottomrule},
      ]{BGROGLSvarmod300886.csv} 
    \caption{FGLS variance models of BEL corresponding to $M_{\text{max}} \in \left\{2, 6, 10, 14, 18, 22\right\}$ derived by adaptive selection from the set of basis functions of the $300$-$886$ OLS proxy function given in Table \ref{tab:BGROOLS3008861} with exponents summing up to at max two. Furthermore, p-values of Breusch-Pagan test, AIC scores and out-of-sample MAEs in $\%$ after each iteration.}
    \label{tab:BGROGLS300886varmod}
  \end{center}
\end{table}

\begin{table}[htb]
  \tiny
  \begin{center}
  \tabcolsep=0.084cm
  \renewcommand{\arraystretch}{0.90}
      \begin{filecontents*}{BGROGLSvarmodvalfig.csv}
m;v.wre.y;v.a.wre.y;v.res.avg.y;v.wre.delta.y;v.res.avg.delta.y;ns.wre.y;ns.a.wre.y;ns.res.avg.y;ns.wre.delta.y;ns.res.avg.delta.y;cr.wre.y;cr.a.wre.y;cr.res.avg.y;cr.wre.delta.y;cr.res.avg.delta.y
0;0.238115981;0.227659528;-15.08300067;8.10267491;45.2475378;0.251724221;0.243215589;-29.59861963;3.984039395;16.34045604;0.153979127;0.150733792;2.705372037;7.37857712;48.64444771
1;0.238179183;0.227719955;-14.68484864;8.66848546;49.20254169;0.245824536;0.237515322;-30.47067152;4.120438564;19.02525601;0.122371466;0.119792309;2.500022003;7.872621436;51.99594953
2;0.239464156;0.228948501;-15.96767262;8.14671415;45.62210033;0.255095858;0.24647326;-30.31511496;4.032066156;16.88319519;0.152702071;0.149483652;2.213307802;7.489019611;49.41161796
3;0.236720123;0.226324967;-15.00599598;7.788807371;42.71151946;0.225553408;0.217929387;-23.74231801;4.423432511;19.58373463;0.165284096;0.161800492;10.24199567;8.116915844;53.56804831
4;0.230831959;0.220695372;-13.12664894;7.684266623;42.1259635;0.206131472;0.199163939;-18.4065654;4.816965125;22.45458424;0.183760339;0.179887321;16.9692705;8.756009062;57.83042014
5;0.231210486;0.221057276;-12.76192044;7.665532418;42.00311728;0.20467789;0.19775949;-18.03924295;4.803147094;22.33433196;0.184505738;0.18061701;17.3519339;8.740124611;57.72550882
6;0.230636006;0.220508024;-12.72065952;7.577363791;41.2791701;0.203356331;0.196482602;-17.69734558;4.762292695;21.91102124;0.186451574;0.182521835;17.42951698;8.636643019;57.0378838
7;0.231974966;0.221788185;-12.12307721;7.66069885;41.73774071;0.20178665;0.194965978;-17.31638361;4.786504158;22.1529715;0.187364886;0.183415898;17.92357256;8.691389838;57.39292767
8;0.23494409;0.224626926;-11.62646291;7.773871832;42.40878908;0.200099268;0.193335632;-16.55448237;4.913710742;23.08930682;0.190262502;0.186252442;19.21484857;8.911689805;58.85863776
9;0.233096374;0.222860349;-11.31799123;7.69157329;41.82883532;0.198175226;0.191476626;-16.23523731;4.837844455;22.52012644;0.190322239;0.18631092;19.10703858;8.762894607;57.86240232
10;0.226880727;0.216917652;-10.12762056;7.459517323;40.45964405;0.194085946;0.187525569;-14.84624408;4.707817397;21.34955772;0.194804319;0.190698534;20.16851621;8.5365753;56.36431802
11;0.228536767;0.218500969;-9.735881107;7.447269101;40.1454614;0.193806187;0.187255266;-14.53331535;4.685744263;20.95656436;0.19733015;0.193171129;20.26300274;8.454532708;55.75288245
12;0.227807415;0.217803645;-9.642365548;7.426287186;40.00647592;0.192914034;0.186393269;-14.29267942;4.686974648;20.96469924;0.198068142;0.193893567;20.42666488;8.444196789;55.68404355
13;0.229804523;0.219713054;-9.032487702;7.513140023;40.6180183;0.193261495;0.186728986;-14.18208705;4.695780085;21.07695615;0.198268388;0.194089592;20.76246048;8.490650945;56.02150368
14;0.230909895;0.220769886;-8.826750475;7.526752978;40.59534387;0.192814;0.186296616;-13.95586097;4.701114966;21.07477057;0.199553771;0.195347884;21.03448684;8.497021025;56.06511839
15;0.230671461;0.220541921;-8.70747815;7.522756806;40.58326872;0.191185753;0.184723407;-13.47829291;4.741780636;21.42099116;0.201684677;0.197433878;21.63828054;8.569190954;56.5375646
16;0.228070362;0.218055045;-9.069861665;7.437282988;40.10657469;0.18887839;0.182494036;-13.31865339;4.730083096;21.46632017;0.201270591;0.19702852;21.63314895;8.556989633;56.41812251
17;0.228042446;0.218028355;-8.949432412;7.420815759;39.95758212;0.18835867;0.181991883;-13.01832535;4.747260047;21.49722638;0.203305529;0.199020568;21.92586342;8.56797004;56.44141515
18;0.228009502;0.217996858;-8.79248094;7.433276412;40.04418546;0.187429515;0.181094134;-12.69102096;4.779954973;21.75418264;0.204456319;0.200147103;22.35519488;8.620893155;56.80039847
19;0.228475679;0.218442564;-8.895281113;7.434513168;40.06604677;0.187983328;0.181629228;-12.79171395;4.785820883;21.77815113;0.204424989;0.200116433;22.27113799;8.627662971;56.84100307
20;0.229510273;0.219431725;-8.88535888;7.442303825;40.08051517;0.188249774;0.181886668;-12.73011869;4.796356618;21.84429256;0.205077589;0.200755279;22.42210582;8.649557362;56.99651708
21;0.230051034;0.21994874;-8.592459234;7.465522246;40.18038247;0.187574565;0.181234282;-12.51091952;4.800445343;21.87045938;0.205662443;0.201327807;22.62861236;8.648409199;57.00999125
22;0.229833019;0.219740299;-8.468185373;7.435732001;39.93576495;0.186753543;0.180441011;-12.20558614;4.801847514;21.80690138;0.207178324;0.202811738;22.92395435;8.639152535;56.93644187
\end{filecontents*}
\pgfplotstabletypeset[
        multicolumn names, 
        col sep=semicolon, 
        display columns/0/.style={column name=$\bm{m}$,dec sep align},
        display columns/1/.style={column name={\textbf{v.mae}},column type={r},fixed,fixed zerofill,precision=3,set thousands separator={,}},
        display columns/2/.style={column name={$\bm{\textbf{\text{v.mae}}^a}$},column type={r},fixed,fixed zerofill,precision=3,set thousands separator={,}},
        display columns/3/.style={column name={\textbf{v.res}},column type={r},fixed,fixed zerofill,precision=0,set thousands separator={,}},
        display columns/4/.style={column name={$\bm{\textbf{\text{v.mae}}^0}$},column type={r},fixed,fixed zerofill,precision=3,set thousands separator={,}},
        display columns/5/.style={column name={$\bm{\textbf{\text{v.res}}^0}$},column type={r},fixed,fixed zerofill,precision=0,set thousands separator={,}},
        display columns/6/.style={column name={\textbf{ns.mae}},column type={r},fixed,fixed zerofill,precision=3,set thousands separator={,}},
        display columns/7/.style={column name={$\bm{\textbf{\text{ns.mae}}^a}$},column type={r},fixed,fixed zerofill,precision=3,set thousands separator={,}},
        display columns/8/.style={column name={\textbf{ns.res}},column type={r},fixed,fixed zerofill,precision=0,set thousands separator={,}},
        display columns/9/.style={column name={$\bm{\textbf{\text{ns.mae}}^0}$},column type={r},fixed,fixed zerofill,precision=3,set thousands separator={,}},
        display columns/10/.style={column name={$\bm{\textbf{\text{ns.res}}^0}$},column type={r},fixed,fixed zerofill,precision=0,set thousands separator={,}},
        display columns/11/.style={column name={\textbf{cr.mae}},column type={r},fixed,fixed zerofill,precision=3,set thousands separator={,}},
        display columns/12/.style={column name={$\bm{\textbf{\text{cr.mae}}^a}$},column type={r},fixed,fixed zerofill,precision=3,set thousands separator={,}},
        display columns/13/.style={column name={\textbf{cr.res}},column type={r},fixed,fixed zerofill,precision=0,set thousands separator={,}},
        display columns/14/.style={column name={$\bm{\textbf{\text{cr.mae}}^0}$},column type={r},fixed,fixed zerofill,precision=3,set thousands separator={,}},
        display columns/15/.style={column name={$\bm{\textbf{\text{cr.res}}^0}$},column type={r},fixed,fixed zerofill,precision=0,set thousands separator={,}},
	    	every head row/.style={before row=\toprule,after row=\midrule},
	    	every row no 1/.style={before row=\midrule},
	    	every nth row={4[+3]}{before row=\midrule},
      	every last row/.style={after row=\bottomrule},
      ]{BGROGLSvarmodvalfig.csv} 
    \caption{Iteration-wise out-of-sample validation figures in adaptive variance model selection of BEL corresponding to $M_{\text{max}} \in \left\{2, 6, 10, 14, 18, 22\right\}$ based on the $150$-$443$ OLS proxy function given in Table \ref{tab:BGROOLS1} with exponents summing up to at max two. Simultaneously type I FGLS regression results.}
    \label{tab:BGROGLSvarmodvalfig}
  \end{center}
\end{table}

\begin{table}[htb]
  \tiny
  \begin{center}
  \tabcolsep=0.084cm
  \renewcommand{\arraystretch}{0.90}
      \begin{filecontents*}{BGROGLSvarmod300886valfig.csv}
m;v.wre.y;v.a.wre.y;v.res.avg.y;v.wre.delta.y;v.res.avg.delta.y;ns.wre.y;ns.a.wre.y;ns.res.avg.y;ns.wre.delta.y;ns.res.avg.delta.y;cr.wre.y;cr.a.wre.y;cr.res.avg.y;cr.wre.delta.y;cr.res.avg.delta.y
0;0.194940084;0.186379627;-8.687718179;6.467518944;32.59585586;0.275281611;0.265976707;-30.18056667;4.600553042;-3.288455438;0.174923745;0.171236972;4.609340526;5.314810996;31.50145176
1;0.198694341;0.189969022;-9.103991102;6.647904047;34.03045663;0.272558278;0.263345427;-31.27127969;4.272142932;-2.528294763;0.165921202;0.16242417;0.829004845;5.005368855;29.57198977
2;0.195546773;0.186959674;-8.955140705;6.526855254;32.96659883;0.275337354;0.266030566;-30.30406755;4.563555133;-2.773790814;0.174680263;0.170998621;4.815708167;5.400683209;32.3459849
3;0.19506455;0.186498627;-8.902372709;6.487445029;33.09943925;0.255471486;0.246836191;-26.53131705;4.350004963;1.079032105;0.174741196;0.17105827;9.112629637;5.916464835;36.72297879
4;0.198038725;0.189342196;-9.254579566;6.305424825;31.85592437;0.238916421;0.230840711;-23.10833259;4.261759232;3.610708541;0.181904756;0.178070848;12.93328022;6.302787142;39.65232135
5;0.198365322;0.189654452;-8.812495647;6.297704461;31.87399702;0.235972149;0.22799596;-22.47030308;4.252332733;3.824726778;0.182257928;0.178416577;13.62979832;6.335728713;39.92482818
6;0.199702422;0.190932835;-8.642398393;6.398950397;32.59446597;0.239872557;0.231764529;-22.94088039;4.292205031;3.904521175;0.183007473;0.179150324;13.44771598;6.388883111;40.29311754
7;0.19918418;0.19043735;-8.76574103;6.364222254;32.31997242;0.240896043;0.232753419;-23.14156071;4.304135305;3.552689937;0.183183243;0.179322388;13.04542602;6.323726275;39.73967667
8;0.199181118;0.190434423;-8.299146592;6.380891839;32.27523242;0.237895617;0.229854412;-22.33772874;4.313197109;3.845187461;0.185697451;0.181783606;14.14815792;6.406636661;40.33107413
9;0.20140655;0.192562129;-8.118843828;6.432412398;32.57366809;0.235939293;0.227964214;-21.79985691;4.312814799;4.501192202;0.187962046;0.184000472;14.8112948;6.520907899;41.11234392
10;0.202849686;0.193941892;-7.878868813;6.472880497;32.76602658;0.234119668;0.226206095;-21.25250339;4.310388404;5.000929199;0.189419206;0.18542692;15.65410122;6.621107298;41.9075338
11;0.203462002;0.194527319;-7.537476045;6.491554576;32.93712307;0.233151278;0.225270438;-20.9912577;4.302825335;5.09187861;0.189472055;0.185478655;15.93701764;6.62845525;42.02015395
12;0.202934187;0.194022683;-7.500157391;6.475882369;32.934436;0.232196758;0.224348182;-20.80992089;4.294407016;5.233209702;0.189497438;0.185503503;16.13475201;6.641098664;42.1778826
13;0.199862388;0.191085776;-7.328324458;6.254377187;31.5628433;0.223345973;0.215796566;-19.17265856;4.25227505;5.327046398;0.192132155;0.18808269;17.41013215;6.614766233;41.90983711
14;0.20024496;0.191451548;-7.322763741;6.246241625;31.4636737;0.221802173;0.214304948;-18.70252607;4.256668744;5.69244856;0.194365881;0.190269337;18.10142042;6.697070344;42.49639505
15;0.200227168;0.191434537;-7.45124739;6.215590882;31.41930924;0.220137747;0.212696783;-18.34562243;4.243488905;6.133471397;0.195836286;0.19170875;18.55636427;6.773081141;43.0354581
16;0.199962741;0.191181723;-7.340974931;6.180325686;31.14642629;0.21813351;0.210760292;-17.94685108;4.238912551;6.149087336;0.196801854;0.192653968;18.72577142;6.753139094;42.82170984
17;0.200434665;0.191632923;-7.413525518;6.196679247;31.28364163;0.218777265;0.211382287;-17.97580117;4.249018951;6.329903169;0.197678169;0.193511813;18.88087798;6.804256269;43.18658232
18;0.199935116;0.19115531;-7.138702583;6.193747164;31.24508551;0.217240717;0.209897676;-17.60746776;4.249709354;6.38485753;0.198094352;0.193919224;19.20066472;6.801082331;43.19299001
19;0.200020698;0.191237134;-7.230191467;6.206838445;31.39730528;0.217764798;0.210404042;-17.8325131;4.237739488;6.403520843;0.196716757;0.192570664;18.88963674;6.786637999;43.12567067
20;0.200002572;0.191219804;-7.214819967;6.228879477;31.67578759;0.218232316;0.210855758;-18.00815128;4.226001944;6.490993475;0.195732696;0.191607344;18.70586261;6.792700826;43.20500736
21;0.200315906;0.191519378;-7.176278411;6.24049988;31.78250316;0.21793089;0.210564521;-17.93687014;4.223673092;6.630448624;0.195711708;0.191586798;18.82509498;6.814140698;43.39241375
22;0.200430635;0.191629069;-6.995540169;6.255897995;31.9533266;0.217220372;0.209878019;-17.74662837;4.223366448;6.810775596;0.195868093;0.191739887;19.05816812;6.843713346;43.61557209
\end{filecontents*}
\pgfplotstabletypeset[
        multicolumn names, 
        col sep=semicolon, 
        display columns/0/.style={column name=$\bm{m}$,dec sep align},
        display columns/1/.style={column name={\textbf{v.mae}},column type={r},fixed,fixed zerofill,precision=3,set thousands separator={,}},
        display columns/2/.style={column name={$\bm{\textbf{\text{v.mae}}^a}$},column type={r},fixed,fixed zerofill,precision=3,set thousands separator={,}},
        display columns/3/.style={column name={\textbf{v.res}},column type={r},fixed,fixed zerofill,precision=0,set thousands separator={,}},
        display columns/4/.style={column name={$\bm{\textbf{\text{v.mae}}^0}$},column type={r},fixed,fixed zerofill,precision=3,set thousands separator={,}},
        display columns/5/.style={column name={$\bm{\textbf{\text{v.res}}^0}$},column type={r},fixed,fixed zerofill,precision=0,set thousands separator={,}},
        display columns/6/.style={column name={\textbf{ns.mae}},column type={r},fixed,fixed zerofill,precision=3,set thousands separator={,}},
        display columns/7/.style={column name={$\bm{\textbf{\text{ns.mae}}^a}$},column type={r},fixed,fixed zerofill,precision=3,set thousands separator={,}},
        display columns/8/.style={column name={\textbf{ns.res}},column type={r},fixed,fixed zerofill,precision=0,set thousands separator={,}},
        display columns/9/.style={column name={$\bm{\textbf{\text{ns.mae}}^0}$},column type={r},fixed,fixed zerofill,precision=3,set thousands separator={,}},
        display columns/10/.style={column name={$\bm{\textbf{\text{ns.res}}^0}$},column type={r},fixed,fixed zerofill,precision=0,set thousands separator={,}},
        display columns/11/.style={column name={\textbf{cr.mae}},column type={r},fixed,fixed zerofill,precision=3,set thousands separator={,}},
        display columns/12/.style={column name={$\bm{\textbf{\text{cr.mae}}^a}$},column type={r},fixed,fixed zerofill,precision=3,set thousands separator={,}},
        display columns/13/.style={column name={\textbf{cr.res}},column type={r},fixed,fixed zerofill,precision=0,set thousands separator={,}},
        display columns/14/.style={column name={$\bm{\textbf{\text{cr.mae}}^0}$},column type={r},fixed,fixed zerofill,precision=3,set thousands separator={,}},
        display columns/15/.style={column name={$\bm{\textbf{\text{cr.res}}^0}$},column type={r},fixed,fixed zerofill,precision=0,set thousands separator={,}},
	    	every head row/.style={before row=\toprule,after row=\midrule},
	    	every row no 1/.style={before row=\midrule},
	    	every nth row={4[+3]}{before row=\midrule},
      	every last row/.style={after row=\bottomrule},
      ]{BGROGLSvarmod300886valfig.csv} 
    \caption{Iteration-wise out-of-sample validation figures in adaptive variance model selection of BEL corresponding to $M_{\text{max}} \in \left\{2, 6, 10, 14, 18, 22\right\}$ based on the $300$-$886$ OLS proxy function given in Table \ref{tab:BGROOLS3008861} with exponents summing up to at max two. Simultaneously type I FGLS regression results.}
    \label{tab:BGROGLSvarmod300886valfig}
  \end{center}
\end{table}

\begin{table}[htb]
  \tiny
  \begin{center}
  \tabcolsep=0.052cm
  \renewcommand{\arraystretch}{0.90}
      \begin{filecontents*}{BGROGLSvalfig1.csv}
k;AIC;v.wre.y;v.a.wre.y;v.res.avg.y;v.wre.delta.y;v.res.avg.delta.y;ns.wre.y;ns.a.wre.y;ns.res.avg.y;ns.wre.delta.y;ns.res.avg.delta.y;cr.wre.y;cr.a.wre.y;cr.res.avg.y;cr.wre.delta.y;cr.res.avg.delta.y
0;437251.0284;4.556602029;4.356506712;-237.5325499;100;38.252875;3.230529781;3.121333346;-0.294727358;100;261.0992348;4.026996295;3.942121469;105.619188;100;367.0131501
10;336389.8192;1.786493749;1.708042958;184.014688;44.08178365;197.8641292;1.401624177;1.354247315;209.1535989;39.15164025;208.6115773;2.290392599;2.242119231;344.2035212;52.033224;343.6614996
20;323883.4565;0.825821311;0.789556793;24.99981902;22.00689618;111.0708207;0.423596777;0.409278612;-28.06417497;10.76354526;43.61536392;0.437154292;0.427940627;27.79617707;16.42363861;99.47571596
30;319957.7126;0.465177282;0.444749824;2.579286012;12.87629255;55.32594776;0.287800024;0.27807198;1.673728897;9.649710426;40.02892784;0.466803658;0.45696509;57.25757948;15.2335059;95.61277843
40;318945.0397;0.401383703;0.38375763;-15.86086729;11.03641503;51.08315063;0.356951391;0.344885933;-37.00587748;7.157898752;15.54667764;0.329536522;0.322591059;2.619761287;10.12722675;55.17231641
50;318206.1487;0.354995983;0.339406947;-23.70882164;9.269891157;34.87461063;0.335766134;0.324416768;-36.12955736;6.611049425;8.062412104;0.339100224;0.331953192;-8.260823214;8.602253018;35.93114625
60;317484.8412;0.322816728;0.308640788;-24.88610365;8.407261975;36.06118395;0.308758554;0.298322082;-35.78686173;5.548107514;10.76896308;0.278547877;0.272677074;-10.92540645;7.243767247;35.63041836
70;317196.9675;0.306295625;0.292845181;-27.91968592;7.631489736;28.31630261;0.34544659;0.333770011;-42.64578977;5.405261151;-0.80126404;0.271695313;0.265968938;-16.82195242;5.898761343;25.0225733
80;316262.5255;0.272205603;0.260252163;-23.86305798;6.946355037;32.23293038;0.320395484;0.309565667;-42.10398485;4.051365554;-0.399459301;0.226522799;0.2217485;-17.14134928;4.898145573;24.56317627
90;316020.5793;0.260422535;0.248986529;-23.15256197;7.142863718;38.52898424;0.297631309;0.287570954;-37.17930935;3.853967455;10.11077406;0.173132138;0.169483125;-4.820598608;6.460587271;42.4694848
100;315871.3835;0.256175863;0.244926342;-22.82657023;7.424014179;40.6825409;0.294012186;0.284074162;-35.27514124;4.0775734;13.84250709;0.186085329;0.182163309;-0.168405309;7.442566777;48.94924302
110;315783.8468;0.256180389;0.244930669;-22.24822803;7.396046455;41.39204528;0.301845517;0.291642716;-36.91514111;3.961868881;12.3336694;0.188594052;0.184619157;-3.230426401;7.012717193;46.01838411
120;315718.7052;0.256523734;0.245258936;-23.21950756;6.923359376;38.18709938;0.296065557;0.286058127;-35.97975336;3.870402617;11.03539078;0.180801588;0.17699093;-1.877711416;6.87229993;45.13743272
130;315675.2945;0.258439609;0.24709068;-24.65179312;6.506280048;35.00984587;0.295237112;0.285257684;-36.01548727;3.760211822;9.254688911;0.187741526;0.183784599;-3.031671376;6.460503069;42.23850481
140;315648.5352;0.252229827;0.241153589;-23.31969563;6.424495108;34.12355989;0.28326172;0.273687077;-33.95934021;3.748626767;9.092452507;0.183876848;0.180001375;-1.224229002;6.398805656;41.82756372
150;315628.8046;0.239469443;0.228953555;-20.5349129;6.467171748;33.96135026;0.26082306;0.252006875;-29.66946586;3.79628727;10.4353345;0.177220593;0.17348541;3.489977964;6.654014631;43.59477832
0;437251.0284;4.556602029;4.356506712;-237.5325499;100;38.252875;3.230529781;3.121333346;-0.294727358;100;261.0992347;4.026996295;3.942121469;105.619188;100;367.0131501
10;332478.5798;2.014081236;1.925636333;258.9438625;49.09828338;213.2773044;2.000195692;1.932586274;298.4738237;44.74521434;238.4158027;2.963625165;2.901162438;445.3778874;58.340649;385.3198665
20;320872.833;0.881075251;0.842384351;50.7439594;22.82074933;115.1407532;0.340858164;0.329336681;16.25364721;13.42804829;66.25897826;0.621945457;0.608837049;83.74523587;20.78984926;133.7505669
30;316187.4947;0.428933329;0.410097462;18.66232844;10.87455902;31.98828397;0.307745695;0.297343459;29.09686015;8.537009775;28.03135288;0.560753494;0.548934796;72.98600944;12.63327443;71.92050217
40;315132.4741;0.36634235;0.350255058;6.302404912;10.24342371;45.08322502;0.254399414;0.245800357;0.595591892;7.853060997;24.9849492;0.40146874;0.3930072;36.15810224;11.22100318;60.54745954
50;314473.1201;0.302750787;0.289456008;3.370148694;9.34635955;45.54238659;0.229319335;0.22156802;0.244203935;7.543078265;28.02497902;0.360883959;0.353277803;34.04009237;10.77586736;61.82086746
60;313643.3287;0.306945581;0.293466596;-17.70799715;7.567266417;27.79261526;0.250801573;0.242324128;-20.58336708;5.807550788;10.52578254;0.266322038;0.260708913;9.470790236;7.67630176;40.57993985
70;313301.2101;0.280337598;0.268027056;-17.16336951;7.767931707;30.31494747;0.221713265;0.214219046;-11.60392207;6.228691729;21.48293211;0.267698074;0.262055946;23.40944218;9.315157756;56.49629636
80;313059.685;0.269706645;0.257862943;-20.37485848;7.092263873;28.19682948;0.23024931;0.22246656;-12.63230878;6.272580771;21.54791638;0.279568628;0.273676311;25.09780907;9.554044961;59.27803423
90;312882.6357;0.262023271;0.250516971;-22.33296337;6.754121363;29.33683235;0.239285587;0.231197399;-17.36557772;5.976790945;19.9127552;0.253424048;0.248082766;19.05526363;9.077365277;56.33359656
100;312099.8218;0.245801318;0.235007377;-18.60403175;6.176713368;28.58519437;0.201926951;0.195101537;-14.43889845;4.81406119;18.35886487;0.221004061;0.216346077;21.13294782;8.305129658;53.93071113
110;311655.6116;0.230739441;0.220606916;-15.54520397;6.445758488;33.09228636;0.188567967;0.182194106;-12.38012005;4.827050709;21.86590748;0.210858662;0.206414508;24.75436904;8.963851966;59.00039657
120;311573.5985;0.235699597;0.225349256;-15.74552594;6.544573;33.95432219;0.209441479;0.202362063;-16.4264567;4.593535739;18.88192862;0.206798237;0.202439663;21.55288754;8.636819553;56.86127287
130;311511.2139;0.237517364;0.227087198;-16.59676082;6.55124472;34.92535486;0.206858234;0.199866135;-15.80207905;4.797036662;21.32857383;0.204277183;0.199971743;22.97039689;9.103672402;60.10104976
140;311460.8045;0.231165445;0.221014213;-16.44690654;6.025924691;31.0560735;0.189008699;0.18261994;-11.83824266;4.72621171;21.27327457;0.216157595;0.211601758;25.26939354;8.852950001;58.38091076
150;311425.5391;0.224394554;0.214540654;-14.32024407;5.904249516;30.78273055;0.176692675;0.170720215;-8.566368431;4.756383755;22.14514339;0.22588917;0.221128226;28.73867462;9.005130545;59.45018644
0;437251.0284;4.556602029;4.356506712;-237.5325499;100;38.252875;3.230529781;3.121333346;-0.294727358;100;261.0992348;4.026996295;3.942121469;105.619188;100;367.0131501
10;328519.0081;2.119935198;2.026841901;287.8657336;50.52443017;221.1344256;2.206256896;2.13168232;329.2227526;46.56271412;248.0999818;3.194136988;3.126815888;480.0195386;60.39630536;398.8967678
20;319481.4898;0.970931262;0.928294491;95.45700051;24.18477677;105.246041;0.43909023;0.424248365;53.37865409;11.8392992;48.77623175;0.820677067;0.8033801;116.5605804;18.0856144;111.958158
30;316528.8986;0.65543461;0.62665233;56.3124325;16.56020786;74.42396668;0.419778235;0.405589143;57.20133594;12.30147527;60.92140732;0.780411444;0.763963134;113.2894277;18.28542569;117.009499
40;314460.0063;0.378608582;0.36198264;19.27285635;10.08875924;42.45190036;0.26829657;0.259227769;19.0225915;8.120161121;27.8101727;0.472715432;0.462752264;54.10740934;11.6082901;62.89499054
50;313842.098;0.324334149;0.310091575;2.236164488;8.422048121;32.52988074;0.228807002;0.221073005;-4.378432783;6.420482616;11.52382066;0.338587951;0.331451716;20.2616819;8.600326721;36.16393535
60;313021.9133;0.296590936;0.283566657;-12.64145613;7.619457513;31.21647302;0.222707161;0.215179347;-12.56824178;6.12336466;16.89822456;0.277078149;0.271238323;14.02123356;8.292178556;43.4876999
70;312692.2015;0.281625555;0.269258455;-16.72050238;7.493674458;26.4026059;0.220639368;0.213181449;-4.64287446;6.762361309;24.08877102;0.326245931;0.319369822;35.1894591;10.46667586;63.92110458
80;312443.2182;0.271044871;0.259142403;-18.81494247;7.171332552;27.37099928;0.218049488;0.21067911;-6.752078776;6.624677215;25.04240018;0.303404845;0.297010145;33.00003414;10.30552273;64.79451309
90;312264.4893;0.260787247;0.249335225;-21.13029786;6.610063215;26.8875839;0.222021553;0.214516913;-10.75726371;6.300482585;22.86915525;0.278284474;0.272419222;27.87852392;9.805828299;61.50494288
100;312186.798;0.261617992;0.250129489;-21.45734684;6.567775577;26.15035416;0.215629587;0.208341004;-9.859981618;6.265494498;23.35625659;0.271593165;0.265868943;27.88054803;9.707153465;61.09678624
110;312108.4811;0.255560789;0.244338278;-21.31908688;6.031360644;22.84690676;0.20277718;0.195923028;-5.012594072;6.323966666;24.76193677;0.288210581;0.282136122;31.35998889;9.753663171;61.13451973
120;312042.7264;0.261380523;0.249902448;-22.8958606;5.989033895;20.46221277;0.200403668;0.193629743;-4.154745166;6.287246728;24.8118654;0.292964314;0.286789663;33.19551606;9.856896812;62.16212662
130;311077.6834;0.226116533;0.216187015;-17.50501334;5.465547124;24.97382891;0.160068746;0.154658198;-4.10456619;5.115201038;23.98281325;0.243983419;0.238841112;32.38594383;9.191897406;60.47332328
140;310918.3543;0.219538953;0.209898279;-15.94468982;5.450716703;25.47792642;0.153409098;0.148223656;-4.064209984;4.819603228;22.96694345;0.232699491;0.227795009;31.2125271;8.858616008;58.24368053
150;310868.4044;0.212448988;0.203119657;-14.43870438;5.375228628;25.34555252;0.14849644;0.143477052;-0.325865963;5.097903576;25.06692814;0.255822562;0.250430728;35.94142527;9.296046661;61.33421937
\end{filecontents*}
\pgfplotstabletypeset[
        multicolumn names, 
        col sep=semicolon, 
        display columns/0/.style={column name=$\bm{k}$,dec sep align},
        display columns/1/.style={column name={\textbf{AIC}},fixed,fixed zerofill,precision=0,set thousands separator={,},dec sep align},
        display columns/2/.style={column name={\textbf{v.mae}},column type={r},fixed,fixed zerofill,precision=3,set thousands separator={,}},
        display columns/3/.style={column name={$\bm{\textbf{\text{v.mae}}^a}$},column type={r},fixed,fixed zerofill,precision=3,set thousands separator={,}},
        display columns/4/.style={column name={\textbf{v.res}},column type={r},fixed,fixed zerofill,precision=0,set thousands separator={,}},
        display columns/5/.style={column name={$\bm{\textbf{\text{v.mae}}^0}$},column type={r},fixed,fixed zerofill,precision=3,set thousands separator={,}},
        display columns/6/.style={column name={$\bm{\textbf{\text{v.res}}^0}$},column type={r},fixed,fixed zerofill,precision=0,set thousands separator={,}},
        display columns/7/.style={column name={\textbf{ns.mae}},column type={r},fixed,fixed zerofill,precision=3,set thousands separator={,}},
        display columns/8/.style={column name={$\bm{\textbf{\text{ns.mae}}^a}$},column type={r},fixed,fixed zerofill,precision=3,set thousands separator={,}},
        display columns/9/.style={column name={\textbf{ns.res}},column type={r},fixed,fixed zerofill,precision=0,set thousands separator={,}},
        display columns/10/.style={column name={$\bm{\textbf{\text{ns.mae}}^0}$},column type={r},fixed,fixed zerofill,precision=3,set thousands separator={,}},
        display columns/11/.style={column name={$\bm{\textbf{\text{ns.res}}^0}$},column type={r},fixed,fixed zerofill,precision=0,set thousands separator={,}},
        display columns/12/.style={column name={\textbf{cr.mae}},column type={r},fixed,fixed zerofill,precision=3,set thousands separator={,}},
        display columns/13/.style={column name={$\bm{\textbf{\text{cr.mae}}^a}$},column type={r},fixed,fixed zerofill,precision=3,set thousands separator={,}},
        display columns/14/.style={column name={\textbf{cr.res}},column type={r},fixed,fixed zerofill,precision=0,set thousands separator={,}},
        display columns/15/.style={column name={$\bm{\textbf{\text{cr.mae}}^0}$},column type={r},fixed,fixed zerofill,precision=3,set thousands separator={,}},
        display columns/16/.style={column name={$\bm{\textbf{\text{cr.res}}^0}$},column type={r},fixed,fixed zerofill,precision=0,set thousands separator={,}},
	    	every head row/.style={before row=\toprule,after row=\midrule},
	    	every row no 0/.style={before row={\multicolumn{18}{l}{\textbf{$\bm{M}_{\text{max}} \bm{= 2}$ in variance model selection}}\\ \\}},
	    	every row no 16/.style={before row={\midrule \multicolumn{18}{l}{\textbf{$\bm{M}_{\text{max}} \bm{= 6}$ in variance model selection}}\\ \\}},
	    	every row no 32/.style={before row={\midrule \multicolumn{18}{l}{\textbf{$\bm{M}_{\text{max}} \bm{= 10}$ in variance model selection}}\\ \\}},
      	every last row/.style={after row=\bottomrule},
      ]{BGROGLSvalfig1.csv} 
    \caption{AIC scores and out-of-sample validation figures of type II FGLS proxy functions of BEL under $150$-$443$ with variance models of varying complexity $M_{\text{max}}$ after each tenth iteration.}
    \label{tab:BGROGLSvalfig1}
  \end{center}
\end{table}

\begin{table}[htb]
\ContinuedFloat
  \tiny
  \begin{center}
  \tabcolsep=0.052cm
  \renewcommand{\arraystretch}{0.90}
      \begin{filecontents*}{BGROGLSvalfig2.csv}
k;AIC;v.wre.y;v.a.wre.y;v.res.avg.y;v.wre.delta.y;v.res.avg.delta.y;ns.wre.y;ns.a.wre.y;ns.res.avg.y;ns.wre.delta.y;ns.res.avg.delta.y;cr.wre.y;cr.a.wre.y;cr.res.avg.y;cr.wre.delta.y;cr.res.avg.delta.y
0;437251.0284;4.556602029;4.356506712;-237.5325499;100;38.252875;3.230529781;3.121333346;-0.294727358;100;261.0992347;4.026996295;3.942121469;105.619188;100;367.0131501
10;326307.761;2.120120829;2.02701938;290.4868417;50.30554296;220.0830522;2.21539127;2.14050794;330.5858051;46.12928692;245.7905529;3.197165751;3.129780815;480.4747054;59.90917703;395.6794532
20;319199.224;1.024266873;0.979287961;99.51834404;26.04903211;136.9274096;0.526962421;0.509150353;75.3941261;18.6394227;98.41172887;1.044493672;1.022479443;154.9209742;27.14242046;177.938577
30;316093.0447;0.701529418;0.670722964;66.63633332;17.57371242;78.94757651;0.503020931;0.486018118;72.50323105;13.7451207;70.42301144;0.901302025;0.882305769;133.2293295;20.20790842;131.1491099
40;314154.5723;0.39321423;0.375946906;23.75426504;10.36346817;44.34701924;0.282463129;0.272915479;25.20197961;8.425846545;31.40327101;0.504996091;0.494352561;61.90672322;12.13144248;68.10801462
50;313561.6307;0.327380056;0.313003726;5.744299701;8.560893432;33.95605853;0.224579312;0.216988216;0.946788944;6.53478662;14.76708497;0.351949338;0.344531491;27.15157907;8.936132529;40.9718751
60;312810.5819;0.29774515;0.284670185;-10.11133864;7.607592299;29.2112768;0.203021789;0.196159368;3.696209581;7.086215723;28.62736222;0.336149752;0.329064905;36.93246451;10.28276529;61.86361714
70;312455.1573;0.288872331;0.276187001;-15.13254118;7.409079067;25.62171138;0.218743174;0.211349348;-1.628602451;6.863402551;24.7341873;0.342643287;0.33542158;38.26682141;10.61202712;64.62961117
80;312235.2423;0.273227446;0.261229134;-17.07394096;7.221694802;27.52256003;0.215009983;0.207742344;-4.09625482;6.737715784;26.10878336;0.322453813;0.315657628;37.08281408;10.66223038;67.28785227
90;312057.3117;0.264441828;0.252829322;-21.66174063;6.679818145;26.96258354;0.221522063;0.214034306;-10.33401508;6.406270672;23.89884628;0.282865687;0.276903879;28.37396849;9.981033661;62.60682985
100;311952.5334;0.255130742;0.243927115;-20.52075919;6.117448059;23.86805065;0.201184174;0.194383867;-4.643538488;6.38110556;25.35380855;0.289961075;0.283849722;31.16340527;9.780112887;61.16075231
110;311897.7479;0.252323441;0.241243093;-20.43526601;5.928715543;21.74495482;0.200158695;0.193393051;-3.960371723;6.235881979;23.8283863;0.293301016;0.287119269;31.7212288;9.582859282;59.50998683
120;311831.6985;0.262676599;0.251141609;-22.73094759;5.962095877;19.0821238;0.198270773;0.191568943;-2.685018582;6.30023498;24.73659;0.30288287;0.296499171;34.47083917;9.877999514;61.89244776
130;310916.1511;0.222937461;0.213147547;-16.59542212;5.36294082;23.08700134;0.154361579;0.149143941;-0.624098365;5.232650473;24.66686229;0.263016999;0.257473532;35.89573302;9.305286704;61.18669368
140;310756.8189;0.215451357;0.205990182;-14.76297749;5.338870744;23.85868245;0.146999798;0.142030999;-0.47986124;4.954405246;23.7503359;0.251249207;0.245953763;34.70532348;8.971631393;58.93552062
150;310713.9804;0.214291071;0.204880848;-14.34572911;5.36813261;24.84780827;0.146033164;0.141097038;-1.323952364;4.856784706;23.47812221;0.244280977;0.239132399;33.79524888;8.905620897;58.59732346
0;437251.0284;4.556602029;4.356506712;-237.5325499;100;38.252875;3.230529781;3.121333346;-0.294727358;100;261.0992347;4.026996295;3.942121469;105.619188;100;367.0131501
10;326124.9119;2.127230172;2.033816529;291.7627205;50.42466897;220.2290411;2.226276366;2.151025104;332.2101042;46.22207626;246.284962;3.209493429;3.141848669;482.3273267;60.01860459;396.4021845
20;318761.6797;1.036016183;0.990521319;110.9602635;25.66835544;113.4813077;0.538474643;0.520273445;75.42791371;13.42906173;63.55749511;0.982621519;0.961911336;144.4235254;20.70817131;132.5531068
30;315994.5333;0.710116601;0.678933056;69.4168981;17.74058286;79.69845429;0.523043268;0.505363673;75.83506659;13.96293692;71.72515998;0.925470186;0.905964551;137.0263816;20.46492122;132.916475
40;314059.7831;0.400608072;0.383016059;26.51096388;10.5289066;45.30786316;0.291525667;0.281671691;28.37946748;8.560279774;32.78490396;0.521459238;0.510468723;65.59314977;12.34126497;69.99858625
50;313482.8762;0.329083486;0.314632352;8.570489932;8.686658207;34.99440047;0.224730468;0.217134263;4.227641535;6.619740338;16.26008927;0.362051523;0.354420758;31.01201864;9.119867306;43.04446637
60;312937.8625;0.315711504;0.301847578;-5.330174024;7.840463057;30.24290365;0.209484762;0.202403883;4.518028737;6.855042519;25.69964361;0.347379298;0.340057772;41.23724987;10.29694452;62.41886474
70;312363.2689;0.269955039;0.258100429;-9.956575579;6.959913819;21.44480305;0.214574974;0.207322039;10.53194821;7.088706504;27.54186404;0.389140404;0.380938702;48.35629054;10.79519592;65.36620637
80;312166.0993;0.259302845;0.247916008;-12.24625081;6.557983632;22.46841523;0.20441456;0.197505062;8.833853355;7.008398316;29.15705659;0.368606662;0.360837739;46.77510439;10.7178979;67.09830763
90;311963.1465;0.233653169;0.223392693;-14.84332046;6.140720349;23.98241138;0.195728449;0.189112553;1.231881545;6.432219109;25.66615058;0.312913652;0.30631854;36.98572988;9.843928146;61.41999892
100;311883.4332;0.241240911;0.230647233;-17.63408724;6.030644316;23.68798888;0.193683494;0.18713672;-0.567410403;6.449235855;26.36320292;0.299093583;0.292789749;34.35002075;9.77732652;61.28063408
110;311829.6846;0.239175427;0.228672451;-17.56975476;5.836347927;21.60106277;0.193055838;0.18653028;0.128702005;6.298267778;24.90805673;0.3027647;0.296383492;35.01180947;9.610113001;59.79116419
120;311766.0547;0.244398383;0.233666049;-18.80881117;5.71254722;17.9797875;0.190835759;0.184385243;3.236802332;6.339902594;25.6339382;0.32091768;0.314153872;39.49744247;9.865951692;61.89457834
130;311044.9821;0.225338554;0.2154432;-15.12679517;5.395746361;23.14105753;0.148416299;0.14339962;0.220091925;5.060813615;24.09648182;0.259450494;0.253982196;34.82028768;8.949782762;58.69667757
140;310693.8336;0.213233337;0.203869563;-13.49014412;5.313723697;23.8962708;0.138802469;0.134110751;0.599227545;4.854685968;23.59417966;0.244696009;0.239538684;33.93371897;8.671922313;56.92867108
150;310643.6246;0.210977111;0.201712415;-13.80651418;5.13118427;22.59981315;0.139387229;0.134675746;1.078389506;4.816062937;23.09325403;0.250107662;0.244836278;34.55367771;8.618033812;56.56854223
0;437251.0284;4.556602029;4.356506712;-237.5325499;100;38.252875;3.230529781;3.121333346;-0.294727358;100;261.0992347;4.026996295;3.942121469;105.619188;100;367.0131501
10;325988.0378;2.127006695;2.033602865;291.7846017;50.41391699;220.4317946;2.226384616;2.151129696;332.2262576;46.2590535;246.4819877;3.210158881;3.142500096;482.4273319;60.0611318;396.683062
20;318925.859;1.033721272;0.988327186;104.6005566;26.16000815;136.9227785;0.569487285;0.550237817;82.7522026;19.04262541;100.6829617;1.097968677;1.074827385;163.2409711;27.62085215;181.1717301
30;315805.0957;0.712160583;0.68088728;70.58399723;17.76253197;79.10160437;0.537359164;0.519195672;78.25320227;14.06281898;72.37934661;0.942722084;0.922852839;139.7746431;20.60291121;133.9007874
40;313972.8085;0.40894227;0.390984275;28.60601616;10.72957272;45.96603063;0.300668986;0.290505953;31.41875196;8.709307179;34.38730363;0.538780813;0.527425221;69.52112168;12.58942223;72.48967335
50;313410.9605;0.348953279;0.333629598;6.757119206;8.949886746;33.96233948;0.223217774;0.2156727;3.372299404;6.618356081;16.18605687;0.356929956;0.349407136;29.6066754;9.081435659;42.42043287
60;312872.8678;0.308034165;0.294507376;-2.16208782;8.204777262;36.91207424;0.203284101;0.196412813;7.867863831;7.490424569;32.55056309;0.350155634;0.342775592;42.65639607;10.85322138;67.33909533
70;312286.496;0.271445942;0.259525862;-8.981090623;6.950063725;21.15163701;0.216973613;0.2096396;11.98821222;7.124493271;27.72947705;0.397539338;0.389160616;50.03461995;10.85609341;65.77588479
80;312090.6783;0.260637668;0.249192214;-11.36780519;6.556890497;22.1367763;0.206509489;0.199529179;10.28915106;7.0513787;29.40226975;0.377401034;0.369446756;48.47839941;10.79281186;67.5915181
90;311893.3609;0.234814317;0.224502852;-15.37686195;6.042682711;22.73382363;0.195647233;0.189034083;1.112042503;6.366716402;24.83126529;0.313518399;0.306910541;36.23423252;9.682917001;59.9534553
100;311814.5884;0.238258719;0.227795999;-16.61988753;5.969924517;22.61193741;0.193733293;0.187184836;1.317551082;6.461571528;26.15791322;0.310553663;0.304008292;36.65100126;9.828822224;61.4913634
110;311760.6615;0.237412336;0.226986783;-17.28093104;5.780039048;20.68360374;0.194175875;0.187612458;1.5146555;6.363799039;25.08772748;0.313126065;0.306526476;36.73794648;9.694263995;60.31101846
120;311696.7031;0.243086774;0.232412038;-18.92240649;5.818039983;18.46906115;0.191434057;0.184963317;2.475970298;6.325122919;25.47597513;0.320146068;0.313398522;39.09628486;9.885181585;62.09628969
130;311654.8729;0.232117943;0.221924884;-17.01430594;5.688449078;17.8959617;0.194529515;0.187954145;8.219546211;6.71391994;28.73835104;0.353025582;0.345585053;46.47904097;10.50886812;66.9978458
140;310747.9174;0.215081002;0.205636091;-14.35498605;5.20570769;22.50903957;0.148062246;0.143057535;5.025874945;5.578097436;27.49843776;0.293370551;0.287187338;41.99710121;9.788236307;64.46966402
150;310590.368;0.207914061;0.198783874;-13.02535308;5.208757852;22.6445265;0.138639092;0.133952896;4.582323843;5.192722181;25.86074062;0.275429488;0.269624409;39.70091037;9.255833578;60.97932715
\end{filecontents*}
\pgfplotstabletypeset[
        multicolumn names, 
        col sep=semicolon, 
        display columns/0/.style={column name=$\bm{k}$,dec sep align},
        display columns/1/.style={column name={\textbf{AIC}},fixed,fixed zerofill,precision=0,set thousands separator={,},dec sep align},
        display columns/2/.style={column name={\textbf{v.mae}},column type={r},fixed,fixed zerofill,precision=3,set thousands separator={,}},
        display columns/3/.style={column name={$\bm{\textbf{\text{v.mae}}^a}$},column type={r},fixed,fixed zerofill,precision=3,set thousands separator={,}},
        display columns/4/.style={column name={\textbf{v.res}},column type={r},fixed,fixed zerofill,precision=0,set thousands separator={,}},
        display columns/5/.style={column name={$\bm{\textbf{\text{v.mae}}^0}$},column type={r},fixed,fixed zerofill,precision=3,set thousands separator={,}},
        display columns/6/.style={column name={$\bm{\textbf{\text{v.res}}^0}$},column type={r},fixed,fixed zerofill,precision=0,set thousands separator={,}},
        display columns/7/.style={column name={\textbf{ns.mae}},column type={r},fixed,fixed zerofill,precision=3,set thousands separator={,}},
        display columns/8/.style={column name={$\bm{\textbf{\text{ns.mae}}^a}$},column type={r},fixed,fixed zerofill,precision=3,set thousands separator={,}},
        display columns/9/.style={column name={\textbf{ns.res}},column type={r},fixed,fixed zerofill,precision=0,set thousands separator={,}},
        display columns/10/.style={column name={$\bm{\textbf{\text{ns.mae}}^0}$},column type={r},fixed,fixed zerofill,precision=3,set thousands separator={,}},
        display columns/11/.style={column name={$\bm{\textbf{\text{ns.res}}^0}$},column type={r},fixed,fixed zerofill,precision=0,set thousands separator={,}},
        display columns/12/.style={column name={\textbf{cr.mae}},column type={r},fixed,fixed zerofill,precision=3,set thousands separator={,}},
        display columns/13/.style={column name={$\bm{\textbf{\text{cr.mae}}^a}$},column type={r},fixed,fixed zerofill,precision=3,set thousands separator={,}},
        display columns/14/.style={column name={\textbf{cr.res}},column type={r},fixed,fixed zerofill,precision=0,set thousands separator={,}},
        display columns/15/.style={column name={$\bm{\textbf{\text{cr.mae}}^0}$},column type={r},fixed,fixed zerofill,precision=3,set thousands separator={,}},
        display columns/16/.style={column name={$\bm{\textbf{\text{cr.res}}^0}$},column type={r},fixed,fixed zerofill,precision=0,set thousands separator={,}},
	    	every head row/.style={before row=\toprule,after row=\midrule},
	    	every row no 0/.style={before row={\multicolumn{18}{l}{\textbf{$\bm{M}_{\text{max}} \bm{= 14}$ in variance model selection}}\\ \\}},
	    	every row no 16/.style={before row={\midrule \multicolumn{18}{l}{\textbf{$\bm{M}_{\text{max}} \bm{= 18}$ in variance model selection}}\\ \\}},
	    	every row no 32/.style={before row={\midrule \multicolumn{18}{l}{\textbf{$\bm{M}_{\text{max}} \bm{= 22}$ in variance model selection}}\\ \\}},
      	every last row/.style={after row=\bottomrule},
      ]{BGROGLSvalfig2.csv} 
    \caption[]{Cont.}
    \label{tab:BGROGLSvalfig2}
  \end{center}
\end{table}

\begin{table}[htb]
  \tiny
  \begin{center}
  \tabcolsep=0.052cm
  \renewcommand{\arraystretch}{0.8}
      \begin{filecontents*}{BGROGLS300886valfig1.csv}
k;AIC;v.wre.y;v.a.wre.y;v.res.avg.y;v.wre.delta.y;v.res.avg.delta.y;ns.wre.y;ns.a.wre.y;ns.res.avg.y;ns.wre.delta.y;ns.res.avg.delta.y;cr.wre.y;cr.a.wre.y;cr.res.avg.y;cr.wre.delta.y;cr.res.avg.delta.y
0;437251.0284;4.556602029;4.356506712;-237.5325499;100;38.252875;3.230529781;3.121333346;-0.294727358;100;261.0992348;4.026996295;3.942121469;105.619188;100;367.0131501
10;336389.8192;1.786493815;1.708043021;184.0147205;44.0817856;197.8641309;1.401624454;1.354247583;209.1536401;39.15164221;208.6115877;2.290392903;2.242119529;344.2035669;52.03322624;343.6615144
20;323883.4565;0.82582082;0.789556323;24.99974648;22.00688946;111.0707903;0.423597109;0.409278933;-28.06428753;10.7635378;43.61529347;0.437153986;0.427940328;27.79603439;16.423626;99.47561539
30;319957.7126;0.465177262;0.444749805;2.579281146;12.87629192;55.32594493;0.28780002;0.278071976;1.673721953;9.649709786;40.02892293;0.466803609;0.456965041;57.25756917;15.2335048;95.61277015
40;318945.0397;0.401383692;0.38375762;-15.86087164;11.03641471;51.08314927;0.356951402;0.344885944;-37.00588102;7.15789865;15.54667708;0.329536521;0.322591057;2.619757202;10.12722656;55.1723153
50;318206.1487;0.354996013;0.339406975;-23.70882845;9.269890935;34.87460099;0.335766169;0.324416802;-36.12956532;6.611049571;8.062401322;0.339100269;0.331953236;-8.260839947;8.602252339;35.93112669
60;317484.8412;0.322816761;0.30864082;-24.88610411;8.407261924;36.06117813;0.308758558;0.298322086;-35.7868613;5.548107907;10.76895814;0.278547931;0.272677126;-10.92541441;7.243767205;35.63040502
70;317196.9675;0.306295639;0.292845194;-27.91968499;7.631489839;28.31630139;0.345446588;0.333770009;-42.64578845;5.405261359;-0.801264874;0.271695323;0.265968948;-16.82195117;5.898761465;25.02257241
80;316262.5255;0.272205601;0.260252161;-23.86305765;6.946355029;32.23293065;0.32039548;0.309565664;-42.10398424;4.051365517;-0.399458744;0.226522795;0.221748496;-17.14134846;4.898145612;24.56317704
90;316020.5793;0.260422538;0.248986532;-23.15256343;7.142863769;38.52898513;0.297631324;0.287570968;-37.17931162;3.853967604;10.11077414;0.173132136;0.169483123;-4.820599518;6.460587453;42.46948624
100;315871.3835;0.256175866;0.244926345;-22.82657009;7.424014299;40.68254141;0.294012187;0.284074163;-35.27514082;4.077573525;13.84250787;0.186085333;0.182163313;-0.168404497;7.442566966;48.9492442
110;315783.8468;0.256180389;0.244930669;-22.24822802;7.396046457;41.39204529;0.301845517;0.291642716;-36.9151411;3.961868884;12.3336694;0.188594052;0.184619157;-3.230426376;7.012717196;46.01838413
120;315718.7052;0.256523759;0.245258961;-23.21950922;6.923359614;38.18709941;0.29606556;0.286058129;-35.97975189;3.870402917;11.03539393;0.180801603;0.176990945;-1.877708182;6.872300705;45.13743765
130;315675.2945;0.258439612;0.247090683;-24.65179333;6.506280119;35.00984613;0.295237113;0.285257684;-36.01548711;3.760211876;9.254689546;0.187741528;0.183784601;-3.031670721;6.460503233;42.23850594
140;315640.9944;0.249977349;0.239000025;-22.8960722;6.441381003;34.41527801;0.284290073;0.27468067;-34.17386717;3.741016693;8.746020228;0.182056652;0.178219542;-1.544336437;6.337876374;41.37555096
150;315621.6471;0.238246041;0.227783877;-20.38146457;6.432532664;33.89415036;0.258439034;0.249703432;-29.05345796;3.821071093;10.83069417;0.177290298;0.173553647;4.351938555;6.740372482;44.23609068
160;315599.383;0.232942382;0.222713119;-19.72486317;6.578189148;34.9056911;0.255512467;0.246875787;-28.08450644;3.920453788;12.15458502;0.183132866;0.179273074;5.659671253;6.987884176;45.89876271
170;315572.9395;0.232177768;0.221982081;-19.40425244;6.615729617;34.97748335;0.254326647;0.24573005;-27.94621624;3.88020842;12.04405675;0.181374694;0.177551957;5.450339607;6.926913606;45.4406126
180;315535.2043;0.224748643;0.214879194;-18.65495233;6.502319649;34.84728045;0.251876234;0.243362465;-27.89805338;3.773496886;11.2127166;0.172197824;0.168568503;5.306800228;6.796506747;44.41757021
190;315523.0107;0.229245487;0.219178567;-18.93348735;6.808704752;36.81024333;0.244377167;0.236116877;-26.117431;4.019676392;15.23483687;0.164424061;0.160958584;8.884209031;7.607250084;50.2364769
200;315506.5162;0.215385241;0.205926969;-18.33734252;6.738491414;36.04051972;0.243445382;0.235216587;-25.86093924;3.969101039;14.12546019;0.164309414;0.160846353;8.725208136;7.386655909;48.71160757
210;315499.5135;0.21432758;0.204915754;-18.05912877;6.703566698;34.97935174;0.233903171;0.225996916;-24.39912989;3.988614597;14.24788782;0.162154446;0.158736804;9.664142423;7.3229347;48.31116013
220;315491.7798;0.216734404;0.207216887;-18.16269217;6.769065992;35.42174205;0.238869473;0.23079535;-25.53328458;3.929801953;13.65968684;0.158815847;0.155468571;8.799098579;7.277277373;47.99206999
224;315490.539;0.208567853;0.199408956;-16.55318111;6.583869232;33.88579153;0.226415996;0.218762818;-22.05215693;3.998696206;13.99535292;0.164737188;0.161265111;12.01007566;7.289529666;48.0575855
0;437251.0284;4.556602029;4.356506712;-237.5325499;100;38.252875;3.230529781;3.121333346;-0.294727358;100;261.0992347;4.026996295;3.942121469;105.619188;100;367.0131501
10;332478.5798;2.014076687;1.925631984;258.9427186;49.09820923;213.2772164;2.000186143;1.932577048;298.4723987;44.74514508;238.4154337;2.963614648;2.901152142;445.3763069;58.34056958;385.3193419
20;320872.8331;0.88107517;0.842384274;50.74394356;22.82074733;115.140743;0.340858146;0.329336662;16.25362304;13.42804555;66.25895971;0.621945282;0.608836877;83.74520014;20.78984496;133.7505368
30;316187.4947;0.428933342;0.410097474;18.66233317;10.87455932;31.98828658;0.307745716;0.29734348;29.09686706;8.537010199;28.03135767;0.560753541;0.548934842;72.98601874;12.63327525;71.92050935
40;315132.4741;0.366342399;0.350255104;6.302449029;10.2434224;45.08321182;0.254399348;0.245800293;0.59567547;7.853062516;24.98497545;0.401468988;0.393007443;36.15819257;11.22100576;60.54749256
50;314473.1201;0.302750759;0.289455982;3.370255056;9.346360686;45.54239753;0.229319229;0.221567918;0.244374616;7.543084899;28.02505428;0.360884444;0.353278277;34.04029568;10.77587872;61.82097535
60;313643.3287;0.306945577;0.293466591;-17.70799968;7.567265944;27.79261334;0.250801551;0.242324107;-20.58336183;5.807550687;10.52578838;0.266322019;0.260708894;9.470799423;7.676302343;40.57994964
70;313301.2101;0.280337601;0.268027059;-17.16338172;7.767930975;30.31493567;0.221713279;0.214219059;-11.6039313;6.228690965;21.48292328;0.267698044;0.262055917;23.40942783;9.315156115;56.49628241
80;313059.685;0.269706632;0.257862931;-20.37487001;7.092263217;28.19681792;0.230249307;0.222466557;-12.63231252;6.272580338;21.5479126;0.27956862;0.273676303;25.09779819;9.554043812;59.27802331
90;312882.6357;0.262023317;0.250517015;-22.33297523;6.754122278;29.33684174;0.239285649;0.231197459;-17.36560072;5.976790306;19.91275345;0.253423968;0.248082688;19.05523874;9.077364501;56.33359291
100;312099.8218;0.245801293;0.235007354;-18.60400281;6.176714185;28.58517702;0.201926775;0.195101367;-14.43881517;4.814068005;18.35890186;0.221004511;0.216346518;21.13305487;8.305139503;53.93077189
110;311655.6116;0.23073943;0.220606906;-15.5451965;6.445758055;33.09227855;0.188567917;0.182194057;-12.38010497;4.827050781;21.86590729;0.210858716;0.20641456;24.75438138;8.963851563;59.00039364
120;311573.5985;0.235699604;0.225349262;-15.74553012;6.544571953;33.95431456;0.20944148;0.202362064;-16.42645841;4.593535215;18.88192347;0.206798241;0.202439666;21.55288206;8.636818223;56.86126394
130;311506.6311;0.23355378;0.223297669;-15.718624;6.705924671;35.59433894;0.206220296;0.199249761;-15.62626016;4.801369066;21.29523998;0.204114335;0.199812328;23.09413439;9.093625019;60.01563454
140;311455.8116;0.226137722;0.216207274;-15.55813342;6.102105149;31.74203719;0.188556355;0.182182886;-11.73902593;4.717000368;21.16968188;0.215379018;0.210839591;25.27940565;8.826710753;58.18811346
150;311419.1109;0.223517207;0.213701834;-14.86913473;5.898928111;31.48008912;0.178008782;0.171991835;-9.927546758;4.711685619;22.03021429;0.213444351;0.2089457;27.29500112;8.971363535;59.25276216
160;311354.8117;0.216827784;0.207306166;-14.53636192;5.536226348;28.97993891;0.159508445;0.154116836;-4.17759442;5.012604648;24.94724361;0.24568611;0.240507916;33.07269244;9.420193493;62.19753047
170;311307.9932;0.197803989;0.189117768;-13.28763138;5.089753799;23.42729662;0.141396977;0.136617561;-3.618456548;4.143838592;18.70500865;0.220934461;0.216277945;26.69524899;7.491082599;49.01871419
180;311265.9158;0.201556371;0.192705371;-14.44545458;5.112490023;24.34922203;0.131855341;0.127398445;-2.770051143;4.432745524;21.63316266;0.218086272;0.213489786;27.26714924;7.867682633;51.67036305
190;311248.0984;0.2075698;0.19845473;-15.80699236;5.287231645;23.48449418;0.143158645;0.138319683;-5.445277498;4.162569179;19.45474624;0.212623229;0.208141884;25.09546028;7.630425622;49.99548401
200;311227.8903;0.201915943;0.193049153;-14.4876308;5.269368564;23.51392899;0.137232128;0.13259349;-3.979850777;4.147744988;19.63024621;0.213157618;0.20866501;26.56373544;7.638970725;50.17383242
210;311196.0674;0.192497387;0.184044196;-13.77998832;5.032330423;19.69797573;0.124932889;0.120709983;3.86879964;4.655401076;22.95530089;0.253066836;0.247733083;32.41442201;7.918602438;51.50092326
220;311164.0417;0.195393534;0.186813164;-15.45429438;5.079394749;21.10536857;0.121708985;0.11759505;1.179299487;4.619921918;23.34749963;0.237242091;0.232241868;30.58747531;8.069913571;52.75567546
230;311148.2801;0.193705961;0.185199698;-14.71669179;5.145850108;22.08723915;0.12196857;0.117845861;0.643404785;4.570970773;23.05587293;0.235562525;0.230597701;29.45905665;7.949287;51.8715248
237;311143.6358;0.196349469;0.187727121;-15.26973808;5.341548115;23.45931655;0.125412693;0.121173568;-0.038382688;4.765150534;24.29920913;0.23522153;0.230263893;29.60425834;8.242727;53.94185017
0;437251.0284;4.556602029;4.356506712;-237.5325499;100;38.252875;3.230529781;3.121333346;-0.294727358;100;261.0992347;4.026996295;3.942121469;105.619188;100;367.0131501
10;331055.5098;2.073279573;1.982235077;273.1800112;50.08506259;215.8653026;2.11285496;2.041437501;315.2851;45.71421544;243.5789286;3.089946804;3.024821664;464.3616866;59.45132755;392.6555152
20;320198.7206;0.924417363;0.883823169;75.8619375;23.1326498;100.8110692;0.374561235;0.361900541;24.67090109;10.92067735;35.22857;0.655300941;0.641489517;81.80581806;15.99863135;92.36348697
30;316044.1087;0.542607901;0.518780211;30.62495165;14.06823775;56.03217162;0.3717458;0.359180271;44.83216733;11.72870466;55.8479245;0.742387738;0.726740832;107.4576184;18.44986128;118.4733756
40;314821.3823;0.384948914;0.368044547;10.95688141;10.62578113;47.34392063;0.256284714;0.247621931;5.770712826;8.118141526;27.76628925;0.42366592;0.414736542;42.51930844;11.68450591;64.51488486
50;314200.8595;0.327023383;0.312662715;2.427845291;9.206000933;40.93550692;0.239742614;0.231638977;-7.579106877;6.713389479;16.53709195;0.336245724;0.329158854;20.92860846;9.102593263;45.04480729
60;313385.7594;0.26918503;0.257364234;-4.83722905;7.830715004;34.06226131;0.220406828;0.212956768;6.012638803;7.505716803;30.52066635;0.364514398;0.356831725;46.04723923;11.22310508;70.55526678
70;312985.8896;0.290434582;0.277680649;-16.63400504;7.315794185;25.93378208;0.210038263;0.202938675;-3.517448996;6.645533104;24.65887533;0.310439061;0.303896104;32.85868397;9.955441542;61.03500829
80;312722.0036;0.280077336;0.267778223;-17.85002471;7.424534057;31.37780925;0.222561849;0.215038946;-7.870115793;6.791560969;26.96625536;0.299697723;0.293381156;32.97434146;10.65213264;67.81071261
90;312545.4012;0.270406323;0.258531896;-22.00438821;7.109566149;32.15415124;0.232711772;0.224845788;-13.31972449;6.634498173;26.44735215;0.273008032;0.267253989;26.8294523;10.44991511;66.59652894
100;312469.3584;0.264574479;0.252956147;-20.69624336;6.800441953;29.05306417;0.224310131;0.216728134;-10.84330466;6.419740188;24.51454007;0.273753594;0.267983837;29.0015568;10.1275498;64.35940153
110;312396.748;0.254197835;0.243035176;-18.75212915;6.135561523;25.13070528;0.201742356;0.194923182;-4.112898312;6.36017237;25.37847332;0.290195928;0.284079625;33.35306789;9.939913235;62.84443952
120;312345.8453;0.247298655;0.236438962;-18.84074954;5.939930761;21.86528761;0.193094908;0.18656803;0.613120611;6.468277361;26.92769496;0.307182296;0.300707981;37.55472191;10.07844097;63.86929626
130;312299.377;0.240460306;0.229900906;-17.11674679;5.784027375;20.88007404;0.191774634;0.185292382;4.394237737;6.562720093;27.99959577;0.329281066;0.322340987;42.61451938;10.36859315;66.21987741
140;312274.4705;0.246691612;0.235858576;-18.07488307;5.81149368;21.67837239;0.192779815;0.186263587;5.353629508;6.870408784;30.71542216;0.337634823;0.330518676;45.36309286;10.94435794;70.72488552
150;312242.9392;0.248932018;0.238000598;-18.66952943;5.950283142;23.93147982;0.192781841;0.186265545;2.734050655;6.871585298;30.9435971;0.324136598;0.317304946;42.83965553;10.98444964;71.04920197
160;312221.8781;0.255427315;0.244210665;-18.65728879;6.162082325;25.45375206;0.198026819;0.191333235;0.55327305;6.858635559;30.2728511;0.324449434;0.317611189;41.97850873;11.09192253;71.69808678
170;311203.8274;0.228013558;0.218000736;-14.06938827;5.956530556;31.20330028;0.161147393;0.155700386;-1.294739088;5.873995268;29.58648666;0.276059622;0.270241262;39.81035451;10.70329622;70.69158026
180;311039.7255;0.222752149;0.212970373;-12.88661743;6.020634919;31.18800254;0.154339326;0.14912244;-1.179356407;5.593627383;28.50380076;0.264610607;0.259033552;38.7128217;10.35572298;68.39597887
190;310996.415;0.222355297;0.212590948;-12.61494976;6.15153553;32.09083189;0.154462967;0.149241902;-1.870357794;5.583589283;28.44396105;0.257946907;0.252510299;37.78784632;10.31123713;68.10216517
200;310968.0267;0.206023505;0.196976338;-9.683184995;6.163042874;32.25674346;0.14410158;0.139230745;3.088391706;5.924426734;30.63685736;0.284999477;0.278992697;42.2475005;10.56769276;69.79596615
210;310953.1156;0.211393881;0.202110884;-10.36010171;5.929593006;29.99905285;0.143201448;0.138361038;2.652669807;5.615429602;28.62036157;0.276240391;0.270418222;41.08911154;10.1529606;67.0568033
220;310927.1869;0.207901179;0.198771557;-11.302775;6.353140659;33.42200716;0.146892049;0.141926892;-1.398256348;5.602219573;28.93506301;0.252069672;0.246756936;37.14477585;10.22533169;67.47809521
230;310918.7399;0.210849963;0.20159085;-10.92571213;6.454048644;33.80200619;0.148621571;0.143597953;-0.929421024;5.702278179;29.4068345;0.258663509;0.253211798;38.19202498;10.37575455;68.5282805
240;310907.7303;0.210106645;0.200880174;-10.85702176;6.558973141;34.50928236;0.152447991;0.147295035;-2.571926695;5.57025797;28.40291462;0.250700255;0.245416381;36.47549487;10.2178046;67.45033619
244;310904.5235;0.208492259;0.199336681;-10.81124475;6.576542185;34.60242988;0.152654553;0.147494615;-2.350457835;5.617254132;28.67175399;0.252094994;0.246781724;36.70675993;10.2588368;67.72897176
\end{filecontents*}
\pgfplotstabletypeset[
        multicolumn names, 
        col sep=semicolon, 
        display columns/0/.style={column name=$\bm{k}$,dec sep align},
        display columns/1/.style={column name={\textbf{AIC}},fixed,fixed zerofill,precision=0,set thousands separator={,},dec sep align},
        display columns/2/.style={column name={\textbf{v.mae}},column type={r},fixed,fixed zerofill,precision=3,set thousands separator={,}},
        display columns/3/.style={column name={$\bm{\textbf{\text{v.mae}}^a}$},column type={r},fixed,fixed zerofill,precision=3,set thousands separator={,}},
        display columns/4/.style={column name={\textbf{v.res}},column type={r},fixed,fixed zerofill,precision=0,set thousands separator={,}},
        display columns/5/.style={column name={$\bm{\textbf{\text{v.mae}}^0}$},column type={r},fixed,fixed zerofill,precision=3,set thousands separator={,}},
        display columns/6/.style={column name={$\bm{\textbf{\text{v.res}}^0}$},column type={r},fixed,fixed zerofill,precision=0,set thousands separator={,}},
        display columns/7/.style={column name={\textbf{ns.mae}},column type={r},fixed,fixed zerofill,precision=3,set thousands separator={,}},
        display columns/8/.style={column name={$\bm{\textbf{\text{ns.mae}}^a}$},column type={r},fixed,fixed zerofill,precision=3,set thousands separator={,}},
        display columns/9/.style={column name={\textbf{ns.res}},column type={r},fixed,fixed zerofill,precision=0,set thousands separator={,}},
        display columns/10/.style={column name={$\bm{\textbf{\text{ns.mae}}^0}$},column type={r},fixed,fixed zerofill,precision=3,set thousands separator={,}},
        display columns/11/.style={column name={$\bm{\textbf{\text{ns.res}}^0}$},column type={r},fixed,fixed zerofill,precision=0,set thousands separator={,}},
        display columns/12/.style={column name={\textbf{cr.mae}},column type={r},fixed,fixed zerofill,precision=3,set thousands separator={,}},
        display columns/13/.style={column name={$\bm{\textbf{\text{cr.mae}}^a}$},column type={r},fixed,fixed zerofill,precision=3,set thousands separator={,}},
        display columns/14/.style={column name={\textbf{cr.res}},column type={r},fixed,fixed zerofill,precision=0,set thousands separator={,}},
        display columns/15/.style={column name={$\bm{\textbf{\text{cr.mae}}^0}$},column type={r},fixed,fixed zerofill,precision=3,set thousands separator={,}},
        display columns/16/.style={column name={$\bm{\textbf{\text{cr.res}}^0}$},column type={r},fixed,fixed zerofill,precision=0,set thousands separator={,}},
	    	every head row/.style={before row=\toprule,after row=\midrule},
	    	every row no 0/.style={before row={\multicolumn{18}{l}{\textbf{$\bm{M}_{\text{max}} \bm{= 2}$ in variance model selection}}\\ \\}},
	    	every row no 24/.style={before row={\midrule \multicolumn{18}{l}{\textbf{$\bm{M}_{\text{max}} \bm{= 6}$ in variance model selection}}\\ \\}},
	    	every row no 49/.style={before row={\midrule \multicolumn{18}{l}{\textbf{$\bm{M}_{\text{max}} \bm{= 10}$ in variance model selection}}\\ \\}},
      	every last row/.style={after row=\bottomrule},
      ]{BGROGLS300886valfig1.csv} 
    \caption{AIC scores and out-of-sample validation figures of type II FGLS proxy functions of BEL under $300$-$886$ with variance models of varying complexity $M_{\text{max}}$ after each tenth and the final iteration.}
    \label{tab:BGROGLS300886valfig1}
  \end{center}
\end{table}

\begin{table}[htb]
\ContinuedFloat
  \tiny
  \begin{center}
  \tabcolsep=0.052cm
  \renewcommand{\arraystretch}{0.88}
      \begin{filecontents*}{BGROGLS300886valfig2.csv}
k;AIC;v.wre.y;v.a.wre.y;v.res.avg.y;v.wre.delta.y;v.res.avg.delta.y;ns.wre.y;ns.a.wre.y;ns.res.avg.y;ns.wre.delta.y;ns.res.avg.delta.y;cr.wre.y;cr.a.wre.y;cr.res.avg.y;cr.wre.delta.y;cr.res.avg.delta.y
0;437251.0284;4.556602029;4.356506712;-237.5325499;100;38.252875;3.230529781;3.121333346;-0.294727358;100;261.0992347;4.026996295;3.942121469;105.619188;100;367.0131501
10;327048.7716;2.13301253;2.039344965;292.2856061;50.56087205;222.2291804;2.232943358;2.157466743;333.2049682;46.68603641;248.7570797;3.221860259;3.153954851;484.185832;60.52366638;399.7379434
20;318964.9079;1.020469942;0.975657765;107.8026094;25.28751252;110.722646;0.506917103;0.489782594;68.86913048;12.75911427;57.39770429;0.931383241;0.91175298;135.737885;19.63425756;124.2664588
30;316261.9278;0.693601217;0.663142916;64.54332291;17.38557287;77.58145833;0.484094197;0.467731134;69.0840223;13.34104124;67.73069491;0.871863641;0.853487842;128.4561309;19.64296994;127.1028035
40;314271.6289;0.392136388;0.374916395;22.79808431;10.37300276;44.33667502;0.277491131;0.268111542;22.77219618;8.321893414;29.91932409;0.493397274;0.482998207;58.78634636;11.9410623;65.93347427
50;313690.9901;0.348655482;0.333344878;1.128604495;8.772356391;32.24517898;0.227816239;0.22011573;-4.781243477;6.439908537;11.9438682;0.335384436;0.328315719;19.46204302;8.633154697;36.1871547
60;312859.8077;0.288840403;0.276156475;-9.556971355;7.474926048;30.43820478;0.204307781;0.197401892;-2.054206917;6.582702905;23.54950642;0.301609973;0.295253103;27.59623688;9.218351755;53.19995022
70;312542.1057;0.285516216;0.272978264;-15.58442604;7.500655673;26.16008352;0.218815975;0.211419688;-2.960974814;6.80216355;24.39207195;0.3341978;0.327154093;37.09500847;10.5483335;64.44805523
80;312337.2699;0.281041877;0.268700408;-18.00888139;7.254222474;26.73924673;0.214558484;0.207306107;-3.722019299;6.834443321;26.63464602;0.32272078;0.315918968;36.86467526;10.65477076;67.22134058
90;312126.2312;0.261427763;0.249947614;-20.63198112;6.672061718;26.78296285;0.220862268;0.213396814;-9.654596765;6.383805548;23.3688844;0.28623655;0.280203697;29.23426851;9.942319069;62.25774968
100;312045.7169;0.267907009;0.256142335;-22.4947258;6.695311677;27.26280547;0.222352923;0.214837082;-11.82416917;6.31743596;23.5418993;0.270219187;0.264523923;26.05069568;9.778905577;61.41676415
110;311961.2362;0.256586396;0.245318848;-21.66823955;5.978973438;22.75808386;0.199545063;0.192800161;-4.921836847;6.315666793;25.11302376;0.284062181;0.278075155;30.59685573;9.694845607;60.63171634
120;311903.4907;0.252212623;0.241137141;-21.30132905;5.891748756;18.51364235;0.193006428;0.18648254;0.542836176;6.410901569;25.96634477;0.310514144;0.303969605;37.13177987;9.976639006;62.55528847
130;311860.0792;0.24369112;0.232989845;-19.27019107;5.885808735;19.80577982;0.190379659;0.183944559;3.09410435;6.561565649;27.77861244;0.322048826;0.315261177;41.26350412;10.34426479;65.94801222
140;311824.0888;0.242948717;0.232280043;-19.61172079;5.879689386;19.49993947;0.189863641;0.183445984;5.14486744;6.757831635;29.8650649;0.335016712;0.327955745;43.81570506;10.69571428;68.53590252
150;311799.5006;0.247329798;0.236468737;-21.26474472;6.011129501;19.72695773;0.185349032;0.179083975;1.586200303;6.452309955;28.18643995;0.309152689;0.302636845;39.85287407;10.36472143;66.45311372
160;310805.5379;0.217534909;0.207982239;-15.72726321;5.451394815;25.07587467;0.140117921;0.135381738;0.233022694;5.233768938;26.64469777;0.254674097;0.249306468;36.96937261;9.596420476;63.38104769
170;310709.8762;0.210140629;0.200912666;-14.98202422;5.47349764;25.07492507;0.136559002;0.131943116;0.253738101;5.077300755;25.91922458;0.24921032;0.243957849;36.14595715;9.3587693;61.81144363
180;310681.7921;0.205572009;0.196544669;-14.18867462;5.303389972;23.52487721;0.135995004;0.131398182;2.249645674;5.06369695;25.5717347;0.26588242;0.26027856;39.37005176;9.492114211;62.69214078
190;310660.8298;0.20014305;0.191354113;-12.5668905;5.284807643;22.71369156;0.143734513;0.138876085;4.750991408;5.162772957;25.64011067;0.298319125;0.292031614;44.11916308;9.842797406;65.00828234
200;310639.3538;0.200985286;0.192159364;-12.75146618;5.41261;22.25160995;0.143167748;0.138328477;4.401148298;5.08805879;25.01276163;0.293476063;0.287290627;43.6246301;9.72590426;64.23624343
210;310605.8913;0.202869191;0.19396054;-12.98060643;5.598822634;22.63556423;0.145487406;0.140569728;6.143179989;5.458776697;27.36788784;0.313795523;0.307181824;46.76456122;10.29414372;67.98926907
220;310524.5208;0.182509651;0.174495054;-13.17552163;4.671998916;11.53617818;0.147816364;0.142819963;-3.228776674;3.743574731;7.091460338;0.221199831;0.216537722;30.14336246;6.238468076;40.46359947
230;310513.2974;0.179325731;0.171450951;-13.81241974;4.667798647;12.86007823;0.153428924;0.148242811;-5.688846786;3.728527802;6.592188379;0.205925008;0.201584838;27.44680813;6.113198821;39.72784329
240;310475.221;0.171897804;0.164349208;-13.75013573;4.346698174;10.45333365;0.130223098;0.125821375;-0.654487221;3.522683223;9.157519349;0.21888877;0.21427537;29.62724532;6.153634105;39.43925189
250;310462.3478;0.170816991;0.163315857;-14.27354991;4.306527685;10.17067067;0.134092164;0.129559661;-2.375855494;3.479756434;7.676902285;0.210889825;0.206445014;28.0544407;5.957678363;38.10719848
258;310442.5491;0.172108606;0.164550754;-14.35072523;4.370849546;10.37226601;0.134024421;0.129494208;-2.101255909;3.504188289;8.230272533;0.214336326;0.209818875;28.22481841;6.063173584;38.55634685
0;437251.0284;4.556602029;4.356506712;-237.5325499;100;38.252875;3.230529781;3.121333346;-0.294727358;100;261.0992347;4.026996295;3.942121469;105.619188;100;367.0131501
10;325845.8559;2.11240019;2.01963778;289.6520154;50.14236648;221.4867684;2.201229439;2.126824798;328.4725438;46.15279936;245.9158341;3.182952417;3.115867048;478.338705;59.92470276;395.7819953
20;318984.6211;1.027407561;0.98229073;104.4070094;25.99121924;135.6580339;0.565979218;0.546848327;82.17735863;18.74772293;99.03692039;1.089346884;1.066387308;161.8979435;27.26066761;178.7575052
30;315896.4266;0.704811041;0.67386048;69.37696169;17.59494292;78.57687557;0.526228597;0.508441333;76.38466629;13.87134966;71.19311737;0.927521957;0.907973078;137.3681036;20.35551398;132.1765547
40;314043.8939;0.403832361;0.38609876;27.53840524;10.60205315;45.48234016;0.296259223;0.286245246;30.05967702;8.629846837;33.61214913;0.53052719;0.519345555;67.74832396;12.46209926;71.30079608
50;313483.1358;0.330336318;0.315830169;9.19167004;8.71513218;35.08263117;0.224787462;0.21718933;5.297245811;6.6430005;16.79674414;0.365278678;0.357579897;32.44198276;9.177303426;43.94148109
60;312939.1882;0.315957522;0.302082793;-4.657342684;7.833350874;30.51730806;0.209914842;0.202819427;5.342354908;6.894737816;26.12554285;0.352021853;0.344602479;42.43968447;10.3820621;63.22287241
70;312359.0497;0.269690404;0.257847415;-9.653717342;6.92691716;20.8575271;0.215694654;0.208403872;11.2852612;7.084368067;27.40504283;0.392857148;0.38457711;49.09101933;10.78097851;65.21080096
80;312165.0922;0.259686456;0.248282773;-11.82622124;6.554613124;22.00331415;0.205510443;0.198563902;9.704916357;7.017784197;29.14298894;0.373070098;0.365207101;47.66071596;10.72083933;67.09878854
90;311963.9539;0.233259966;0.223016757;-14.75965455;6.12988928;23.93794608;0.195968781;0.189344761;1.315307714;6.433201796;25.62144554;0.313270169;0.306667543;37.10142542;9.838138582;61.40756324
100;311882.2354;0.237057991;0.226647998;-16.77101995;5.755680142;20.31697115;0.189904849;0.183485799;1.532917975;6.21779562;24.22944627;0.30478413;0.29836036;35.7346742;9.430719507;58.43120249
110;311826.6873;0.23946774;0.228951927;-17.93954099;5.733055092;20.59397058;0.190034854;0.18361141;1.22056106;6.305080319;25.36260983;0.302875603;0.296492058;35.68063107;9.588013651;59.82267984
120;311768.7987;0.244620476;0.23387839;-19.55963267;5.762467695;17.96327824;0.189034078;0.182644461;3.476296725;6.424978271;26.60774484;0.319266777;0.312537763;39.23167821;9.924056546;62.36312633
130;311716.26;0.223971861;0.214136524;-16.15193575;5.502105274;15.09350873;0.189633076;0.183223212;9.89985401;6.402650161;26.75383569;0.349865016;0.3424911;46.37653549;9.992725175;63.23051716
140;311004.9937;0.215750429;0.206276121;-13.28444712;5.222026054;21.1210959;0.141635526;0.136848046;6.381813785;5.36106922;26.395894;0.291353698;0.285212993;41.86430616;9.415563759;61.87838637
150;310659.749;0.203073797;0.194156161;-11.87498744;5.093739362;21.22572794;0.133031771;0.128535111;7.041302472;5.157583096;25.75055505;0.284165683;0.278176476;41.51962975;9.129248883;60.22888232
160;310610.6718;0.201279919;0.192441058;-11.76590062;5.033072663;20.89937814;0.137444965;0.132799133;8.396519166;5.360002357;26.67033512;0.303441711;0.297046234;44.91717157;9.567643778;63.19098753
170;310586.0936;0.195977593;0.187371575;-10.90979253;4.994256484;21.29494244;0.136465603;0.131852874;10.30566811;5.547700755;28.11894028;0.316207579;0.309543043;47.04808947;9.820552382;64.86136163
180;310550.4635;0.192639595;0.184180159;-12.18682124;4.987258899;21.30576318;0.135019039;0.130455206;0.716085653;4.264288842;19.81720727;0.241356618;0.236269675;35.05804499;8.200150585;54.1591666
190;310535.2642;0.195829088;0.187229592;-13.83974912;5.087153308;21.33613794;0.139461385;0.134747395;-2.619643121;4.048696618;18.16478113;0.216767967;0.212199265;31.28838166;7.884258136;52.07280591
200;310510.7017;0.181907831;0.173919663;-11.19780624;4.965380192;20.70563716;0.130945292;0.126519157;0.400366138;3.991793756;17.91234674;0.231001259;0.22613257;34.06853477;7.809721232;51.58051537
210;310467.1709;0.185009344;0.176884978;-11.96200602;5.010620834;19.54279748;0.131434918;0.126992233;0.255125782;3.967117261;17.36846649;0.230539563;0.225680605;34.01040905;7.7405631;51.12374976
220;310463.1987;0.181296994;0.173335649;-11.50785527;5.058544185;20.04461143;0.129645544;0.125263343;2.130454827;4.180698266;19.29145873;0.246285634;0.241094804;36.40333796;8.110089147;53.56434186
230;310453.5226;0.18118625;0.173229769;-11.01152159;5.408654945;22.8050715;0.137515147;0.132866942;0.870872531;4.404884917;20.29600282;0.246209757;0.241020527;36.29300215;8.436191046;55.71813244
240;310440.4117;0.181756489;0.173774967;-11.30445205;5.398113559;22.69371819;0.137562399;0.132912597;1.172388418;4.457279983;20.77909586;0.250496821;0.245217235;36.9218778;8.558900375;56.52858524
250;310430.8643;0.181042945;0.173092756;-11.05743768;5.50899178;23.23252298;0.137650178;0.132997409;1.442655671;4.524875882;21.34115352;0.251138844;0.245845726;37.14726721;8.638284957;57.04576506
252;310425.3469;0.184539166;0.176435447;-11.12281926;5.515265264;23.32765839;0.138154482;0.133484667;1.462134951;4.547966055;21.52114979;0.252904057;0.247573735;37.17531418;8.700002058;57.23432902
0;437251.0284;4.556602029;4.356506712;-237.5325499;100;38.252875;3.230529781;3.121333346;-0.294727358;100;261.0992347;4.026996295;3.942121469;105.619188;100;367.0131501
10;325796.003;2.115488145;2.022590132;290.0737782;50.2034756;221.7017933;2.20564924;2.131095204;329.1320769;46.23777867;246.3686291;3.188543701;3.121340487;479.1789712;60.02062423;396.4155235
20;318939.8645;1.026335653;0.981265893;112.1603804;25.96451962;135.3986751;0.666205253;0.643686581;98.47371051;20.2431832;107.3205425;1.199260121;1.173983963;179.0084127;28.60636123;187.8552447
30;315848.9664;0.708408849;0.677300296;70.09149138;17.68113801;79.06155891;0.532490955;0.514492014;77.43872945;14.00457615;72.01733418;0.936479917;0.916742235;138.7800005;20.52617282;133.3586052
40;314001.0194;0.407297967;0.389412179;28.26818413;10.71179157;46.26635619;0.298593468;0.28850059;30.82056044;8.709784365;34.42726969;0.535638055;0.524348701;68.92170883;12.58880385;72.52841808
50;313412.7903;0.347765839;0.332494302;9.809593192;9.024954608;36.06144821;0.223379013;0.215828489;5.118875472;6.61555817;16.97926769;0.36398833;0.356316745;32.2868219;9.225069805;44.14721412
60;312897.1771;0.316193177;0.302308099;-3.901079134;7.865789221;31.29184165;0.210543172;0.203426517;6.243058577;6.982679615;27.04451655;0.358152559;0.350603971;43.84202786;10.54916802;64.64348583
70;312316.6093;0.271195895;0.259286795;-8.943831646;6.968515851;21.73576778;0.217048991;0.209712431;12.12006219;7.18462442;28.40819882;0.399100933;0.390689298;50.47245685;10.96123112;66.76059347
80;312120.0259;0.2602707;0.248841361;-11.20249574;6.564648803;22.84257929;0.206575241;0.199592708;10.44002317;7.118739116;30.0936354;0.37870886;0.370727018;48.92774957;10.89618525;68.5813618
90;311920.2548;0.234517315;0.224218891;-15.22094638;6.091260666;23.62116118;0.19561831;0.189006137;1.105070497;6.427190364;25.55571526;0.31308445;0.306485738;36.52011961;9.790672018;60.97076437
100;311841.5199;0.238025335;0.227572863;-16.46905106;6.034037465;23.46736921;0.193748471;0.187199501;1.340657322;6.531269426;26.88561479;0.310554955;0.304009556;36.98093058;9.948700354;62.52588805
110;311784.4892;0.240985165;0.230402717;-17.68730058;5.899747113;23.56491317;0.191522808;0.185049068;0.855584157;6.553940595;27.7163351;0.303710052;0.29730892;36.32722628;10.00408671;63.18797722
120;311736.7377;0.240738853;0.230167221;-18.1963982;5.808883551;20.81506829;0.188825303;0.182442743;2.070458067;6.39455385;26.69046175;0.309901508;0.303369881;38.13278106;9.924130136;62.75278474
130;311689.9175;0.227443471;0.217455683;-15.81940174;5.653450754;18.16255469;0.187397042;0.181062759;7.999622608;6.468049037;27.59011623;0.338696244;0.331557726;44.79806821;10.0996663;64.38856184
140;310925.4862;0.212763244;0.203420113;-12.71124664;5.205856442;22.09388491;0.140299509;0.135557189;6.671402702;5.430473847;27.08507145;0.292657809;0.286489619;42.5130463;9.548353936;62.92671505
150;310603.7991;0.202095435;0.193220762;-11.29381233;5.130515844;22.28226259;0.133484594;0.128972628;7.443329981;5.286446432;26.6279421;0.28865844;0.282574542;42.29565617;9.320827796;61.48026829
160;310559.0449;0.200469315;0.191666051;-11.25131422;5.06252002;21.78800499;0.138813113;0.134121035;8.950997815;5.507496527;27.59885422;0.310107035;0.303571076;46.02096728;9.791400527;64.66882369
170;310531.7317;0.188937405;0.180640545;-10.31301561;4.998967892;22.1603719;0.134015603;0.129485688;7.923210846;5.193866165;26.00513555;0.296936202;0.290677838;44.25030894;9.437621262;62.33223365
180;310503.1597;0.193351381;0.184860688;-12.251651;5.2215962;24.38145287;0.132285291;0.127813863;3.799954347;5.136816994;26.04159542;0.26998763;0.264297246;40.25344115;9.462277897;62.49508222
190;310481.0966;0.194488154;0.185947542;-13.4074137;5.112914165;22.09994263;0.140263684;0.135522575;-2.386274002;4.124123584;18.72961953;0.219517225;0.214890579;31.84411394;8.018587868;52.96000746
200;310453.8217;0.189433763;0.181115106;-12.68481986;5.164001171;20.51061071;0.134641793;0.130090712;-0.847944728;4.033050546;17.95602303;0.22422708;0.219501167;32.95078531;7.836102277;51.75475307
210;310411.8523;0.184813754;0.176697977;-11.95906657;5.038287212;19.86593383;0.132419327;0.127943368;0.23165001;4.019071646;17.66518761;0.231176769;0.226304381;34.11473557;7.804839493;51.54827317
220;310406.3572;0.184507618;0.176405284;-11.65923128;5.066941689;19.96764459;0.132009554;0.127547446;0.843192854;4.061975315;18.07860592;0.239326849;0.234282686;35.47482341;7.980770452;52.71023648
224;310403.5012;0.183678666;0.175612734;-11.63495212;5.112494379;19.88594032;0.132461713;0.127984322;0.90449566;4.076223075;18.0339253;0.239132618;0.234092549;35.27457151;7.934403864;52.40400115
\end{filecontents*}
\pgfplotstabletypeset[
        multicolumn names, 
        col sep=semicolon, 
        display columns/0/.style={column name=$\bm{k}$,dec sep align},
        display columns/1/.style={column name={\textbf{AIC}},fixed,fixed zerofill,precision=0,set thousands separator={,},dec sep align},
        display columns/2/.style={column name={\textbf{v.mae}},column type={r},fixed,fixed zerofill,precision=3,set thousands separator={,}},
        display columns/3/.style={column name={$\bm{\textbf{\text{v.mae}}^a}$},column type={r},fixed,fixed zerofill,precision=3,set thousands separator={,}},
        display columns/4/.style={column name={\textbf{v.res}},column type={r},fixed,fixed zerofill,precision=0,set thousands separator={,}},
        display columns/5/.style={column name={$\bm{\textbf{\text{v.mae}}^0}$},column type={r},fixed,fixed zerofill,precision=3,set thousands separator={,}},
        display columns/6/.style={column name={$\bm{\textbf{\text{v.res}}^0}$},column type={r},fixed,fixed zerofill,precision=0,set thousands separator={,}},
        display columns/7/.style={column name={\textbf{ns.mae}},column type={r},fixed,fixed zerofill,precision=3,set thousands separator={,}},
        display columns/8/.style={column name={$\bm{\textbf{\text{ns.mae}}^a}$},column type={r},fixed,fixed zerofill,precision=3,set thousands separator={,}},
        display columns/9/.style={column name={\textbf{ns.res}},column type={r},fixed,fixed zerofill,precision=0,set thousands separator={,}},
        display columns/10/.style={column name={$\bm{\textbf{\text{ns.mae}}^0}$},column type={r},fixed,fixed zerofill,precision=3,set thousands separator={,}},
        display columns/11/.style={column name={$\bm{\textbf{\text{ns.res}}^0}$},column type={r},fixed,fixed zerofill,precision=0,set thousands separator={,}},
        display columns/12/.style={column name={\textbf{cr.mae}},column type={r},fixed,fixed zerofill,precision=3,set thousands separator={,}},
        display columns/13/.style={column name={$\bm{\textbf{\text{cr.mae}}^a}$},column type={r},fixed,fixed zerofill,precision=3,set thousands separator={,}},
        display columns/14/.style={column name={\textbf{cr.res}},column type={r},fixed,fixed zerofill,precision=0,set thousands separator={,}},
        display columns/15/.style={column name={$\bm{\textbf{\text{cr.mae}}^0}$},column type={r},fixed,fixed zerofill,precision=3,set thousands separator={,}},
        display columns/16/.style={column name={$\bm{\textbf{\text{cr.res}}^0}$},column type={r},fixed,fixed zerofill,precision=0,set thousands separator={,}},
	    	every head row/.style={before row=\toprule,after row=\midrule},
	    	every row no 0/.style={before row={\multicolumn{18}{l}{\textbf{$\bm{M}_{\text{max}} \bm{= 14}$ in variance model selection}}\\ \\}},
 	    	every row no 27/.style={before row={\midrule \multicolumn{18}{l}{\textbf{$\bm{M}_{\text{max}} \bm{= 18}$ in variance model selection}}\\ \\}},
 	    	every row no 54/.style={before row={\midrule \multicolumn{18}{l}{\textbf{$\bm{M}_{\text{max}} \bm{= 22}$ in variance model selection}}\\ \\}},
      	every last row/.style={after row=\bottomrule},
      ]{BGROGLS300886valfig2.csv} 
    \caption[]{Cont.}
    \label{tab:BGROGLS300886valfig2}
  \end{center}
\end{table}

\begin{table}[htb]
  \tiny
  \begin{center}
  \tabcolsep=0.026cm
  \renewcommand{\arraystretch}{0.90}
      \begin{filecontents*}{BGROGLSallvalfig.csv}
k;Mmax;AIC;v.wre.y;v.a.wre.y;v.res.avg.y;v.wre.delta.y;v.res.avg.delta.y;ns.wre.y;ns.a.wre.y;ns.res.avg.y;ns.wre.delta.y;ns.res.avg.delta.y;cr.wre.y;cr.a.wre.y;cr.res.avg.y;cr.wre.delta.y;cr.res.avg.delta.y
150;2;315980.418;0.239464156;0.228948501;-15.96767262;8.14671415;45.62210033;0.255095858;0.24647326;-30.31511496;4.032066156;16.88319519;0.152702071;0.149483652;2.213307802;7.489019611;49.41161796
150;6;311949.2027;0.230636006;0.220508024;-12.72065952;7.577363791;41.2791701;0.203356331;0.196482602;-17.69734558;4.762292695;21.91102124;0.186451574;0.182521835;17.42951698;8.636643019;57.0378838
150;10;311363.2967;0.226880727;0.216917652;-10.12762056;7.459517323;40.45964405;0.194085946;0.187525569;-14.84624408;4.707817397;21.34955772;0.194804319;0.190698534;20.16851621;8.5365753;56.36431802
150;14;311160.6866;0.230909895;0.220769886;-8.826750475;7.526752978;40.59534387;0.192814;0.186296616;-13.95586097;4.701114966;21.07477057;0.199553771;0.195347884;21.03448684;8.497021025;56.06511839
150;18;311048.4659;0.228009502;0.217996858;-8.79248094;7.433276412;40.04418546;0.187429515;0.181094134;-12.69102096;4.779954973;21.75418264;0.204456319;0.200147103;22.35519488;8.620893155;56.80039847
150;22;310974.0209;0.229833019;0.219740299;-8.468185373;7.435732001;39.93576495;0.186753543;0.180441011;-12.20558614;4.801847514;21.80690138;0.207178324;0.202811738;22.92395435;8.639152535;56.93644187
224;2;315614.7932;0.195546773;0.186959674;-8.955140705;6.526855254;32.96659883;0.275337354;0.266030566;-30.30406755;4.563555133;-2.773790814;0.174680263;0.170998621;4.815708167;5.400683209;32.3459849
224;6;311554.4018;0.199702422;0.190932835;-8.642398393;6.398950397;32.59446597;0.239872557;0.231764529;-22.94088039;4.292205031;3.904521175;0.183007473;0.179150324;13.44771598;6.388883111;40.29311754
224;10;311287.0343;0.202849686;0.193941892;-7.878868813;6.472880497;32.76602658;0.234119668;0.226206095;-21.25250339;4.310388404;5.000929199;0.189419206;0.18542692;15.65410122;6.621107298;41.9075338
224;14;310980.0188;0.20024496;0.191451548;-7.322763741;6.246241625;31.4636737;0.221802173;0.214304948;-18.70252607;4.256668744;5.69244856;0.194365881;0.190269337;18.10142042;6.697070344;42.49639505
224;18;310880.8341;0.199935116;0.19115531;-7.138702583;6.193747164;31.24508551;0.217240717;0.209897676;-17.60746776;4.249709354;6.38485753;0.198094352;0.193919224;19.20066472;6.801082331;43.19299001
224;22;310832.2798;0.200430635;0.191629069;-6.995540169;6.255897995;31.9533266;0.217220372;0.209878019;-17.74662837;4.223366448;6.810775596;0.195868093;0.191739887;19.05816812;6.843713346;43.61557209
150;2;315628.8046;0.239469443;0.228953555;-20.5349129;6.467171748;33.96135026;0.26082306;0.252006875;-29.66946586;3.79628727;10.4353345;0.177220593;0.17348541;3.489977964;6.654014631;43.59477832
150;6;311425.5391;0.224394554;0.214540654;-14.32024407;5.904249516;30.78273055;0.176692675;0.170720215;-8.566368431;4.756383755;22.14514339;0.22588917;0.221128226;28.73867462;9.005130545;59.45018644
150;10;310868.4044;0.212448988;0.203119657;-14.43870438;5.375228628;25.34555252;0.14849644;0.143477052;-0.325865963;5.097903576;25.06692814;0.255822562;0.250430728;35.94142527;9.296046661;61.33421937
150;14;310713.9804;0.214291071;0.204880848;-14.34572911;5.36813261;24.84780827;0.146033164;0.141097038;-1.323952364;4.856784706;23.47812221;0.244280977;0.239132399;33.79524888;8.905620897;58.59732346
150;18;310643.6246;0.210977111;0.201712415;-13.80651418;5.13118427;22.59981315;0.139387229;0.134675746;1.078389506;4.816062937;23.09325403;0.250107662;0.244836278;34.55367771;8.618033812;56.56854223
150;22;310590.368;0.207914061;0.198783874;-13.02535308;5.208757852;22.6445265;0.138639092;0.133952896;4.582323843;5.192722181;25.86074062;0.275429488;0.269624409;39.70091037;9.255833578;60.97932715
224;2;315490.539;0.208567853;0.199408956;-16.55318111;6.583869232;33.88579153;0.226415996;0.218762818;-22.05215693;3.998696206;13.99535292;0.164737188;0.161265111;12.01007566;7.289529666;48.0575855
237;6;311143.6358;0.196349469;0.187727121;-15.26973808;5.341548115;23.45931655;0.125412693;0.121173568;-0.038382688;4.765150534;24.29920913;0.23522153;0.230263893;29.60425834;8.242727;53.94185017
244;10;310904.5235;0.208492259;0.199336681;-10.81124475;6.576542185;34.60242988;0.152654553;0.147494615;-2.350457835;5.617254132;28.67175399;0.252094994;0.246781724;36.70675993;10.2588368;67.72897176
258;14;310442.5491;0.172108606;0.164550754;-14.35072523;4.370849546;10.37226601;0.134024421;0.129494208;-2.101255909;3.504188289;8.230272533;0.214336326;0.209818875;28.22481841;6.063173584;38.55634685
252;18;310425.3469;0.184539166;0.176435447;-11.12281926;5.515265264;23.32765839;0.138154482;0.133484667;1.462134951;4.547966055;21.52114979;0.252904057;0.247573735;37.17531418;8.700002058;57.23432902
224;22;310403.5012;0.183678666;0.175612734;-11.63495212;5.112494379;19.88594032;0.132461713;0.127984322;0.90449566;4.076223075;18.0339253;0.239132618;0.234092549;35.27457151;7.934403864;52.40400115
\end{filecontents*}
\pgfplotstabletypeset[
        multicolumn names, 
        col sep=semicolon, 
        display columns/0/.style={column name=$\bm{k}$,dec sep align},
        display columns/1/.style={column name={$\bm{M_\textbf{\text{max}}}$},column type={r},fixed,precision=0,set thousands separator={,}},
        display columns/2/.style={column name={\textbf{AIC}},column type={r},fixed,fixed zerofill,precision=0,set thousands separator={,}},
        display columns/3/.style={column name={\textbf{v.mae}},column type={r},fixed,fixed zerofill,precision=3,set thousands separator={,}},
        display columns/4/.style={column name={$\bm{\textbf{\text{v.mae}}^a}$},column type={r},fixed,fixed zerofill,precision=3,set thousands separator={,}},
        display columns/5/.style={column name={\textbf{v.res}},column type={r},fixed,fixed zerofill,precision=0,set thousands separator={,}},
        display columns/6/.style={column name={$\bm{\textbf{\text{v.mae}}^0}$},column type={r},fixed,fixed zerofill,precision=3,set thousands separator={,}},
        display columns/7/.style={column name={$\bm{\textbf{\text{v.res}}^0}$},column type={r},fixed,fixed zerofill,precision=0,set thousands separator={,}},
        display columns/8/.style={column name={\textbf{ns.mae}},column type={r},fixed,fixed zerofill,precision=3,set thousands separator={,}},
        display columns/9/.style={column name={$\bm{\textbf{\text{ns.mae}}^a}$},column type={r},fixed,fixed zerofill,precision=3,set thousands separator={,}},
        display columns/10/.style={column name={\textbf{ns.res}},column type={r},fixed,fixed zerofill,precision=0,set thousands separator={,}},
        display columns/11/.style={column name={$\bm{\textbf{\text{ns.mae}}^0}$},column type={r},fixed,fixed zerofill,precision=3,set thousands separator={,}},
        display columns/12/.style={column name={$\bm{\textbf{\text{ns.res}}^0}$},column type={r},fixed,fixed zerofill,precision=0,set thousands separator={,}},
        display columns/13/.style={column name={\textbf{cr.mae}},column type={r},fixed,fixed zerofill,precision=3,set thousands separator={,}},
        display columns/14/.style={column name={$\bm{\textbf{\text{cr.mae}}^a}$},column type={r},fixed,fixed zerofill,precision=3,set thousands separator={,}},
        display columns/15/.style={column name={\textbf{cr.res}},column type={r},fixed,fixed zerofill,precision=0,set thousands separator={,}},
        display columns/16/.style={column name={$\bm{\textbf{\text{cr.mae}}^0}$},column type={r},fixed,fixed zerofill,precision=3,set thousands separator={,}},
        display columns/17/.style={column name={$\bm{\textbf{\text{cr.res}}^0}$},column type={r},fixed,fixed zerofill,precision=0,set thousands separator={,}},
	    	every head row/.style={before row=\toprule,after row=\midrule},
	    	every row no 0/.style={before row={\multicolumn{19}{l}{\textbf{Type I algorithm under 150-443}}\\ \\}},
	    	every row no 6/.style={before row={\midrule \multicolumn{19}{l}{\textbf{Type I algorithm under 300-886}}\\ \\}},
	    	every row no 12/.style={before row={\midrule \multicolumn{19}{l}{\textbf{Type II algorithm under 150-443}}\\ \\}},
	    	every row no 18/.style={before row={\midrule \multicolumn{19}{l}{\textbf{Type II algorithm under 300-886}}\\ \\}},
      	every last row/.style={after row=\bottomrule},
      	every row 23 column 2/.style={postproc cell content/.append style={/pgfplots/table/@cell content/.add={\cellcolor{green!26!white}}{},}},
      	every row 21 column 3/.style={postproc cell content/.append style={/pgfplots/table/@cell content/.add={\cellcolor{green!26!white}}{},}},
      	every row 21 column 4/.style={postproc cell content/.append style={/pgfplots/table/@cell content/.add={\cellcolor{green!26!white}}{},}},
      	every row 11 column 5/.style={postproc cell content/.append style={/pgfplots/table/@cell content/.add={\cellcolor{green!26!white}}{},}},
      	every row 21 column 6/.style={postproc cell content/.append style={/pgfplots/table/@cell content/.add={\cellcolor{green!26!white}}{},}},
      	every row 21 column 7/.style={postproc cell content/.append style={/pgfplots/table/@cell content/.add={\cellcolor{green!26!white}}{},}},
      	every row 19 column 8/.style={postproc cell content/.append style={/pgfplots/table/@cell content/.add={\cellcolor{green!26!white}}{},}},
      	every row 19 column 9/.style={postproc cell content/.append style={/pgfplots/table/@cell content/.add={\cellcolor{green!26!white}}{},}},
      	every row 19 column 10/.style={postproc cell content/.append style={/pgfplots/table/@cell content/.add={\cellcolor{green!26!white}}{},}},
      	every row 21 column 11/.style={postproc cell content/.append style={/pgfplots/table/@cell content/.add={\cellcolor{green!26!white}}{},}},
      	every row 6 column 12/.style={postproc cell content/.append style={/pgfplots/table/@cell content/.add={\cellcolor{green!26!white}}{},}},
      	every row 0 column 13/.style={postproc cell content/.append style={/pgfplots/table/@cell content/.add={\cellcolor{green!26!white}}{},}},
      	every row 0 column 14/.style={postproc cell content/.append style={/pgfplots/table/@cell content/.add={\cellcolor{green!26!white}}{},}},
      	every row 0 column 15/.style={postproc cell content/.append style={/pgfplots/table/@cell content/.add={\cellcolor{green!26!white}}{},}},
      	every row 6 column 16/.style={postproc cell content/.append style={/pgfplots/table/@cell content/.add={\cellcolor{green!26!white}}{},}},
      	every row 6 column 17/.style={postproc cell content/.append style={/pgfplots/table/@cell content/.add={\cellcolor{green!26!white}}{},}},
      	every row 0 column 2/.style={postproc cell content/.append style={/pgfplots/table/@cell content/.add={\cellcolor{red!15!white}}{},}},
      	every row 12 column 3/.style={postproc cell content/.append style={/pgfplots/table/@cell content/.add={\cellcolor{red!15!white}}{},}},
      	every row 12 column 4/.style={postproc cell content/.append style={/pgfplots/table/@cell content/.add={\cellcolor{red!15!white}}{},}},
      	every row 12 column 5/.style={postproc cell content/.append style={/pgfplots/table/@cell content/.add={\cellcolor{red!15!white}}{},}},
      	every row 0 column 6/.style={postproc cell content/.append style={/pgfplots/table/@cell content/.add={\cellcolor{red!15!white}}{},}},
      	every row 0 column 7/.style={postproc cell content/.append style={/pgfplots/table/@cell content/.add={\cellcolor{red!15!white}}{},}},
      	every row 6 column 8/.style={postproc cell content/.append style={/pgfplots/table/@cell content/.add={\cellcolor{red!15!white}}{},}},
      	every row 6 column 9/.style={postproc cell content/.append style={/pgfplots/table/@cell content/.add={\cellcolor{red!15!white}}{},}},
      	every row 0 column 10/.style={postproc cell content/.append style={/pgfplots/table/@cell content/.add={\cellcolor{red!15!white}}{},}},
      	every row 20 column 11/.style={postproc cell content/.append style={/pgfplots/table/@cell content/.add={\cellcolor{red!15!white}}{},}},
      	every row 20 column 12/.style={postproc cell content/.append style={/pgfplots/table/@cell content/.add={\cellcolor{red!15!white}}{},}},
      	every row 17 column 13/.style={postproc cell content/.append style={/pgfplots/table/@cell content/.add={\cellcolor{red!15!white}}{},}},
      	every row 17 column 14/.style={postproc cell content/.append style={/pgfplots/table/@cell content/.add={\cellcolor{red!15!white}}{},}},
      	every row 17 column 15/.style={postproc cell content/.append style={/pgfplots/table/@cell content/.add={\cellcolor{red!15!white}}{},}},
      	every row 20 column 16/.style={postproc cell content/.append style={/pgfplots/table/@cell content/.add={\cellcolor{red!15!white}}{},}},
      	every row 20 column 17/.style={postproc cell content/.append style={/pgfplots/table/@cell content/.add={\cellcolor{red!15!white}}{},}},
      ]{BGROGLSallvalfig.csv} 
    \caption{AIC scores and out-of-sample validation figures of all derived FGLS proxy functions of BEL under $150$-$443$ and $300$-$886$ after the final iteration. Highlighted in green and red respectively the best and worst AIC scores and validation figures.}
    \label{tab:BGROGLSallvalfig}
  \end{center}
\end{table}

\begin{table}[htb]
  \tiny
  \begin{center}
  \tabcolsep=0.028cm
  \renewcommand{\arraystretch}{0.90}
      \begin{filecontents*}{BGROMARSRun2.csv}
k;Kmax;thresh;o;pmethod;glm;v.wre.y;v.a.wre.y;v.res.avg.y;v.wre.delta.y;v.res.avg.delta.y;ns.wre.y;ns.a.wre.y;ns.res.avg.y;ns.wre.delta.y;ns.res.avg.delta.y;cr.wre.y;cr.a.wre.y;cr.res.avg.y;cr.wre.delta.y;cr.res.avg.delta.y
148;206;0;6;s;inv.g, id;0.2649869;0.253350458;-24.39297853;10.31675184;54.96081478;0.574579965;0.555158356;-39.94207534;16.23385524;-55.78751494;0.822106601;0.804779504;79.70472426;17.65716201;63.85928466
49;50;0;3;n;inv.g, log;0.370155775;0.353901023;-0.039670371;9.167857661;19.45886862;0.704766007;0.680943928;-11.7480176;29.47665975;-102.3952496;0.524696325;0.513637584;25.15456613;16.89112053;-65.49266589
60;66;0;4;s;inv.g, id;0.324447416;0.310199867;-11.4657355;8.517012149;15.85747109;1.71154997;1.653697182;151.0792258;44.50388044;132.2984236;0.916616941;0.897297901;101.9561809;19.87732276;83.17537873
45;50;0;4;b;inv.g, id;0.347172568;0.331927084;-2.423797702;8.685681215;10.56640169;0.446589637;0.431494281;-36.22430845;22.70159423;-124.8002523;0.5105819;0.499820641;34.78870639;15.78544609;-53.78723743
45;50;0;4;b;inv.g, id;0.347172568;0.331927084;-2.423797702;8.685681215;10.56640169;0.446589637;0.431494281;-36.22430845;22.70159423;-124.8002523;0.5105819;0.499820641;34.78870639;15.78544609;-53.78723743
17;19;0;4;b;inv.g, id;0.833953936;0.797332287;25.42862193;24.67287084;123.9714418;0.479871475;0.463651147;-3.564239924;41.35560943;-243.4478781;0.763100336;0.747016882;108.3815503;21.39785391;-131.5020879
70;81;0;4;b;inv.g, id;0.334916629;0.320209343;-21.57997665;10.87222641;52.14548304;0.553994357;0.535268571;-34.73436493;14.07257616;-37.99807417;0.875006509;0.856564469;101.8945383;18.25001254;98.63082907
33;34;0;3;n;inv.g, id;0.425555165;0.406867644;-9.892778741;10.87146431;21.39602316;1.564502882;1.511620492;108.1039888;52.38438238;0.934549863;0.661798801;0.647850425;31.90140957;20.99725932;-75.26802939
45;50;0;3;b;pois, log;0.378795812;0.362161648;0.152148763;9.555706651;27.75239617;0.480199452;0.463968038;-42.63704333;24.87811973;-139.2511157;0.510291544;0.499536405;27.75822666;16.93803421;-68.85584568
31;34;0;3;b;pois, log;0.476157716;0.455248071;-12.53355661;12.75206424;45.75223741;0.59319155;0.573140845;-53.90801587;31.14774619;-175.1506278;0.660652269;0.646728058;18.22223784;23.08755417;-103.0203741
45;50;0;4;b;inv.g, id;0.347172568;0.331927084;-2.423797702;8.685681215;10.56640169;0.446589637;0.431494281;-36.22430845;22.70159423;-124.8002523;0.5105819;0.499820641;34.78870639;15.78544609;-53.78723743
59;66;0;3;b;pois, log;0.428209663;0.438950808;39.67307353;16.67388629;97.92197616;0.760185829;0.734490483;-11.83438995;22.51095659;-40.72495242;0.809246349;0.792190301;68.06156425;18.40289736;39.17100178
134;144;1.56E-05;5;n;gaus, log;0.273381043;0.261375986;-22.1818519;10.25462049;54.45915527;1.024539249;0.989908386;-1.258910411;28.19171531;-22.97676229;1.515447042;1.483506784;178.9338366;32.61591529;157.2159847
45;50;0;4;s;inv.g, id;0.347172568;0.331927084;-2.423797702;8.685681215;10.56640169;0.446589637;0.431494281;-36.22430845;22.70159423;-124.8002523;0.5105819;0.499820641;34.78870639;15.78544609;-53.78723743
60;66;0;4;s;inv.g, id;0.324447416;0.310199867;-11.4657355;8.517012149;15.85747109;1.71154997;1.653697182;151.0792258;44.50388044;132.2984236;0.916616941;0.897297901;101.9561809;19.87732276;83.17537873
45;50;0;4;b;inv.g, id;0.347172568;0.331927084;-2.423797702;8.685681215;10.56640169;0.446589637;0.431494281;-36.22430845;22.70159423;-124.8002523;0.5105819;0.499820641;34.78870639;15.78544609;-53.78723743
45;50;0;4;b;inv.g, id;0.347172568;0.331927084;-2.423797702;8.685681215;10.56640169;0.446589637;0.431494281;-36.22430845;22.70159423;-124.8002523;0.5105819;0.499820641;34.78870639;15.78544609;-53.78723743
146;159;9.38E-06;5;n;gaus, log;0.278767118;0.266525541;-24.05649659;10.00798708;53.16868503;1.024915771;0.990272181;-0.018011048;26.7787886;-11.18886972;1.498203488;1.466626662;173.9495799;31.70221813;162.7787212
76;97;3.75E-05;4;b;inv.g, log;0.343786826;0.328690021;-16.8534216;10.67638543;51.88687788;0.538494544;0.520292673;-36.56902523;11.87391304;-24.33100457;0.803504852;0.786569814;87.7070729;16.58402024;99.94509356
107;113;0;4;n;gaus, log;0.321222585;0.307116649;-19.85161205;11.97578707;63.32138401;0.996614888;0.962927908;8.371041868;25.69380191;0.090416647;1.52859749;1.496380066;190.5475695;32.14846013;182.2669443
45;50;0;4;s;pois, id;0.353317145;0.337801832;-2.987854995;8.890844215;18.34617947;0.449235331;0.434050547;-36.46199616;23.63373261;-131.0573195;0.503599711;0.492985612;36.11568338;16.07862599;-58.47963991
31;34;0;4;s;pois, id;0.436922954;0.417736236;-11.2283421;11.25430795;31.64369734;0.548162605;0.529633941;-44.68160163;28.44362834;-157.2990484;0.647667596;0.634017056;28.74343133;21.37355725;-83.87401543
72;82;3.13E-05;4;b;inv.g, inv;0.365076353;0.349044655;-15.92347548;11.18101429;52.91298716;0.57922525;0.559646625;-49.20540094;14.5283204;-51.0416725;0.700233485;0.685475042;65.47140775;14.61865798;63.63513619
45;50;0;4;b;inv.g, id;0.347172568;0.331927084;-2.423797702;8.685681215;10.56640169;0.446589637;0.431494281;-36.22430845;22.70159423;-124.8002523;0.5105819;0.499820641;34.78870639;15.78544609;-53.78723743
125;144;0;5;f;inv.g, inv;0.282947075;0.270521942;-20.10979577;10.33566634;53.9600262;0.629562382;0.608282291;-63.47073237;17.24480258;-76.06203064;0.674672803;0.660453088;44.55875634;14.73725402;31.96745807
45;50;0;4;s;gaus, log;0.381945489;0.365173012;-0.615773129;9.915701952;31.84706015;0.468802769;0.452956579;-40.66919967;25.48711892;-143.6555924;0.495064252;0.48463005;31.56984616;16.86756513;-71.41654652
114;144;1.88E-05;5;s;inv.g,$1/\mu^2$;0.313214082;0.299459826;-12.12987651;9.414339681;40.05076629;0.708198863;0.684260749;-77.37001984;20.11501115;-96.77724768;0.625636842;0.612450632;36.0446334;14.09473181;16.63740556
45;50;0;4;b;gaus, log;0.381945489;0.365173012;-0.615773129;9.915701952;31.84706015;0.468802769;0.452956579;-40.66919967;25.48711892;-143.6555924;0.495064252;0.48463005;31.56984616;16.86756513;-71.41654652
45;50;0;4;f;gaus, log;0.386108183;0.369152909;-1.490260635;10.09517355;34.11982907;0.46814499;0.452321034;-40.51161885;25.70898565;-145.0714369;0.496036365;0.485581674;31.68501526;17.07729132;-72.87480276
64;66;0;4;n;inv.g,$1/\mu^2$;0.419777331;0.401343534;-2.678496877;11.50641666;38.50908422;0.839695634;0.811312745;2.838835128;25.96900496;-38.44657181;1.297904662;1.270549426;146.3110545;29.11018191;105.0256476
148;175;0;6;s;inv.g,$1/\mu^2$;0.310635528;0.296994505;-15.92450509;10.44713266;52.02481316;0.575690471;0.556231326;-54.64268073;14.56463185;-57.25744547;0.611080368;0.598200957;29.93201085;12.84426992;27.31724611
77;81;0;4;n;inv.g,$1/\mu^2$;0.38725998;0.370254127;-10.85289959;11.51923944;51.9910527;1.029007686;0.994225784;-28.03182061;25.83103109;-31.50107099;1.279396639;1.252431487;148.1527018;26.70033957;144.6834514
45;50;0;4;s;gaus, log;0.381945489;0.365173012;-0.615773129;9.915701952;31.84706015;0.468802769;0.452956579;-40.66919967;25.48711892;-143.6555924;0.495064252;0.48463005;31.56984616;16.86756513;-71.41654652
33;34;0;3;n;inv.g,$1/\mu^2$;0.563966197;0.539200594;-14.01120004;15.69276772;63.70039467;0.827485428;0.799515261;-53.55790702;38.64515569;-185.3163452;0.744593762;0.728900361;-1.832766073;26.33841542;-133.5912043
148;175;0;6;s;inv.g,$1/\mu^2$;0.310635528;0.296994505;-15.92450509;10.44713266;52.02481316;0.575690471;0.556231326;-54.64268073;14.56463185;-57.25744547;0.611080368;0.598200957;29.93201085;12.84426992;27.31724611
148;175;4.69E-06;5;f;inv.g, inv;0.295572929;0.282593353;-19.53916889;10.41629722;52.92438799;0.548902145;0.530348484;-54.06056713;18.25957442;-87.2433075;0.664326586;0.650324933;32.36584963;16.30707037;-0.816890746
\end{filecontents*}
\pgfplotstabletypeset[
        multicolumn names, 
        col sep=semicolon, 
        display columns/0/.style={column name={$\bm{k}$},dec sep align},
        display columns/1/.style={column name={$\bm{K}_{\textbf{\text{max}}}$},column type={r},fixed,fixed zerofill,precision=0},
        display columns/2/.style={column name={$\bm{t}_{\textbf{\text{min}}}$},sci subscript,precision=1},
        display columns/3/.style={column name={\textbf{o}},column type={r},fixed,fixed zerofill,precision=0},
        display columns/4/.style={column name={\textbf{ p}},string type,column type={r}},
        display columns/5/.style={column name={\textbf{glm}},string type},
        display columns/6/.style={column name={v.mae},column type={r},fixed,fixed zerofill,precision=3,set thousands separator={,}},
        display columns/7/.style={column name={$\text{v.mae}^a$},column type={r},fixed,fixed zerofill,precision=3,set thousands separator={,}},
        display columns/8/.style={column name={v.res},column type={r},fixed,fixed zerofill,precision=0,set thousands separator={,}},
        display columns/9/.style={column name={$\text{v.mae}^0$},column type={r},fixed,fixed zerofill,precision=3,set thousands separator={,}},
        display columns/10/.style={column name={$\text{v.res}^0$},column type={r},fixed,fixed zerofill,precision=0,set thousands separator={,}},
        display columns/11/.style={column name={ns.mae},column type={r},fixed,fixed zerofill,precision=3,set thousands separator={,}},
        display columns/12/.style={column name={$\text{ns.mae}^a$},column type={r},fixed,fixed zerofill,precision=3,set thousands separator={,}},
        display columns/13/.style={column name={ns.res},column type={r},fixed,fixed zerofill,precision=0,set thousands separator={,}},
        display columns/14/.style={column name={$\text{ns.mae}^0$},column type={r},fixed,fixed zerofill,precision=3,set thousands separator={,}},
        display columns/15/.style={column name={$\text{ns.res}^0$},column type={r},fixed,fixed zerofill,precision=0,set thousands separator={,}},
        display columns/16/.style={column name={cr.mae},column type={r},fixed,fixed zerofill,precision=3,set thousands separator={,}},
        display columns/17/.style={column name={$\text{cr.mae}^a$},column type={r},fixed,fixed zerofill,precision=3,set thousands separator={,}},
        display columns/18/.style={column name={cr.res},column type={r},fixed,fixed zerofill,precision=0,set thousands separator={,}},
        display columns/19/.style={column name={$\text{cr.mae}^0$},column type={r},fixed,fixed zerofill,precision=3,set thousands separator={,}},
        display columns/20/.style={column name={$\text{cr.res}^0$},column type={r},fixed,fixed zerofill,precision=0,set thousands separator={,}},
	    	every head row/.style={before row=\toprule,after row=\midrule},
	    	every row no 0/.style={before row={\multicolumn{22}{l}{$\bm{\textbf{\text{Sobol set}}^2}$}\\ \\}},
	    	every row no 4/.style={before row={\midrule \multicolumn{22}{l}{\textbf{Sobol set and nested simulations set}}\\ \\}},
	    	every row no 8/.style={before row={\midrule \multicolumn{22}{l}{\textbf{Sobol set and capital region set}}\\ \\}},
	    	every row no 12/.style={before row={\midrule \multicolumn{22}{l}{\textbf{Nested simulations set and Sobol set}}\\ \\}},
	    	every row no 16/.style={before row={\midrule \multicolumn{22}{l}{$\bm{\textbf{\text{Nested simulations set}}^2}$}\\ \\}},
	    	every row no 20/.style={before row={\midrule \multicolumn{22}{l}{\textbf{Nested simulations set and capital region set}}\\ \\}},
	    	every row no 24/.style={before row={\midrule \multicolumn{22}{l}{\textbf{Capital region set and Sobol set}}\\ \\}},
	    	every row no 28/.style={before row={\midrule \multicolumn{22}{l}{\textbf{Capital region set and nested simulations set}}\\ \\}},
	    	every row no 32/.style={before row={\midrule \multicolumn{22}{l}{$\bm{\textbf{\text{Capital region set}}^2}$}\\ \\}},
      	every last row/.style={after row=\bottomrule},
      	every row 0 column 6/.style={postproc cell content/.append style={/pgfplots/table/@cell content/.add={\cellcolor{green!26!white}}{},}},
      	every row 0 column 7/.style={postproc cell content/.append style={/pgfplots/table/@cell content/.add={\cellcolor{green!26!white}}{},}},
      	every row 1 column 8/.style={postproc cell content/.append style={/pgfplots/table/@cell content/.add={\cellcolor{green!26!white}}{},}},
      	every row 2 column 9/.style={postproc cell content/.append style={/pgfplots/table/@cell content/.add={\cellcolor{green!26!white}}{},}},
      	every row 14 column 9/.style={postproc cell content/.append style={/pgfplots/table/@cell content/.add={\cellcolor{green!26!white}}{},}},
      	every row 3 column 10/.style={postproc cell content/.append style={/pgfplots/table/@cell content/.add={\cellcolor{green!26!white}}{},}},
      	every row 4 column 10/.style={postproc cell content/.append style={/pgfplots/table/@cell content/.add={\cellcolor{green!26!white}}{},}},
      	every row 10 column 10/.style={postproc cell content/.append style={/pgfplots/table/@cell content/.add={\cellcolor{green!26!white}}{},}},
      	every row 13 column 10/.style={postproc cell content/.append style={/pgfplots/table/@cell content/.add={\cellcolor{green!26!white}}{},}},
      	every row 15 column 10/.style={postproc cell content/.append style={/pgfplots/table/@cell content/.add={\cellcolor{green!26!white}}{},}},
      	every row 16 column 10/.style={postproc cell content/.append style={/pgfplots/table/@cell content/.add={\cellcolor{green!26!white}}{},}},
      	every row 23 column 10/.style={postproc cell content/.append style={/pgfplots/table/@cell content/.add={\cellcolor{green!26!white}}{},}},
      	every row 3 column 11/.style={postproc cell content/.append style={/pgfplots/table/@cell content/.add={\cellcolor{green!26!white}}{},}},
      	every row 4 column 11/.style={postproc cell content/.append style={/pgfplots/table/@cell content/.add={\cellcolor{green!26!white}}{},}},
      	every row 10 column 11/.style={postproc cell content/.append style={/pgfplots/table/@cell content/.add={\cellcolor{green!26!white}}{},}},
      	every row 13 column 11/.style={postproc cell content/.append style={/pgfplots/table/@cell content/.add={\cellcolor{green!26!white}}{},}},
      	every row 15 column 11/.style={postproc cell content/.append style={/pgfplots/table/@cell content/.add={\cellcolor{green!26!white}}{},}},
      	every row 16 column 11/.style={postproc cell content/.append style={/pgfplots/table/@cell content/.add={\cellcolor{green!26!white}}{},}},
      	every row 23 column 11/.style={postproc cell content/.append style={/pgfplots/table/@cell content/.add={\cellcolor{green!26!white}}{},}},
      	every row 3 column 12/.style={postproc cell content/.append style={/pgfplots/table/@cell content/.add={\cellcolor{green!26!white}}{},}},
      	every row 4 column 12/.style={postproc cell content/.append style={/pgfplots/table/@cell content/.add={\cellcolor{green!26!white}}{},}},
      	every row 10 column 12/.style={postproc cell content/.append style={/pgfplots/table/@cell content/.add={\cellcolor{green!26!white}}{},}},
      	every row 13 column 12/.style={postproc cell content/.append style={/pgfplots/table/@cell content/.add={\cellcolor{green!26!white}}{},}},
      	every row 15 column 12/.style={postproc cell content/.append style={/pgfplots/table/@cell content/.add={\cellcolor{green!26!white}}{},}},
      	every row 16 column 12/.style={postproc cell content/.append style={/pgfplots/table/@cell content/.add={\cellcolor{green!26!white}}{},}},
      	every row 23 column 12/.style={postproc cell content/.append style={/pgfplots/table/@cell content/.add={\cellcolor{green!26!white}}{},}},
      	every row 17 column 13/.style={postproc cell content/.append style={/pgfplots/table/@cell content/.add={\cellcolor{green!26!white}}{},}},
      	every row 18 column 14/.style={postproc cell content/.append style={/pgfplots/table/@cell content/.add={\cellcolor{green!26!white}}{},}},
      	every row 19 column 15/.style={postproc cell content/.append style={/pgfplots/table/@cell content/.add={\cellcolor{green!26!white}}{},}},
      	every row 25 column 16/.style={postproc cell content/.append style={/pgfplots/table/@cell content/.add={\cellcolor{green!26!white}}{},}},
      	every row 27 column 16/.style={postproc cell content/.append style={/pgfplots/table/@cell content/.add={\cellcolor{green!26!white}}{},}},
      	every row 32 column 16/.style={postproc cell content/.append style={/pgfplots/table/@cell content/.add={\cellcolor{green!26!white}}{},}},
        every row 25 column 17/.style={postproc cell content/.append style={/pgfplots/table/@cell content/.add={\cellcolor{green!26!white}}{},}},
      	every row 27 column 17/.style={postproc cell content/.append style={/pgfplots/table/@cell content/.add={\cellcolor{green!26!white}}{},}},
      	every row 32 column 17/.style={postproc cell content/.append style={/pgfplots/table/@cell content/.add={\cellcolor{green!26!white}}{},}},
        every row 33 column 18/.style={postproc cell content/.append style={/pgfplots/table/@cell content/.add={\cellcolor{green!26!white}}{},}},
        every row 30 column 19/.style={postproc cell content/.append style={/pgfplots/table/@cell content/.add={\cellcolor{green!26!white}}{},}},
        every row 34 column 19/.style={postproc cell content/.append style={/pgfplots/table/@cell content/.add={\cellcolor{green!26!white}}{},}},
        every row 35 column 20/.style={postproc cell content/.append style={/pgfplots/table/@cell content/.add={\cellcolor{green!26!white}}{},}},
      	every row 5 column 6/.style={postproc cell content/.append style={/pgfplots/table/@cell content/.add={\cellcolor{red!15!white}}{},}},
      	every row 5 column 7/.style={postproc cell content/.append style={/pgfplots/table/@cell content/.add={\cellcolor{red!15!white}}{},}},
      	every row 11 column 8/.style={postproc cell content/.append style={/pgfplots/table/@cell content/.add={\cellcolor{red!15!white}}{},}},
      	every row 5 column 9/.style={postproc cell content/.append style={/pgfplots/table/@cell content/.add={\cellcolor{red!15!white}}{},}},
      	every row 5 column 10/.style={postproc cell content/.append style={/pgfplots/table/@cell content/.add={\cellcolor{red!15!white}}{},}},
      	every row 2 column 11/.style={postproc cell content/.append style={/pgfplots/table/@cell content/.add={\cellcolor{red!15!white}}{},}},
      	every row 2 column 12/.style={postproc cell content/.append style={/pgfplots/table/@cell content/.add={\cellcolor{red!15!white}}{},}},
      	every row 2 column 13/.style={postproc cell content/.append style={/pgfplots/table/@cell content/.add={\cellcolor{red!15!white}}{},}},
      	every row 14 column 11/.style={postproc cell content/.append style={/pgfplots/table/@cell content/.add={\cellcolor{red!15!white}}{},}},
      	every row 14 column 12/.style={postproc cell content/.append style={/pgfplots/table/@cell content/.add={\cellcolor{red!15!white}}{},}},
      	every row 14 column 13/.style={postproc cell content/.append style={/pgfplots/table/@cell content/.add={\cellcolor{red!15!white}}{},}},
      	every row 7 column 14/.style={postproc cell content/.append style={/pgfplots/table/@cell content/.add={\cellcolor{red!15!white}}{},}},
      	every row 5 column 15/.style={postproc cell content/.append style={/pgfplots/table/@cell content/.add={\cellcolor{red!15!white}}{},}},
      	every row 19 column 16/.style={postproc cell content/.append style={/pgfplots/table/@cell content/.add={\cellcolor{red!15!white}}{},}},
      	every row 19 column 17/.style={postproc cell content/.append style={/pgfplots/table/@cell content/.add={\cellcolor{red!15!white}}{},}},
      	every row 19 column 18/.style={postproc cell content/.append style={/pgfplots/table/@cell content/.add={\cellcolor{red!15!white}}{},}},
      	every row 12 column 19/.style={postproc cell content/.append style={/pgfplots/table/@cell content/.add={\cellcolor{red!15!white}}{},}},
      	every row 19 column 20/.style={postproc cell content/.append style={/pgfplots/table/@cell content/.add={\cellcolor{red!15!white}}{},}},
      ]{BGROMARSRun2.csv} 
    \caption{Settings and out-of-sample validation figures of best performing MARS models derived in a two-step approach sorted by first and second step validation sets. Highlighted in green and red respectively the best and worst validation figures.}
    \label{tab:BGROMARSRun2}
  \end{center}
\end{table}

\begin{table}[htb]
  \tiny
  \begin{center}
  \tabcolsep=0.15cm
  \renewcommand{\arraystretch}{0.90}
      \begin{filecontents*}{BGROMARSbest.csv}
k;h;coef
0;1;15397.13441
1;$h\left(X_8-0.104892\right)$;7901.887111
2;$h\left(0.104892-X_8\right)$;-8165.642986
3;$h\left(0.205577-X_1\right) \cdot h\left(0.104892-X_8\right)$;688.8259158
4;$h\left(X_6-1.17224\right)$;265.0823072
5;$h\left(1.17224-X_6\right)$;-280.9423711
6;$h\left(X_{15}-53.8706\right)$;-2.1117534
7;$h\left(53.8706-X_{15}\right)$;1.1626033
8;$h\left(X_7--0.147599\right)$;-60.9019674
9;$h\left(-0.147599-X_7\right)$;-334.7732451
10;$h\left(X_8--0.0456197\right)$;3183.070278
11;$h\left(0.205577-X_1\right) \cdot h\left(0.104892-X_8\right) \cdot h\left(X_{15}-64.6262\right)$;-9.4811822
12;$h\left(0.205577-X_1\right) \cdot h\left(0.104892-X_8\right) \cdot h\left(64.6262-X_{15}\right)$;29.8518484
13;$h\left(X_1-0.945371\right)$;-64.875516
14;$h\left(0.945371-X_1\right)$;124.4512543
15;$h\left(X_6-1.56058\right) \cdot h\left(0.104892-X_8\right)$;-815.2037149
16;$h\left(1.56058-X_6\right) \cdot h\left(0.104892-X_8\right)$;1085.796192
17;$h\left(1.44218-X_2\right)$;-60.2308942
18;$h\left(X_1--1.61447\right) \cdot h\left(1.56058-X_6\right) \cdot h\left(0.104892-X_8\right)$;-233.1411908
19;$h\left(-1.61447-X_1\right) \cdot h\left(1.56058-X_6\right) \cdot h\left(0.104892-X_8\right)$;415.9235436
20;$h\left(X_8-0.0159508\right) \cdot h\left(53.8706-X_{15}\right)$;8.9407891
21;$h\left(0.0159508-X_8\right) \cdot h\left(53.8706-X_{15}\right)$;47.9861185
22;$h\left(X_9-0.247192\right)$;47.7215432
23;$h\left(0.247192-X_9\right)$;-82.5804328
24;$h\left(0.993896-X_{12}\right)$;-63.6091725
25;$h\left(X_1-0.0195594\right) \cdot h\left(0.0159508-X_8\right) \cdot h\left(53.8706-X_{15}\right)$;-12.5776126
26;$h\left(0.0195594-X_1\right) \cdot h\left(0.0159508-X_8\right) \cdot h\left(53.8706-X_{15}\right)$;-42.2487134
27;$h\left(X_7--0.147599\right) \cdot h\left(X_8--0.191689\right)$;2124.930651
28;$h\left(X_7--0.147599\right) \cdot h\left(-0.191689-X_8\right)$;1510.412509
29;$h\left(X_3-0.323352\right) \cdot h\left(0.104892-X_8\right)$;948.861628
30;$h\left(0.323352-X_3\right) \cdot h\left(0.104892-X_8\right)$;-577.6149869
31;$h\left(X_1--1.26627\right) \cdot h\left(X_7--0.147599\right)$;101.1484964
32;$h\left(-1.26627-X_1\right) \cdot h\left(X_7--0.147599\right)$;-9.9980626
33;$h\left(X_{14}-0.684998\right)$;109.7603493
34;$h\left(0.684998-X_{14}\right)$;-37.8867519
35;$h\left(1.17224-X_6\right) \cdot h\left(X_8--0.12538\right)$;216.6201595
36;$h\left(1.17224-X_6\right) \cdot h\left(-0.12538-X_8\right)$;2076.177089
37;$h\left(0.945371-X_1\right) \cdot h\left(X_8-0.0019988\right)$;-156.7930408
38;$h\left(0.945371-X_1\right) \cdot h\left(0.0019988-X_8\right)$;1262.557717
39;$h\left(X_1--1.58818\right) \cdot h\left(X_6-1.56058\right) \cdot h\left(0.104892-X_8\right)$;137.6024823
40;$h\left(1.56058-X_6\right) \cdot h\left(0.104892-X_8\right) \cdot h\left(X_{15}-76.9327\right)$;-4.8694112
41;$h\left(1.56058-X_6\right) \cdot h\left(0.104892-X_8\right) \cdot h\left(76.9327-X_{15}\right)$;2.1135646
42;$h\left(0.205577-X_1\right) \cdot h\left(X_2-1.43028\right) \cdot h\left(0.104892-X_8\right)$;24003.06762
43;$h\left(0.205577-X_1\right) \cdot h\left(1.43028-X_2\right) \cdot h\left(0.104892-X_8\right)$;-161.877976
44;$h\left(X_1-0.945371\right) \cdot h\left(X_8--0.0165546\right)$;-224.1771427
45;$h\left(X_1-0.945371\right) \cdot h\left(-0.0165546-X_8\right)$;-987.468527
\end{filecontents*}
\pgfplotstabletypeset[
        multicolumn names, 
        col sep=semicolon, 
        display columns/0/.style={column name=$\bm{k}$,dec sep align},
        display columns/1/.style={column type=l,column name=$\bm{h_k\left(X\right)}$,string type},
        display columns/2/.style={column name=$\bm{\widehat{\beta}_{\textbf{\text{MARS}},k}}$,fixed,fixed zerofill,precision=2,set thousands separator={,},dec sep align},
	    	every head row/.style={before row=\toprule,after row=\midrule},
	    	every row no 1/.style={before row=\midrule},
	    	every nth row={10[+1]}{before row=\midrule},
      	every last row/.style={after row=\bottomrule},
      ]{BGROMARSbest.csv} 
    \caption{Best MARS model of BEL derived in a two-step approach with the final coefficients.}
    \label{tab:BGROMARSbest}
  \end{center}
\end{table}

\begin{table}[htb]
  \tiny
  \begin{center}
  \tabcolsep=0.106cm
  \renewcommand{\arraystretch}{0.90}
      \begin{filecontents*}{KRregressors1.csv}
k,r1,r2,r3,r4,r5,r6,r7,r8,r9,r10,r11,r12,r13,r14,r15
0,0,0,0,0,0,0,0,0,0,0,0,0,0,0,0
1,0,0,0,0,0,0,0,1,0,0,0,0,0,0,0
2,1,0,0,0,0,0,0,0,0,0,0,0,0,0,0
3,0,0,0,0,0,1,0,0,0,0,0,0,0,0,0
4,0,0,0,0,0,0,0,0,0,0,0,0,0,0,1
5,0,0,0,0,0,0,1,0,0,0,0,0,0,0,0
6,1,0,0,0,0,0,0,1,0,0,0,0,0,0,0
7,0,0,0,0,0,0,0,2,0,0,0,0,0,0,0
8,2,0,0,0,0,0,0,0,0,0,0,0,0,0,0
9,0,0,0,0,0,1,0,1,0,0,0,0,0,0,0
10,0,0,0,0,0,0,0,1,0,0,0,0,0,0,1
11,1,0,0,0,0,0,0,0,0,0,0,0,0,0,1
12,0,1,0,0,0,0,0,0,0,0,0,0,0,0,0
13,1,0,0,0,0,1,0,0,0,0,0,0,0,0,0
14,0,0,0,0,0,0,0,0,1,0,0,0,0,0,0
15,0,0,0,0,0,0,0,0,0,0,0,1,0,0,0
16,0,0,0,0,0,0,1,1,0,0,0,0,0,0,0
17,1,0,0,0,0,0,1,0,0,0,0,0,0,0,0
18,1,0,0,0,0,0,0,1,0,0,0,0,0,0,1
19,0,0,0,0,0,0,0,2,0,0,0,0,0,0,1
20,0,0,0,0,0,0,0,0,0,0,0,0,0,1,0
21,1,0,0,0,0,0,0,2,0,0,0,0,0,0,0
22,0,0,0,0,0,0,0,0,0,0,0,0,0,0,2
23,2,0,0,0,0,0,0,0,0,0,0,0,0,0,1
24,0,0,0,0,0,0,0,0,0,0,0,0,1,0,0
25,0,0,1,0,0,0,0,0,0,0,0,0,0,0,0
26,0,0,1,0,0,0,0,1,0,0,0,0,0,0,0
27,1,0,1,0,0,0,0,0,0,0,0,0,0,0,0
\end{filecontents*}
\pgfplotstabletypeset[
        multicolumn names, 
        col sep=comma, 
        display columns/0/.style={column name=$\bm{k}$,dec sep align},
        display columns/1/.style={column type=C,column name=$\bm{r_{k}^{1}}$},
        display columns/2/.style={column type=C,column name=$\bm{r_{k}^{2}}$},
        display columns/3/.style={column type=C,column name=$\bm{r_{k}^{3}}$},
        display columns/4/.style={column type=C,column name=$\bm{r_{k}^{4}}$},
        display columns/5/.style={column type=C,column name=$\bm{r_{k}^{5}}$},
        display columns/6/.style={column type=C,column name=$\bm{r_{k}^{6}}$},
        display columns/7/.style={column type=C,column name=$\bm{r_{k}^{7}}$},
        display columns/8/.style={column type=C,column name=$\bm{r_{k}^{8}}$},
        display columns/9/.style={column type=C,column name=$\bm{r_{k}^{9}}$},
        display columns/10/.style={column type=C,column name=$\bm{r_{k}^{10}}$},
        display columns/11/.style={column type=C,column name=$\bm{r_{k}^{11}}$},
        display columns/12/.style={column type=C,column name=$\bm{r_{k}^{12}}$},
        display columns/13/.style={column type=C,column name=$\bm{r_{k}^{13}}$},
        display columns/14/.style={column type=C,column name=$\bm{r_{k}^{14}}$},
        display columns/15/.style={column type=C,column name=$\bm{r_{k}^{15}}$},
	    	every head row/.style={before row=\toprule,after row=\midrule},
	    	every row no 0/.style={before row={\multicolumn{17}{l}{\textbf{$\bm{K}_{\text{max}} \bm{= 16}$ in adaptive basis function selection}}\\ \\}},
	    	every row no 17/.style={before row={\midrule \multicolumn{17}{l}{\textbf{$\bm{K}_{\text{max}} \bm{= 27}$ in adaptive basis function selection}}\\ \\}},
      	every last row/.style={after row=\bottomrule},
      ]{KRregressors1.csv} 
    \caption{Basis function sets of LC and LL proxy functions of BEL corresponding to $K_{\text{max}} \in \left\{16, 27\right\}$ derived by adaptive OLS selection.}
    \label{tab:KRregressors1}
  \end{center}
\end{table}

\begin{table}[htb]
  \tiny
  \begin{center}
  \tabcolsep=0.106cm
  \renewcommand{\arraystretch}{0.90}
      \begin{filecontents*}{KRregressors2.csv}
k,r1,r2,r3,r4,r5,r6,r7,r8,r9,r10,r11,r12,r13,r14,r15
0,0,0,0,0,0,0,0,0,0,0,0,0,0,0,0
1,1,0,0,0,0,0,0,0,0,0,0,0,0,0,0
2,0,1,0,0,0,0,0,0,0,0,0,0,0,0,0
3,0,0,1,0,0,0,0,0,0,0,0,0,0,0,0
4,0,0,0,1,0,0,0,0,0,0,0,0,0,0,0
5,0,0,0,0,1,0,0,0,0,0,0,0,0,0,0
6,0,0,0,0,0,1,0,0,0,0,0,0,0,0,0
7,0,0,0,0,0,0,1,0,0,0,0,0,0,0,0
8,0,0,0,0,0,0,0,1,0,0,0,0,0,0,0
9,0,0,0,0,0,0,0,0,1,0,0,0,0,0,0
10,0,0,0,0,0,0,0,0,0,1,0,0,0,0,0
11,0,0,0,0,0,0,0,0,0,0,1,0,0,0,0
12,0,0,0,0,0,0,0,0,0,0,0,1,0,0,0
13,0,0,0,0,0,0,0,0,0,0,0,0,1,0,0
14,0,0,0,0,0,0,0,0,0,0,0,0,0,1,0
15,0,0,0,0,0,0,0,0,0,0,0,0,0,0,1
16,1,0,0,0,0,0,0,1,0,0,0,0,0,0,0
17,0,0,0,0,0,1,0,1,0,0,0,0,0,0,0
18,0,0,0,0,0,0,0,1,0,0,0,0,0,0,1
19,1,0,0,0,0,0,0,0,0,0,0,0,0,0,1
20,1,0,0,0,0,1,0,0,0,0,0,0,0,0,0
21,0,0,0,0,0,0,1,1,0,0,0,0,0,0,0
22,1,0,0,0,0,0,1,0,0,0,0,0,0,0,0
\end{filecontents*}
\pgfplotstabletypeset[
        multicolumn names, 
        col sep=comma, 
        display columns/0/.style={column name=$\bm{k}$,dec sep align},
        display columns/1/.style={column type=C,column name=$\bm{r_{k}^{1}}$},
        display columns/2/.style={column type=C,column name=$\bm{r_{k}^{2}}$},
        display columns/3/.style={column type=C,column name=$\bm{r_{k}^{3}}$},
        display columns/4/.style={column type=C,column name=$\bm{r_{k}^{4}}$},
        display columns/5/.style={column type=C,column name=$\bm{r_{k}^{5}}$},
        display columns/6/.style={column type=C,column name=$\bm{r_{k}^{6}}$},
        display columns/7/.style={column type=C,column name=$\bm{r_{k}^{7}}$},
        display columns/8/.style={column type=C,column name=$\bm{r_{k}^{8}}$},
        display columns/9/.style={column type=C,column name=$\bm{r_{k}^{9}}$},
        display columns/10/.style={column type=C,column name=$\bm{r_{k}^{10}}$},
        display columns/11/.style={column type=C,column name=$\bm{r_{k}^{11}}$},
        display columns/12/.style={column type=C,column name=$\bm{r_{k}^{12}}$},
        display columns/13/.style={column type=C,column name=$\bm{r_{k}^{13}}$},
        display columns/14/.style={column type=C,column name=$\bm{r_{k}^{14}}$},
        display columns/15/.style={column type=C,column name=$\bm{r_{k}^{15}}$},
	    	every head row/.style={before row=\toprule,after row=\midrule},
	    	every row no 0/.style={before row={\multicolumn{17}{l}{\textbf{$\bm{K}_{\text{max}} \bm{= 15}$ in risk factor wise basis function selection}}\\ \\}},
	    	every row no 16/.style={before row={\midrule \multicolumn{17}{l}{\textbf{$\bm{K}_{\text{max}} \bm{= 22}$ in combined risk factor wise and adaptive basis function selection}}\\ \\}},
      	every last row/.style={after row=\bottomrule},
      ]{KRregressors2.csv} 
    \caption{Basis function sets of LC and LL proxy functions of BEL corresponding to $K_{\text{max}} \in \left\{15, 22\right\}$ derived by risk factor wise or combined risk factor wise and adaptive OLS selection.}
    \label{tab:KRregressors2}
  \end{center}
\end{table}

\begin{table}[htb]
  \tiny
  \begin{center}
  \tabcolsep=0.004cm
  \renewcommand{\arraystretch}{0.80}
      \begin{filecontents*}{KR.csv}
iter;n.bw;c.kerorder;v.wre.y;v.a.wre.y;v.res.avg.y;v.wre.delta.y;v.res.avg.delta.y;ns.wre.y;ns.a.wre.y;ns.res.avg.y;ns.wre.delta.y;ns.res.avg.delta.y;cr.wre.y;cr.a.wre.y;cr.res.avg.y;cr.wre.delta.y;cr.res.avg.delta.y
16;0.1;2;0.546153;0.52217;-43.894639;12.628891;49.69111;0.703398;0.679622;-86.303467;12.307277;-7.10918;0.549314;0.537736;-34.54401;12.154774;44.650277
16;0.2;2;0.400395;0.382812;-25.81661;11.129815;47.063761;0.519889;0.502316;-51.353788;10.841666;7.13512;0.443222;0.433881;4.520303;12.734376;63.009212
16;0.3;2;0.369697;0.353463;-24.51917;10.836127;45.387092;0.452493;0.437198;-36.825653;10.698888;18.689146;0.440188;0.430911;4.949341;12.480295;60.46414
27;0.2;2;0.393065;0.375804;-25.682309;10.622731;42.622467;0.512082;0.494773;-50.548208;10.734276;3.365106;0.434516;0.425358;3.821685;12.152289;57.734999
16;0.1;4;2.79803;2.67515;-154.51802;84.44613;-407.17321;8.04807;7.77603;-558.4392;246.6812;-825.48584;5.0434;4.9371;-95.7696;128.2946;-362.8163
16;0.1;2;0.375365;0.358882;-10.654574;11.969555;56.822028;0.570465;0.551183;-68.237897;9.71328;-15.152759;0.412035;0.403351;-22.152362;8.718015;30.932777
16;0.2;2;0.342821;0.327766;-6.367537;11.378924;58.728241;0.448152;0.433003;-48.927597;8.352881;1.776718;0.372189;0.364344;4.669775;10.265173;55.374091
27;0.1;2;210.299;201.064;-30682.204;5209.354;-30588.582;131.043;126.614;-18981.015;3669.754;-18901.784;4.08631;4.00019;-82.28107;92.00606;-3.04998
27;0.2;2;2726.47;2606.74;400253.73;67487.06;400306.02;3502.24;3383.85;422442.71;98081.04;422480.61;1.85169;1.81266;-25.17703;41.20744;12.72291
16;0.1;2;0.570197;0.545157;-43.056718;14.009036;55.104361;0.645925;0.624092;-72.190383;12.40994;11.579233;0.50032;0.489775;-12.018227;13.873249;71.751388
16;0.2;2;1.62557;1.55419;38.41462;41.43294;73.12942;1.94348;1.87779;265.63984;57.26095;285.96318;2.56754;2.51342;384.06585;61.39052;404.38919
27;0.1;2;0.563064;0.538338;-41.910183;13.937522;56.023719;0.64253;0.620811;-71.829842;12.327896;11.712597;0.497017;0.486542;-11.994305;13.791471;71.548133
15;0.1;2;0.528086;0.504896;-36.41342;12.54782;40.502121;1.05207;1.0165;-38.39929;22.32939;24.12479;0.509355;0.498619;-29.35237;10.963527;33.171708
15;0.2;2;0.412718;0.394594;-30.607283;9.608547;33.155126;1.13878;1.10029;3.15227;26.42195;52.52321;1.18362;1.15867;96.90285;27.4225;146.27379
15;0.3;2;0.395583;0.378212;-29.622255;9.011983;23.267757;0.960228;0.927771;15.831505;23.39627;54.330054;0.463552;0.453782;-5.723145;11.015387;32.775404
15;0.4;2;0.347164;0.331919;-22.379244;8.513386;18.485394;1.11272;1.07511;12.47924;28.2151;38.95241;0.469432;0.459538;-1.70692;10.65764;24.766255
15;0.5;2;0.343464;0.328381;-17.508742;9.455628;36.857923;1.23971;1.19781;6.49072;29.56431;46.46592;0.508641;0.497921;-21.632198;10.826473;18.343003
15;0.6;2;0.332587;0.317982;-17.305581;10.442177;49.587277;1.15877;1.1196;21.19324;26.91621;73.69464;0.457765;0.448117;-2.298486;11.099905;50.202909
15;0.7;2;0.32949;0.315021;-16.389063;9.689764;40.5701;1.16544;1.12605;17.94471;27.70462;60.51241;0.443527;0.434179;-14.454873;9.477018;28.112828
15;0.8;2;0.327269;0.312898;-16.11549;9.99128;45.014855;1.20812;1.16728;29.29205;28.83994;76.03093;1.15648;1.1321;100.97171;26.4803;147.71059
15;0.9;2;0.318113;0.304144;-20.442145;11.522969;60.60937;1.14118;1.1026;40.36144;27.20808;107.02149;1.13705;1.11309;110.96987;28.8628;177.62992
15;1;2;0.323692;0.309478;-21.961282;10.108442;49.352881;1.18943;1.14923;51.99575;28.61697;108.91845;1.13091;1.10708;106.31185;27.34395;163.23455
16;0.1;2;0.525867;0.502774;-40.378112;12.795798;42.794735;1.19847;1.15796;2.15335;27.50517;70.93473;0.505804;0.495144;-20.09142;12.160221;48.689964
16;0.2;2;0.405211;0.387417;-26.122455;11.451967;49.75666;1.15732;1.1182;26.73014;27.61899;88.2178;0.437583;0.42836;2.407587;12.356754;63.89524
16;0.3;2;0.355647;0.34003;-26.922973;8.662343;29.374182;1.06931;1.03317;40.86694;26.75642;82.77263;0.437886;0.428657;0.855796;10.713671;42.761488
16;0.4;2;0.331618;0.317056;-19.080318;8.262673;21.920994;1.1578;1.11867;26.72118;29.54357;53.33103;0.446196;0.436791;3.752961;10.230724;30.36281
16;0.5;2;0.3249;0.310632;-16.155644;9.248019;35.946338;1.34131;1.29597;29.71293;32.8505;67.42345;1.21866;1.19298;100.69204;27.23035;138.40256
16;0.1;4;0.45449;0.434532;-25.816821;12.527693;33.966003;0.735633;0.710767;-68.245179;16.474619;-22.853818;0.58504;0.572709;5.488605;15.074536;50.879967
16;0.2;4;3.29075;3.14624;-103.53207;159.5738;891.49089;7.49635;7.24297;-14.40687;328.50786;966.22463;8.06275;7.89282;176.44003;295.20366;1157.07152
16;0.1;6;3.30783;3.16257;-32.40544;83.83065;67.61665;5.74221;5.54812;-95.803;158.11561;-10.17237;6.62283;6.48324;-53.258;148.38886;32.37263
16;0.2;6;3.32259;3.17669;-71.11734;85.00992;-216.6332;9.37358;9.05674;72.75723;267.88489;-87.15009;13.1821;12.9042;245.8579;303.5575;85.9506
16;0.1;8;3.93818;3.76524;145.9072;104.71814;-118.74321;10.7096;10.3476;-190.8777;307.8494;-469.9196;8.84002;8.6537;-311.63099;205.15039;-590.67286
16;0.2;8;8.53467;8.15989;396.66907;285.7741;-639.23591;7.78528;7.52213;70.13264;347.27891;-980.1638;12.3658;12.1052;1365.1071;389.9667;314.8107
22;0.1;2;0.50472;0.482556;-37.005067;12.248323;43.947203;1.0705;1.03432;-41.19888;22.33681;25.36192;0.515155;0.504297;-29.589864;11.397877;36.970944
22;0.2;2;0.419866;0.401428;-28.237043;10.410045;38.719923;1.06502;1.02902;-2.51835;24.76444;50.04715;1.19519;1.17;106.47572;28.6507;159.04122
22;0.3;2;0.389415;0.372315;-28.956863;8.710499;22.50589;0.894971;0.86472;6.342132;21.715586;43.413423;0.450252;0.440762;-3.212756;10.934308;33.858535
22;0.4;2;0.347533;0.332272;-21.224716;8.345253;16.355926;1.05091;1.01538;2.74076;26.73844;25.92994;0.485668;0.475432;-4.478034;11.070894;18.711144
22;0.5;2;0.326944;0.312587;-13.792873;9.151808;31.728337;1.16542;1.12603;-1.72065;28.47758;29.40909;0.465972;0.456151;-15.094556;9.843389;16.035192
22;0.6;2;0.332701;0.318091;-17.252494;9.824264;45.795457;1.09274;1.05581;11.01592;25.16919;59.67241;0.447417;0.437987;-1.078739;10.98689;47.577749
22;0.7;2;0.322846;0.308668;-15.372658;9.412172;39.487384;1.22554;1.18411;25.8166;29.48404;66.28518;1.16888;1.14425;98.57349;26.19243;139.04207
22;0.8;2;0.315293;0.301447;-14.747563;9.774146;45.80198;1.18731;1.14718;31.99686;28.23564;78.15494;1.12426;1.10056;105.54112;26.24104;151.6992
22;0.9;2;0.311829;0.298136;-19.322662;10.879246;57.895544;1.14584;1.10711;38.94512;27.06034;101.77187;1.12306;1.09939;110.9219;28.33974;173.74864
22;1;2;0.312224;0.298513;-21.069047;9.700039;48.141463;1.12912;1.09095;40.89074;26.74107;95.70979;1.11892;1.09534;107.27837;27.09108;162.09742
27;0.2;2;0.397957;0.380481;-26.15439;10.728211;45.397323;1.15484;1.1158;26.20136;27.52539;83.36161;0.43607;0.426879;0.693237;11.768933;57.853487
27;0.3;2;0.378723;0.362092;-28.238232;8.851802;24.465349;0.898229;0.867868;6.632808;21.765886;44.944926;0.458446;0.448783;-1.913562;11.098039;36.398556
27;0.4;2;0.348714;0.333401;-20.908727;8.611455;17.13483;1.05331;1.01771;2.2423;26.74263;25.89439;0.480345;0.470221;-3.615492;11.025928;20.036602
15;0.1;2;0.446074;0.426485;-48.937764;10.404897;40.03169;1.21763;1.17648;-100.3442;21.71011;-25.76621;0.784103;0.767577;-104.355905;10.912104;-29.777914
15;0.2;2;0.35558;0.339965;-34.053773;8.207957;13.451077;1.58795;1.53427;-145.4052;39.61998;-112.29181;0.595605;0.583051;-54.214498;11.159007;-21.101112
15;0.3;2;0.319296;0.305275;-36.359887;7.388496;16.890117;1.91258;1.84793;133.95309;48.40141;172.81163;0.59511;0.582567;-35.777876;10.683074;3.080665
15;0.4;2;0.341575;0.326575;-39.805014;8.165098;33.090485;1.82544;1.76374;-164.04919;42.36417;-105.54515;0.42549;0.416522;-49.136521;6.153075;9.367515
15;0.5;2;0.328635;0.314204;-39.939971;7.96633;33.74775;2.19874;2.12441;-219.43217;52.52594;-160.13591;0.41476;0.406018;-44.592136;5.588761;14.704122
15;0.6;2;0.298452;0.285346;-33.458885;7.281068;29.219342;0.938441;0.906721;7.785074;19.453453;56.071839;0.32594;0.31907;-27.74969;5.46956;20.53708
15;0.7;2;0.311144;0.297481;-39.78496;6.60746;22.956395;0.94252;0.910661;-12.821589;19.099614;35.528303;0.362262;0.354627;-39.994438;4.790072;8.355455
15;0.8;2;0.293681;0.280784;-38.155042;5.489087;8.17114;0.855845;0.826917;4.372327;19.107488;36.307045;0.323491;0.316673;-29.427392;4.748958;2.507327
22;0.1;2;731.512;699.389;2738.327;85172.21;479611.586;1564.87;1511.98;-111627.94;127409.54;365230.93;492.488;482.109;-19403.855;76575.305;457455.013
22;0.2;2;0.3404518;0.32550145;-33.75420158;7.82817242;-0.00576539;0.832117;0.80399;-15.081102;20.555568;4.275871;0.417942;0.409133;-24.754119;8.3514;-5.397146
22;0.3;2;98.0342;93.7292;14396.3148;147.7161;-250.4256;101.6892;98.2519;15174.2926;147.0882;513.1608;100;97.8924;15028.145;100;367.0132
22;0.4;2;98.0519;93.7461;14398.9183;147.2773;-247.8221;113.994;110.141;13157.8;495.292;-1503.332;100;97.8924;15028.145;100;367.0132
22;0.5;2;100;95.6087;14684.9932;100;38.2528;118.946;114.926;14983.71;650.687;322.579;100;97.8924;15028.145;100;367.0132
22;0.6;2;99.72123;95.34215;14644.05655;105.9587;-2.68385;100.5922;97.1921;15004.1511;119.8722;343.0192;100;97.8924;15028.145;100;367.0132
22;0.7;2;100;95.6087;14684.9933;100;38.2529;100;96.6199;14922.2311;100;261.0992;100;97.8924;15028.145;100;367.0132
22;0.8;2;0.294007;0.281096;-38.860703;5.476624;8.579633;152.426;147.274;22622.276;4263.712;22655.325;0.309118;0.302603;-34.87707;4.512311;-1.828196
16;0.1;2;0.745556;0.712816;-56.155358;17.719796;46.077873;1.52663;1.47503;-52.10592;31.57767;35.73585;0.732031;0.716603;-58.928896;14.539019;28.912872
16;0.5;2;1.22239;1.16871;-77.72782;29.4151;16.38742;2.60252;2.51455;300.80804;81.57244;380.53182;10.4536;10.2333;1418.5586;242.197;1498.2824
27;0.1;2;0.639425;0.611346;-38.142914;15.559134;31.232585;1.30328;1.25923;13.07598;32.01972;68.06002;0.594331;0.581804;-2.074805;14.679435;52.909232
27;0.5;2;0.354532;0.338963;-15.736624;11.683017;52.703082;1.34251;1.29713;25.19347;33.47736;79.24171;1.39588;1.36646;116.55623;32.1523;170.60448
16;0.1;4;0.713599;0.682262;-33.352287;17.063592;46.674028;1.273469;1.230424;-0.743778;30.711794;64.891075;0.665411;0.651386;-22.840822;14.717063;42.79403
16;0.5;4;1.84597;1.76491;-138.64435;39.02283;50.1561;2.29465;2.21709;18.23405;51.39049;192.64304;7.09252;6.94304;769.07277;157.1516;943.48176
27;0.1;4;0.659467;0.630508;-37.737584;15.375377;32.276491;1.31625;1.27176;7.32072;31.96216;62.94333;0.579838;0.567617;-15.330448;13.791497;40.292164
27;0.5;4;0.39072;0.373562;-13.112013;13.097079;67.244719;1.25692;1.21444;16.09142;30.87012;82.05669;0.519679;0.508726;-9.977855;13.052395;55.987414
16;0.1;6;1.82838;1.74809;-165.10336;38.33927;99.95468;1.95002;1.88411;-178.37356;28.80045;72.29301;1.54607;1.51348;-190.23321;23.66757;60.43336
16;0.5;6;1.82801;1.74773;-6.24862;56.04693;271.44298;1.08098;1.04444;80.29642;65.40029;343.59657;1.66022;1.62523;225.1853;73.96078;488.48544
\end{filecontents*}
\pgfplotstabletypeset[
        multicolumn names, 
        col sep=semicolon, 
        display columns/0/.style={column name={$\bm{k}$},dec sep align},
        display columns/1/.style={column name={\textbf{ bw}},column type={r},fixed,fixed zerofill,precision=1},
        display columns/2/.style={column name={\textbf{ o}},column type={r},fixed,fixed zerofill,precision=0},
        display columns/3/.style={column name={\textbf{v.mae}},column type={r},fixed,fixed zerofill,precision=2,set thousands separator={,}},
        display columns/4/.style={column name={$\bm{\textbf{\text{v.mae}}^a}$},column type={r},fixed,fixed zerofill,precision=2,set thousands separator={,}},
        display columns/5/.style={column name={\textbf{v.res}},column type={r},fixed,fixed zerofill,precision=0,set thousands separator={,}},
        display columns/6/.style={column name={$\bm{\textbf{\text{v.mae}}^0}$},column type={r},fixed,fixed zerofill,precision=0,set thousands separator={,}},
        display columns/7/.style={column name={$\bm{\textbf{\text{v.res}}^0}$},column type={r},fixed,fixed zerofill,precision=0,set thousands separator={,}},
        display columns/8/.style={column name={\textbf{ns.mae}},column type={r},fixed,fixed zerofill,precision=2,set thousands separator={,}},
        display columns/9/.style={column name={$\bm{\textbf{\text{ns.mae}}^a}$},column type={r},fixed,fixed zerofill,precision=2,set thousands separator={,}},
        display columns/10/.style={column name={\textbf{ns.res}},column type={r},fixed,fixed zerofill,precision=0,set thousands separator={,}},
        display columns/11/.style={column name={$\bm{\textbf{\text{ns.mae}}^0}$},column type={r},fixed,fixed zerofill,precision=0,set thousands separator={,}},
        display columns/12/.style={column name={$\bm{\textbf{\text{ns.res}}^0}$},column type={r},fixed,fixed zerofill,precision=0,set thousands separator={,}},
        display columns/13/.style={column name={\textbf{cr.mae}},column type={r},fixed,fixed zerofill,precision=2,set thousands separator={,}},
        display columns/14/.style={column name={$\bm{\textbf{\text{cr.mae}}^a}$},column type={r},fixed,fixed zerofill,precision=2,set thousands separator={,}},
        display columns/15/.style={column name={\textbf{cr.res}},column type={r},fixed,fixed zerofill,precision=0,set thousands separator={,}},
        display columns/16/.style={column name={$\bm{\textbf{\text{cr.mae}}^0}$},column type={r},fixed,fixed zerofill,precision=0,set thousands separator={,}},
        display columns/17/.style={column name={$\bm{\textbf{\text{cr.res}}^0}$},column type={r},fixed,fixed zerofill,precision=0,set thousands separator={,}},
	    	every head row/.style={before row=\toprule,after row=\midrule},
	    	every row no 0/.style={before row={\multicolumn{19}{l}{\textbf{LC regression with gaussian kernel and LOO-CV}}\\ \\}},
	    	every row no 5/.style={before row={\midrule \multicolumn{19}{l}{\textbf{LL regression with gaussian kernel and LOO-CV}}\\ \\}},
	    	every row no 9/.style={before row={\midrule \multicolumn{19}{l}{\textbf{LC regression with gaussian kernel and AIC}}\\ \\}},
	    	every row no 12/.style={before row={\midrule \multicolumn{19}{l}{\textbf{LC regression with Epanechnikov kernel and LOO-CV}}\\ \\}},
	    	every row no 46/.style={before row={\midrule \multicolumn{19}{l}{\textbf{LL regression with Epanechnikov kernel and LOO-CV}}\\ \\}},
	    	every row no 62/.style={before row={\midrule \multicolumn{19}{l}{\textbf{LC regression with uniform kernel and LOO-CV}}\\ \\}},
      	every last row/.style={after row=\bottomrule},
      	every row 53 column 3/.style={postproc cell content/.append style={/pgfplots/table/@cell content/.add={\cellcolor{green!26!white}}{},}},
      	every row 53 column 4/.style={postproc cell content/.append style={/pgfplots/table/@cell content/.add={\cellcolor{green!26!white}}{},}},
      	every row 71 column 5/.style={postproc cell content/.append style={/pgfplots/table/@cell content/.add={\cellcolor{green!26!white}}{},}},
      	every row 61 column 6/.style={postproc cell content/.append style={/pgfplots/table/@cell content/.add={\cellcolor{green!26!white}}{},}},
      	every row 55 column 7/.style={postproc cell content/.append style={/pgfplots/table/@cell content/.add={\cellcolor{green!26!white}}{},}},
      	every row 6 column 8/.style={postproc cell content/.append style={/pgfplots/table/@cell content/.add={\cellcolor{green!26!white}}{},}},
      	every row 6 column 9/.style={postproc cell content/.append style={/pgfplots/table/@cell content/.add={\cellcolor{green!26!white}}{},}},
      	every row 66 column 10/.style={postproc cell content/.append style={/pgfplots/table/@cell content/.add={\cellcolor{green!26!white}}{},}},
      	every row 6 column 11/.style={postproc cell content/.append style={/pgfplots/table/@cell content/.add={\cellcolor{green!26!white}}{},}},
      	every row 6 column 12/.style={postproc cell content/.append style={/pgfplots/table/@cell content/.add={\cellcolor{green!26!white}}{},}},
      	every row 61 column 13/.style={postproc cell content/.append style={/pgfplots/table/@cell content/.add={\cellcolor{green!26!white}}{},}},
      	every row 61 column 14/.style={postproc cell content/.append style={/pgfplots/table/@cell content/.add={\cellcolor{green!26!white}}{},}},
      	every row 43 column 15/.style={postproc cell content/.append style={/pgfplots/table/@cell content/.add={\cellcolor{green!26!white}}{},}},
      	every row 61 column 16/.style={postproc cell content/.append style={/pgfplots/table/@cell content/.add={\cellcolor{green!26!white}}{},}},
      	every row 61 column 17/.style={postproc cell content/.append style={/pgfplots/table/@cell content/.add={\cellcolor{green!26!white}}{},}},
      	every row 8 column 3/.style={postproc cell content/.append style={/pgfplots/table/@cell content/.add={\cellcolor{red!15!white}}{},}},
      	every row 8 column 4/.style={postproc cell content/.append style={/pgfplots/table/@cell content/.add={\cellcolor{red!15!white}}{},}},
      	every row 8 column 5/.style={postproc cell content/.append style={/pgfplots/table/@cell content/.add={\cellcolor{red!15!white}}{},}},
      	every row 54 column 6/.style={postproc cell content/.append style={/pgfplots/table/@cell content/.add={\cellcolor{red!15!white}}{},}},
      	every row 54 column 7/.style={postproc cell content/.append style={/pgfplots/table/@cell content/.add={\cellcolor{red!15!white}}{},}},
      	every row 8 column 8/.style={postproc cell content/.append style={/pgfplots/table/@cell content/.add={\cellcolor{red!15!white}}{},}},
      	every row 8 column 9/.style={postproc cell content/.append style={/pgfplots/table/@cell content/.add={\cellcolor{red!15!white}}{},}},
      	every row 8 column 10/.style={postproc cell content/.append style={/pgfplots/table/@cell content/.add={\cellcolor{red!15!white}}{},}},
      	every row 54 column 11/.style={postproc cell content/.append style={/pgfplots/table/@cell content/.add={\cellcolor{red!15!white}}{},}},
      	every row 8 column 12/.style={postproc cell content/.append style={/pgfplots/table/@cell content/.add={\cellcolor{red!15!white}}{},}},
      	every row 54 column 13/.style={postproc cell content/.append style={/pgfplots/table/@cell content/.add={\cellcolor{red!15!white}}{},}},
      	every row 54 column 14/.style={postproc cell content/.append style={/pgfplots/table/@cell content/.add={\cellcolor{red!15!white}}{},}},
      	every row 54 column 15/.style={postproc cell content/.append style={/pgfplots/table/@cell content/.add={\cellcolor{red!15!white}}{},}},
      	every row 54 column 16/.style={postproc cell content/.append style={/pgfplots/table/@cell content/.add={\cellcolor{red!15!white}}{},}},
      	every row 54 column 17/.style={postproc cell content/.append style={/pgfplots/table/@cell content/.add={\cellcolor{red!15!white}}{},}},
      ]{KR.csv} 
    \caption{Settings and out-of-sample validation figures of LC and LL proxy functions of BEL using basis function sets from Tables \ref{tab:KRregressors1} and \ref{tab:KRregressors2}. Highlighted in green and red respectively the best and worst validation figures.}
    \label{tab:KR}
  \end{center}
\end{table}

\end{document}